\documentclass{aa}
\usepackage{graphicx}
\usepackage{txfonts} 
\usepackage{subcaption}
\usepackage{placeins} 
\usepackage{textcomp} 
\usepackage{gensymb} 

\newcommand{\hi}{\text{H\textsc{i}}}
\newcommand{\hii}{\text{H\textsc{ii}}}

\usepackage{xcolor}

\usepackage{romannum} 

\usepackage{natbib}
\usepackage{hyperref}
\addto\extrasenglish{
    
}
\addto\extrasenglish{
    
}

\addto\extrasenglish{
    
}

\def\pasa{PASA}

\def\aap{A\&A}
\def\aaps{A\&A Suppl.\ Ser.}
\def\apj{ApJ}
\def\apjs{ApJ\ Suppl.}
\def\apjl{ApJ\ Let.}
\def\aj{AJ}
\def\mnras{MNRAS}

\def\pasp{PASP}

\title{ViCTORIA project: The LOFAR HBA Virgo Cluster Survey\thanks{Full Tables 3 and 4 are available at the CDS via anonymous ftp to \url{cdsarc.cds.unistra.fr} (\url{130.79.128.5}) or via \url{https://cdsarc.cds.unistra.fr/cgi-bin/qcat?J/A+A/}}}

\begin{document}
\pagenumbering{arabic} 
 \author{H.~W.~Edler
          \inst{1}
          \and
          F.~de~Gasperin\inst{1,2}
          \and
          T.~W.~Shimwell\inst{3,4}
          \and
          M.~J.~Hardcastle\inst{5}
          \and
          A.~Boselli\inst{6}
          \and
          V.~Heesen\inst{1}
          \and              
          H.~McCall\inst{7}
          \and
          D.~J.~Bomans\inst{8}
          \and     
          M.~Brüggen\inst{1}
          \and
          E.~Bulbul\inst{9}
          \and
          K.~T.~Chy\.zy\inst{10}
          \and
          A.~Ignesti\inst{11}
          \and
          A.~Merloni\inst{9}
          \and
          F.~Pacaud\inst{7}
          \and
          T.~H.~Reiprich\inst{7}
          \and
          I.~D.~Roberts\inst{3}
          \and
          H.~J.~A.~Rottgering\inst{3}
          \and
          R.~J.~van~Weeren\inst{3}
          }

   \institute{Hamburger Sternwarte, University of Hamburg,
              Gojenbergsweg 112, D-21029, Hamburg, Germany\\
              \email{henrik.edler@hs.uni-hamburg.de}
              \and
              INAF - Istituto di Radioastronomia,
              via P. Gobetti 101, Bologna, Italy
              \and
              Leiden Observatory, Leiden University, PO Box 9513, 2300 RA Leiden, The Netherlands
              \and
              ASTRON, Netherlands Institute for Radio Astronomy, Oude Hoogeveensedijk 4, 7991 PD, Dwingeloo, The Netherlands 
              \and
              Centre for Astrophysics Research, University of Hertfordshire, College Lane, Hatfield AL10 9AB, UK
              \and
              Aix Marseille Univ, CNRS, CNES, LAM, Marseille, France
              \and
              Argelander-Institut für Astronomie (AIfA), Universität Bonn, Auf
dem Hügel 71, 53121 Bonn, Germany
              \and
              Ruhr University Bochum, Faculty of Physics and Astronomy, Astronomical Institute (AIRUB), Universitätsstrasse 150, 44801 Bochum, Germany
              \and
              Max Planck Institute for Extraterrestrial Physics, Giessenbachstrasse 1, 85748 Garching, Germany
              \and
              Astronomical Observatory, Jagiellonian University, ul. Orla 171, 30-244 Krak\'ow, Poland 
              \and 
              INAF - Astronomical Observatory of Padova, vicolo dell'Osservatorio 5, IT-35122 Padova, Italy 
              }

   \date{Received March 20, 2023; accepted June 2, 2023}

  \abstract
   {The Virgo cluster is the nearest ($d=16.5$\,Mpc) massive ($M\geq 10^{14}\,M_\odot$) galaxy cluster and thus a prime target to study astrophysical processes in dense large-scale environments. 
   In the radio band, we can probe the non-thermal components of the interstellar medium (ISM), intracluster-medium (ICM) and of active galactic nuclei (AGN). This allows an investigation of the impact of the environment on the evolution of galaxies and the contribution of AGN to the ICM-heating. 
   With the ViCTORIA (Virgo Cluster multi-Telescope Observations in Radio of Interacting galaxies and AGN) project, we are carrying out multiple wide-field surveys of the Virgo cluster at different frequencies.
   }
   {We aim to investigate the impact of the environment on the evolution of galaxies and the contribution of AGN to the ICM-heating, from the inner cluster regions out to beyond the virial radius.}
   {We perform a survey of the cluster at $120-168\,$MHz using the LOw-Frequency ARray (LOFAR). We image a $132\,\mathrm{deg}^2$ region of the cluster, reaching an order of magnitude greater sensitivity than existing wide-field radio surveys of this field at three times higher spatial resolution compared to other low-frequency observations.
   We developed a tailored data processing strategy to subtract the bright central radio galaxy M\,87 from the data. This allowed us to correct for the systematic effects due to ionospheric variation as a function of time and direction.}
   {In the final mosaic with a resolution of $9''\times5''$, we reach a median noise level of $140\,\mathrm{{\mu}Jy\,beam^{-1}}$ inside the virial radius and $280\,\mathrm{{\mu}Jy\,beam^{-1}}$ for the full area. We detect 112 Virgo member galaxies and 114 background galaxies.
   In at least 18 cases, the radio morphology of the cluster member galaxies shows clear signs of ram-pressure stripping. This includes three previously unreported candidates.
   In addition, we reveal for the first time 150\,kpc long tails from a previous epoch of AGN activity for NGC\,4472 (M\,49).
   While no cluster-scale diffuse radio sources are discovered, we find the presence of an extended radio signature of the $W'$-group. This feature is coincident with an X-ray filament detected with SRG/eROSITA in the outskirts of the cluster. We speculate that this emission is synchrotron radiation which could be related to shocks or turbulence from accretion processes.} 
   {   The data published in this paper serve as a valuable resource for future studies. In the follow-up work of the ViCTORIA project, we will rely on them for an analysis of environmental effects on the radio-properties of star-forming galaxies in Virgo.}

\maketitle

\section{Introduction} 
The Virgo cluster is the nearest galaxy cluster to us; its center is coincident with the bright radio galaxy M\,87 \citep[Virgo \,A, NGC\,4486, e.g.][]{Owen2000,deGasperin2012} at a distance of 16.5\,Mpc \citep{Gavazzi1998TheDeterminations,Mei2007,Cantiello2018TheBeyond}. Due to its proximity, it spans an enormous angular size with a virial radius of $3.3\degree$ (974\,kpc) \citep{Simionescu2017}. Estimates of the virial mass $M_\mathrm{vir}$ ($\approx M_{200}$) of the cluster are around $M_{vir}\approx1.0\mathrm{-}1.4\times10^{14}M_\odot$ \citep{Urban2011X-rayRadius,Ferrarese2012,Simionescu2017}.
The system is considered to be dynamically young and in the process of formation, as indicated by the high fraction of spiral galaxies, the significant deviation from a spherical symmetry \citep{Binggeli1987} and the properties of intracluster stars \citep{Aguerri2005,Arnaboldi2005}. Further indication of the young age of the cluster is the presence of pronounced substructure in the galaxy distribution as well as in the X-rays, such as the sub-clusters and -groups associated with the massive ellipticals M\,49, M\,60 and M\,86 \citep{Bohringer1994TheImages}. All this makes the Virgo cluster (for conciseness, we will also refer to the cluster simply as ``Virgo'') an exceptional target for studies on the evolution of clusters and their member galaxies.

An example of a phenomenon that can be studied in rich and dynamic environments such as Virgo is ram-pressure stripping, a perturbation which affects galaxies that move at high velocity with respect to the intracluster-medium (ICM). This effect removes part of the interstellar medium (ISM) of star-forming galaxies, eventually causing a quenching of the star-formation rate (SFR) and turning them quiescent \citep[e.g.][]{Gunn1972,Sarazin1986,Boselli2022RamEnvironments}. 
The synchrotron-emitting cosmic rays in the ISM of star-forming galaxies are accelerated in supernovae of short-lived massive stars, which gives rise to the particularly tight radio-SFR relation \citep{vanderKruit1971ObservationsMHz.,vanderKruit1973High-resolutionDiscussion,CalistroRivera2017The2.5,Gurkan2018LOFAR/H-ATLAS:Relation}. For galaxies that experience ram-pressure stripping, an excess of radio emission can often be found \citep{Gavazzi1991MultifrequencyPhotometry,Gavazzi1999OnGalaxies,Miller2001TheClusters,Murphy2009,Ignesti2022WalkGalaxies}.
Integrated flux density measurements and high-resolution studies of individual cluster galaxies can help to understand the cause of this excess. 
Cosmic ray-electrons in the stripped ISM of galaxies likely show a steep spectrum due to radiative aging. Low-frequency ($\lesssim1\,$GHz) observations are required to detect the oldest and most distant parts of those stripped tails. Indeed, it was recently demonstrated in a number of studies utilizing 144\,MHz observations with the Low-Frequency Array High-band Antenna (LOFAR HBA) that low-frequency observations are well suited in discovering ram-pressure stripped tails  \citep{Ignesti2022Gasp,Ignesti2022WalkGalaxies,Roberts2021LoTSSClusters,Roberts2021II,Roberts2022LoTSSCluster}. 

Contrary to star-forming galaxies, where the radio luminosity is dominated by aforementioned star-formation, in early type galaxies, active galactic nuclei (AGN) are responsible for the radio emission. The relativistic plasma supplied by AGN can halt star-formation of the host galaxy \citep{Gaspari2012MechanicalGalaxies} and in clusters even influence the thermodynamical properties of the intracluster-medium (ICM) via AGN-feedback \citep{Fabian2012}. 
For Virgo, interactions between the AGN-jets and the surrounding ICM have previously been observed for  M\,87 \citep{Forman2007,Million2010} as well as numerous other systems at the center of the less massive sub-structures \citep{Kraft2011,Dunn2010TheGalaxies,Paggi2014,Su2019} 
or in non-central galaxies  \citep{Finoguenov2008,Birzan2008RadiativePower,Kraft2011,Million2010,Paggi2014,Su2019}.
As the relativistic electrons accelerated by the AGN lose energy over time, their radio spectrum also becomes steeper. Thus, we require observations at low-frequencies to unveil the emission of past phases of nuclear activity.

Furthermore, in the radio band, the non-thermal component of the ICM can generate radio halos and radio relics if there is sufficient particle acceleration and magnetic field strength \citep{Brunetti2014,vanWeeren2019}. Such emission is prominent in nearby clusters like Coma and Abell\,1367. Contrary to these clusters of higher richness, Virgo is not known to host cluster-scale diffuse radio emission -- nevertheless, deeper observations have the chance to reveal (or put stringent limits on) the presence of such a source, which could possibly be linked to merging sub-systems. Again, it is critical to observe at low-frequency due to the steep-spectrum nature of diffuse radio sources. This is the case for Virgo in particular, given that clusters of lower mass are expected to host steeper radio halos \citep{Cassano2010TheFrequency}.

Until now, the most sensitive wide-field surveys that cover Virgo are the TIFR GMRT Sky-Survey \citep[TGSS,][]{Intema2017TheADR1}, the NRAO VLA Sky Survey \citep[NVSS,][]{Condon1998TheSurvey} and more recently, the Rapid ASKAP Continuum Sky Survey \citep[RACS,][]{McConnell2020TheResults} and the Karl G. Jansky Very Large Array Sky Survey \citep{Lacy2020}. All of these are subject to strong imaging and calibration artifacts in proximity to M\,87, and only TGSS with a frequency of 150\,MHz is a low-frequency survey. 
In addition to the wide-field surveys, a number of radio studies of samples of Virgo galaxies exist \citep{Vollmer2007TheGalaxies,Vollmer2010,Vollmer2013,Wezgowiec2007TheSpirals,Wezgowiec2012,Murphy2009,Capetti2009}. However, those are limited to a small number of objects (<20).

We aim to significantly increase the radio coverage of the Virgo Cluster with the ViCTORIA (Virgo Cluster multi-Telescope Observations in Radio of Interacting galaxies and AGN) project. The project includes data from the low- and high-band systems of LOFAR as well as MeerKAT and it will greatly improve the sensitivity and resolution of the wide-field coverage in radio continuum between $42-1700$\,MHz and in the 21\,cm line. Polarization data from the L-band observations will also be included. This work represents the first data-release of ViCTORIA, which consists of data taken with the LOFAR HBA at 144\,MHz.
LOFAR \citep{vanHaarlem2013} is a radio interferometer based in the Netherlands and operating at frequencies between 10 and 240\,MHz. The LOFAR Surveys Key Science Project is currently carrying out the LOFAR Two-metre Sky Survey \citep[LoTSS,][]{Shimwell2017,Shimwell2019,Shimwell2022TheRelease}, imaging the Northern Sky at more than an order of  magnitude greater sensitivity and significantly higher resolution than previous large sky surveys. Due to its low declination and the presence of the extremely bright source M\,87 (1250\,Jy at 144\,MHz), which severely complicates calibration and imaging of the field, Virgo is not yet included in the published footprint of LoTSS. With this work, we built up on the significant advances in low-frequency radio interferometric calibration techniques developed for LoTSS \citep{Tasse2021TheImaging} to extend the coverage of sensitive LOFAR surveys to the Virgo region by employing a partly custom-developed calibration strategy to address the severe image fidelity issues associated with M\,87. 

A wealth of multi-wavelength data exists on Virgo, which will be complemented by the LOFAR observations presented in this paper and the further releases of the ViCTORIA project. For example, the classical reference for galaxies in the Virgo field, derived from optical observations,  is \citet{Binggeli1985}, who published the photographic Virgo Cluster Catalog (VCC), which includes 2096 galaxies in a $140\,\mathrm{deg^2}$ area. 
A digital successor to the VCC is the Extended Virgo Cluster Catalog (EVCC) \citep{Kim2014}, which, based on SDSS Data Release 7 photometric and spectroscopic data \citep{Abazajian2009TheSurvey}, provides an updated cluster membership and morphology categorization over galaxies residing in a five times larger area than the VCC. The deepest optical survey of the cluster, although only covering $104\,\mathrm{deg^2}$, is the Next Generation Virgo Cluster Survey \citep[NGVS,][]{Ferrarese2012} conducted with the MegaCam instrument at the CFHT. Other auxiliary data include the Virgo Environmental Survey Tracing Ionised Gas Emission \citep[VESTIGE,][]{Boselli2018b} which has mapped the cluster in $H{\alpha}+[NII]$ also using MegaCam and following the NGVS footprint. In the far-infrared, data of the Herschel Reference Survey \citep[HRS,][]{Boselli2010TheSurvey} and the Herschel Virgo Cluster Survey \citep[HeViCS,][]{Davies2010} exists. In the near- and far-ultraviolet, the cluster was observed by the GALEX Ultraviolet Virgo Cluster Survey \citep[GUViCS,][]{Boselli2011}. Combined, these data give an outstanding repertoire of star-formation tracers over the entire cluster region.
Other data include the sample from the ACS Virgo Cluster Survey \citep[ACSVCS,][]{Cote2004} of 100 early-type Virgo galaxies. In infrared and X-rays, this sample has also been studied with Spitzer and Chandra \citep{Gallo2010Amuse-virgo.Accretion}. 
Samples of late-type galaxies were subject to dedicated studies X-rays with Chandra \citep{Soria2022}, in molecular gas with ALMA \citep[][]{Brown2021VERTICO:Survey} and in atomic gas \citep[\hi{},][]{Chung2009} with the VLA. 
The properties of the ICM were analyzed based on X-ray observations with ROSAT \citep{Bohringer1994TheImages}, XMM-Newton \citep{Urban2011X-rayRadius} and Suzaku \citep{Simionescu2017}. An ongoing study beyond the virial radius is currently being conducted with the extended ROentgen Survey with an Imaging Telescope Array (eROSITA, \citep[][McCall et al.\ in prep.]{Merloni2012,Predehl2021}.

With this paper, we will extend the multi-wavelength coverage of Virgo by providing wide-field images of the cluster and radio-measurements of all LOFAR-detected Virgo galaxies. The detailed scientific interpretation of the radio data will follow in upcoming studies, where we will analyze the environmental impact on galaxy evolution and the radio galaxy population in the Virgo cluster.

Throughout this paper, we assume a flat $\mathrm{{\Lambda}CDM}$ cosmology with $\Omega_\mathrm{m}=0.3$ and $H_0=70\,\mathrm{km\,s^{-1}\,Mpc^{-1}}$. At the distance of M\,87, for which we adopt a value of 16.5\,Mpc \citep{Mei2007,Cantiello2018TheBeyond}, one arcsecond corresponds to 80\,pc. 
This paper is arranged as follows: In \autoref{sec:obsanddata}, we present the observations and the data reduction strategy of the survey. In section \autoref{sec:cat}, we report details on the catalog of Virgo galaxies we provide, and in \autoref{sec:discussion}, we highlight individual objects and findings in the LOFAR maps. We conclude in \autoref{sec:conclusion} .

\section{Observations and Data Reduction} \label{sec:obsanddata}
\begin{figure*}
    \centering
    \includegraphics[width=0.925\textwidth]{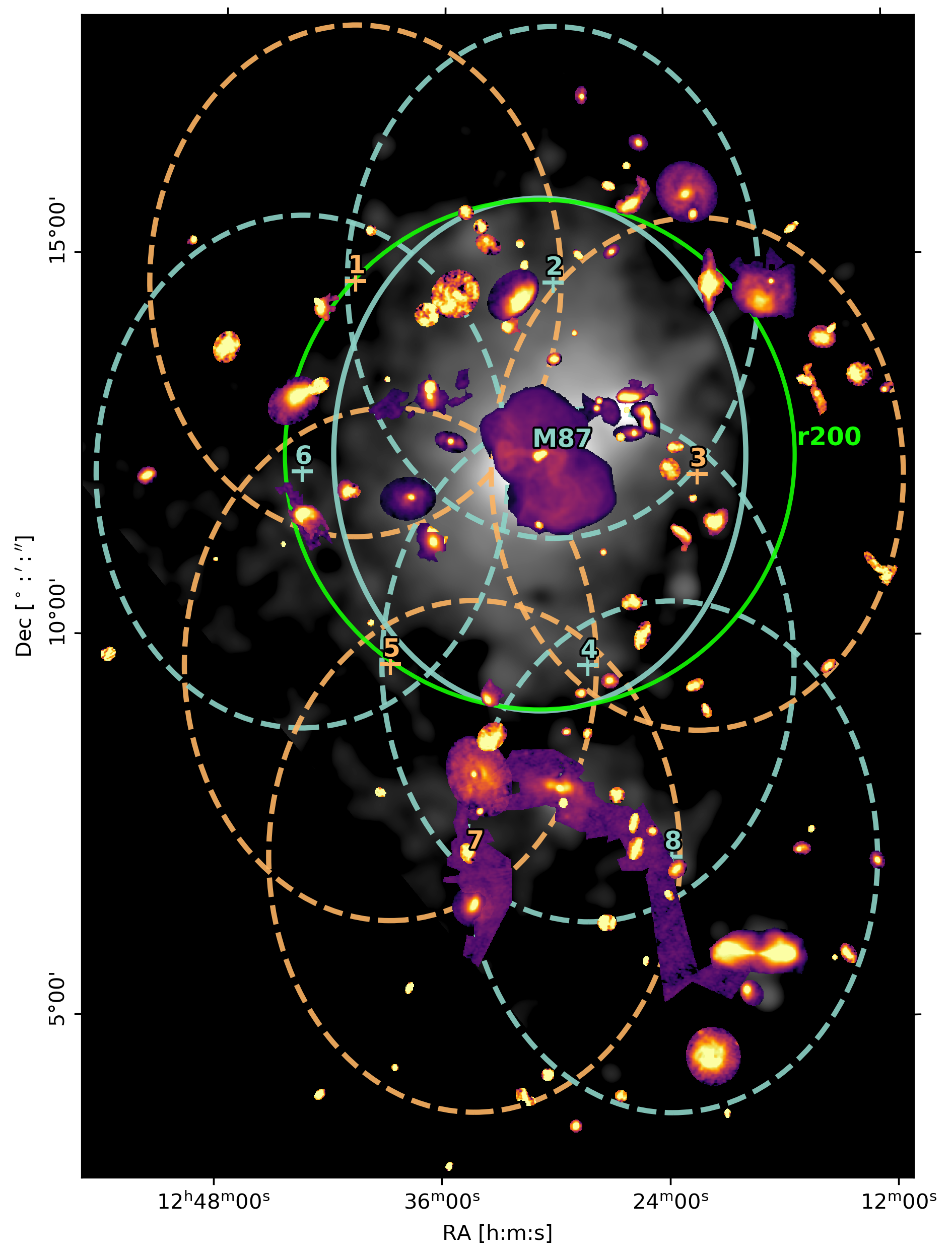}
    \caption{Overview of the LOFAR HBA Virgo cluster survey. As cutouts, we show the radio-detected galaxies from the $20''$-mosaic, for displaying purposes, their sizes are magnified by a factor of six and on different color-scales. The central points and the half-maximum primary beam ellipses of the nine survey pointings are displayed in alternating colors (orange/blue). The green circle marks $r_{200}$ and the greyscale background image is the eROSITA X-ray map in the $0.3-2\,$keV band (Mc Call et al. in prep.) where we masked the southeastern emission in the direction of the NPS.}
    \label{fig:fields}
\end{figure*}
\begin{figure}
    \centering
    \includegraphics[width=1.0\linewidth]{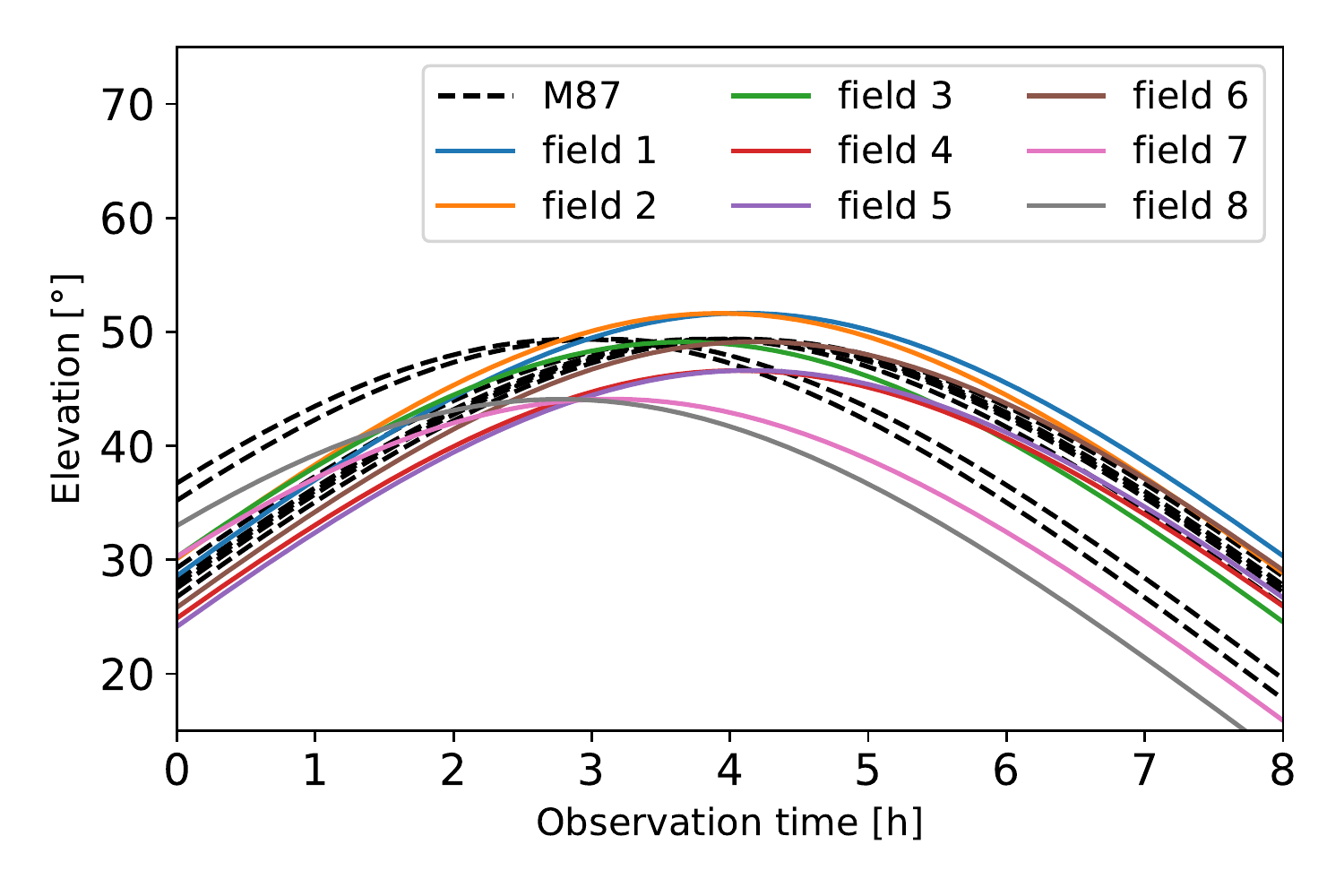}
    \caption{Elevation as a function of observation time for the Virgo cluster observations.}
    \label{fig:elev}
\end{figure}
\begin{figure}
    \centering
    \includegraphics[width=0.99\linewidth]{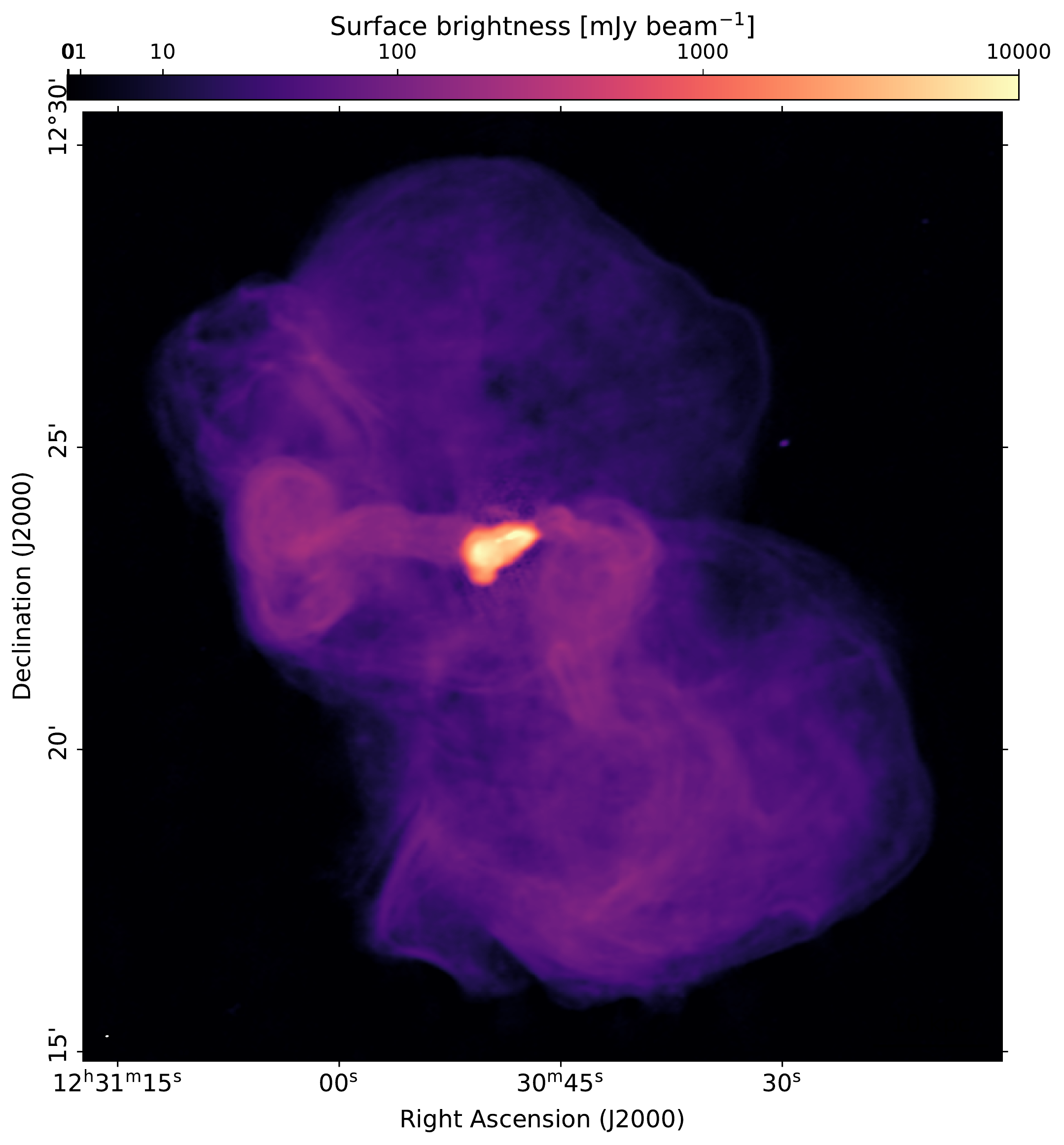}
    \caption{LOFAR image of M\,87 at 144\,MHz. The ellipse in the bottom left corner indicated the primary beam size of $5.6''\times4.1''$. The noise level is 350\,$\mathrm{{\mu}Jy\,beam^{-1}}$, corresponding to a dynamic range of 44000.}
    \label{fig:m87}
\end{figure}

The LOFAR HBA Virgo Cluster Survey covers a 132\,deg$^2$ region of the Virgo cluster between a declination of 3\textdegree\ and 18\textdegree. The survey footprint displayed in \autoref{fig:fields}  is composed of nine different pointings listed in \autoref{tab:pointings}. In total, eight observations of 8\,h each were conducted (project: LC11\_010) between March 18, 2019 and April 2, 2019, book-ended by 10\,min calibrator scans of 3C\,196 and 3C\,295. The observations were taken in dual-beam mode, where one beam was always pointing towards M\,87, and another on one of the eight outer fields. The outer fields lie on the grid of LoTSS. The total observation time of M\,87 is only 56 instead of 64\,h since the observation parallel to field 1 is missing in the LOFAR Long Term Archive (LTA). Data were taken with the LOFAR HBA in a setup identical to observations taken for LoTSS with a nearly continuous frequency range from 120 to 168\,MHz, a frequency resolution of 3.07\,kHz and a time resolution of 1\,s. After initial flagging of the raw data, the data are averaged to a resolution of 12.2\,kHz in frequency and stored in the LTA.

\subsection{Challenges of the Virgo field}
The presence of M\,87 in the cluster center makes radio calibration and imaging of the Virgo field particularly challenging. It is not only one of the brightest  sources on the radio sky \citep[1.25\,kJy at 144\,MHz,][]{Scaife2012ATelescopes}, but also highly extended with an angular size of $15'$. This complex morphology makes an accurate deconvolution of the source hard to achieve, and the large extent means that the primary beam variation across M\,87 is non-negligible, in fact it is $>20\%$ for some pointings.
Another difference to observations of LoTSS-DR2 is the declination of the LOFAR Virgo field which is between $+3\degree$ and $+18\degree$; this translates to rather low-elevation observations in the range $\sim 25\degree-50\degree$ (see \autoref{fig:elev}). This causes an elongated primary beam shape, an increased thermal noise level due to the reduced sensitivity of the dipoles and greater susceptibility to ionospheric disturbances due to the higher air mass compared to observations at more favorable declinations. Furthermore, the LOFAR primary beam model is expected to be less accurate at lower elevations which affects the accuracy of the flux density scale.
Lastly, the majority of the Virgo cluster galaxies are extended objects with a low radio-continuum surface brightness and as such, are difficult to fully deconvolve.

\begin{table*} 
\centering
\caption{Virgo cluster pointings.}
\begin{tabular}{c c c c c} 
\hline\hline
 field & RA (J2000) [h:m:s] & DEC (J2000) [\textdegree:$':''$] & observation time [h] & comments \\ [0.5ex]\hline 

 M87 & 12:30:49.42 & +12:23:28.0 & 56 & \\
 1 & 12:40:47.46 & +14:39:45.5 & 8 & severe ionosphere, high noise level\\ 
 2 & 12:30:05.30 & +14:38:54.9 & 8 &\\
 3 & 12:22:22.72 & +12:07:42.3 & 8 &\\
 4 & 12:28:16.58 & +09:37:33.7 & 8 &\\
 5 & 12:38:46.71 & +09:38:23.2 & 8 &\\
 6 & 12:43:33.66 & +12:09:22.3 & 8 & dynamic ionosphere \\ 
 7 & 12:34:18.36 & +07:07:25.6 & 8 & bright sources\\ 
 8 & 12:23:52.29 & +07:06:36.4 & 8 & bright sources\\ [1ex] 
 \hline
\end{tabular}
\label{tab:pointings}
\end{table*}

\subsection{Data reduction}
To address the aforementioned difficulties, we develop a strategy specifically tuned towards the Virgo cluster field and the targets of interest. The main difference to the default calibration approach for LOFAR HBA, as it is used e.g. for LoTSS-DR2, is an additional step in which we accurately subtract M\,87 from the $uv$-data. Afterwards, direction-dependent calibration can mostly be carried out similar to LOFAR HBA observations of normal fields.

The data reduction is split into a series of steps which are implemented in various pipelines: First, for pre-processing, second to reduce the data of the calibrator scans, third to subtract M\,87 from the data and lastly, for full direction-dependent calibration. Afterwards, an extraction and re-calibration procedure can be applied for selected targets \citep[see][]{vanWeeren2021LOFARHETDEX}.
For the first three steps, we make use of the Library for Low Frequencies (LiLF\footnote{\url{https://github.com/revoltek/LiLF}}), and for the direction-dependent calibration and extraction, we use ddf-pipeline\footnote{https://github.com/mhardcastle/ddf-pipeline} \citep{Tasse2021TheImaging}. 

In the pre-processing step, the data of all calibrator and target observations were downloaded from the LTA. Subsequently, all baselines containing international LOFAR stations were removed and the data averaged down to 48.8\,kHz in frequency (four channels per sub-band) and 4\,s in time. 
The strategy used to find the calibrator solutions is described in detail in \citet{deGasperin2019}; however, we will quickly summarize the key steps. The pipeline derived the polarization alignment delays as well as the Faraday rotation and bandpass solutions. These were then applied to the data in physical order together with the primary beam model. Lastly, we solved for scalar phases which describe the ionospheric and clock delays.

\subsubsection{Peeling pipeline}

\begin{figure*}
    \centering
    \includegraphics[width=0.8\linewidth]{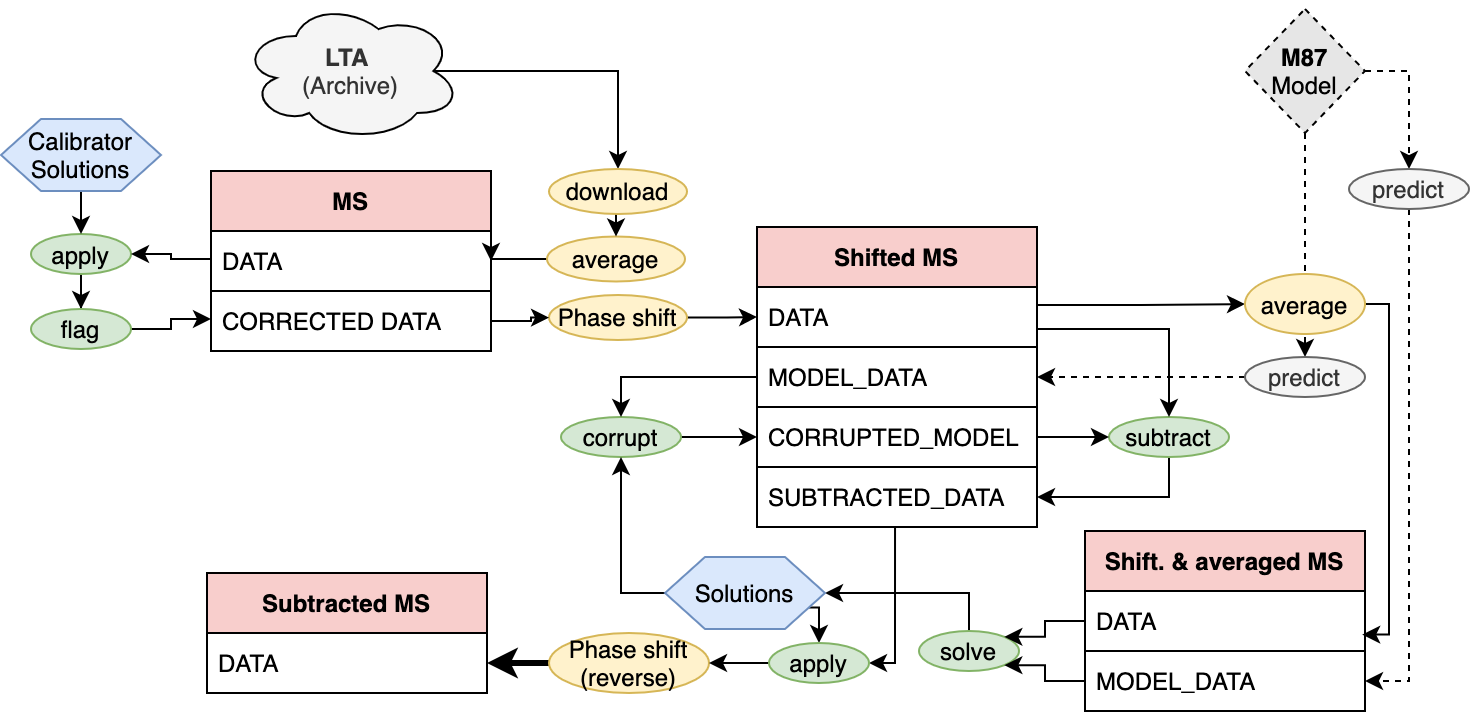}
    \caption{Scheme of the peeling strategy.}
    \label{fig:peelpipeline}
\end{figure*}

The purpose of this pipeline is to subtract M\,87 from the $uv$-data of each observation of the Virgo field. A key requirement for this is a high-quality model of the source, which did not exist at 144\,MHz prior to this project. Starting from the $21''\times15''$ image of \citet{deGasperin2012}, we performed multiple rounds of self-calibration using an 8\,h-observation with M\,87 at the phase center. In each iteration, we solved for scalar-phases at full time- and frequency-resolution and for full-Jones matrices at a resolution of 64\,s and 0.49\,MHz using \texttt{DP3}\footnote{https://github.com/lofar-astron/DP3}. For imaging, we used the multi-scale and multi-frequency deconvolution mode of \texttt{WSCLEAN} \citep{Offringa2014,Offringa2016}. Since M\,87 is extremely bright, of complex morphology and highly resolved, deconvolution is particularly delicate. With improving solutions during self-calibration, less conservative deconvolution parameters were used. A high-resolution image of a single 8\,h observation, shown in \autoref{fig:m87}, was created using a \textit{Briggs}-value of $-1.2$ and the final model of M\,87 that is used for the peeling is created with a \textit{Briggs}-weighting of $-0.6$.

The following peeling procedure was carried out for each observation. The first step of the peeling pipeline is to apply the polarization alignment, bandpass and scalar-phase solutions from the corresponding calibrator scan with the least amount of flagged data to the target field together with the primary beam model at the phase center. Next, we perform `A-Team clipping'. The purpose of this procedure is to flag the parts of the data where the bright sources Cassiopeia A, Cygnus A and Taurus A (which form the infamous `A-team' together with `Virgo A') reside in a side-lobe of the primary beam. If their predicted amplitude is above 5\,Jy for any time, frequency and polarization, this part of the data is flagged. In addition, we flag the parts of the data affected by radio-frequency interference using \texttt{AOFLAGGER} \citep{Offringa2012ADetection} as well as all data taken below $20\degree$ elevation. Afterwards, time steps that contain more than 50\% of flagged data are discarded. 

Next, the data are phase-shifted towards the location of M\,87, if they are not already centered on the source, and corrected for the difference between the primary beam at the original phase center and M\,87. From this phase-shifted data set, a smaller data set is created by averaging down to a resolution of 16\,s in time and 0.39\,MHz in frequency. This speeds up computation and suppresses the contribution of sources away from M\,87 by time- and frequency-smearing them. The model-$uv$-data of M\,87 are predicted to this small data set using \texttt{WSCLEAN} together with the image domain gridder \citep[\texttt{idg},][]{vanderTol2019ImageVisibilities}. Due to a large angular size of $15'$, the primary beam variations across M\,87 are non-negligible, especially for the pointings where M\,87 is close to the edge of the primary beam. Thus, we found that it is important to adjust the predicted visibilities with the direction-dependent component of the primary beam during prediction. It is then solved against the model data for a complex scalar gain on the full time resolution of the averaged data set. In the solver the solutions are smoothed in frequency with a 1\,MHz kernel. From the resulting calibration solutions, we only apply the phases to the data, which encompass the ionospheric delays. The amplitudes, however, are used to identify bad parts of the data by flagging all time and frequency intervals where the amplitudes are more than a factor of five from unity. We perform a second, slow full-Jones calibration on time scales of 256\,s. In this solve step, we smooth the calibration solutions with a 2\,MHz kernel in frequency. After deriving the solutions towards M\,87 from the small data set, we predict the M\,87 model to the large data set at 48.8\,kHz bandwidth and 4\,s time resolution, again using \texttt{idg} to include the direction-dependent component of the primary beam. The predicted visibilities are then corrupted with the phase and slow full-Jones solutions. We create a new data set by subtracting the corrupted M\,87 model from the data. We then correct the subtracted data using the scalar phases towards M\,87, which is intended to correct the clock delays as well as to pre-correct the ionospheric delays in the direction of the Virgo-cluster. The pre-corrected subtracted data column is then phase-shifted back to the original phase-center, averaged to a resolution of 8\,s in time and 98\,kHz in frequency and concatenated into frequency blocks of 1.95\,MHz. Lastly, we correct for the difference of the primary beam at the phase center and at the location of M\,87.
The resulting measurement sets now have M\,87 subtracted and are thus prepared in a format that is suited for the following direction-dependent calibration. A schematic overview of the peeling strategy is presented in \autoref{fig:peelpipeline}.

For the observations centered on M\,87, the subtraction procedure was slightly altered. The phase-shifting and additional beam-correction steps are not necessary due to the greater apparent brightness. Furthermore, the calibration solutions are derived at a higher time resolution of 4\,s for the Jones scalar, and 64\,s for the full-Jones matrix.

For field 8, the subtraction of M\,87 is particularly challenging due to the presence of NGC\,4261 (=3C\,270; $S_{144}=73$\,Jy) in the center of the field. To obtain robust solutions for this pointing, we used a parallel solve on both NGC\,4261 and M\,87, where we employed a model of two Gaussian components obtained from the TGSS for NGC\,4261. In addition, we solved for a diagonal instead of a full-Jones matrix and used a 4\,MHz kernel to smooth the solutions in frequency to reduce the effective number of free parameters. The solutions for M\,87 were used for subtraction and to pre-calibrate the phases, the solutions for NGC\,4261 were discarded. Stable calibration solutions were not obtained for all time- and frequency windows, which lead to a high ratio of flagged data for this field (55\%). 

\subsubsection{Direction-dependent calibration} \label{sec:ddcal}
 For direction-dependent calibration, the \texttt{ddf-pipeline} \citep{Shimwell2019,Tasse2021TheImaging} is used, which is a framework that is based on the \texttt{DDFacet} imager \citep{Tasse2018} and the \texttt{killMS} solver \citep{Tasse2014ApplyingProblem,Tasse2014NonlinearInterferometry}. The \texttt{ddf-pipeline} algorithm was also used for the second data release of LoTSS. Since it is described in detail in \citet{Tasse2021TheImaging}, only a very brief summary is provided here, mainly focusing on the differences in processing compared to LoTSS-DR2. The algorithm starts with a sparse selection of the data which consists of every fourth frequency sub-band. In a series of direction-independent and direction-dependent calibration steps, a sky model of a $8.3\degree\times 8.3\degree$ square region of the target field is obtained. Using this model, calibration is carried out on all sub-bands, again in a series of direction-independent and direction-dependent self-calibration cycles. The direction-dependent steps are carried out in a facet-based approach, i.e. the sky model is split into 45 discrete directions which are solved in parallel. An important difference to the LoTSS-calibration is that no amplitude solutions are applied to the data - this is because we experienced divergent amplitude solutions in a small number of facets with a low flux density. This divergence is likely caused by a reduced quality of the data due to residuals of M\,87 and/or a low signal-to-noise ratio due to the low elevation of the observations. As a consequence of the missing amplitude corrections, artifacts exist around a number of bright sources, mainly further away from the phase center, hinting that they may be at least partly caused by errors in the primary beam model. This, however, does not notably affect the science quality of the images regarding studies of Virgo cluster objects, since the vast majority of these sources are of low-surface brightness.
 During the various imaging steps, it is challenging but also critical to accurately deconvolve faint and extended structures to obtain a complete sky model for calibration and accurate surface brightness values - \texttt{ddf---pipeline} strongly optimizes deconvolution for these targets, using multiple \texttt{CLEAN}-iterations with progressively improving masks. By using a mask based on lower resolution images for the deep high-resolution deconvolution, it is possible to further improve deconvolution for extended structures. Nevertheless, a number of faint Virgo galaxies is not fully picked up in the masking procedure and hence, not accurately deconvolved. To address this, we manually add all detected galaxies of the optical VCC and EVCC catalogs to the masks created by \texttt{ddf-pipeline} and perform one additional deep \texttt{CLEAN}-iteration for both the low- and high-resolution images. For the seven observations pointed on M\,87, we follow the procedure used for the LoTSS deep-fields \citep{Tasse2021TheImaging}. A single 8\,h observation is fully calibrated to derive an accurate sky model. This model is then used to jointly calibrate and deconvolve all 56\,h of data. 
 
 For three of the eight outer fields, a slightly different strategy was necessary: Field 6 is affected by the presence of 3C\,275.1 outside of $8.3''$ square region used for imaging and calibration. We address this by carrying out an additional initial imaging step with a larger field of view to subtract sources outside of the square normally used for calibration. The two southern pointings on field 7 and field 8 are particularly challenging to calibrate, as they have a bright source in the field-of-view (FoV) (NGC\,4261), and the bright quasar 3C\,273 (116\,Jy, \citet{Jacobs2011NewSky}) in the side-lobes. The subtraction of sources outside the primary beam, which was used for field 6, did not yield satisfactory results for fields 7 and 8. Instead, the image size used for direction-dependent calibration of these two fields was increased to allow for direction-dependent solutions towards 3C\,273. Since this strongly increased memory requirements and computation time, the data set was phase-shifted beforehand such that the image size is only increased by as much as necessary to cover 3C\,270, without increasing the coverage in the opposite direction. The size of the square region used for calibration of fields 7 and 8 is $9.5\degree$.
 
Per field, a number of wide-field imaging products are created from the calibrated data sets: We image Stokes $I$  at angular resolutions of $9''\times5''$, $20''$ and $1'$, hereinafter, we refer to these resolutions as \textit{high}, \textit{low} and \textit{very low}.  Compact source-subtracted images were created at $1'$ and $4'$ resolution. In addition, we produced Stokes $Q$ and $U$ dirty-images in a frequency-spacing of 97.6\,kHz at $20''$ and $4'$ resolution and Stokes $V$ dirty images at $20''$ resolution. No flux density scale correction was applied to the per-field data products.

\subsubsection{Flux-scale alignment and mosaicing} 
Uncertainties of the beam model of LOFAR HBA currently limit the accuracy of the flux density scale when it is directly transferred from a calibrator source by introducing a per-field flux density scaling $f_i\neq 1$. Thus, \citet{Shimwell2022TheRelease} correct the flux density scale in a post-processing step. 
By cross-matching sources between LOFAR and the 151\,MHz sixth Cambridge survey of radio sources \citep[6C][]{Hales1988TheFormula} and between LOFAR and the 1.4\,GHz NVSS, they derive the median flux density ratios $F_{\mathrm{6C},i}$ and $F_{\mathrm{NVSS},i}$ of each field $i$ and both surveys. Under the assumption that the flux density scale offsets are direction-independent for a given field, both median ratios should be equally offset by a factor of $f_i$ from the value found for a perfectly accurate flux density scale. Thus, the ratio of the median flux density ratios per field ${F_{\mathrm{6C},i}}/{F_{\mathrm{NVSS},i}}$ should be approximately independent of the offset $f_i$ and thus, constant between the fields if there is no spatial variation in the NVSS and 6C scales. Considering the median ratio of the median ratios found across all fields, a value of ${F_{\mathrm{6C}}}/{F_{\mathrm{NVSS}}} = 5.724$ was found in \citet{Shimwell2022TheRelease}. Taking into account the frequency difference between LOFAR and 6C and using the median spectral index of sources in $6C$ and $NVSS$ of $\alpha=-0.783$, the LOFAR observations are aligned with the 6C flux density scale if the median flux density ratio equals $F_{\mathrm{NVSS}} = 5.724 \times (144/151)^{-0.783} = 5.936$. Consequently, the flux density scale correction factors $f_i^{-1}$ can be determined from the measured median flux density scale ratio $F_{\mathrm{NVSS},i}$ of each field as $f_i^{-1} = 5.936/F_{\mathrm{NVSS},i}$.
We followed the procedure developed in \citet{Shimwell2022TheRelease} with two minor modifications: First, since a flux cut of 30\,mJy is applied in the cross-matching with NVSS prior to the re-scaling and we found correction factors strongly different from unity for some fields, we iteratively repeated the calculation of the correction factors $f_i$, each time updating the flux cut to $30\,\mathrm{mJy}\times f_i^{-1}$. Second, we estimated the correction factors from the $20''$-resolution maps since in the high-resolution maps, ionospheric smearing lead to a systematic over-correction of $10-20\%$ for the fields were we have lower quality (fields 1, 6, 7, 8). For the remaining fields with good image quality, the difference between the correction factors estimated from the high- and low-resolution maps is between 1\% and 8\%.
To derive these factors, we created a LOFAR source catalog for each Virgo field using the Python Blob Detector and Source Finder \citep[\texttt{PyBDSF},][]{Mohan2015} and only considered sources with no neighbor within $30''$, a major axis of less than $25''$ and a significance of at least 5$\sigma$.
The final correction factors we found were applied to the fields during mosaicing, ultimately placing our survey on the scale of the 6C survey which in turn is aligned with the flux density scale of \citet{Roger1973TheSources}.   
In the mosaicing procedure, pixels were weighted according to the primary beam response as well as the central noise level of the corresponding field. We also excluded all areas with a primary beam attenuation factor below 0.3. In the final mosaic, we restore the clean-components of M\,87 that were subtracted in the peeling pipeline.
 \begin{table*}[]
     \caption{Imaging parameters}
     \centering
     \begin{tabular}{c c c c c c}
          resolution & subtracted & [$''$$\times$$''$] & robust weighting & min. $uv$ [km] & max. $uv$ [km]   \\ \hline
          high & & $9\times5$ &  $-0.5$ & 0.1 &  120.6  \\ 
          low & & $20\times20$ & $-0.25$ & 0.1 & 25.8  \\
          very low & & $60\times60$ & $-0.2$ & 0.04  & 7.0  \\
          very low & yes & $60\times60$ & $-0.2$ & 0.04 & $\sim 7.0$ ($1'$-taper)  \\ \hline
     \end{tabular}
     \label{tab:mosaics}
 \end{table*}
Wide-field mosaics are created for the high-, low- and very-low-resolution Stokes $I$ maps as well as for the source-subtracted images. The image parameters of all Virgo field mosaics are presented in \autoref{tab:mosaics}; the image files are available online in fits format\footnote{\url{https://lofar-surveys.org}}.
 
 \begin{figure}
     \centering
     \includegraphics[width=1.0\linewidth]{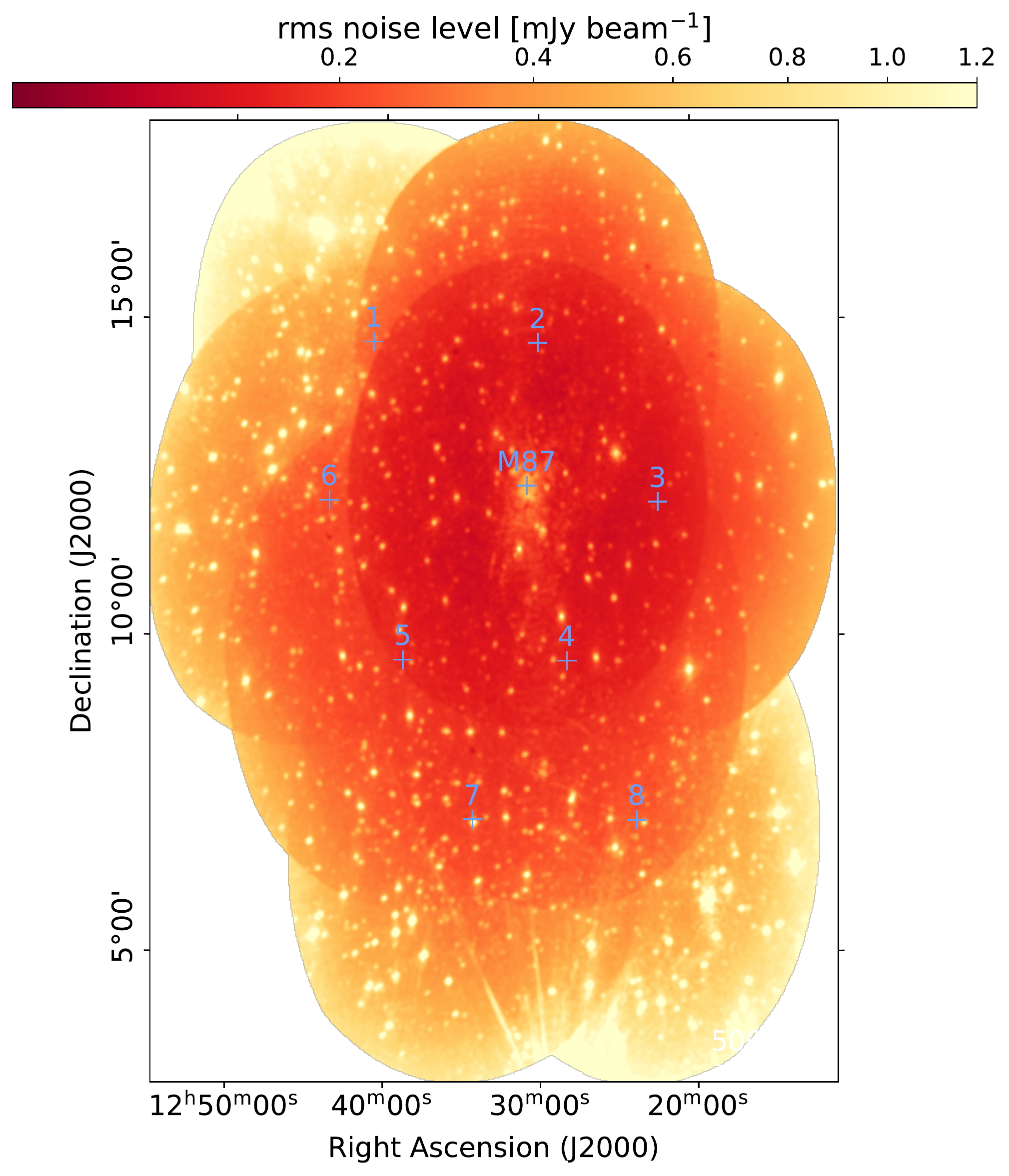}
     \caption{RMS noise map of the $9''\times 5''$ mosaic. Blue annotations indicate the pointing centers of the nine LOFAR fields.}
     \label{fig:noisemap}
 \end{figure}

A map of the root-mean-square (RMS) noise level of the high-resolution Stokes $I$ mosaic is displayed in \autoref{fig:noisemap}; the median noise level is $140\,\mathrm{{\mu}Jy\,beam^{-1}}$ within the virial radius and $280\,\mathrm{{\mu}Jy\,beam^{-1}}$ across the full footprint. In the vicinity of M\,87, the noise level is $\sim170\,\mathrm{{\mu}Jy\,beam^{-1}}$, only 40\% deeper than what we find for a single 8\,h observation, which indicates that we are limited by dynamic range in the central few square degrees. Beyond $2^\circ$ separation from M\,87, the sensitivity ratio between 8\,h and 56\,h of data approaches the expected ratio of $\sqrt{56\,\mathrm{h}/8\,\mathrm{h}}\approx2.65$. Remaining artifacts related to M\,87 express as wave-like patterns, the most dominant of those originate from M\,87 and cross the image towards north and south, extending for multiple degrees. In some of the outer regions of the survey, the noise level is strongly increased. This is apparent most drastically in the northeast, were the observation of field 1 was affected by particularly unfortunate ionospheric conditions with rapid high-amplitude variations of the ionospheric parameters. The noise level for the region that is exclusively covered by this pointing is $\sim800\,\mathrm{{\mu}Jy\,beam^{-1}}$. Fortunately, only very few Virgo cluster galaxies populate that area. Similarly, the eastern field 6 to the south of field 1 was also affected by a dynamic ionosphere which increased the noise level to  $250\,\mathrm{{\mu}Jy\,beam^{-1}}$. Another region where the image quality is reduced is the southwestern field 8, where the presence of the bright sources 3C\,270 and 3C\,273 as well as a high flag ratio and low elevation increase the noise level to $\sim400\,\mathrm{{\mu}Jy\,beam^{-1}}$. Especially in the lower-resolution mosaics, this area also shows the presence of larger-scale calibration artifacts.
 
\begin{figure}
     \centering
     \includegraphics[width=1.0\linewidth]{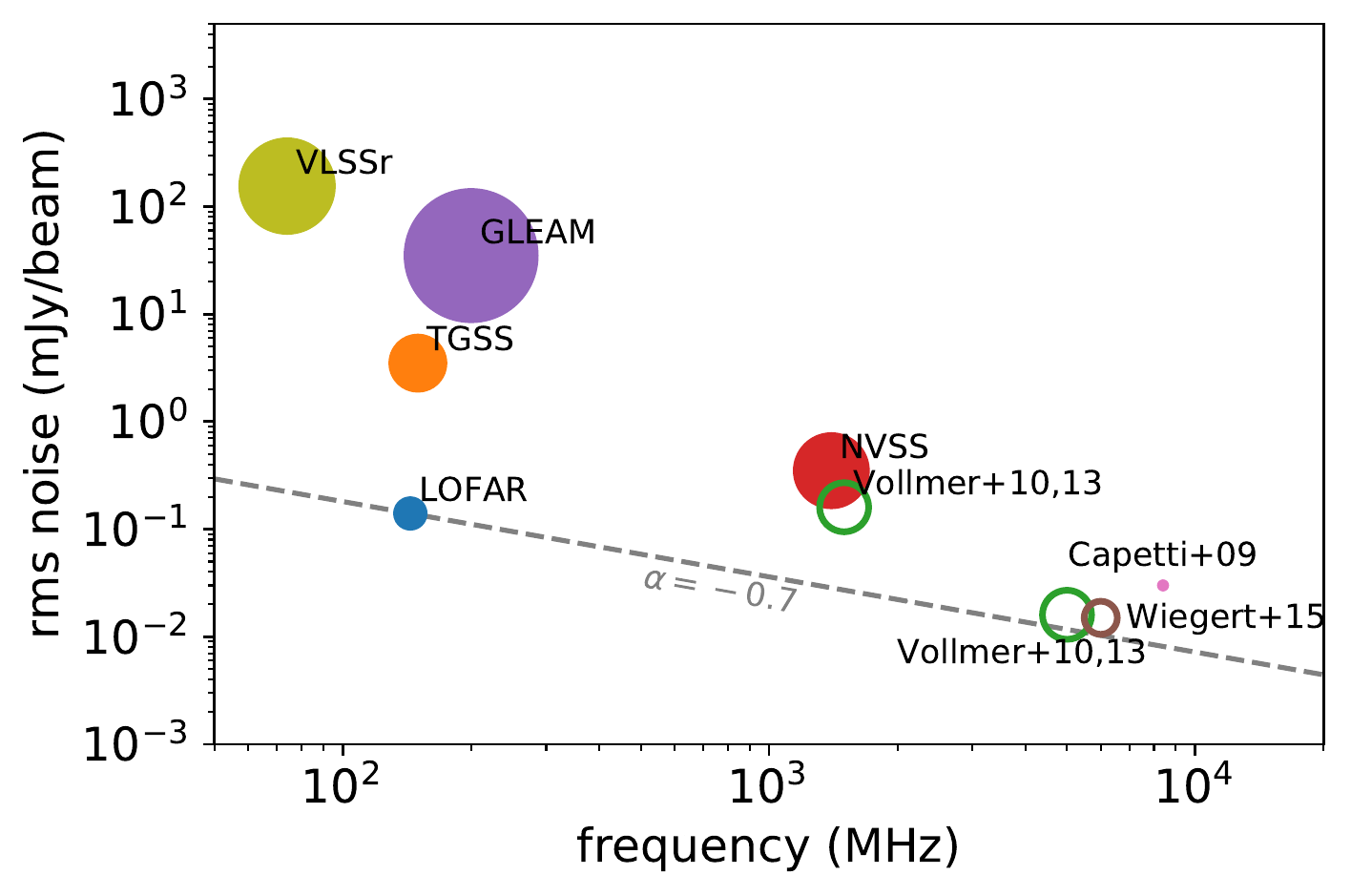}
     \caption{Noise-level comparison of different Virgo cluster surveys. The circles show different wide-field surveys (filled circles) and targeted observations of samples of Virgo-galaxies (hollow circles). The marker size is proportional to the angular resolution of the surveys. The quoted noise level for LOFAR refers to the median noise inside the virial radius. The dashed line shows an interpolation of this noise level assuming a spectral index of $-0.7$.}
     \label{fig:rmssurveys}
 \end{figure}

 In \autoref{fig:rmssurveys}, we compare the noise level of the LOFAR HBA Virgo cluster survey to further surveys and targeted observations of the Virgo field. The quoted noise level of $\sim150\,\mathrm{{\mu}Jy\,beam^{-1}}$ corresponds to the region just inside the area covered by the deeper M\,87 pointing which approximately coincides with the Virgo cluster virial radius. 
 
\subsection{Flux density scale}
For LoTSS-DR2, the global uncertainty of the flux density scale was estimated to be 10\%, with an additional positional variation of 10\% \citep{Shimwell2022TheRelease}. Due to a number of reasons, we expect the uncertainty to be higher for the Virgo field: first, due to the low declination of Virgo, the LOFAR beam model should be less accurate, which translates to a higher systematic uncertainty on the flux densities. Indeed, \citet{Shimwell2022TheRelease} found a declination-dependency of the flux density scale that becomes increasingly severe at lower declination. Second, the presence of M\,87 in the field does not only cause image artifacts in our LOFAR data, but also and even more severely in the NVSS and other all-sky surveys, where no tailored strategy was used to address this bright source. This decreases the quality of reference radio catalogs within a few degrees around M\,87. And lastly, the direction-dependent variation of the flux density scale should be higher for simple statistical reasons --- since our survey consists of only nine pointings, the degree of overlap is lower than in LoTSS, where the mosaicing of multiple pointings was shown to reduce the scatter of flux density measurements \citep{Shimwell2022TheRelease}.

To assess the systematic uncertainty of the flux density scale of the LOFAR HBA Virgo cluster survey, we created a source catalog of the final low-resolution mosaic using PyBDSF. We first compared the LOFAR catalog to NVSS which was also used for the flux density scale alignment. Repeating the source selection and matching as described in \autoref{sec:ddcal}, we found a median ratio of $S_\mathrm{LOFAR}/S_\mathrm{NVSS} = 5.42$, which is 9\% different from the value of 5.94 we used for the alignment.

We then cross-matched the LOFAR catalog with sources of the TGSS and the Galactic and extragalactic all-sky MWA survey \citep[GLEAM][]{Hurley-Walker2017GaLacticCatalogue} which are nearly at the same frequency as LOFAR HBA. We searched for unique matches in a radius of $10''$ for TGSS and $50''$ for GLEAM. To avoid complex sources where the flux density measurements might not be accurate, we only kept sources that are isolated in LOFAR within $30''$ for TGSS and within $2'$ for GLEAM and have an extension in LOFAR of less than $25''$. Furthermore, we consider only sources with a signal-to-noise ratio of five and above.
This resulted in 576 matched sources with TGSS and 68 with GLEAM. We extrapolated the TGSS $150$\,MHz flux densities to $144$\,MHz using a factor of $(144/150)^{-0.783} = 1.03$. For GLEAM, we used the 143\,MHz catalog and neglected the frequency difference as well as the <3\% systematic uncertainty arising due to the Baars flux density scale \citep{Baars1977TheCalibrators.} of GLEAM \citep{Hurley-Walker2017GaLacticCatalogue,Perley2017AnGHz}. Compared to TGSS, the median flux scale ratio is $1.05$ with a standard deviation of $0.23$, and for GLEAM, it is $1.02$ with a standard deviation of $0.19$. This independent test shows that the flux density scale is aligned accurately, although with a considerable scatter. This scatter is partly caused by the direction-dependent flux scale uncertainty in the LOFAR maps; another contribution stems from outliers in the cross-matching procedure. Additionally, also the reference surveys add considerably to the scatter. By cross-matching 211 sources in the Virgo field between TGSS and GLEAM, we find that the flux densities of those two surveys show a median flux ratio of 1.01 (expected: $(143/150)^{-0.78} = 1.04$) with a standard deviation of 0.29. 
Given the comparison with both TGSS and GLEAM, we assume that the systematic uncertainty of the flux density scale of our LOFAR HBA survey of the Virgo cluster is $f_\mathrm{sys} = 20\%$, which includes both the uncertainty of the absolute flux density scale and the direction-dependent variation.  

During the flux-scale alignment, we used the low-resolution mosaic since it is less affected by poor ionospheric conditions. Indeed, repeating the procedure for the high-resolution images, we find that the median flux density is significantly lower for the observations with poor ionospheric conditions. The most extreme case is field 1, where the difference in median flux between the high-resolution and the low-resolution image is $28\%$. A significant difference of $\sim 20\%$ is also observed for the southern fields 7 and 8, where bright sources limit the quality of the calibration. This indicates that for regions close to the edge of the survey footprint which are exclusively covered by one of those fields, particular care must be taken when working with the high-resolution images.

\section{LOFAR Virgo Cluster Catalog}\label{sec:cat}

\begin{figure*}
\centering
    \includegraphics[width=1.0\textwidth]{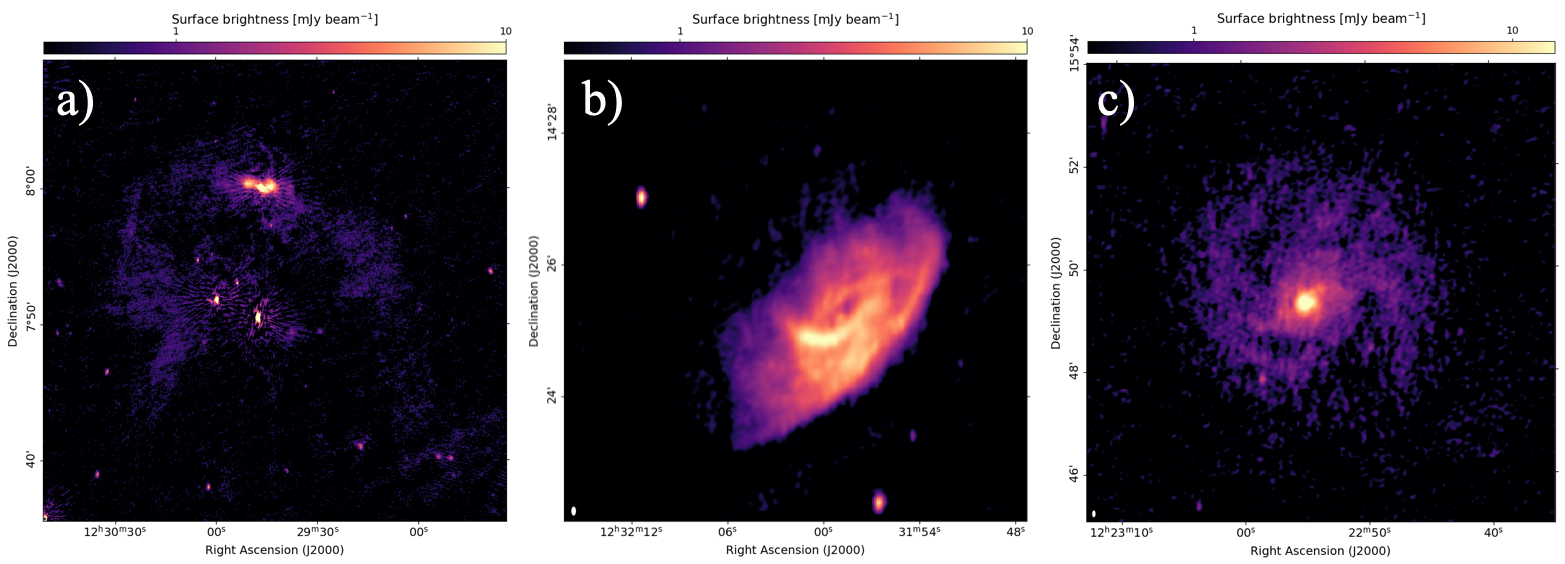}
\caption{LOFAR maps of selected Virgo cluster galaxies. The panels show a)  NGC\,4472 (M\,49), b) NGC\,4501 (M\,88) and c) NGC\,4321 (M\,100). The ellipse in the bottom left corner of the panels indicates the synthesized beam size of $9''\times5''$.}
\label{fig:examples}
\end{figure*}

To create the catalog of all Virgo cluster galaxies that are detected in the 144\,MHz LOFAR Virgo cluster survey, we considered all galaxies in the EVCC, which fully covers our LOFAR footprint. For the EVCC, a radial velocity cut of $v < 3000\mathrm{\,km\,s^{-1}}$ is employed to select possible cluster members, and a redshift-based infall model is used to identify certain cluster members \citep{Kim2014}. Of the 1589 galaxies, 991 are in the footprint of our LOFAR survey. The classical VCC includes a high number of background objects which do not fall into the radial velocity cut. Since the VCC is still used as a common reference for galaxies in the Virgo field, we also compiled a radio catalog of the background VCC-galaxies which are not part of the Virgo cluster.

Since the radio continuum emission of the galaxies is mostly faint, extended and often not well described by a Gaussian profile (see \autoref{fig:examples}), it is challenging to accurately identify all emission attributed to galaxies using an automatized source-finder. Thus, we measured the flux densities manually. We visually inspected the high- and low-resolution images of every galaxy in the EVCC and VCC with a $z$-band magnitude of 16 and below in the footprint of the LOFAR Virgo field. If emission is detected, we manually determine an elliptical region around the emission, if possible fully enclosing the $3\sigma$ contours in the low-resolution mosaic. In cases where radio morphology is particularly complex or background sources are super-imposed on the emission, we instead use a more complex polygon region to trace the emission. To check for misidentifications, the optical images of the DESI Legacy Imaging Survey DR9 \citep{Dey2019} were inspected together with the radio contours. 

We measure the integrated flux density in the regions, and the corresponding uncertainty $\sigma_S$ is calculated according to:
\begin{equation}
    \sigma_S = \sqrt{\sigma^2_\mathrm{rms} N_\mathrm{beam}+ \left(f_\mathrm{sys} S\right)^2},   
\end{equation}
where $f_\mathrm{sys} = 0.2$ is the systematic uncertainty of the flux density scale, $N_\mathrm{beam}$ the number of synthesized beams covering the region and $\sigma_\mathrm{rms}$ the local statistical uncertainty as measured from the RMS of the residual maps. 
We only consider sources with a statistical significance above $4\sigma$ in the low-resolution mosaic. This results in 112 Virgo cluster galaxies that are detected in the LOFAR maps out of the 991 EVCC galaxies in the survey footprint. The detection fraction strongly depends on the optical brightness; it is above 50\% for the 171 galaxies brighter than a z-band magnitude of 12. In contrast, none of the objects fainter than a z-band magnitude of 14.6 is radio-detected, although those constitute for more than half of the EVCC objects in the LOFAR footprint. The optically bright objects that are not detected in LOFAR are mostly ellipticals and lenticulars without strong AGN.
Additionally, we detect 114 background galaxies which are listed in the VCC. 

\begin{figure}
    \centering
    \includegraphics[width=1.0\linewidth]{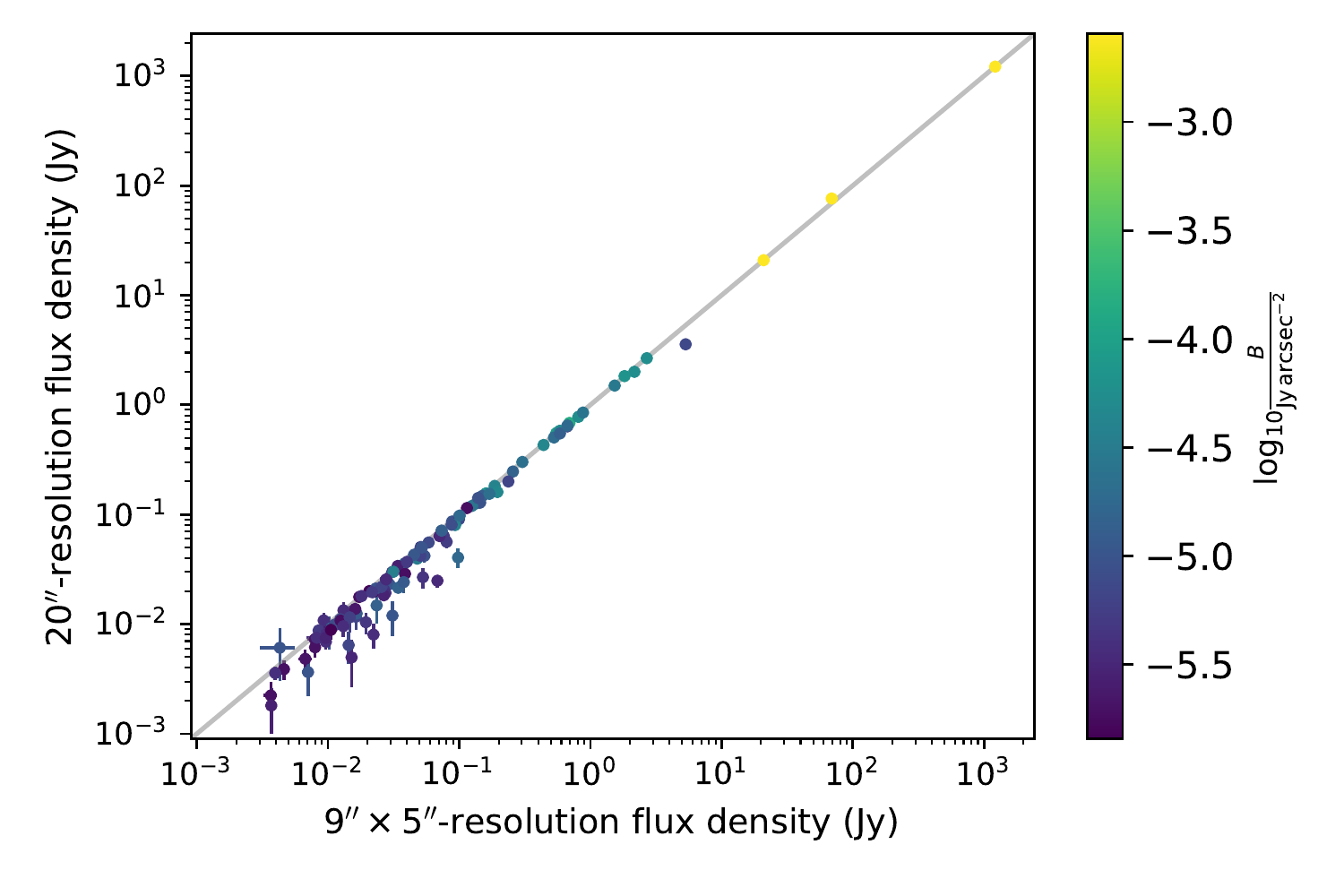}
    \caption{Flux densities of Virgo cluster members measured from the high-resolution ($9''\times 5''$, $x$-axis) and the low-resolution ($20''\times 20''$, $y$-axis) mosaic using identical regions. The color-scale corresponds to the logarithm of the average surface brightness.}
    \label{fig:hivslow}
\end{figure}

\begin{figure}
    \centering
    \includegraphics[width=1.0\linewidth]{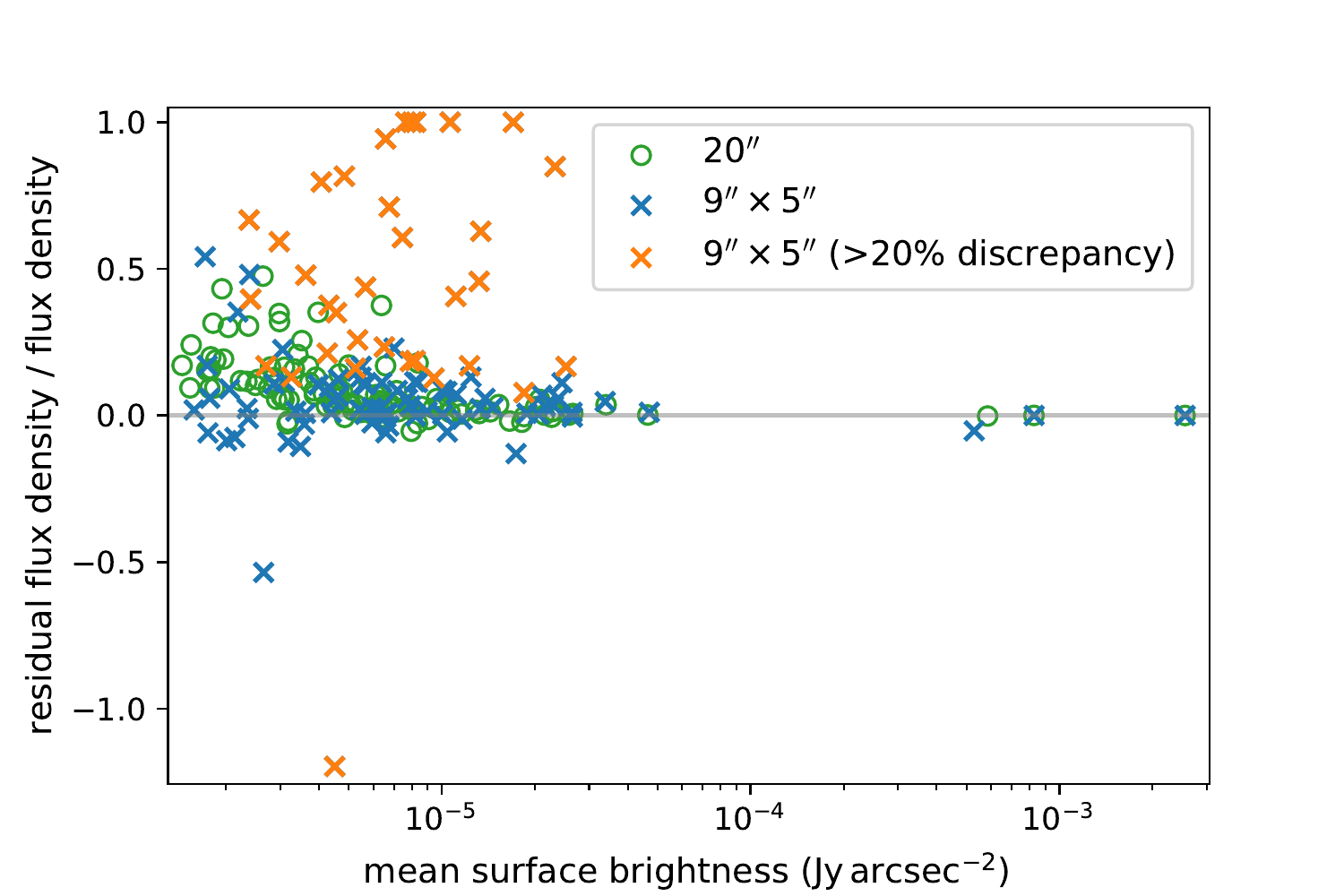}
    \caption{Ratio between the flux density measured from the residual and the restored images ($y$-axis) as a function of mean surface brightness ($x$-axis). Circles mark measurements from the low- and crosses from the high-resolution mosaic. For the orange crosses, the difference between the flux density in the low- and high-resolution measurements is above 20\%.}
    \label{fig:residual_flux_ratio}
\end{figure}

Since the galaxies are mostly low-surface brightness extended sources, particular care must be taken during the deconvolution. If sources are not fully deconvolved, emission coming from the side-lobes of the point-spread function may remain and their flux density will be overestimated.
This problem is mostly affecting the high-resolution mosaic, where the objects are more strongly resolved. To assess the completeness of the deconvolution, we compare the flux densities of the sources in the high- and low-resolution images using identical regions. This comparison is presented in \autoref{fig:hivslow}. 
Particularly for fainter sources, the flux density estimated from the high-resolution image is systematically above the measurement at low-resolution, even after we perform an additional round of deep deconvolution with manual clean-masks for all objects in the catalog (see \autoref{sec:ddcal}). 
A further measure for the completeness of the deconvolution is the integrated flux density measured from the residual images using the source regions --- if a source is not or barely deconvolved, the flux density in the residual image will be close or equal to the flux density in the restored image, while for a perfect deconvolution in the absence of noise and systematics, the integrated flux in the residual image should be zero. In \autoref{fig:residual_flux_ratio}, we show the flux density ratio between the residual and restored mosaics for the Virgo cluster galaxies as a function of the mean surface brightness and mark the galaxies where the difference between the high- and low-resolution images is larger than 20\%. At high-resolution, for 13 out of 112 galaxies the residual flux ratio is at least 50\%, and for all these cases but one, there is a difference above 20\% between the integrated flux densities at the different resolutions. Contrary, at low-resolution, the residual flux density ratio is always below 50\%. Thus, we decided to report the flux density measurements of the low-resolution images. 
Two galaxies in \autoref{fig:residual_flux_ratio} show a residual flux density ratio $< -0.5$. The orange cross is VCC\,144, which is located in a high noise region an not significantly detected in the high-resolution map, causing this fluctuation. The blue cross is VCC\,758, which is located directly next to an extended radio feature that we discuss in \autoref{sec:extended}. Around this feature, a bowl of negative surface brightness is present, caused by the incomplete deconvolution of the large-scale emission. 
\begin{small}
\begin{table*}[h]
\centering
\caption{LOFAR catalog of Virgo cluster galaxies.}\
\begin{tabular}{cccccccccccc}
\hline\hline
VCC & NGC & IC & RA J2000 & Dec J2000 & Type & $m_\mathrm{r}$ & $S$ & $A$ & $d$ & $L_{144}$ \\
 &  &  &  [h:m:s] &  [$\degree$:$'$:$''$] &  & [mag]  & [Jy] &  [$''\times''$] & [Mpc] & [W\,Hz$^{-1}$] \\
(1) & (2) & (3) & (4) & (5) & (6) & (7) & (8) & (9) & (10) & (11) \\
\hline
49 & 4168 &  & 12:12:17.2 & +13:12:19 & E & 11.00 & (2.5$\pm$0.5)e-2 & 4291 & 32.0 & (3.1$\pm$0.7)e+21 \\
66 & 4178 & 3042 & 12:12:46.1 & +10:51:55 & SBc & 11.24 & (5.6$\pm$1.2)e-2 & 14149 & 16.5 & (1.8$\pm$0.4)e+21 \\
73 & 4180 &  & 12:13:03.0 & +07:02:20 & Sb & 12.23 & (1.6$\pm$0.3)e-1 & 7735 & 32.0 & (2.0$\pm$0.4)e+22 \\
89 & 4189 & 3050 & 12:13:47.4 & +13:25:34 & SBc & 11.51 & (9.0$\pm$1.8)e-2 & 17831 & 32.0 & (1.1$\pm$0.2)e+22 \\
120 & 4197 &  & 12:14:38.6 & +05:48:23 & Sc & 12.65 & (7.1$\pm$1.5)e-2 & 8902 & 32.0 & (8.7$\pm$1.8)e+21 \\
\hline
\end{tabular}
\tablefoot{This table is available in its entirety at the CDS and contains the following information. The first three columns list the index of the galaxies in the VCC, NGC and IC. The fourth to seventh column lists the optical coordinates,  the morphological classification and the $r$-band magnitude of the galaxies as in \citet{Kim2014}. Column eight shows the integrated flux density as measured from the $20''$-resolution mosaics, column nine quotes the source area, the tenth the distance and the eleventh the 144\,MHz radio luminosity.}
\label{tab:lvcs}
\end{table*}
\end{small}
In \autoref{tab:lvcs}, we list the measured flux densities of the LOFAR-detected Virgo cluster member galaxies together with the position, morphological classification and $r$-band magnitude as provided in \cite{Kim2014}. Additionally, we specify the area of the region used for the flux-density measurements as well as redshift independent distance measurements. We assume the distances of the galaxies to be identical to the mean distance of the sub-structure they belong to. We follow the sub-structure distances and -membership-criteria defined in \citep{Boselli2014TheEvolution}, which report the clusters \emph{A, C} and the low-velocity cloud (LVC) at $d=17$\,Mpc, the cluster \emph{B} and $W^\prime$-cloud at 23\,Mpc and the $W$ and $M$-clouds at 32\,Mpc. However, to be consistent with the VESTIGE \citep{Boselli2018b} and NGVS \citep{Ferrarese2012} projects, we assume a distance of $d=16.5$\,Mpc instead of 17\,Mpc for clusters $A$, $B$ and the LVC. 
Based on these distances, we also calculated the radio luminosity for all galaxies in the catalog. The catalog is made available online at the CDS and together with cutout-images of all detected galaxies, also at the LOFAR Surveys web page\footnote{\url{https://lofar-surveys.org/virgo_survey.html}}. We report the LOFAR flux density measurements and the auxiliary data also for the LOFAR-detected VCC background galaxies, which can be found in \autoref{tab:lvcs_bg}. Furthermore, while not suited for the analysis of nearby galaxies, the full wide-field source catalog obtained with PyBDSF from the low-resolution mosaic is still highly valuable for studies of background objects. Thus, we also provide this catalog online.
\begin{small}
\begin{table*}[h]
\centering
\caption{LOFAR catalog of VCC galaxies in the background of the Virgo cluster.}
\begin{tabular}{ccccccccccc} 
\hline\hline
VCC & NGC & IC & RA J2000 & Dec J2000 & Type & $m_\mathrm{r}$ & $S$ & $A$ & $d$ &  $L_{144}$  \\
 &  &  & [h:m:s] &  [$\degree$:$'$:$''$] &  & [mag]  & [Jy] &  [$''\times''$] & [Mpc] & [W\,Hz$^{-1}$] \\
(1) & (2) & (3) & (4) & (5) & (6) & (7) & (8) & (9) & (10) & (11)\\
\hline
76 &  & 3046 & 12:13:07.8 & +12:55:05 & Sc(s)I & 14.16 & (1.3$\pm$0.3)e-2 & 3420 & 119 & (2.2$\pm$0.5)e+22 \\
121 &  &  & 12:14:42.5 & +12:59:24 & E7/S017 & 14.87 & (3.5$\pm$1.1)e-3 & 1463 & 267 & (3.0$\pm$0.9)e+22 \\
123 &  &  & 12:14:45.7 & +13:19:35 & Sa & 14.57 & (7.1$\pm$1.6)e-3 & 1346 & 306 & (8.0$\pm$1.8)e+22 \\
129 &  & 3060 & 12:15:02.0 & +12:32:49 & Sab & 14.39 & (4.9$\pm$1.5)e-3 & 2544 & 86 & (4.3$\pm$1.3)e+21 \\
134 &  & 3062 & 12:15:05.3 & +13:35:42 & ScI & 14.00 & (4.2$\pm$1.3)e-3 & 1852 & 116 & (6.8$\pm$2.1)e+21 \\
\hline
\end{tabular}
\tablefoot{This table is available in its entirety at the CDS and contains the following information. The first, second and third column list the indices of the galaxies in the VCC, NGC and IC. The fourth to seventh columns list the optical coordinates, the morphological classification and the $r$-band magnitude of the galaxies as reported in \citet{Binggeli1985}. Column eight shows the integrated flux density as measured from the low-resolution mosaics. The ninth column quotes the source area, the tenth column the radial-velocity inferred distance and the eleventh column the resulting radio luminosity.}
\label{tab:lvcs_bg} 
\end{table*}
\end{small}

\begin{figure}
    \centering
    \includegraphics[width=1.0\linewidth]{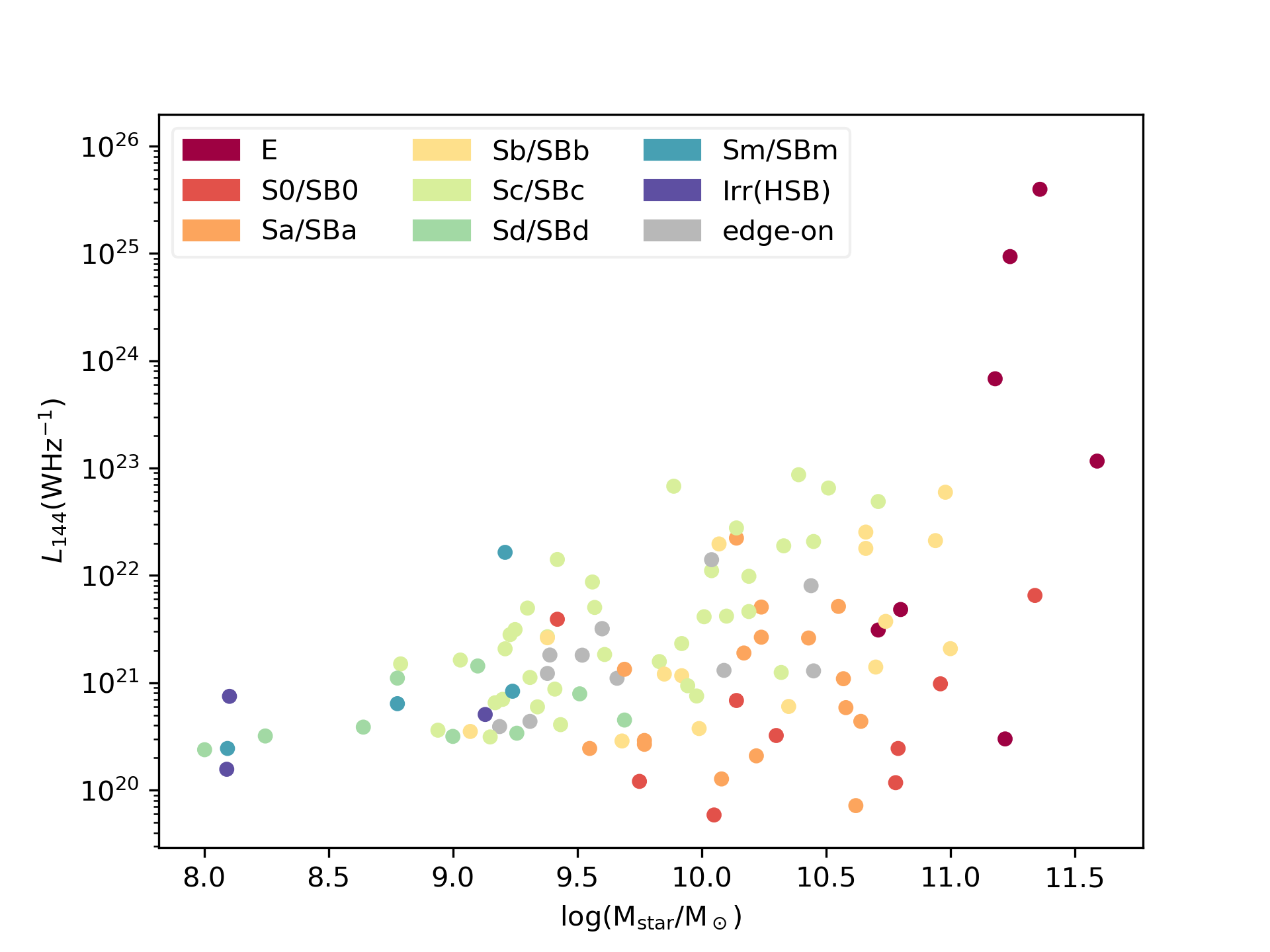}
    \caption{Stellar mass ($x$-axis) vs. 144\,MHz radio luminosity ($y$-axis) for galaxies in the LOFAR Virgo cluster catalog. The marker color corresponds to the Hubble-type.}
    \label{fig:population}
\end{figure}

The Hubble morphological classification is shown together with the stellar masses and radio-luminosities of the LOFAR Virgo cluster catalog galaxies in \autoref{fig:population}. Stellar masses were taken from \citet{Boselli2015HGalaxies}, or, if not available, from \citep{Boselli2014TheEvolution}. In both cases, the calibration of \citet{Zibetti2009}, based on the Chabrier initial-mass function \citep{Chabrier2003}, was used. 
The radio-brightest and most massive objects are radio galaxies in giant ellipticals. For spiral galaxies, later-type objects are on average brighter than earlier-type spirals of similar mass.

\section{Discussion}\label{sec:discussion}
A dedicated scientific analysis of the radio data presented here, including the study of the radio-SFR-relation and the ram-pressure stripped objects in Virgo, will be subject of a follow-up work by our team. However, we will present highlights and a brief initial discussion of radio sources that are noteworthy either due to their environmental interaction or because they otherwise belong to class of objects that is of particular interest in the radio continuum, such as giant ellipticals or dwarf galaxies.
See \autoref{sec:appendixlvcs} for an image atlas of all the galaxies.

\subsection{Galaxies} \label{sec:discuss_galaxies}

\subsubsection*{VCC 144}
VCC\,144 is discussed in \citet{Brosch1997VCCCluster} as a blue compact dwarf galaxy with an extended \hi{}-envelop which is currently undergoing an intense starburst. They name it as an example of galaxy formation currently taking place in the Southern extension of the Virgo cluster. It is the optically faintest galaxy in our radio catalog.  It is also part of the HERSCHEL far-infrared selected star-forming dwarf galaxy survey of \cite{Grossi2016Star-formingDust}

\subsubsection*{VCC 241 (IC 3105)}
This irregular galaxy shows low-surface brightness emission in our LOFAR maps. In the south of the galaxy, the radio emission appears to extend towards west. This feature is coincident with extra-planar \hii-regions traced by VESTIGE (Boselli priv. comm.). Thus, it is likely that the low-frequency radio continuum emission distribution traces cosmic ray electrons advected due to ram-pressure and/or tracing the star-formation in the stripped gas.

\subsubsection*{VCC 307 (NGC 4254, M 99)}
M\,99 is the radio-brightest late-type galaxy in the Virgo cluster and oriented close to face-on. The high star-formation rate, the disturbed morphology with a peculiar spiral arm in the west and the long \hi-tail \citep{Haynes2007NGCSurvey} is thought to originate from an encounter with another massive cluster member in the past \citep{Vollmer2005NGCCluster,Chyzy2007,Duc2008,Chyzy2008MagneticFields,Boselli2018}. However, even though the galaxy is located in the cluster outskirts at a distance of $\approx 1$\,Mpc to M\,87, ram pressure stripping also appears to play a role \citep{Murphy2009}. In the LOFAR images and previous low-frequency studies \citep{Kantharia2008Low4254}, the radio emission extends beyond the optical disk towards the north for at least 10\,kpc, this, together with the steep surface brightness gradient towards the south, is interpreted as a sign of interaction with the ICM while the galaxy is moving rapidly towards the cluster center.

\subsubsection*{VCC 345 (NGC 4261, 3C 270)} 
This bright Fanaroff-Riley type \Romannum{1} radio galaxy sits at the center of a poor group which is located $\sim12$\,Mpc behind the Virgo Cluster core in a structure know as the $W$-cloud \citep{deVaucouleurs1961StructureGalaxies.}. It is well studied at radio wavelengths \citep{Dunn2010TheGalaxies,Kolokythas2015NewObservations,Grossova2022VeryGalaxies} and shows clear signs of interaction between the AGN and the surrounding medium \citep{OSullivan2011InteractionGalaxies}.

\subsubsection*{VCC 596 (NGC 4321, M 100)}
M\,100 is another grand-design spiral galaxy in the Virgo cluster outskirts. In the high-resolution LOFAR map, the supernova SN1979C \citep{Urbanik1986ASpirals.} is visible in the southeast of the galaxy as a faint point-source with a background-subtracted flux density of $3.9\pm0.5$\,mJy. The galaxy does not show particularly asymmetric radio emission or other clear signs of perturbance, thus, it is currently likely not undergoing significant ram-pressure stripping.

\subsubsection*{VCC 630 (NGC 4330)}
The radio emission of this edge-on galaxy shows the presence of a tail in the south-west that is caused by ram-pressure stripping \citep{Murphy2009,Vollmer2012Ram4330} and also visible in the LOFAR maps. Tails in the same region are also present in atomic hydrogen \citep{Chung2007Tails,Chung2009} and H$\alpha$ \citep{Fossati2018,Sardaneta2022}.

\subsubsection*{VCC 664 (IC 3258)}
To our knowledge, we detect this late-type spiral galaxy for the first time in the radio continuum. It is among the optically faintest Virgo cluster members in our catalog. The $3\sigma$ contours in the low-resolution map are elongated towards the west. The orientation of the tail opposite to the cluster center and the lack of any associated stellar component in the optical image suggest that the galaxy is suffering a ram pressure stripping event. This could also explain the \hi{}-deficiency of the object \citep{Koppen2018RamApproach}.

\subsubsection*{VCC 699 (IC 3268)}
This galaxy is another Virgo star-forming dwarf galaxy from \citet{Grossi2016Star-formingDust} sample. The LOFAR detection is slightly offset from the optical image but towards the projected center of the Virgo cluster.  If real, this would 
imply an orbit moving away from the cluster center.  In view of the relative  small mass and very high star-formation rate, interpretation as an galactic outflow would be even more compelling.

\subsubsection*{VCC 763 (NGC 4374, M 84)}
The radio galaxy M\,84 is the second-brightest radio source in the central Virgo cluster after M\,87 and a well-studied example for AGN-feedback in non-central galaxies \citep{Finoguenov2001ITALChandra/ITALCluster,Birzan2004AGalaxies,Finoguenov2008,Laing2011DeepLobes,Bambic2023AGNRadius}. The radio lobes/tails are deflected after 5\,kpc, a clear sign of interaction with the ICM. They coincide with Chandra-detected X-ray cavities \citep{Finoguenov2001ITALChandra/ITALCluster,Finoguenov2008,Bambic2023AGNRadius}. In the 144\,MHz LOFAR images, the source is not significantly more extended than at GHz frequencies.

\subsubsection*{VCC 836 (NGC 4388)}
This object hosts a Seyfert 2-type nucleus and is the only spiral galaxy in the Virgo cluster where an AGN contributes a large fraction of the total radio emission. The sub-parsec nuclear jets are oriented perpendicular to the disk \citep{Giroletti2009THEVLBI} and give rise to a bi-conical outflow \citep{Damas-Segovia2016CHANG-ES4388} that is barely resolved even in our high-resolution maps. Furthermore, the galaxy, which is believed to be post-core passage and moving to the southwest, is undergoing a ram-pressure stripping event \citep{Murphy2009} which generates a spectacular \hi{}-tail \citep{Oosterloo2005ACluster} that can also be partly traced in H$\alpha$ \citep{Yoshida2002}. In the LOFAR maps, this imprints in a strong intensity gradient on the leading-side and an extension of the emission beyond the stellar component towards the trailing-side of the galaxy. Still, \citet{Damas-Segovia2016CHANG-ES4388} found that the leading-edge polarized outflow is able to resist the ram-pressure.

\subsubsection*{VCC 865 (NGC 4396)}
The presence of a tail in neutral hydrogen \citep{Chung2007Tails} is a sign of ram-pressure stripping also acting on this galaxy. However, a lack of compression in the \hi{}-contours \citep{Chung2009} and a radio-to-infrared deficit \citep{Murphy2009} on the leading edge make the interpretation somewhat unclear. In the low-resolution LOFAR map, the radio emission shows a strong gradient towards southeast with a tail-like extension in the opposite direction, thus, favoring the scenario that the galaxy suffering from ram-pressure stripping while falling into the cluster center. Within the optical disk, the LOFAR images reveal enhanced radio emission in the leading-half of the galaxy, coincident with a blue region of current or recent star-formation in the optical. This may be an example of ram-pressure induced star-formation due to gas compression on the leading edge \citep[e.g.][]{Boselli2021Astrophysics3476,Roberts2022LoTSSIC3949}

\subsubsection*{VCC 873 (NGC 4402)}
This edge-on spiral is located at a distance of $\approx 700$\,kpc from M\,87 and shows a truncated radio profile to the southeast with an extension in the opposite direction \citep[see also][]{Murphy2009}. Furthermore, signs of ram-pressure stripping are also visible in the \hi{} and the dust components of the galaxy \citep{Crowl2005Dense4402}. We speculate that in the LOFAR images, a low-surface brightness patch to the northwest of the galaxy might constitute part of a radio continuum tail of the galaxy, which was not observed previously.

\subsubsection*{VCC 881 (NGC 4406, M 86)} 
While close in projection, the massive elliptical galaxy and it's surrounding group, which is extended in the X-rays \citep{Bohringer1994TheImages}, likely lie around 2\,Mpc behind M\,87 \citep{Cantiello2018TheBeyond}. M\,86 is rapidly falling towards the cluster center from behind, as indicated by the blue-shifted spectrum. The galaxy was first discovered in the radio continuum at 4.9\,GHz as faint point source in \citet{Dunn2010TheGalaxies}, and recently also detected at 1.5\,GHz by \citet{Grossova2022VeryGalaxies}. In the latter work, it was concluded that the radio emission could be dominated by star-formation instead of nuclear activity; furthermore, no signs of X-ray cavities were observed.

\subsubsection*{VCC 1043 (NGC 4438)}
NGC\,4438 is a unique object in the Virgo cluster which shows strong disruption of the stellar component. This was caused by a gravitational interaction with NGC\,4435 and NGC\,4406 (M\,86), to which it is connected by filaments visible in H$\alpha$ emission \citep{Kenney2008AEllipticals}. NGC\,4438 shows peculiar radio emission - a central point source corresponds to the LINER-type AGN \citep{Decarli2007TheCluster} and unresolved inner lobes \citep{Hota2007NGCWavelengths}, accompanied by a radio-bubble in the west \citep{Wezgowiec2007TheSpirals,Vollmer2009Astrophysics4438}.

\subsubsection*{VCC 1226 (NGC 4472, M 49)} 
Being optically brighter than M\,87, M\,49 is the BCG of a southern sub-cluster falling into the Virgo core almost perpendicular to the line-of-sight \citep{Mei2007}. It lies at a projected distance of 1.3\,Mpc ($1.3 r_\mathrm{vir}$) from the cluster center. At GHz frequencies, it hosts a slightly extended, double-lobed radio source of low power \citep[e.g.][]{Dunn2010TheGalaxies}. X-ray observations with the XMM-Newton and Chandra satellites \citep{Kraft2011,Gendron-Marsolais2017Uplift4472,Su2019} revealed the presence of cavities in the thermal plasma, excavated by the AGN. These inner cavities connect to $\sim30\,\mathrm{kpc}$ long X-ray filaments. They are thought to originate from a previous nuclear outbreak, which can only be revealed in the radio by deep, low-frequency observations. With our LOFAR survey, we report the discovery of the radio tails which correspond to the X-ray features. The tails extend far beyond the X-ray cavities for a projected distance of 150\,kpc ($\widehat{=} 0.52\degree$) and are detected at a surface-brightness significance of $2 - 4 \sigma$ even in the high-resolution image (panel a) of \autoref{fig:examples}). Due to the ram pressure they experience from the Virgo ICM, they are bent southwards, giving rise to a wide-angle tail morphology. Follow-up studies of the spectral aging along the tails, using 54\,MHz observations of LOFAR LBA, will allow us to constrain both the duty cycle of the AGN and the infall-history of the M\,49 sub-cluster. 

\subsubsection*{VCC 1316 (NGC 4486, M 87)} 
M\,87 is the famous radio galaxy at the center of the Virgo cluster and among the brightest radio sources on the sky. It consists of an inner pair of jets which are also visible in the X-ray and optical and form a cocoon with an extension of $\sim6\,$kpc (see \autoref{fig:m87}). This emission is embedded in an highly extended halo with a size of $\sim75\,$kpc, which contains the outer jets with a prominent {smoking gun} morphology. The extended emission was studied in detail in \citet{Owen2000} at 300\,MHz using the VLA, which for the past 20 years remained the highest quality published image of the large-scale structure of M\,87. In the early commissioning stage of LOFAR, the source was studied from 20 to 160\,MHz, although with limited resolution and image fidelity owed to the incompleteness of the instrument and the lack of sophisticated calibration strategies \citep{deGasperin2012}. A high-fidelity LOFAR LBA image is provided in \citet{deGasperin2020ATeam}. M\,87 and the filaments permeating the extended halo of the source will be subject to a dedicated multi-frequency study combining our LOFAR data with unpublished VLA and MeerKAT observations (de Gasperin et al. in prep.).

\subsubsection*{VCC 1401 (NGC 4501, M 88)}
The radio map \autoref{fig:examples} b) of this bright, highly inclined spiral galaxy shows significant asymmetry with a strong gradient towards the southwest. This is caused by the ram-pressure stripping the galaxy experiences during the infall onto the central Virgo cluster. In our LOFAR 144\,MHz image, the radio emission in the northeast extends further from the optical disk than at 1.4\,GHz \citep{Vollmer2010}, which is in agreement with a larger age of the advected cosmic-ray electrons.  

\subsubsection*{VCC 1450 (IC 3476)}
\citet{Boselli2021Astrophysics3476} reported the presence of ionized gas tails for VCC\,1450, caused by a recent onset of ram-pressure. This is in agreement with the LOFAR images, where a strong gradient on the leading (eastern) side and a tail on the trailing side is visible. To the best of our knowledge, this is the first time this tail is reported in the radio continuum.

\subsubsection*{VCC 1516 (NGC 4522)}
This galaxy is observed at high inclination and shows strongly asymmetric radio emission with a tail towards the northwest \citep{Vollmer2004N4522,Murphy2009}. It is undergoing active ram-pressure stripping and also shows extra-planar  and UV-emission \citep{Kenney1999,Kenney2004,Vollmer201212Gals}.

\subsubsection*{VCC 1532 (IC 800)}
The LOFAR maps of VCC\,1532 reveal a gradient of the radio emission at the edge of the galaxy facing the cluster center, and a tail in the opposite direction. To our best knowledge, the galaxy was not previously reported to show signs of ram-pressure stripping. Thus, we note it as a new candidate for ongoing ram-pressure stripping. 

\subsubsection*{VCC 1575 (IC 3521)}
This is another Virgo star-forming dwarf galaxy from \citet{Grossi2016Star-formingDust} sample. The LOFAR emission appears to extent beyond the optical image of the galaxy, which implies an galactic outflow, consistent with the high star-formation rate of this low mass galaxy. 

\subsubsection*{VCC 1632 (NGC 4552, M 89)}
M\,89 hosts a nuclear point source surrounded by two lobes reminiscent of ears of 5 kpc extent each. Those ears correspond to X-ray cavities revealed by Chandra studies \citep{Machacek2006,Allen2006TheGalaxies,Kraft2017StrippedMedium}.

\subsubsection*{VCC 1686 (IC 3583)}
VCC 1686 has projected location very close to VCC 1690, but both galaxies differ by
more than 1300 km s$^{-1}$ in radial velocity, making a tidal interaction between the two galaxies very unlikely.  Still,  there is spur of LOFAR emission in the south of VCC\,1686 pointing towards VCC\,1690.  This feature has also a counterpart in 
at least two independent H$\alpha$ maps of VCC\,1686.  The origin of the H$\alpha$ feature is unclear, but the radio emission is more extended than the optical image of the galaxies over the whole quadrant, so maybe this is a large-scale outflow of VCC 1686. This would be consistent with the high star-formation rate of this dwarf galaxy. Note, that the two quite bright radio sources are likely background and not physically connected to VCC 1686. 
VCC 1686 also belongs to the \citet{Grossi2016Star-formingDust} sample.

\subsubsection*{VCC 1690 (NGC 4569, M 90)} 
M\,90 is one of the brightest spiral galaxy in the Virgo cluster. As traced by tails of ionized gas to the east \citep{Boselli2016Spectacular4569}, it is currently undergoing a ram-pressure stripping event. Observations in the radio continuum revealed symmetric bubbles extending up to 24\,kpc perpendicular to the disk \citep{ChyZy2006AstrophysicsWind}. Our LOFAR map detects both radio lobes and the southwestern ridge of the emission connecting to the star-forming disk that likely resulting from gas and magnetic field compression by the ambient ICM. Since M\,90 does not show signs of strong nuclear activity, it was concluded that the lobes are likely powered by a nuclear starburst event. Detailed analysis of the stellar emission in the innermost region by spectral synthesis by \citet{Gabel2002NGC4569} resulted in an age of the starburst of $5-6$\,Myr and $5 \times 10^4$ of O and B stars, sufficient to create a large scale outflow/wind. 

\subsubsection*{VCC 1727 (NGC 4579, M 58)} 
The massive barred spiral galaxy M\,58 host a Seyfert 2-type low-luminosity AGN \citep{Contini2004The4579}. Despite its significant \hi{}-deficiency, it was found to have the radio properties of a normal star-forming galaxy by \citep{Vollmer2013} even after subtracting the AGN contribution. In the LOFAR images, a central point source from which swirling features of \,5kpc scale extend can be observed. Since their direction is counterclockwise, opposite to the spiral arms, they most likely correspond to the AGN jets. The central structure is embedded in a low-surface brightness emission with a size similar to the optical disk which traces the star formation activity.

\subsubsection*{VCC 1791 (IC 3617)}
VCC 1791 also belongs to the \citet{Grossi2016Star-formingDust} sample. It has a relatively high starformation rate (but less than that of VCC 1686) but also a 
relatively low mass.  Therefore the extent of the radio emission beyond the optical 
body of the galaxy is suggestive of an outflow. The brightest peak of the radio 
emission is coinciting with three very blue compact knots.  

\subsubsection*{VCC 1932 (NGC 4634)}
The low-resolution LOFAR map of this edge-on spiral galaxy which forms a close pair with NGC\,4633 and shows extra-planar radio emission to the west, which could be interpreted as a radio tail. In the same direction, a star forming object that was likely created from material stripped (or tidally ejected) from VCC\,1932 is located \citep{Stein2018}. However, the galaxy is not known to show a tail at any other wavelength, thus, interpretation of the radio morphology remains somewhat inconclusive.

\subsubsection*{VCC 1972 (NGC 4647)}
The spiral galaxy VCC\,1972 lies close to M\,60 in an eastern sub-cluster of Virgo. The radio surface brightness and the molecular gas distribution \citep{YoungTHE4647} of this galaxy are asymmetric, likely due to ram-pressure exerted by the ICM around M\,60 and/or a disturbance of the gravitational well. 

\subsubsection*{VCC 1978 (NGC 4649, M 60)} 
This giant elliptical galaxy is the most massive galaxy in the Virgo Cluster and dominates a small, X-ray bright sub-cluster \citep{Bohringer1994TheImages}. Deep Chandra and VLA observations revealed the presence of X-ray cavities coincident with the radio jets of the central AGN \citep{Shurkin2007AGN-Induced4649,Dunn2010TheGalaxies,Paggi2014,Grossova2022VeryGalaxies}, placing it among the Virgo galaxies that are prime examples of AGN feedback in non-central galaxies. In the 144\,MHz LOFAR images, the inner radio source of 5\,kpc extension which is also visible at GHz-frequencies lies, is embedded in low-surface brightness diffuse emission of 75\,kpc size.

\subsection{Extended emission}\label{sec:extended}
\begin{figure*}
    \centering
    \includegraphics[width=1.0\textwidth]{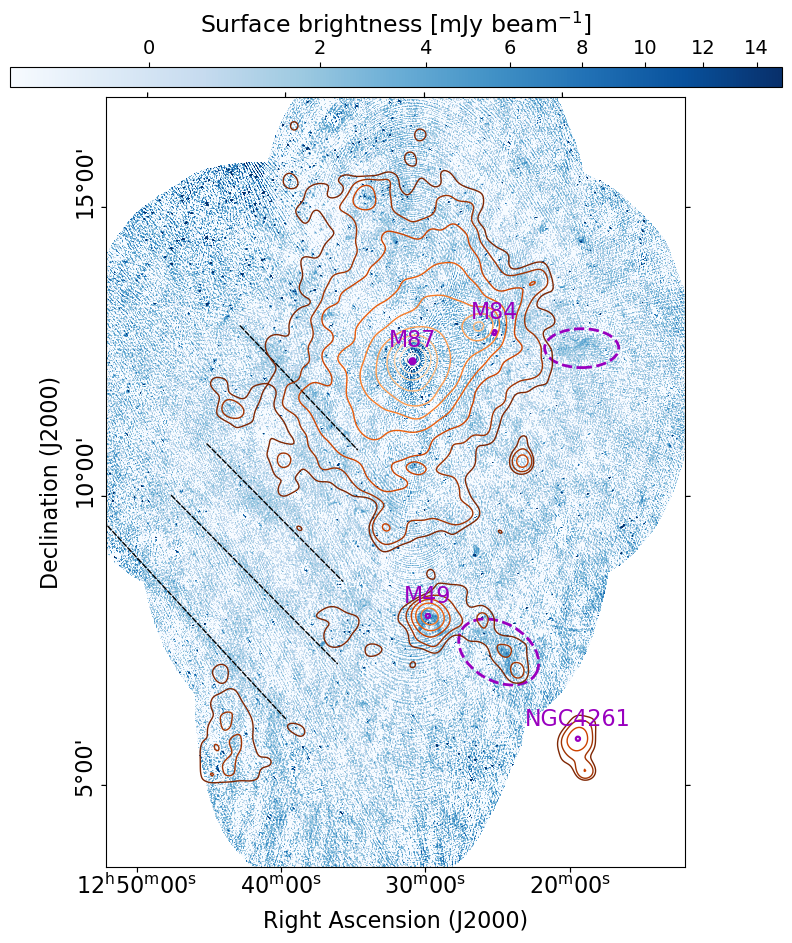}
    \caption{Compact source-subtracted mosaic at $1'$-resolution. Due to poor quality of the very low resolution images, field 1 and 8 are excluded. Black dashed lines correspond large-scale Galactic emission of the North Polar Spur (NPS), purple circles mark the position of the giant elliptical galaxies M\,49 (NGC\,4472), M\,84 (NGC\,4374) and M\,87 (NGC\,4486). The dashed purple ellipses highlight candidate extra-galactic emission, which could be related to the wider Virgo environment, and orange contours mark the eROSITA compact source-subtracted X-ray surface brightness (McCall et al.\ in prep.).}
    \label{fig:vlow}
\end{figure*}

\begin{figure}
    \centering
    \includegraphics[width=1.0\linewidth]{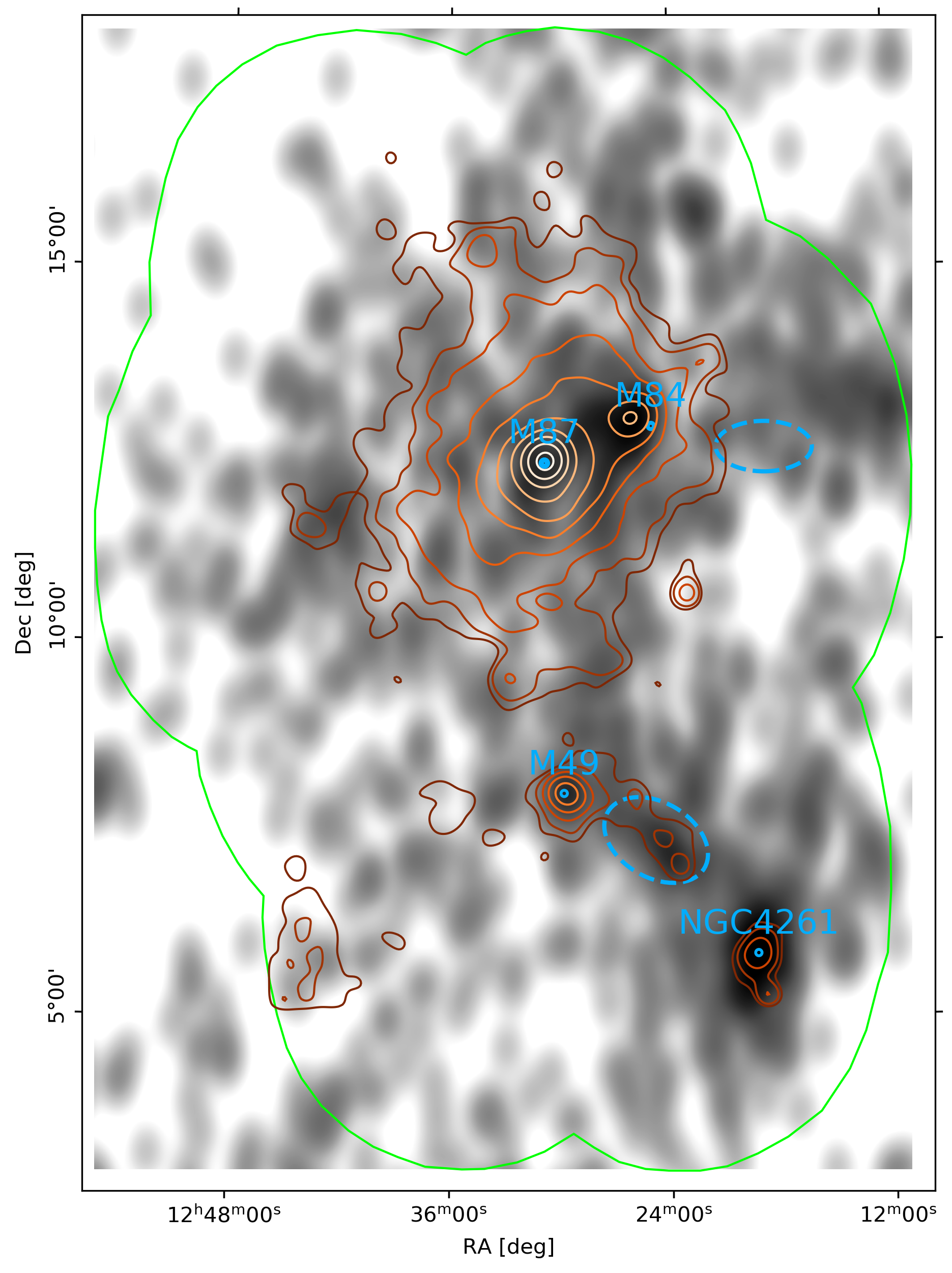}
    \caption{Projected galaxy density of the EVCC. Orange contours show the eROSITA source-subtracted X-ray surface brightness (McCall et al.\ in prep.), the green outline the footprint of the LOFAR observations and the blue circles and dotted ellipses mark individual galaxies and the extended emission as in \autoref{fig:vlow}.}
    \label{fig:vlow_density}
\end{figure}

For the following discussion of emission on large angular scales, it must be considered that radio interferometers have a reduced sensitivity to features beyond a certain extent, for LOFAR, the flux density loss at a scale of $18'$ is around $20\%$ \citep{Bruno2023}, and more extended sources will be attenuated more drastically.
In \autoref{fig:vlow}, we display the very low resolution source-subtracted mosaic which is sensitive to large-scale diffuse radio emission in the Virgo cluster environment. Due to poor quality of the subtracted images, fields 1 and 8 are excluded from the mosaic. Among the most prominent large-scale features are at least four sets of diagonal stripes that extend from northeast towards southwest and are marked by black dashed lines. Given their location and orientation, they are certainly associated with the North Polar Spur (Loop \Romannum{1}), a Galactic spherical structure which is debated to be either local \citep[$d \sim 100$\,pc][]{Salter1983Loop-IEnvironment} or originating from the Galactic center as indicated by studies of the coincident X-ray emission \citep[see also \autoref{fig:fields},][]{Sofue2000BIPOLARSIMULATIONS,Predehl2020DetectionHalo}. Further large-scale emission with an extent of $\approx1\degree$ is marked in \autoref{fig:vlow} by two dashed ellipses. 
The emission in the SW follows a similar orientation as the tails of M\,49, which might indicate that it is part of those. However, the emission does not show a clear connection to M\,49. Furthermore, if the emission were indeed due to a tail of this giant elliptical galaxy, it would follow a highly unexpected surface brightness trend with a complete fainting and a subsequent re-brightening. While this is not fully unheard of for cluster radio galaxies \citep{deGasperin2017MultifrequencySource,Cuciti2018,Edler2022}, it is a rare phenomenon. 
Thus, we consider it unlikely that the southwestern extended feature is causally connected the AGN in M\,49.
However, we note that directly coincident with the radio emission is the $W'$-cloud \citep{deVaucouleurs1961StructureGalaxies.}, a group of galaxies $\sim5.5$\,Mpc behind M\,49 that is located in a filament connecting the cluster to the background $W$-cloud \citep{Binggeli1993TheRevisited.,Mei2007}. In \autoref{fig:vlow_density}, we show that the elongated structure of the filament is apparent both in the galaxy density distribution and eROSITA (McCall et al.\ in prep.) X-ray surface brightness distribution and has a similar morphology and location as the radio emission. The displayed galaxy density was obtained from the EVCC by smoothing the galaxy distribution with a Gaussian kernel of a width $\sigma=14'$. All these aspects combined point towards the idea that the emission originates from phenomena related to the ICM/Intra-group medium, either due to turbulent re-acceleration processes or accretion shocks. This could explain the similar morphology of the X-ray and radio signal.
An alternative explanation is that the radio emission traces a past phase of nuclear activity of a galaxy in the group, such as the dominant elliptical NGC\,4365. However, at present, NGC\,4365 is not associated with a compact radio source. The only radio-detected galaxy nearby is NGC\,4370, although there is no clear connection to the diffuse emission.  The extended radio source coincident with the $W'$-group/filament will be subject of a forthcoming multi-frequency follow-up study.
The second extended and elongated feature in the northeast is of similar size, but does not directly coincide with an over-density of galaxies in the EVCC or with an increase in surface brightness in the ROSAT \citep{Bohringer1994TheImages} or eROSITA (McCall et al.\ in prep.) X-ray images. However, it is located between the Virgo core and the $M$-cloud of galaxies, another concentration of galaxies in the wider Virgo environment \citep{deVaucouleurs1961StructureGalaxies.,Mei2007}.

The three small circles in \autoref{fig:vlow} mark the location of extended emission possibly associated with giant elliptical galaxies in the cluster. The extended tails of M\,49 are discussed in \autoref{sec:discuss_galaxies}. Around M\,87 there is extended emission on a scale of $\sim 1^\circ$ (four times larger than M\,87). Given the cool-core nature of the Virgo cluster and the extent of $200-300$\,kpc of the candidate diffuse emission, a tentative possibility is that this emission is caused by a radio mini-halo \citep{Gitti2004,Giacintucci2017OccurrenceClusters,Giacintucci2019ExpandingClusters,vanWeeren2019} caused by the sloshing of gas in the Virgo-core \citep{Gatuzz2018MeasuringCluster}. However, due to the presence of strong systematics directly next to M\,87, we cannot conclude with certainty that the emission is physical.
Around the bright (21\,Jy) radio galaxy M\,84, a circular halo of emission with an embedded negative hole is most likely a calibration- and imaging artifact.

\section{Summary} \label{sec:conclusion}
In this work, we presented the LOFAR HBA Virgo cluster survey, which is the first data release of the ViCTORIA project and represents the deepest published wide-field radio survey of the Virgo cluster field. This advance was made possible by the general progress in low-frequency radio calibration techniques as well as the development of a specifically tailored subtraction procedure which mitigates the dynamic range limitations due to the extremely bright source M\,87 and is introduced in this work. Within the virial radius of the cluster, where we have increased exposure time and pointing overlap, we reach a median noise level of $140\,\mathrm{{\mu}Jy\,beam^{-1}}$ at a resolution of $9''\times5''$, while across the full survey area, the median noise is twice as high. 
We use this data to create a catalog of the radio properties of 112 LOFAR-detected certain and possible Virgo cluster galaxies ($v_{\mathrm{rad}}<3000\,\mathrm{km\,s^{-1}}$). The detected objects include at least 18 cases of galaxies exhibiting a radio-morphology indicative of ongoing ram-pressure stripping. Of those, we report VCC\,664 (IC\,3258), VCC\,1532 (IC\,800) and VCC\,1932 (NGC\,4632) as new ram-pressure stripping candidates.
Further, for the giant elliptical galaxy VCC\,1226 (NGC\,4472, M\,49), we revealed the presence of old radio tails with an extent of $0.5\degree$/150\,kpc. Due to the interaction with the ICM, the tails are bend towards south and assume a wide-angle tail morphology. 
The image cutouts and the catalog of the Virgo galaxies as well as the full mosaics are made available online\footnote{\url{https://lofar-surveys.org/virgo_survey.html}}. 

We also investigated the presence of large-scale diffuse emission in the Virgo cluster. While no radio emission attributable to the radio halo or -relic phenomena is found, extended emission coincident with the $W'$-group in a filament between Virgo and the background $W$-cloud is detected.  The scale of this feature is $\approx 1\deg$ and emission with a similar extent and orientation is also present in the eROSITA/SRG X-ray map of the cluster. Thus, we speculate that this radio emission may be caused by accretion-processes due to shocks or turbulence.

This work is the first part of ViCTORIA, a project aiming to drastically improve the multi-frequency radio coverage of the Virgo cluster. Further planned radio surveys are being conducted at 54\,MHz with the LOFAR low-band antenna system as well as in the L-band using MeerKAT, this also includes the 21\,cm-line.  
In a forthcoming work of the ViCTORIA-project, we will use the LOFAR data presented here to analyze the impact of the cluster environment on the evolution of star-forming galaxies in Virgo. Further, we will employ the multi-frequency data that will be provided by ViCTORIA for detailed spectral studies on the interacting radio tail we unveiled for VCC\,1226 and the extended emission coincident with the $W'$-filament.

\begin{acknowledgements}

HE acknowledges support by the Deutsche Forschungsgemeinschaft (DFG, German Research Foundation) under project number 427771150. MB acknowledges support from the Deutsche Forschungsgemeinschaft under Germany's Excellence Strategy - EXC 2121 "Quantum Universe" - 390833306.
AI acknowledges financial support from the European Research Council (ERC) programme (grant agreement No. 833824). DJB acknowledges funding from the German Science Foundation DFG, via the Collaborative Research Center SFB1491 ‘Cosmic Interacting Matters - From Source to Signal'. AI acknowledges the INAF founding program 'Ricerca Fondamentale 2022' (PI A. Ignesti).  RJvW acknowledges support from the ERC Starting Grant ClusterWeb 804208. E.B. acknowledges financial support from the European Research Council (ERC) Consolidator Grant under the European Union’s Horizon 2020 research and innovation programme (grant agreement CoG DarkQuest No 101002585). 
LOFAR is the Low Frequency Array designed and constructed by ASTRON. It has observing, data processing, and data storage facilities in several countries, which are owned by various parties (each with their own funding sources), and which are collectively operated by the ILT foundation under a joint scientific policy. The ILT resources have benefited from the following recent major funding sources: CNRS-INSU, Observatoire de Paris and Université d’Orléans, France; BMBF, MIWF-NRW, MPG, Germany; Science Foundation Ireland (SFI), Department of Business, Enterprise and Innovation (DBEI), Ireland; NWO, The Netherlands; The Science and Technology Facilities Council, UK; Ministry of Science and Higher Education, Poland; The Istituto Nazionale di Astrofisica (INAF), Italy. This research made use of the Dutch national e-infrastructure with support of the SURF Cooperative (e-infra 180169) and NWO (grant 2019.056).
The Jülich LOFAR Long Term Archive and the German LOFAR network are both coordinated and operated by the Jülich Supercomputing Centre (JSC), and computing resources on the supercomputer JUWELS at JSC were provided by the Gauss Centre for Supercomputing e.V. (grant CHTB00) through the John von Neumann Institute for Computing (NIC). This research made use of the University of Hertfordshire high-performance computing facility and the LOFAR-UK computing facility located at the University of Hertfordshire and supported by STFC [ST/P000096/1], and of the Italian LOFAR IT computing infrastructure supported and operated by INAF, and by the Physics Department of Turin university (under an agreement with Consorzio Interuniversitario per la Fisica Spaziale) at the C3S Supercomputing Centre, Italy. The data are published via the SURF Data Repository service which is supported by the EU funded DICE project (H2020-INFRAEOSC-2018-2020 under Grant Agreement no. 101017207).
This work is based on data from eROSITA, the soft X-ray instrument aboard SRG, a joint Russian-German science mission supported by the Russian Space Agency (Roskosmos), in the interests of the Russian Academy of Sciences represented by its Space Research Institute (IKI), and the Deutsches Zentrum f{\"{u}}r Luft und Raumfahrt (DLR). The SRG spacecraft was built by Lavochkin Association (NPOL) and its subcontractors and is operated by NPOL with support from the Max Planck Institute for Extraterrestrial Physics (MPE).
The development and construction of the eROSITA X-ray instrument was led by MPE, with contributions from the Dr. Karl Remeis Observatory Bamberg \& ECAP (FAU Erlangen-Nuernberg), the University of Hamburg Observatory, the Leibniz Institute for Astrophysics Potsdam (AIP), and the Institute for Astronomy and Astrophysics of the University of T{\"{u}}bingen, with the support of DLR and the Max Planck Society. The Argelander Institute for Astronomy of the University of Bonn and the Ludwig Maximilians Universit{\"{a}}t Munich also participated in the science preparation for eROSITA.
The eROSITA data shown here were processed using the {\tt eSASS} software system developed by the German eROSITA consortium.
\end{acknowledgements}

\begin{appendix}

\onecolumn
\section{Images of Virgo Cluster Galaxies}
\label{sec:appendixlvcs}
\FloatBarrier

\vspace{-0.5cm}
\begin{figure}[h]
    \centering
    \begin{subfigure}[b]{\textwidth}
        \includegraphics[width=\textwidth]{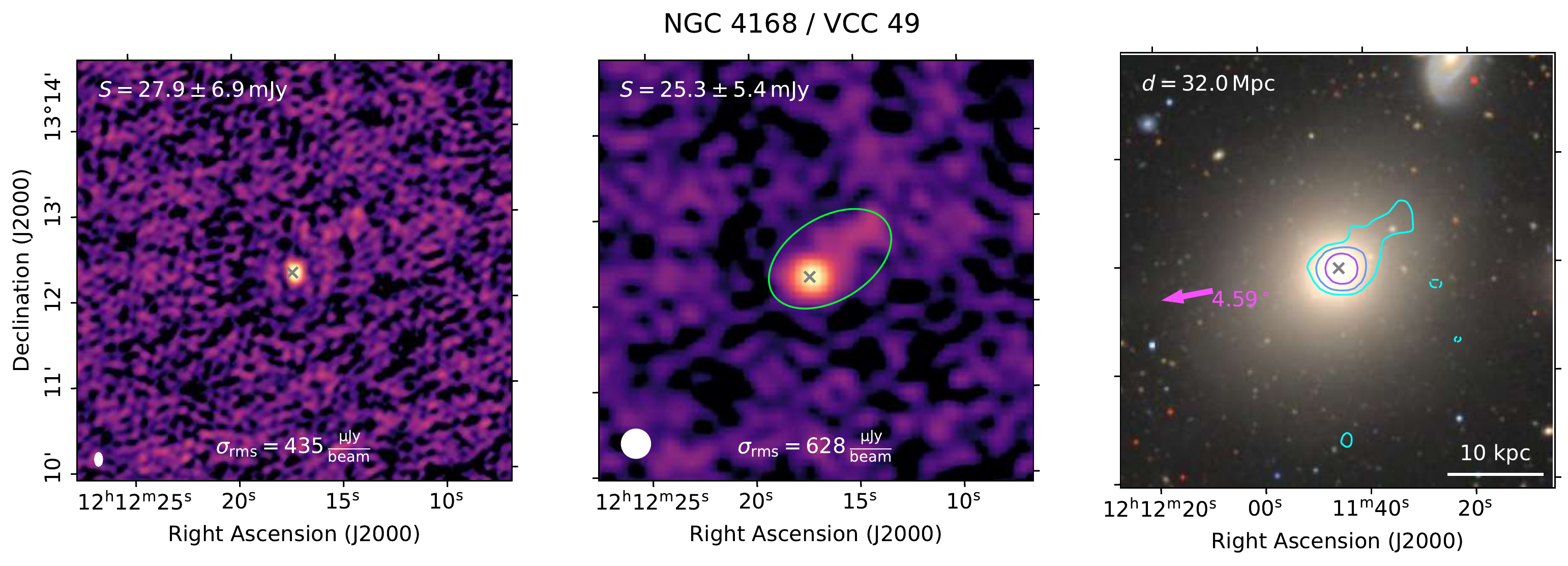}
        \caption{}
    \end{subfigure}
     \hfill
    \begin{subfigure}[b]{\textwidth}
        \includegraphics[width=\textwidth]{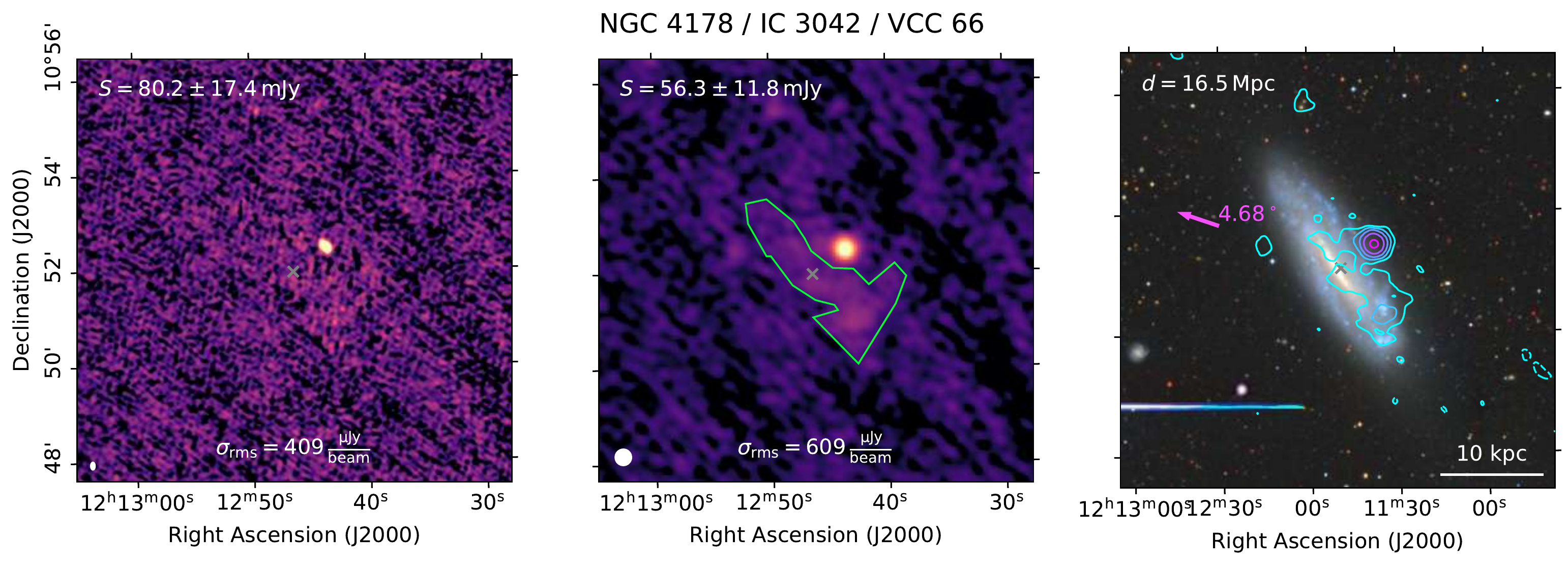}
        \caption{}
    \end{subfigure} 
    \caption{LOFAR images of the galaxy VCC\,144 at $9''\times 5''$ resolution (left panel), $20''$ resolution (center panel) and the corresponding optical image of the DESI Legacy Imaging Survey DR9 \citep{Dey2019}  with the $20''$ LOFAR contours starting from $3\sigma$ and increasing in powers of two (right panel). The region outlined in green in the central panel marks the area used for the flux density measurement at high- and low-resolution, the measured flux density is displayed at the top left and the background RMS $\sigma_\mathit{rms}$ of the maps at the bottom. In the right panel, the pink arrow marks the direction of and distance to the cluster center (M\,87), and the redshift-independent distance $d$ is reported in the top left.}
    \label{fig:144first}
\end{figure}

\begin{figure}
    \centering
    \begin{subfigure}[b]{\textwidth}
        \includegraphics[width=\textwidth]{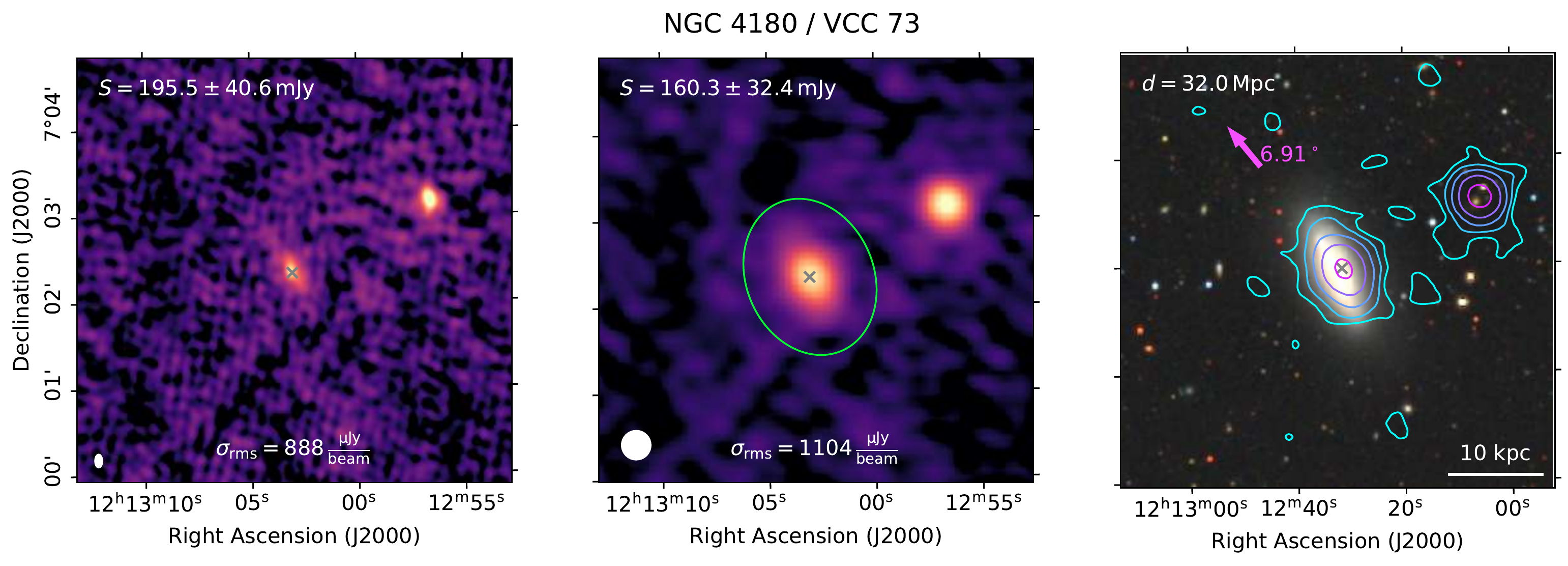}
        \caption{}
    \end{subfigure}
     \hfill
    \begin{subfigure}[b]{\textwidth}
        \includegraphics[width=\textwidth]{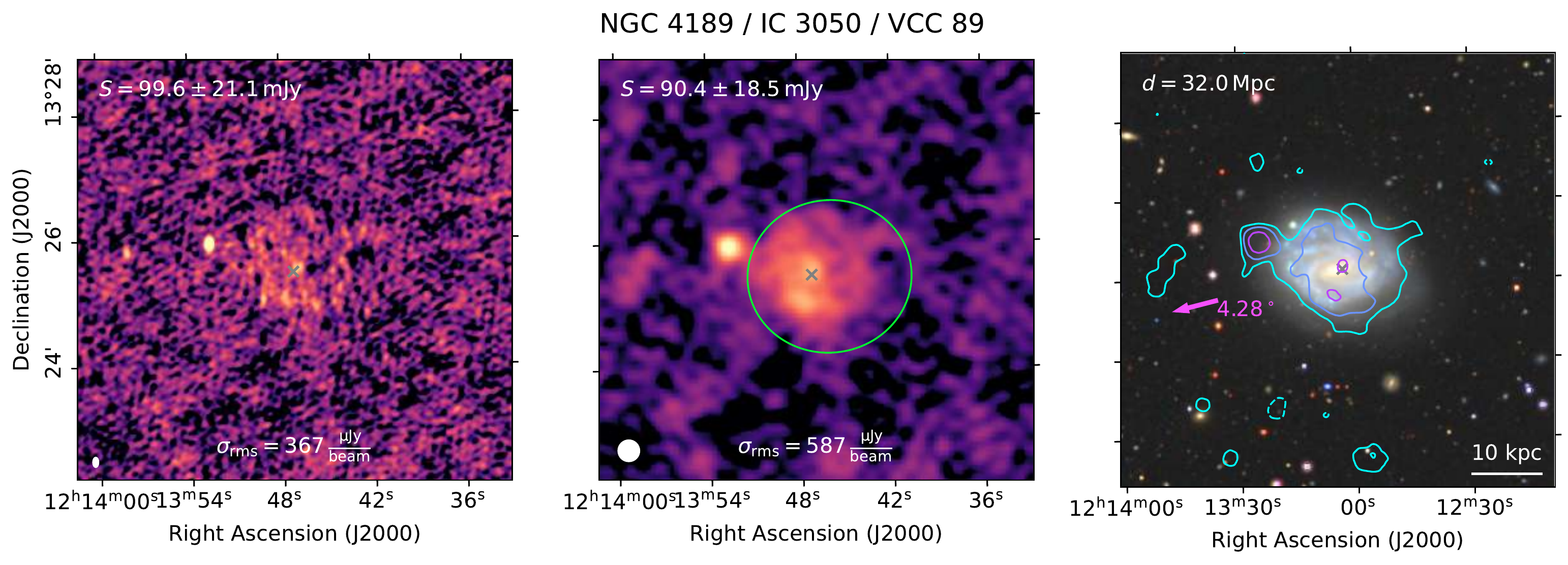}
        \caption{}
    \end{subfigure} 
     \hfill
    \begin{subfigure}[b]{\textwidth}
        \includegraphics[width=\textwidth]{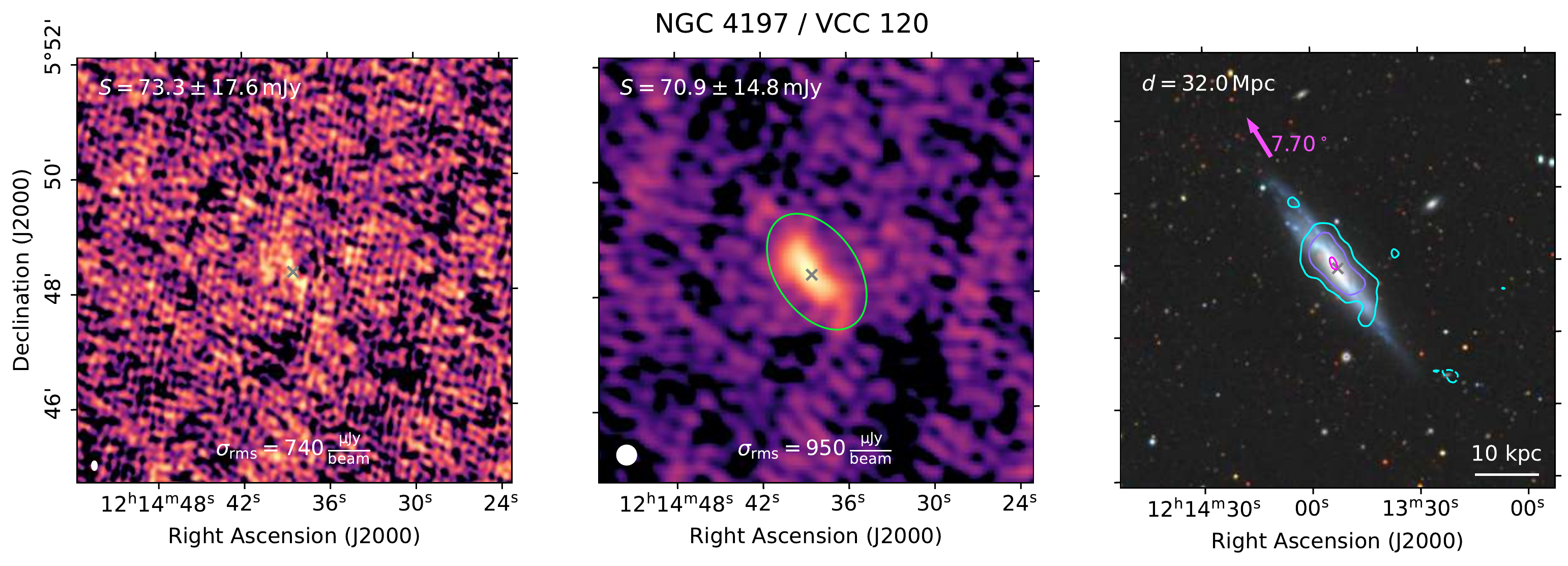}
        \caption{}
    \end{subfigure} 
    \caption{Same as \autoref{fig:144first}.}
\end{figure}

\begin{figure}
    \centering
    \begin{subfigure}[b]{\textwidth}
        \includegraphics[width=\textwidth]{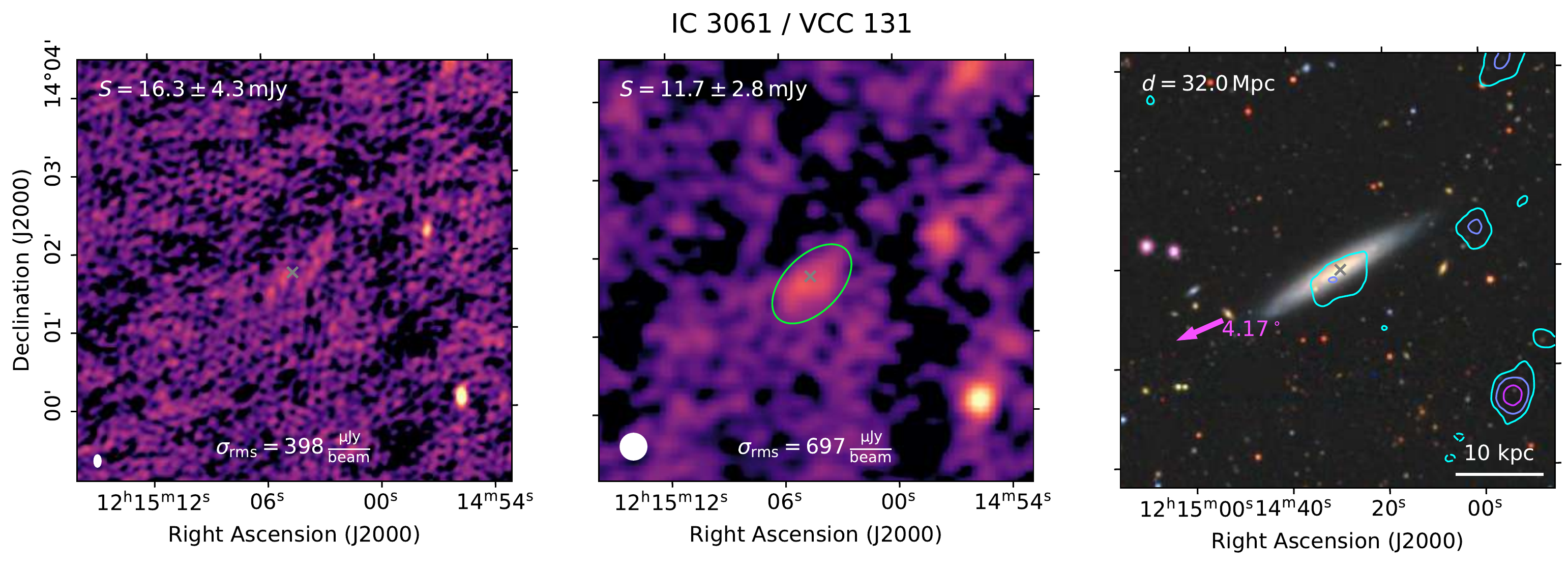}
        \caption{}
    \end{subfigure}
     \hfill
    \begin{subfigure}[b]{\textwidth}
        \includegraphics[width=\textwidth]{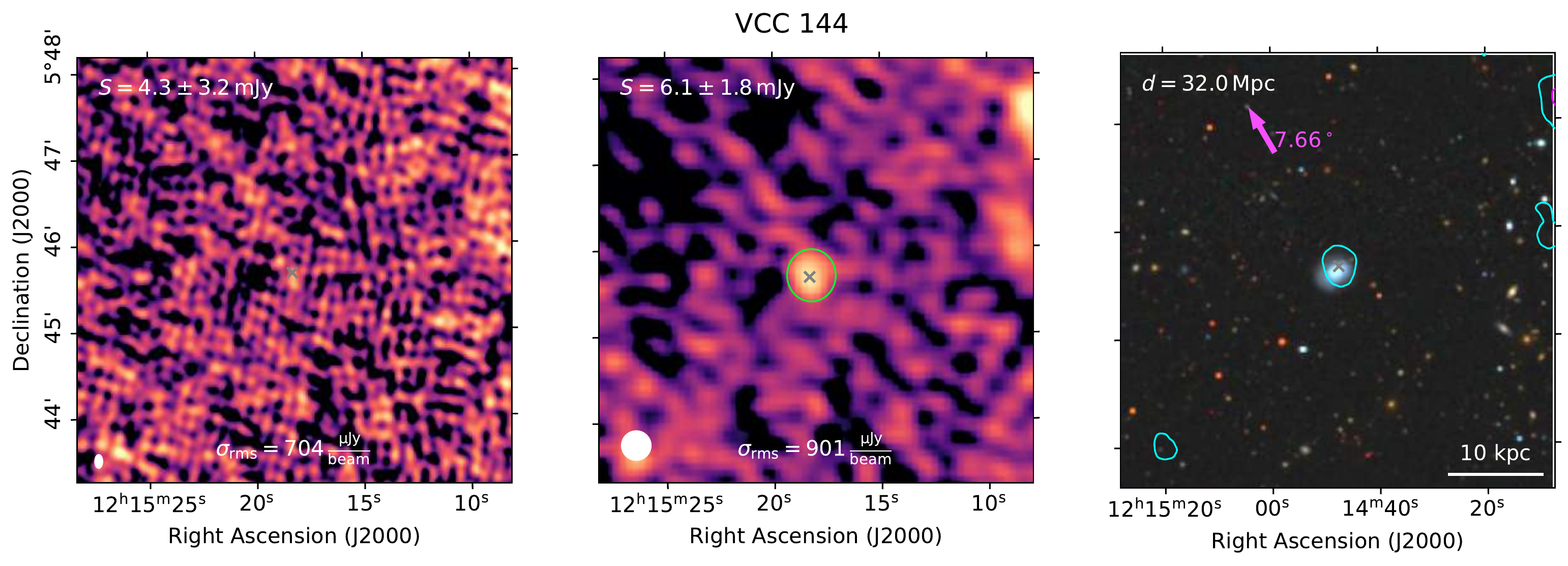}
        \caption{}
    \end{subfigure} 
     \hfill
    \begin{subfigure}[b]{\textwidth}
        \includegraphics[width=\textwidth]{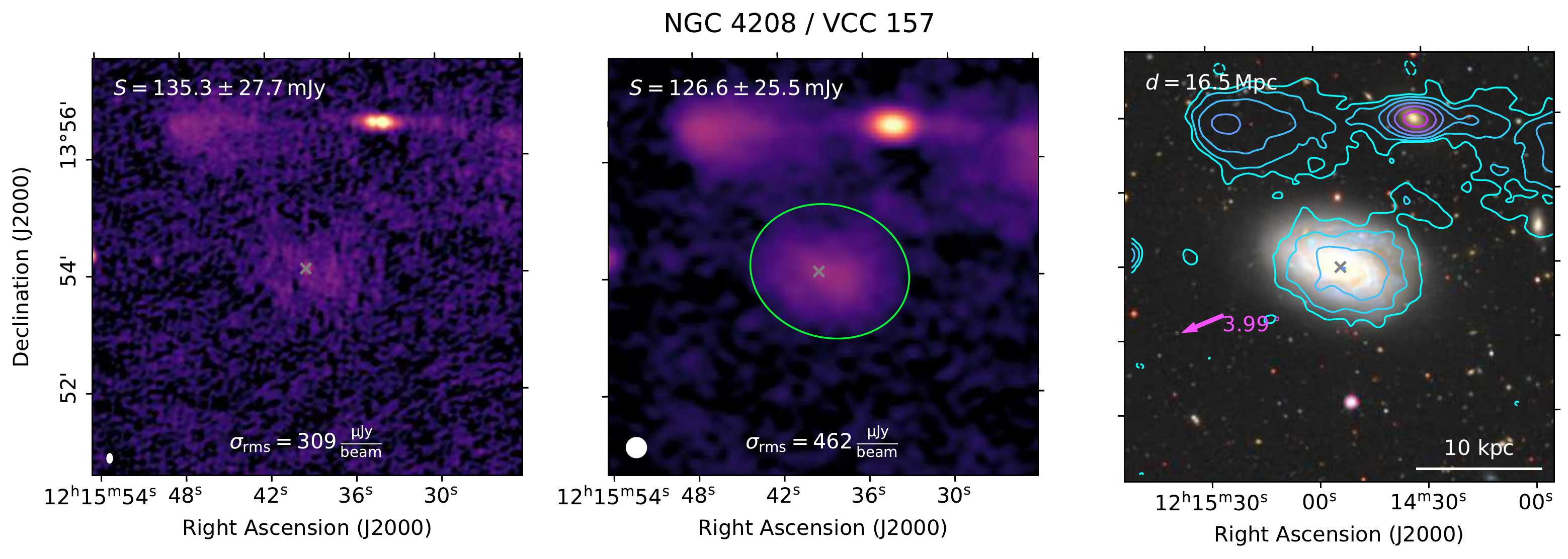}
        \caption{}
    \end{subfigure} 
    \caption{Same as \autoref{fig:144first}.}
\end{figure}

\begin{figure}
    \centering
    \begin{subfigure}[b]{\textwidth}
        \includegraphics[width=\textwidth]{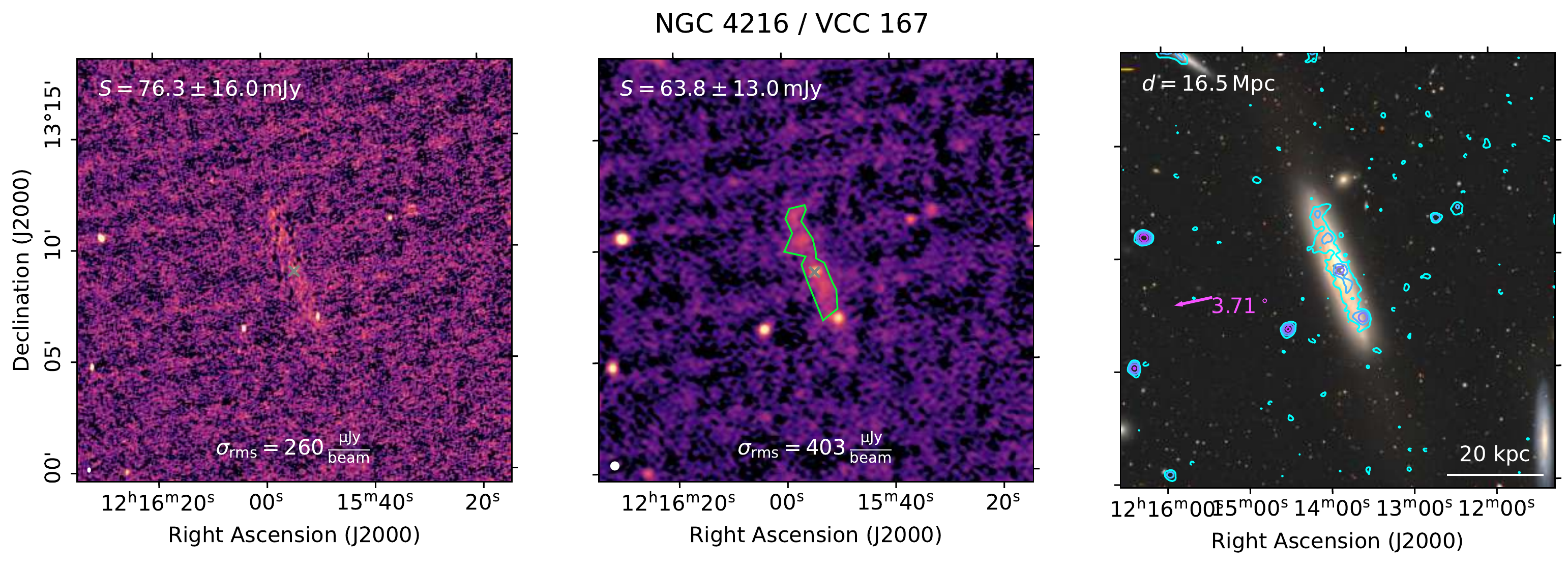}
        \caption{}
    \end{subfigure}
     \hfill
    \begin{subfigure}[b]{\textwidth}
        \includegraphics[width=\textwidth]{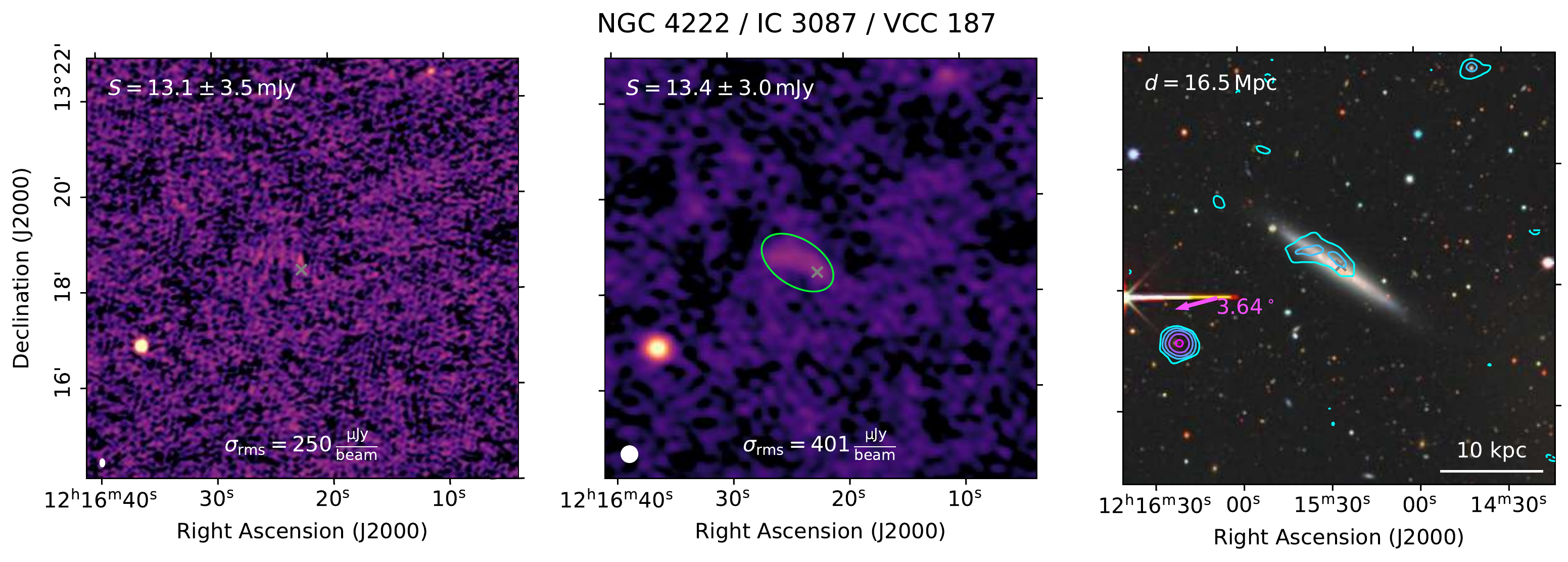}
        \caption{}
    \end{subfigure} 
     \hfill
    \begin{subfigure}[b]{\textwidth}
        \includegraphics[width=\textwidth]{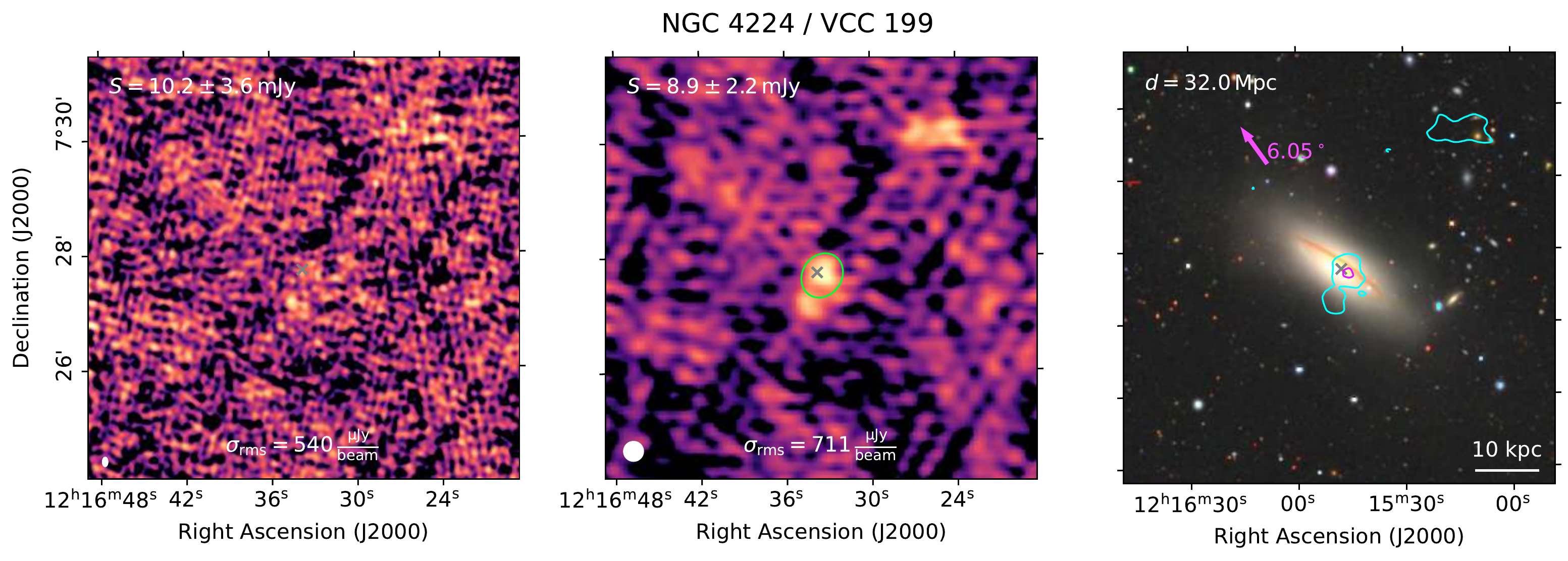}
        \caption{}
    \end{subfigure} 
    \caption{Same as \autoref{fig:144first}.}
\end{figure}

\begin{figure}
    \centering
    \begin{subfigure}[b]{\textwidth}
        \includegraphics[width=\textwidth]{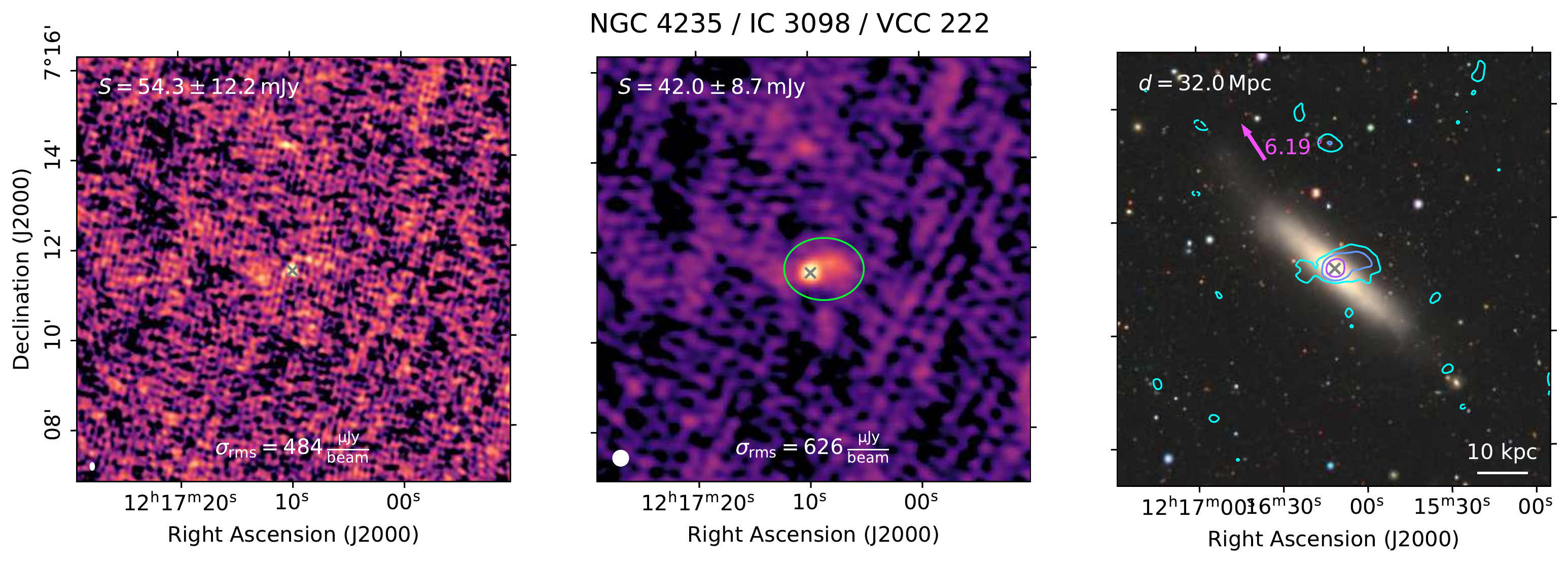}
        \caption{}
    \end{subfigure}
     \hfill
    \begin{subfigure}[b]{\textwidth}
        \includegraphics[width=\textwidth]{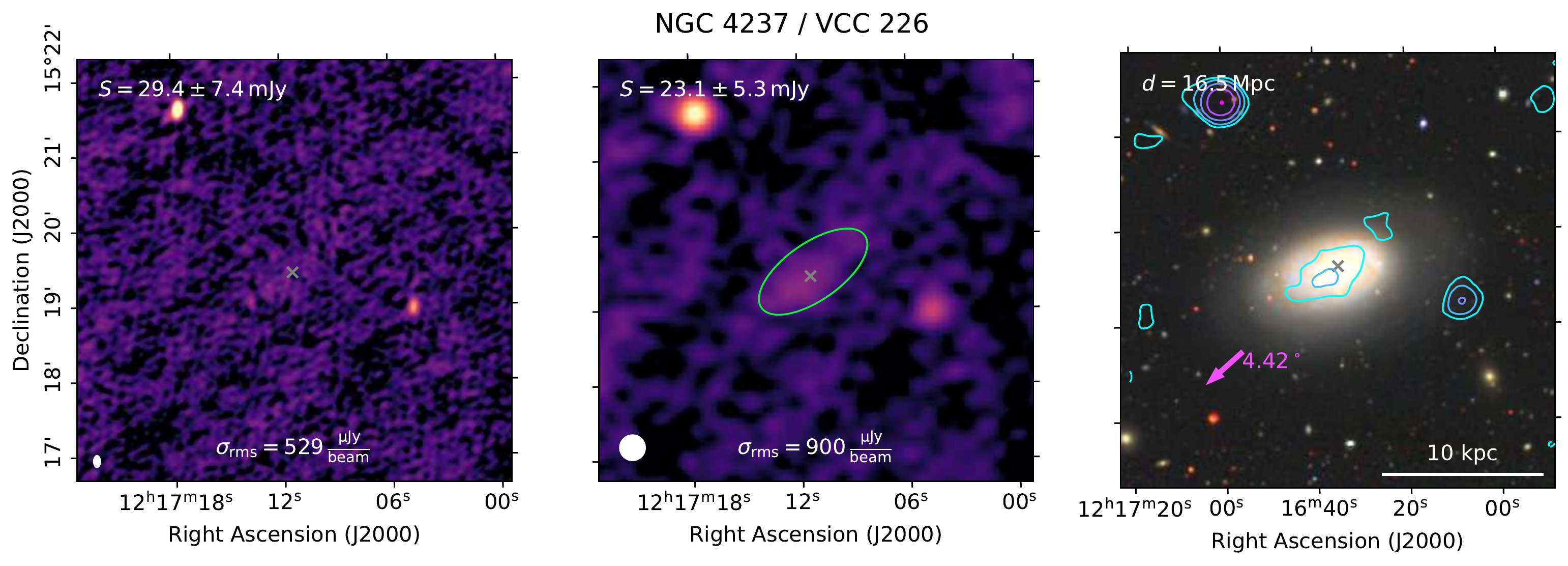}
        \caption{}
    \end{subfigure} 
     \hfill
    \begin{subfigure}[b]{\textwidth}
        \includegraphics[width=\textwidth]{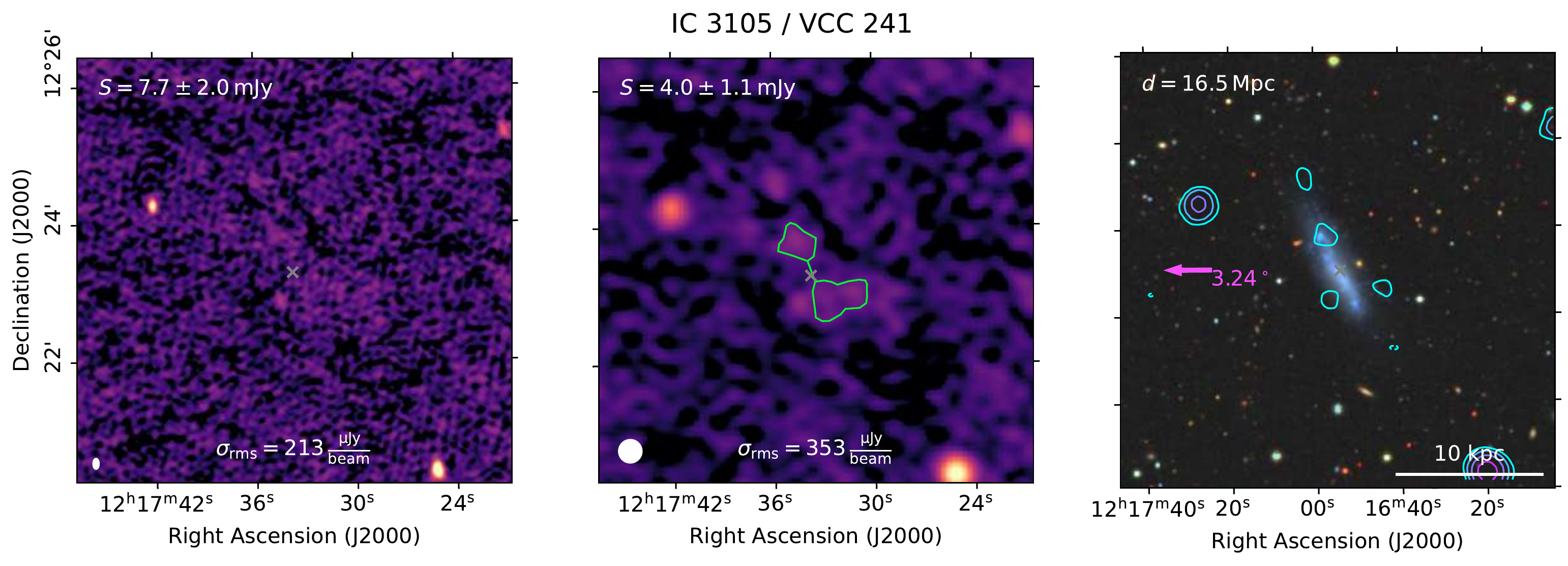}
        \caption{}
    \end{subfigure} 
    \caption{Same as \autoref{fig:144first}.}
\end{figure}

\begin{figure}
    \centering
    \begin{subfigure}[b]{\textwidth}
        \includegraphics[width=\textwidth]{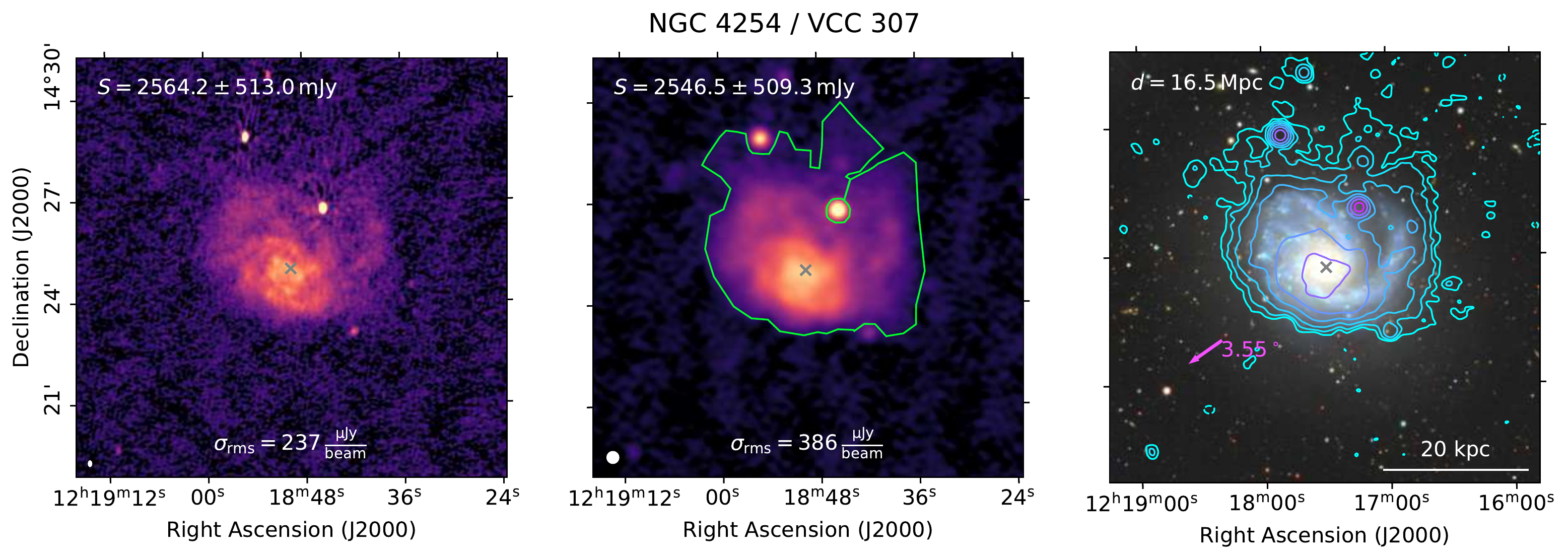}
        \caption{}
        \label{fig:307}
    \end{subfigure}
     \hfill
    \begin{subfigure}[b]{\textwidth}
        \includegraphics[width=\textwidth]{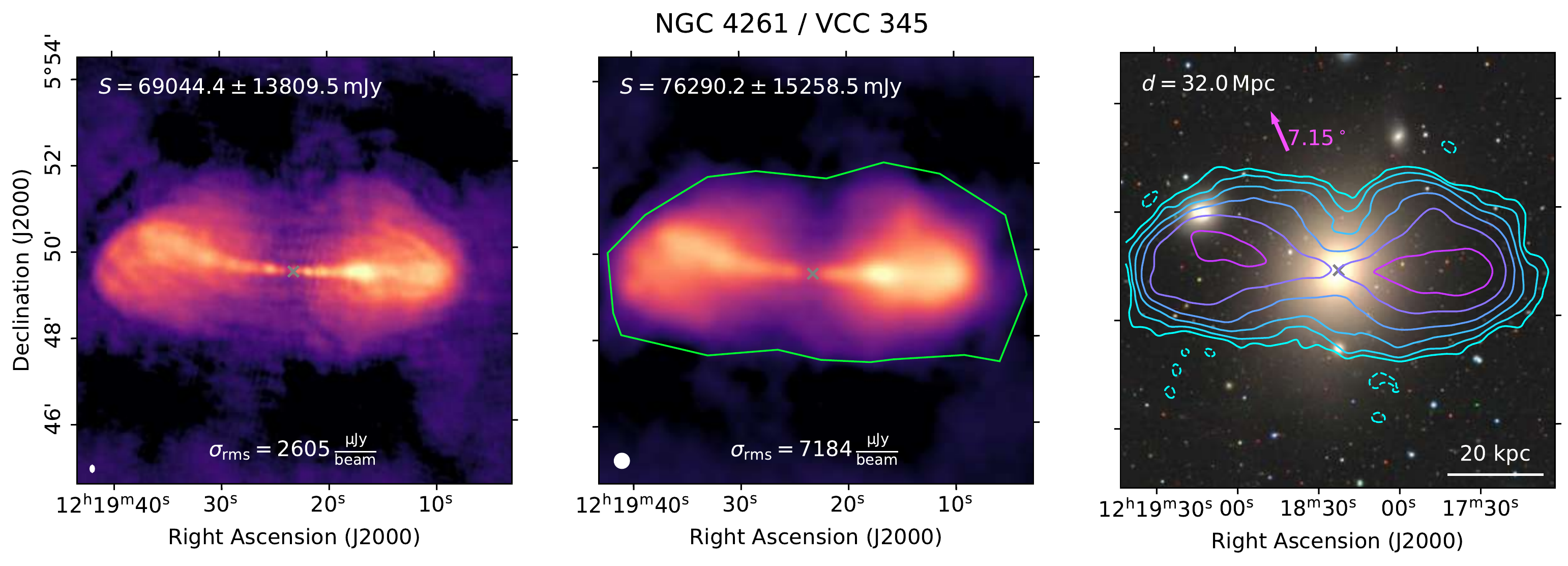}
        \caption{}
        \label{fig:345}
    \end{subfigure} 
     \hfill
    \begin{subfigure}[b]{\textwidth}
        \includegraphics[width=\textwidth]{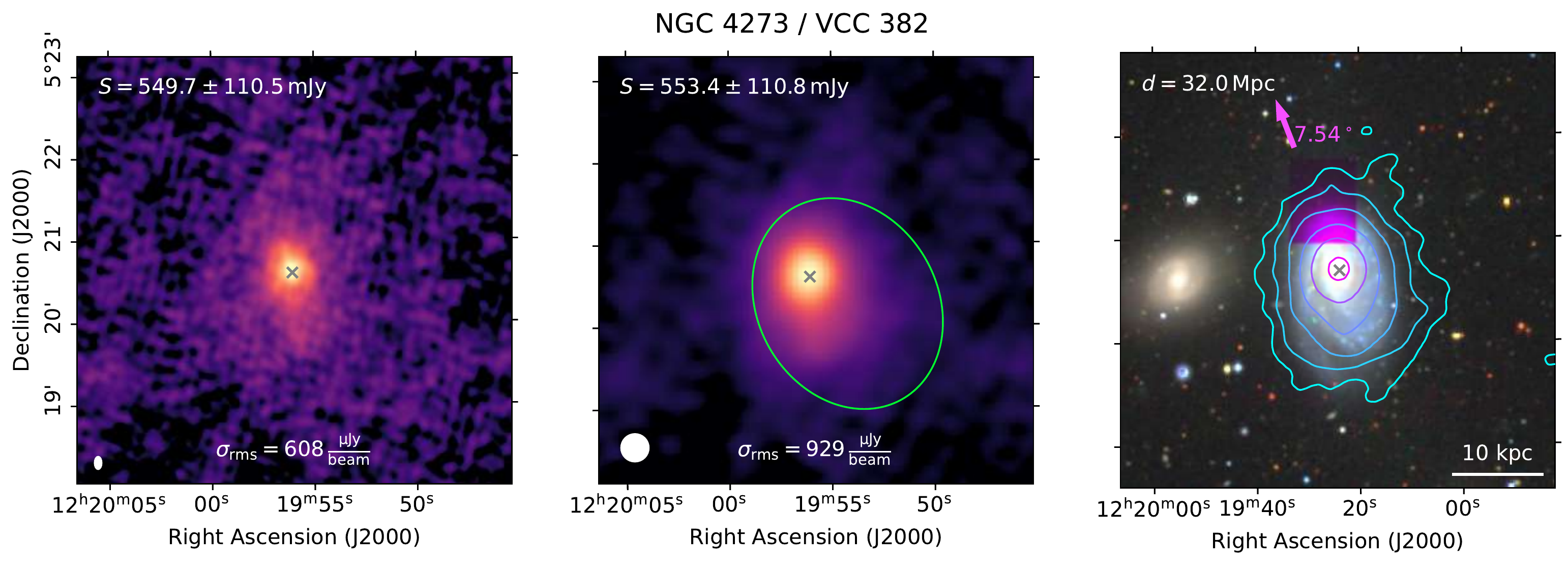}
        \caption{}
    \end{subfigure} 
    \caption{Same as \autoref{fig:144first}.}
    \end{figure}

\begin{figure}
    \centering
    \begin{subfigure}[b]{\textwidth}
        \includegraphics[width=\textwidth]{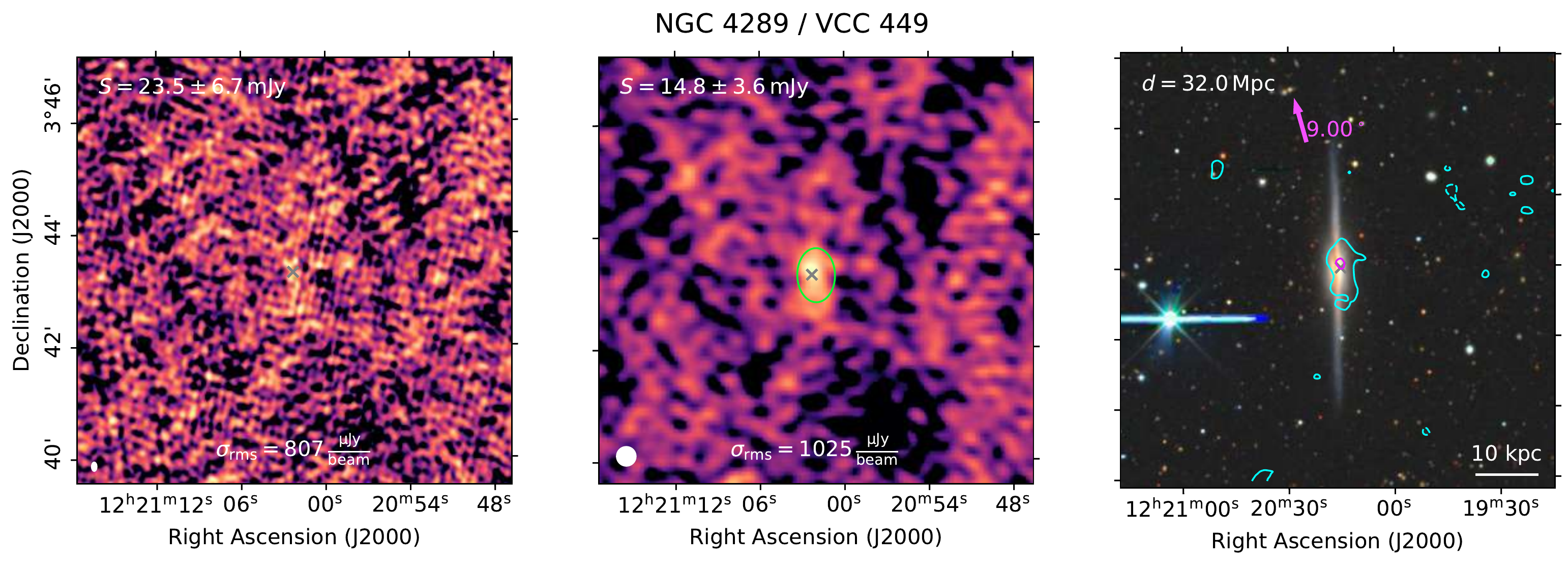}
        \caption{}
    \end{subfigure}
     \hfill
    \begin{subfigure}[b]{\textwidth}
        \includegraphics[width=\textwidth]{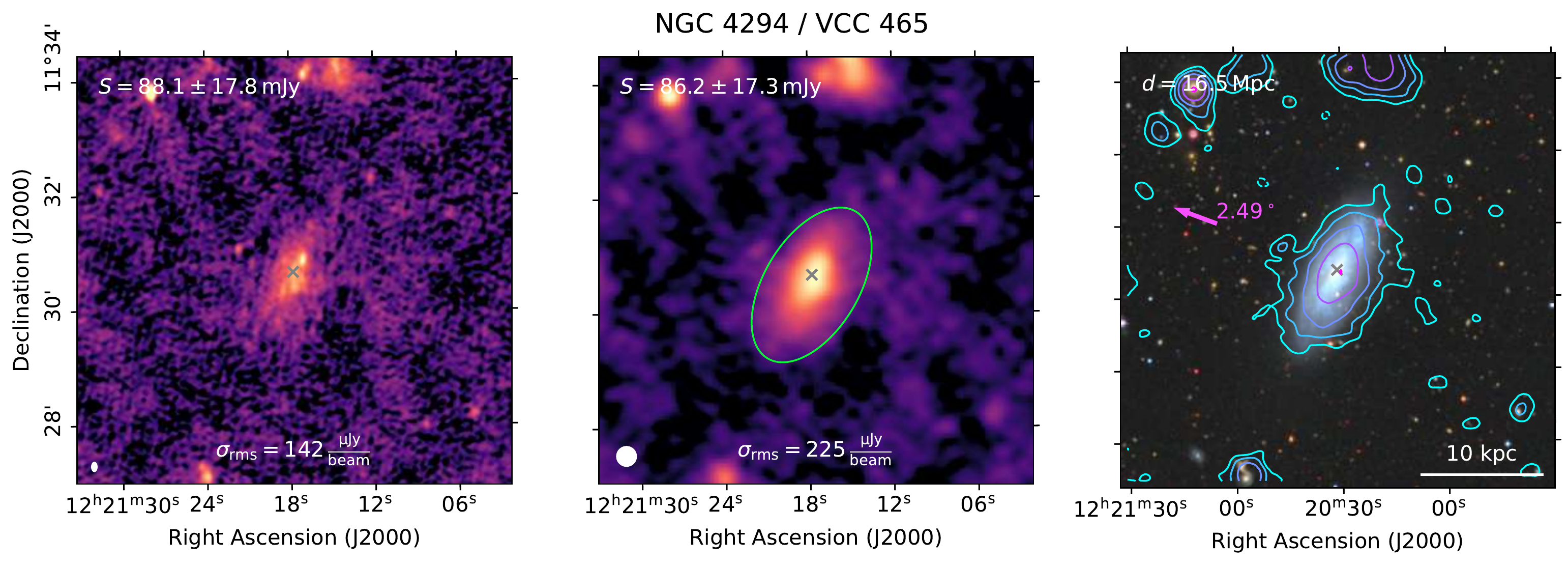}
        \caption{}
    \end{subfigure} 
     \hfill
    \begin{subfigure}[b]{\textwidth}
        \includegraphics[width=\textwidth]{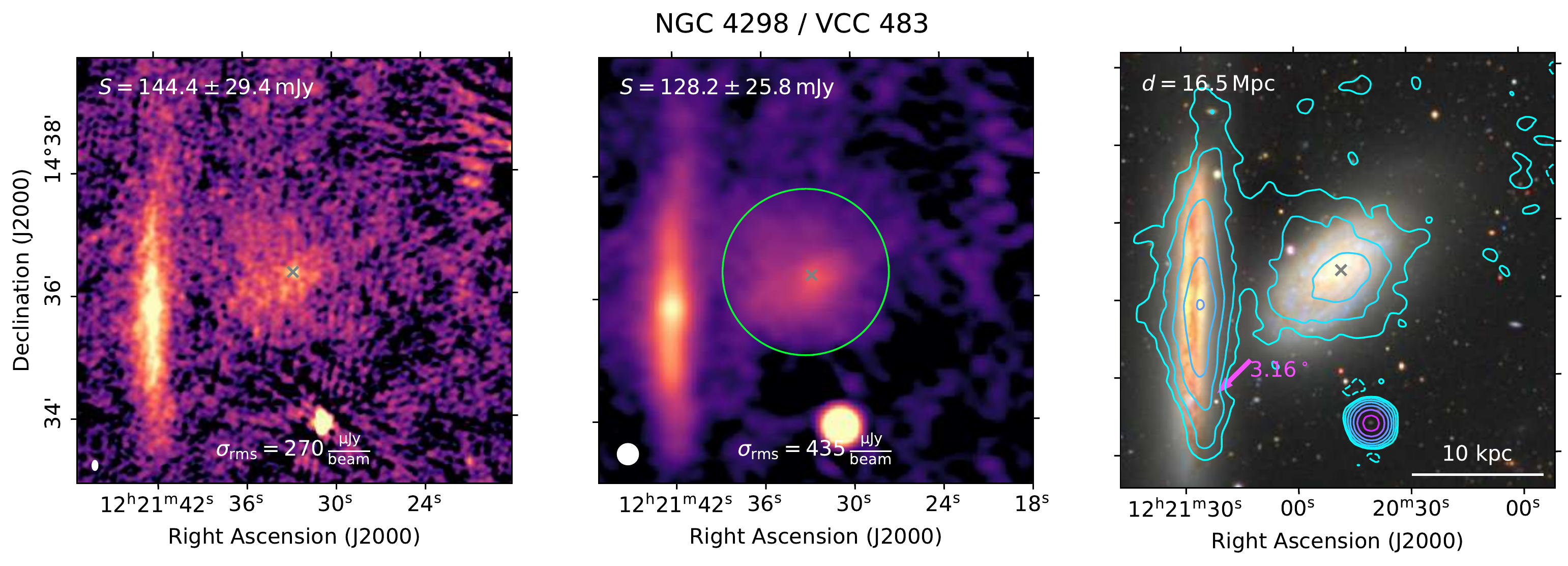}
        \caption{}
    \end{subfigure} 
    \caption{Same as \autoref{fig:144first}.}
\end{figure}

\begin{figure}
    \centering
    \begin{subfigure}[b]{\textwidth}
        \includegraphics[width=\textwidth]{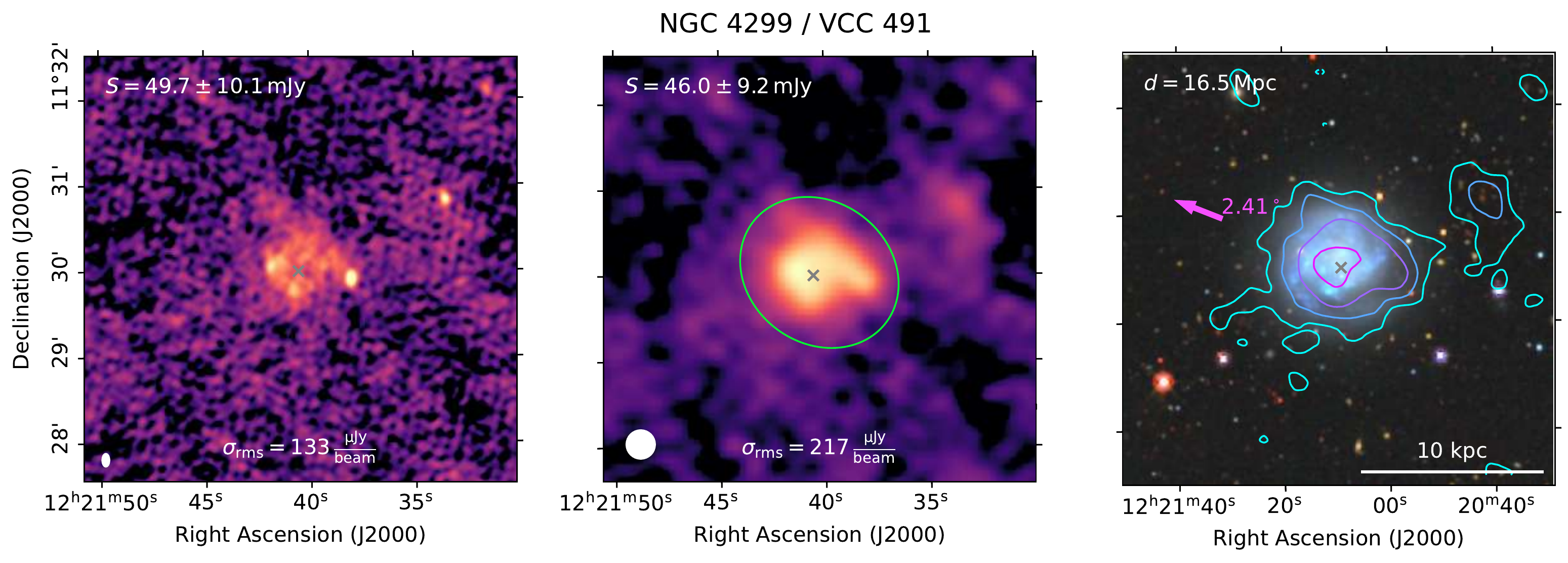}
        \caption{}
    \end{subfigure}
     \hfill
    \begin{subfigure}[b]{\textwidth}
        \includegraphics[width=\textwidth]{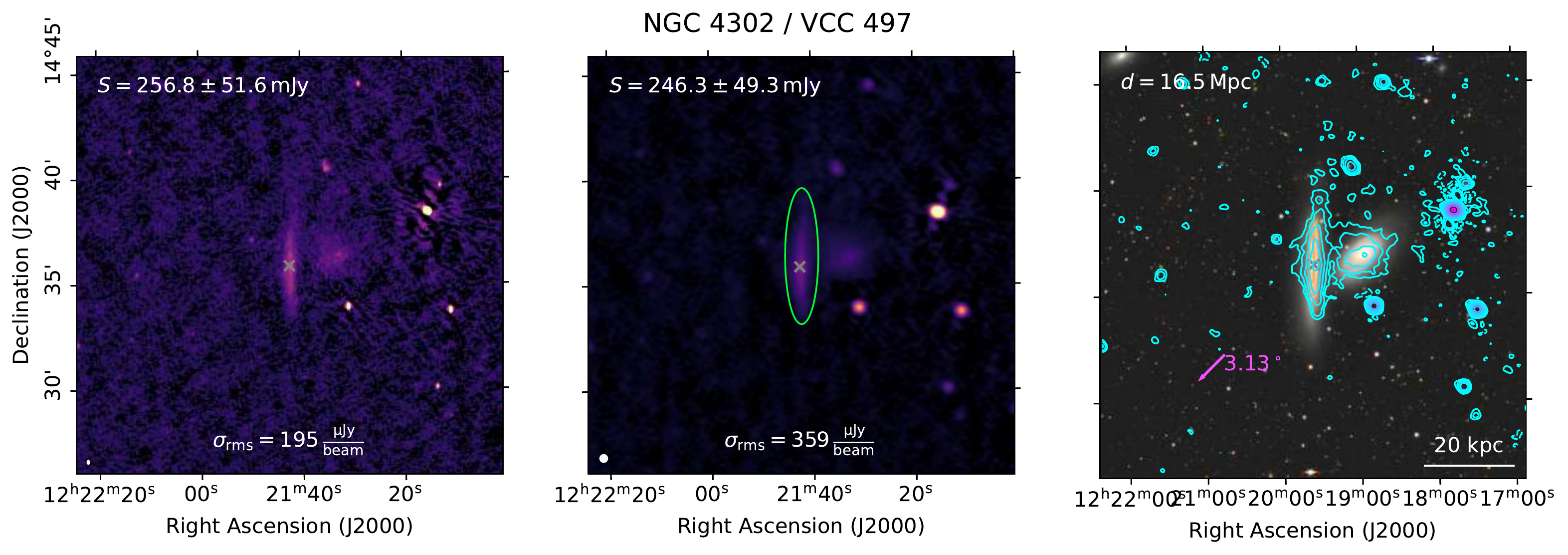}
        \caption{}
    \end{subfigure} 
     \hfill
    \begin{subfigure}[b]{\textwidth}
        \includegraphics[width=\textwidth]{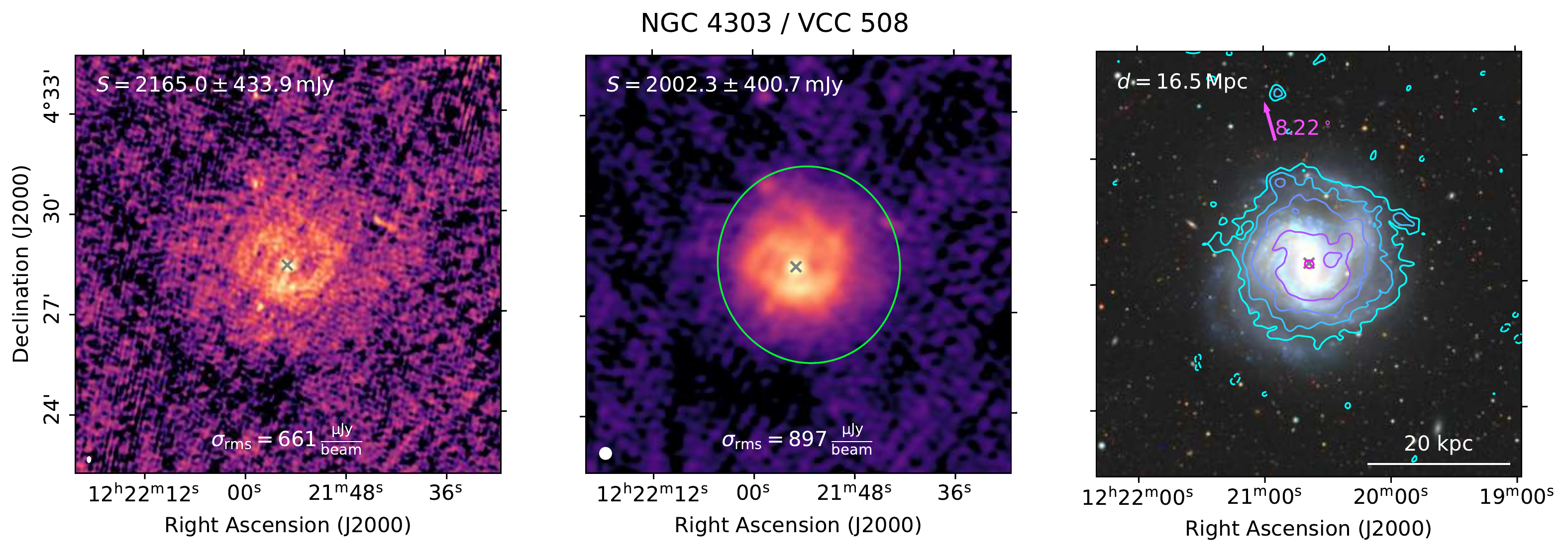}
        \caption{}
    \end{subfigure} 
    \caption{Same as \autoref{fig:144first}.}
\end{figure}

\begin{figure}
    \centering
    \begin{subfigure}[b]{\textwidth}
        \includegraphics[width=\textwidth]{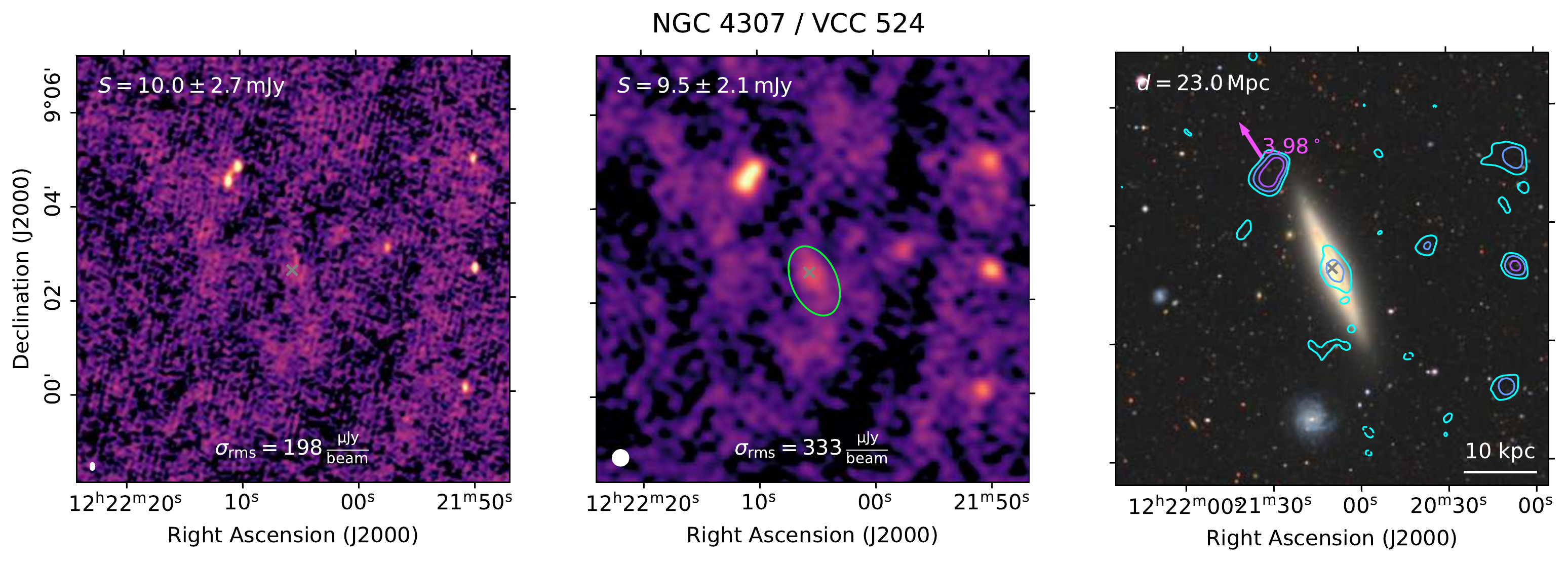}
        \caption{}
    \end{subfigure}
     \hfill
    \begin{subfigure}[b]{\textwidth}
        \includegraphics[width=\textwidth]{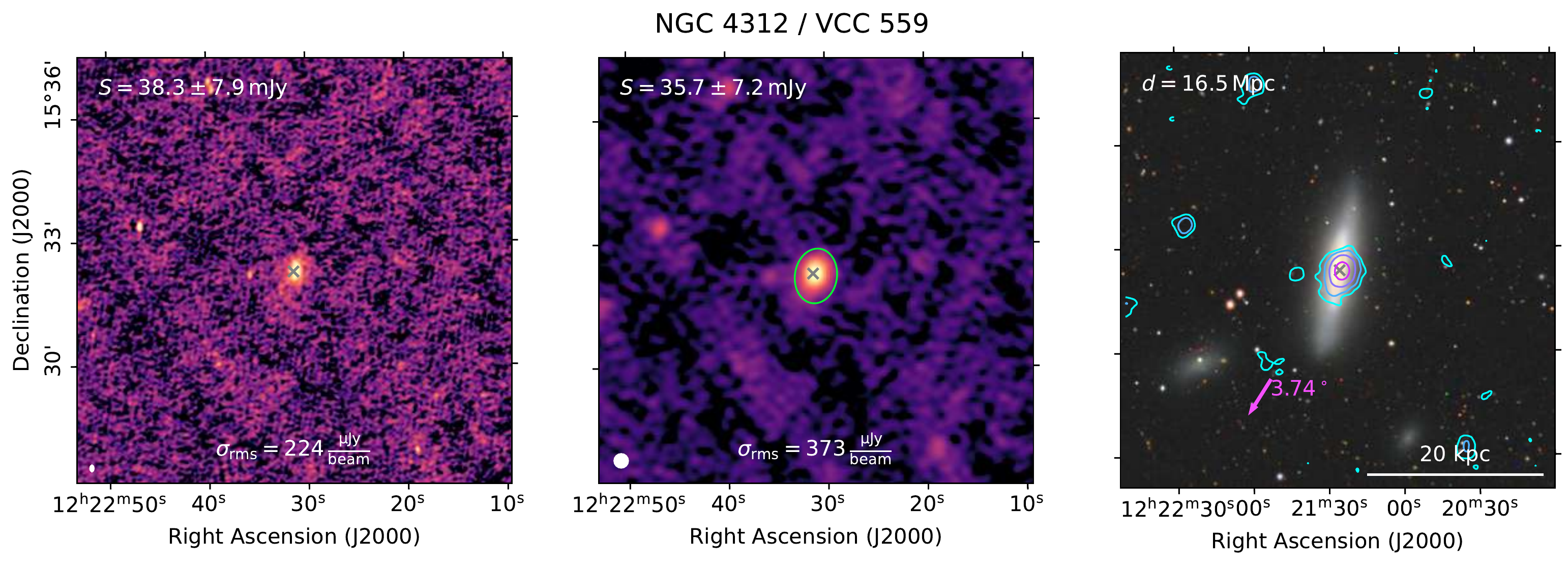}
        \caption{}
    \end{subfigure} 
     \hfill
    \begin{subfigure}[b]{\textwidth}
        \includegraphics[width=\textwidth]{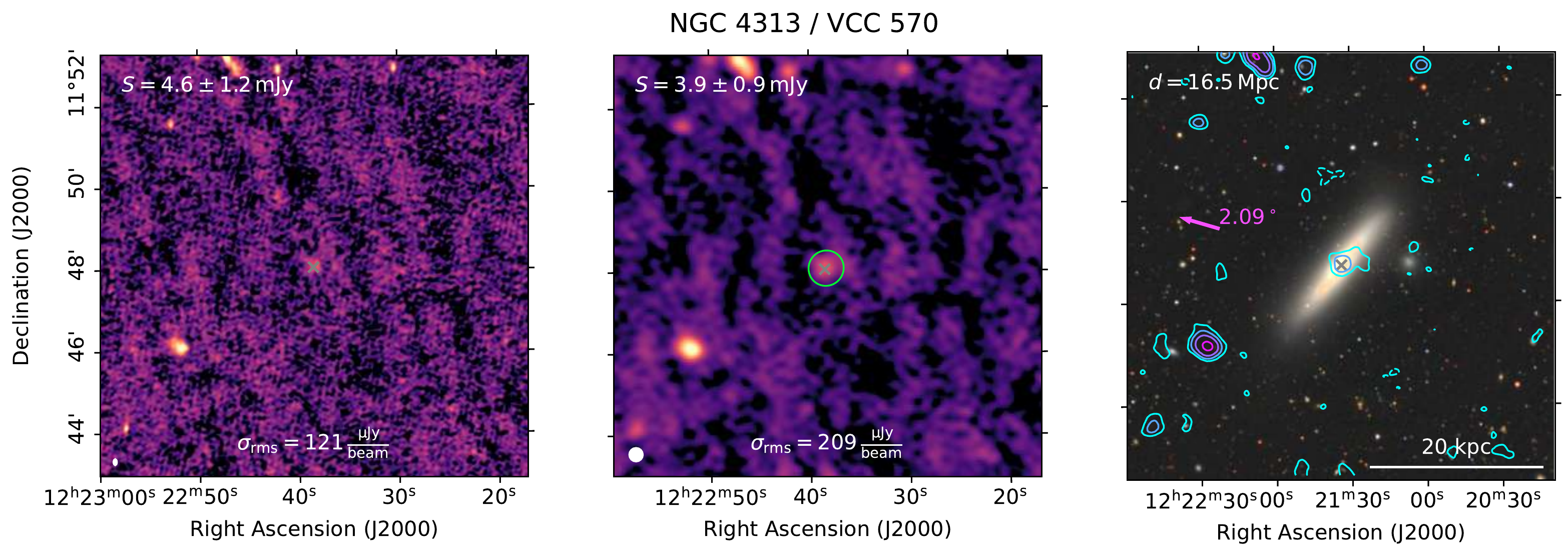}
        \caption{}
    \end{subfigure} 
    \caption{Same as \autoref{fig:144first}.}
\end{figure}

\begin{figure}
    \centering
    \begin{subfigure}[b]{\textwidth}
        \includegraphics[width=\textwidth]{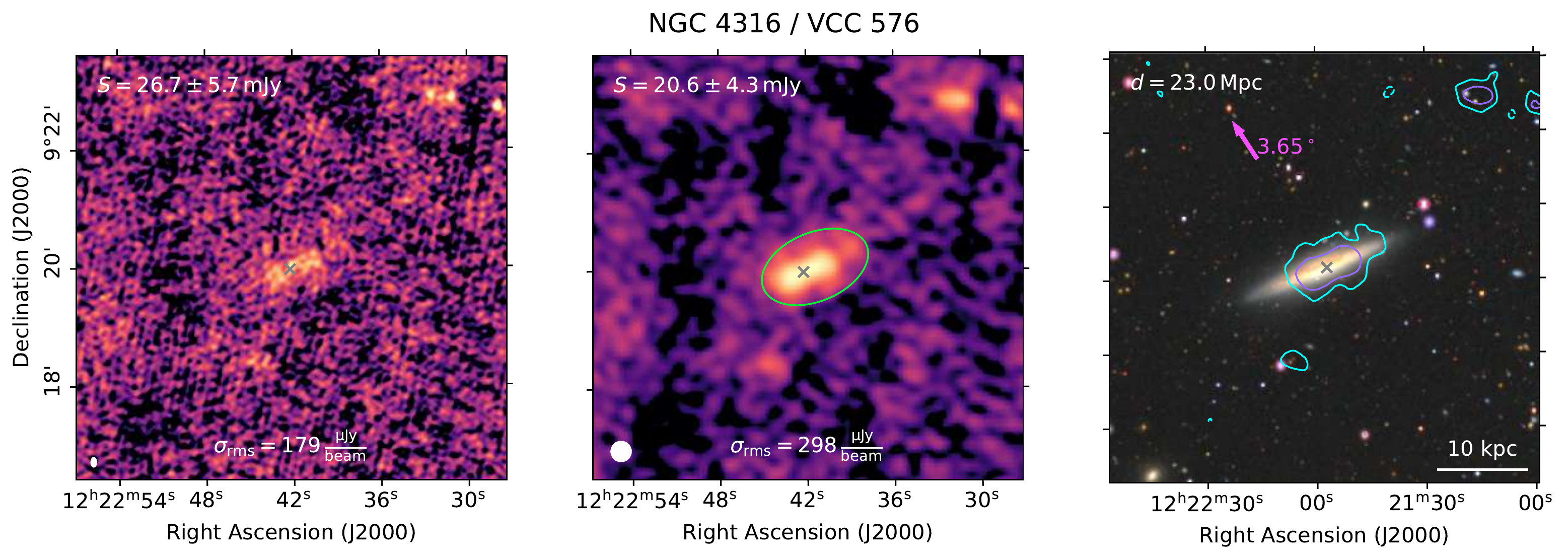}
        \caption{}
    \end{subfigure}
     \hfill
    \begin{subfigure}[b]{\textwidth}
        \includegraphics[width=\textwidth]{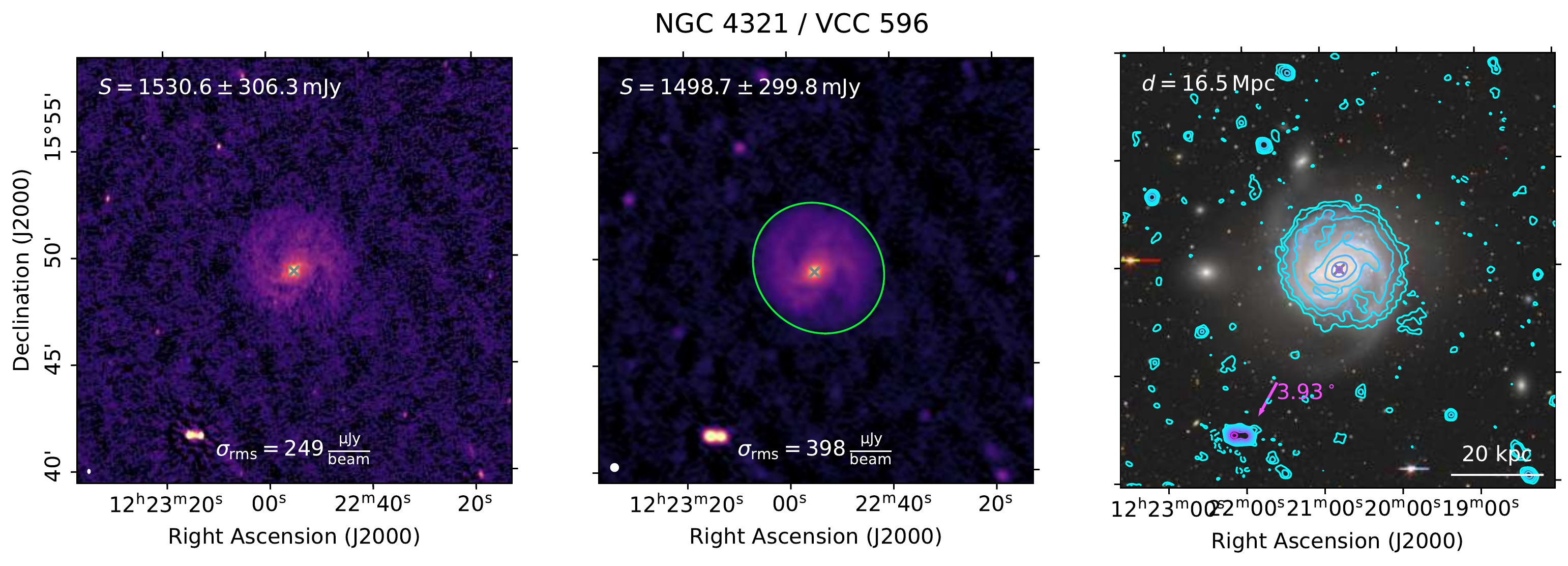}
        \caption{}
        \label{fig:596}
    \end{subfigure} 
     \hfill
    \begin{subfigure}[b]{\textwidth}
        \includegraphics[width=\textwidth]{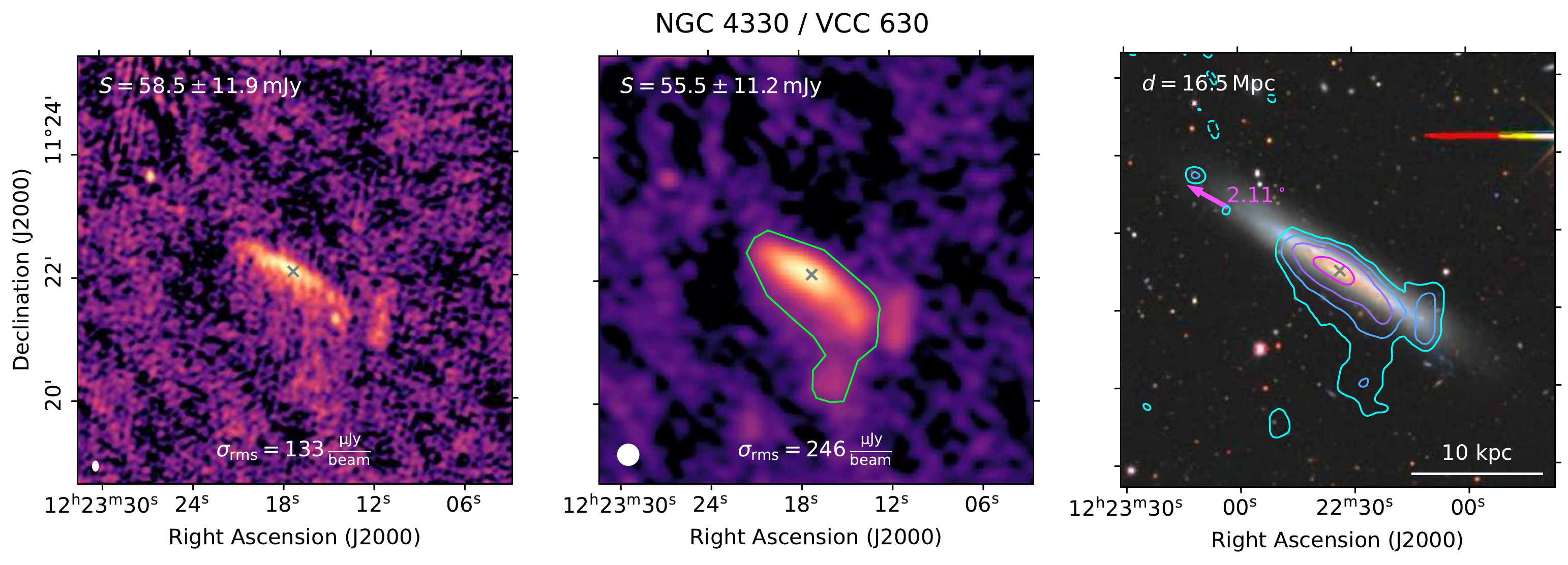}
        \caption{}
        \label{fig:630}
    \end{subfigure} 
    \caption{Same as \autoref{fig:144first}.}
\end{figure}

\begin{figure}
    \centering
    \begin{subfigure}[b]{\textwidth}
        \includegraphics[width=\textwidth]{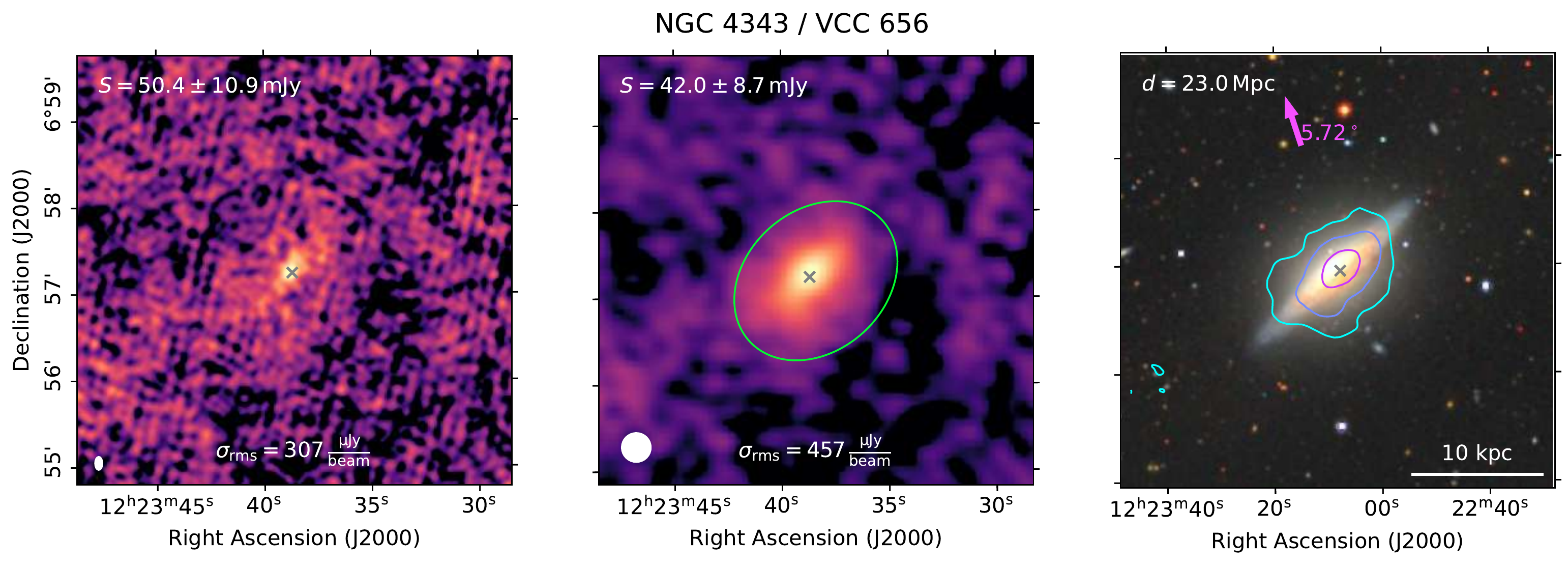}
        \caption{}
    \end{subfigure}
     \hfill
    \begin{subfigure}[b]{\textwidth}
        \includegraphics[width=\textwidth]{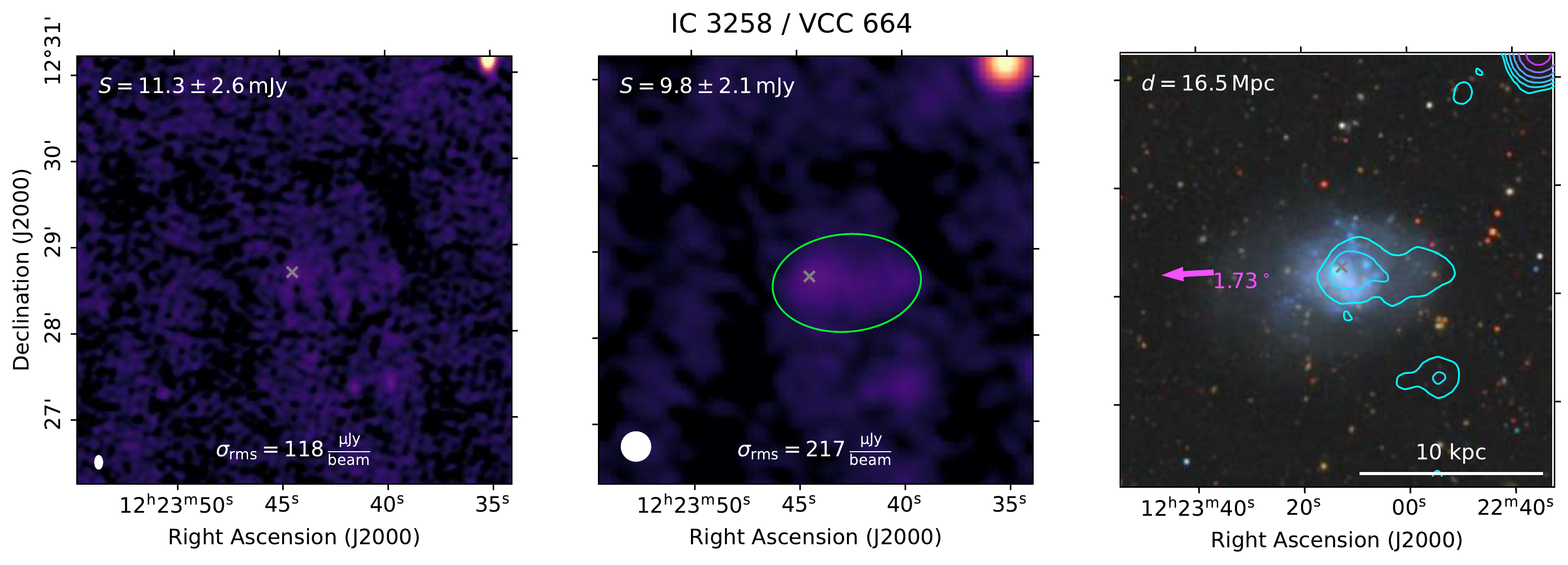}
        \caption{}
        \label{fig:664}
    \end{subfigure} 
     \hfill
    \begin{subfigure}[b]{\textwidth}
        \includegraphics[width=\textwidth]{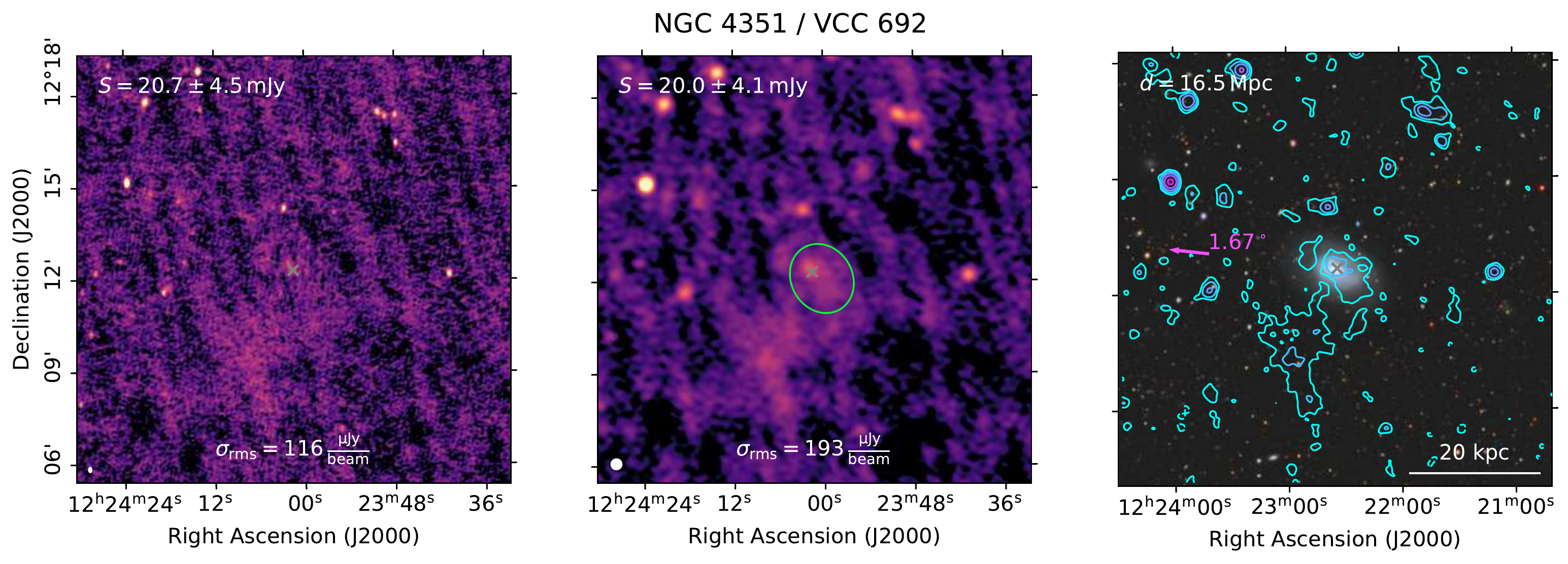}
        \caption{}
    \end{subfigure} 
    \caption{Same as \autoref{fig:144first}.}
\end{figure}

\begin{figure}
    \centering
    \begin{subfigure}[b]{\textwidth}
        \includegraphics[width=\textwidth]{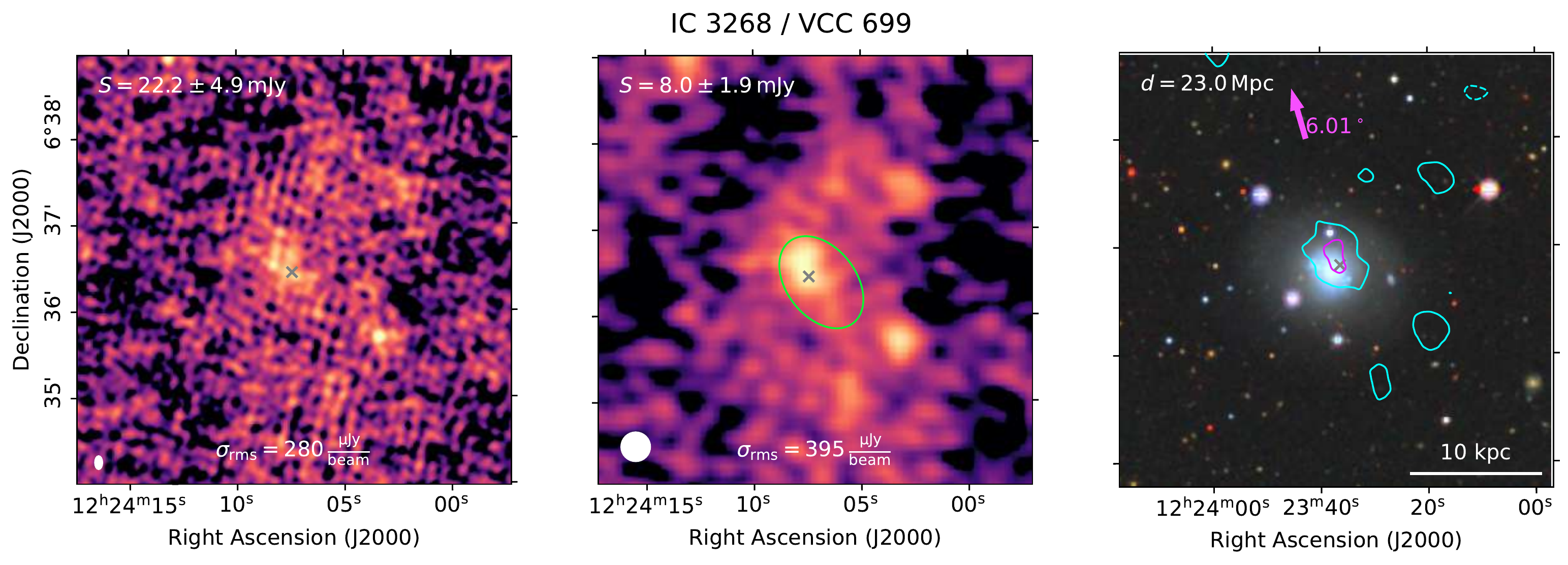}
        \caption{}
    \end{subfigure}
     \hfill
    \begin{subfigure}[b]{\textwidth}
        \includegraphics[width=\textwidth]{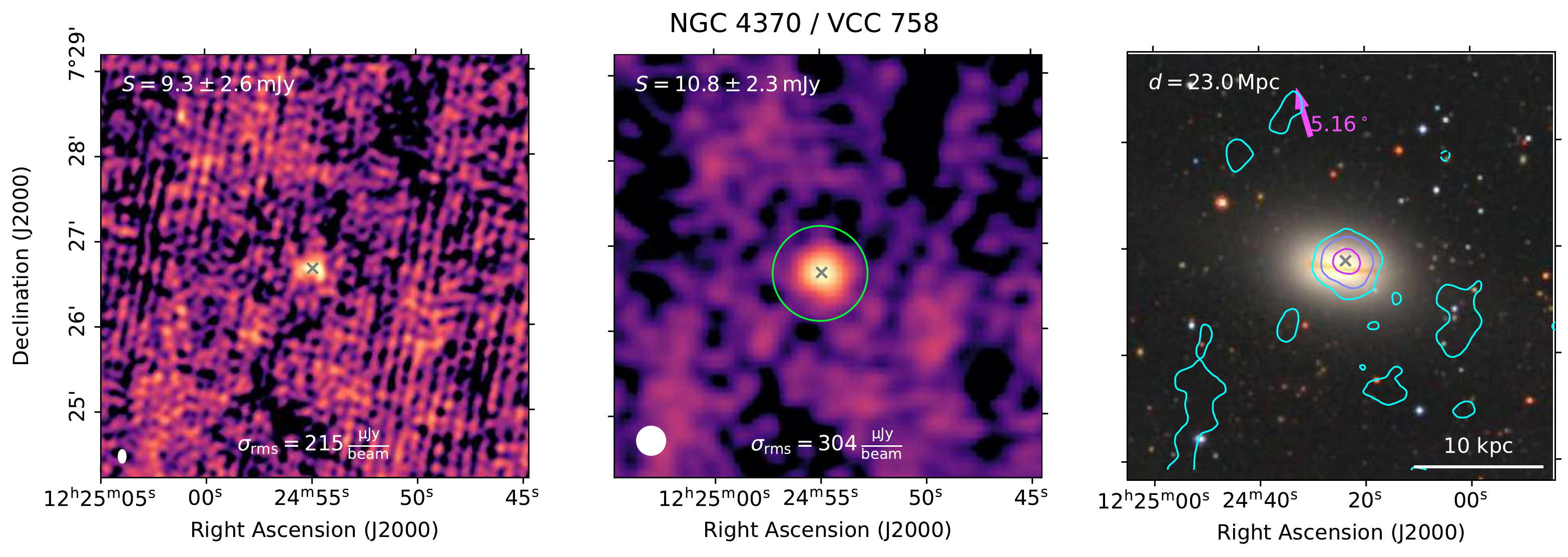}
        \caption{}
    \end{subfigure} 
     \hfill
    \begin{subfigure}[b]{\textwidth}
        \includegraphics[width=\textwidth]{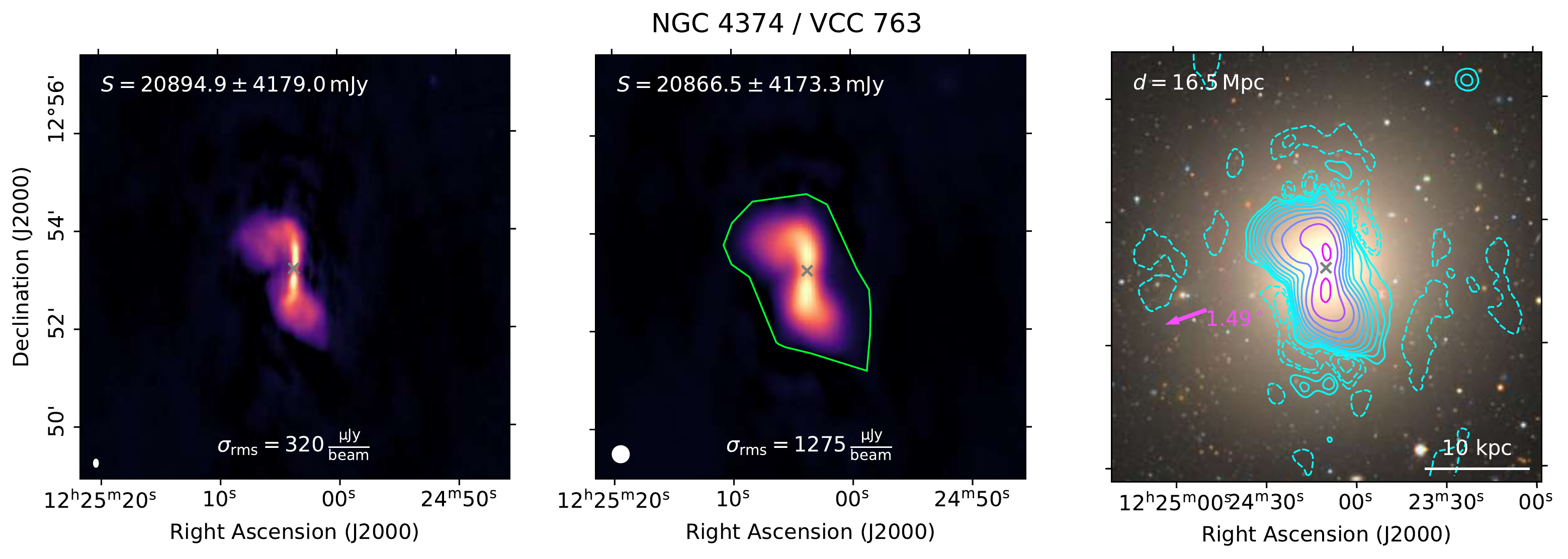}
        \caption{}
        \label{fig:763}
    \end{subfigure} 
    \caption{Same as \autoref{fig:144first}.}
\end{figure}

\begin{figure}
    \centering
    \begin{subfigure}[b]{\textwidth}
        \includegraphics[width=\textwidth]{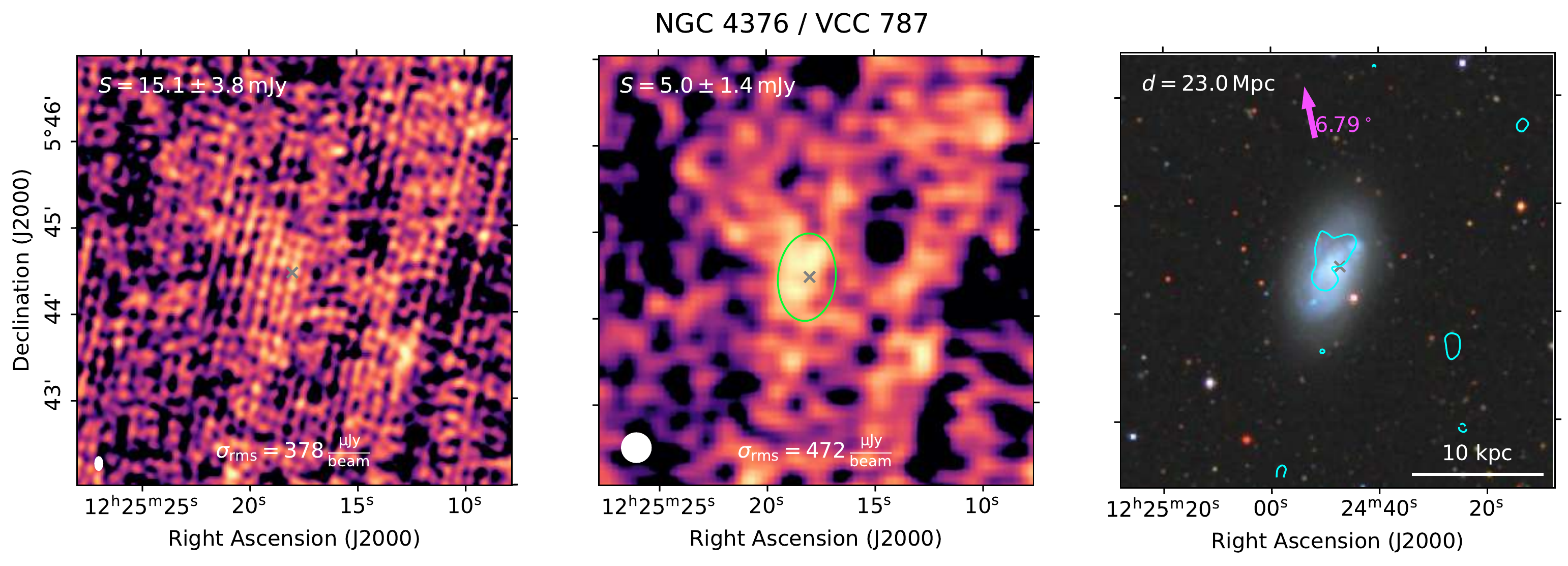}
        \caption{}
    \end{subfigure}
     \hfill
    \begin{subfigure}[b]{\textwidth}
        \includegraphics[width=\textwidth]{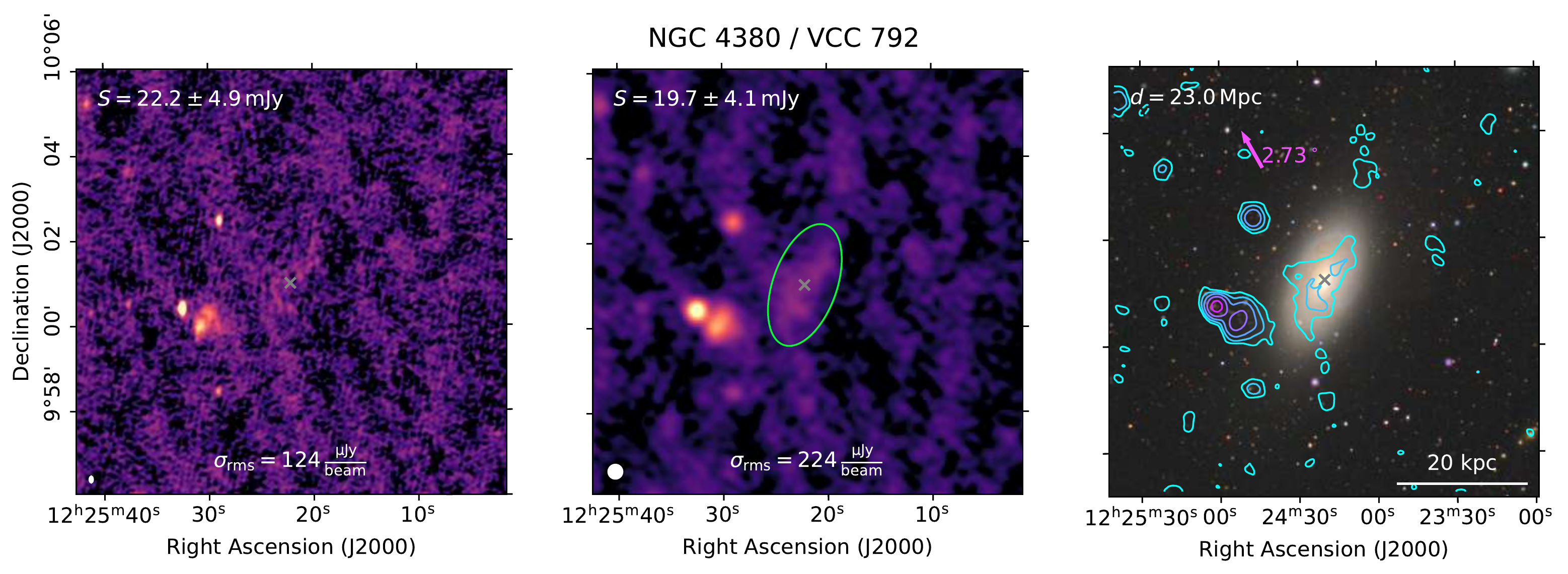}
        \caption{}
    \end{subfigure} 
     \hfill
    \begin{subfigure}[b]{\textwidth}
        \includegraphics[width=\textwidth]{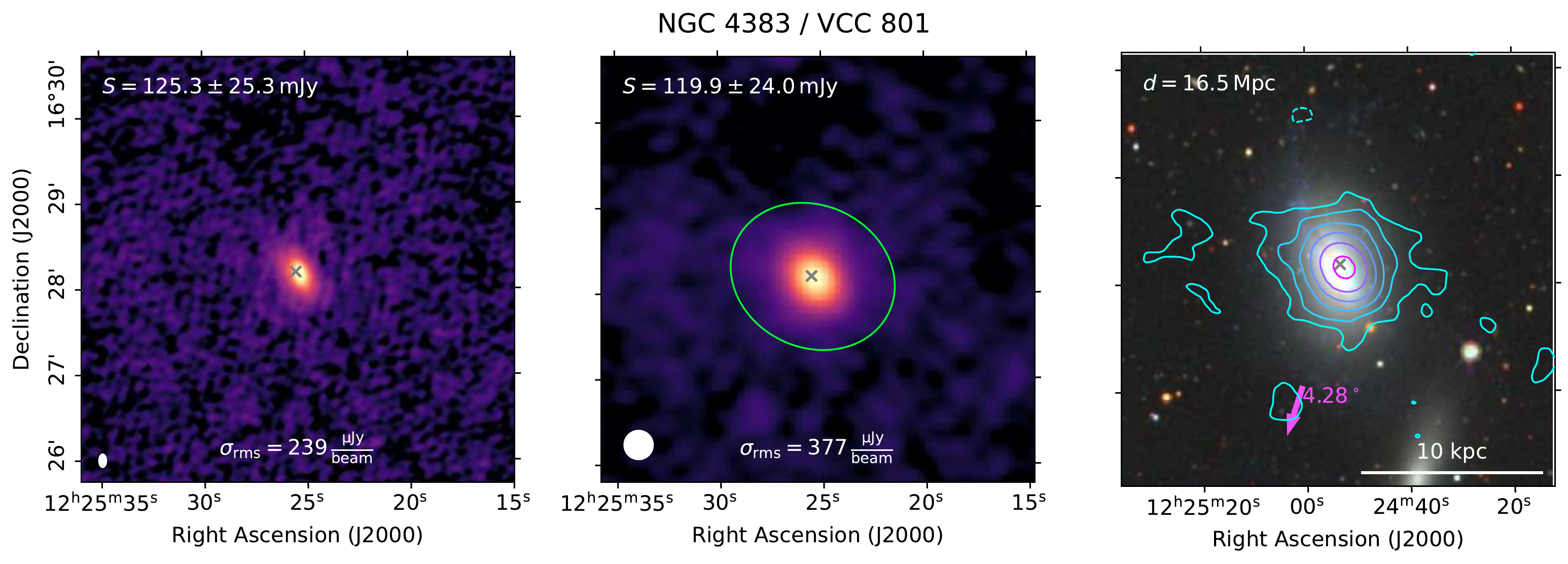}
        \caption{}
    \end{subfigure} 
    \caption{Same as \autoref{fig:144first}.}
\end{figure}

\begin{figure}
    \centering
    \begin{subfigure}[b]{\textwidth}
        \includegraphics[width=\textwidth]{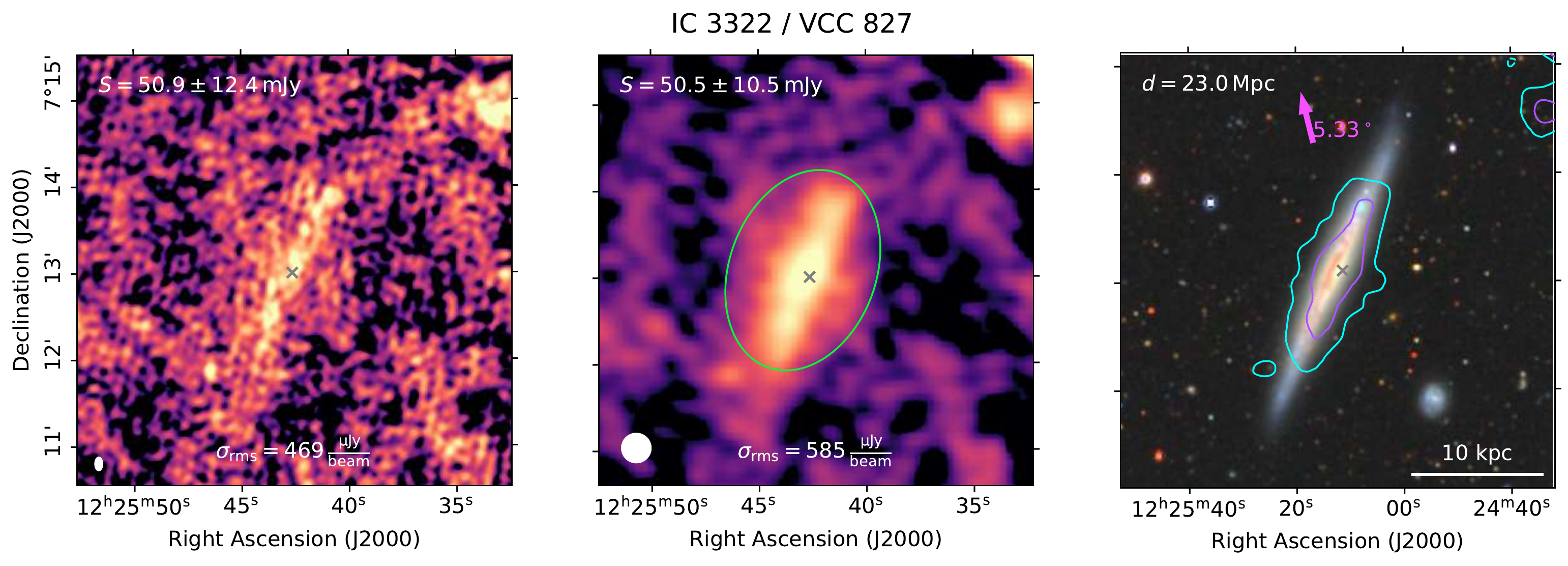}
        \caption{}
    \end{subfigure}
     \hfill
    \begin{subfigure}[b]{\textwidth}
        \includegraphics[width=\textwidth]{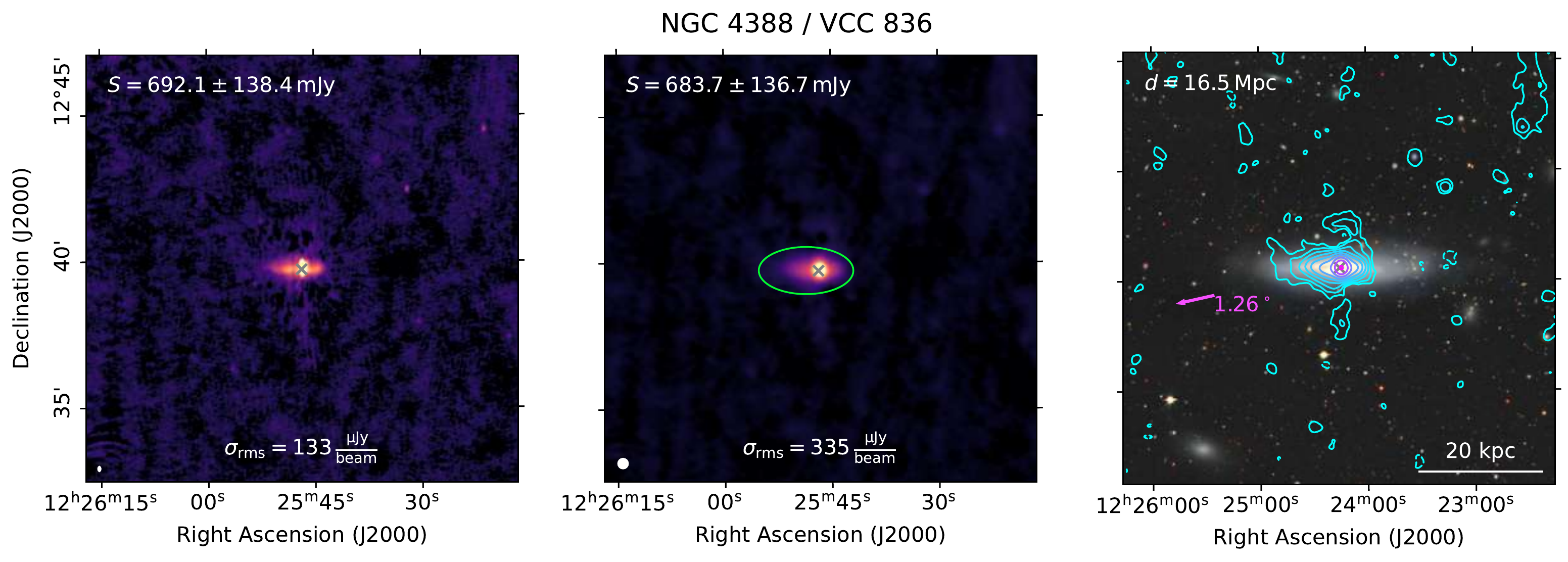}
        \caption{}
        \label{fig:836}
    \end{subfigure} 
     \hfill
    \begin{subfigure}[b]{\textwidth}
        \includegraphics[width=\textwidth]{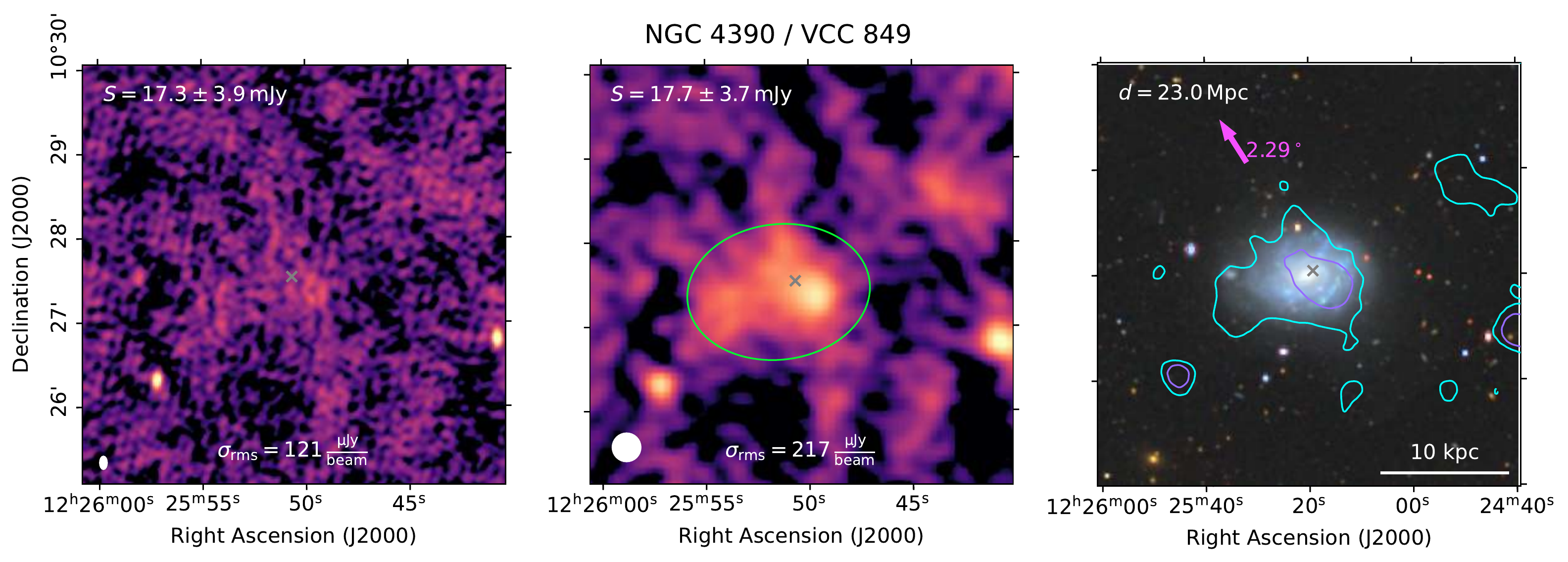}
        \caption{}
    \end{subfigure} 
    \caption{Same as \autoref{fig:144first}.}
\end{figure}

\begin{figure}
    \centering
    \begin{subfigure}[b]{\textwidth}
        \includegraphics[width=\textwidth]{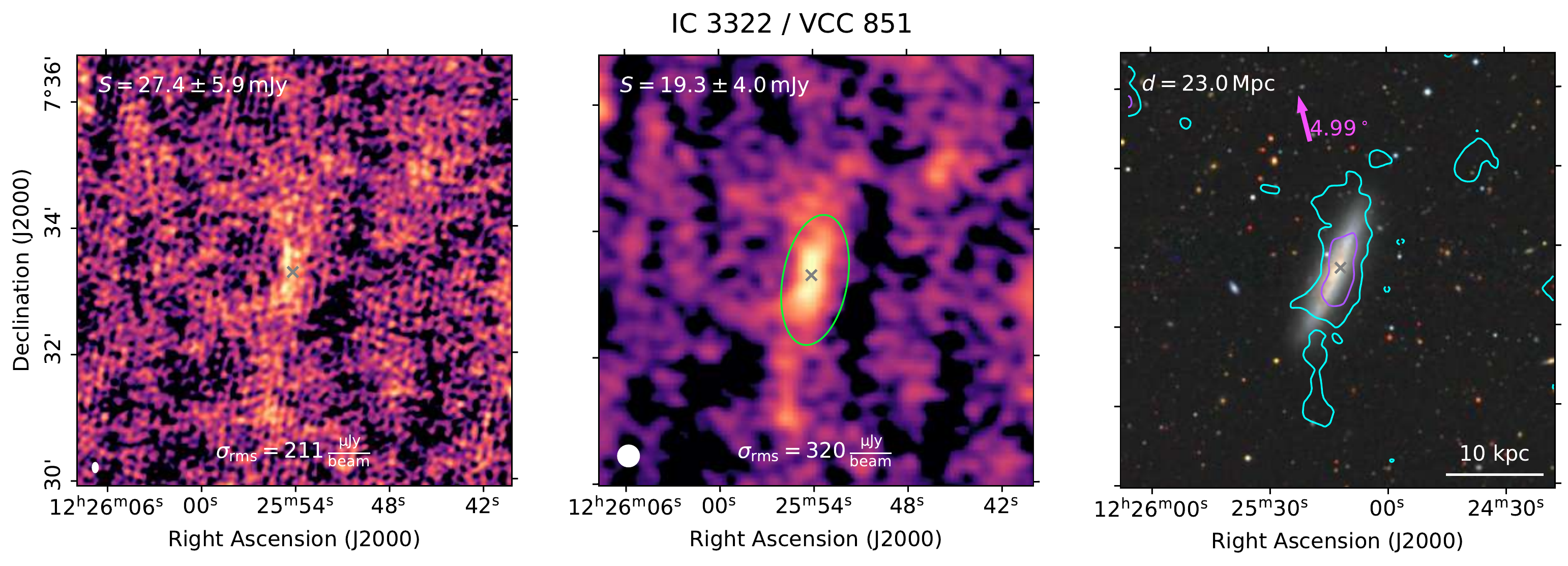}
        \caption{}
    \end{subfigure}
     \hfill
    \begin{subfigure}[b]{\textwidth}
        \includegraphics[width=\textwidth]{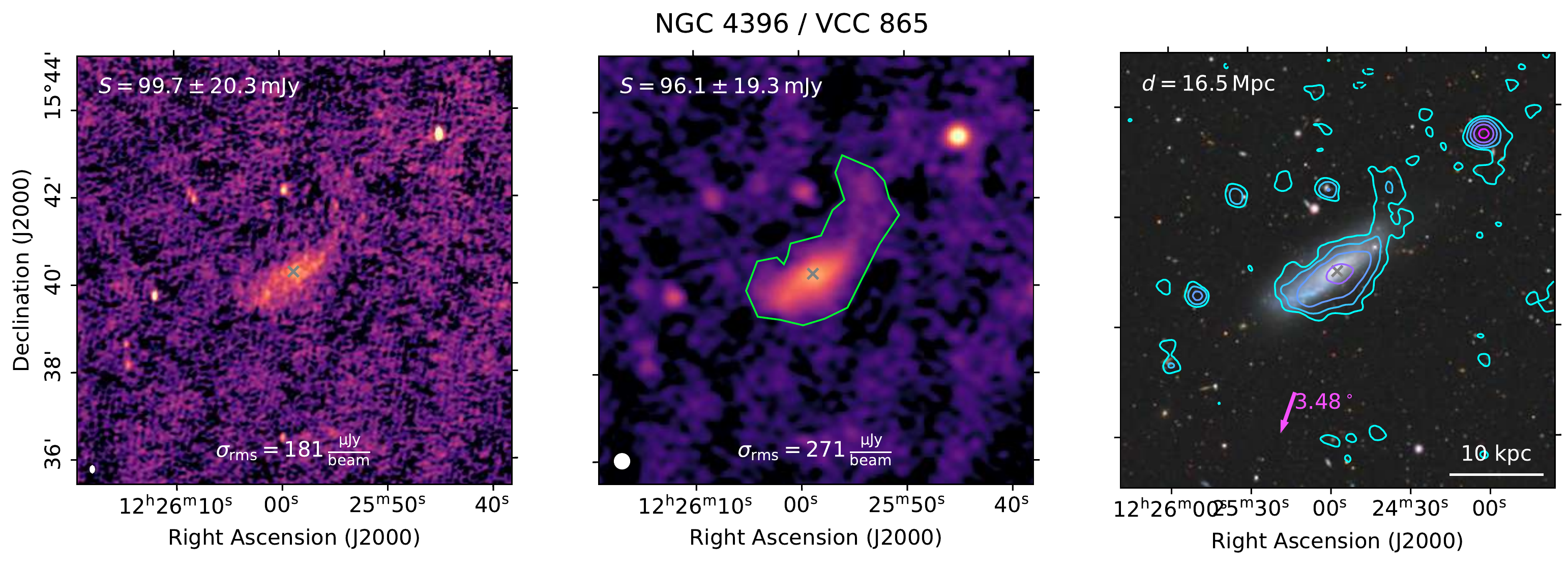}
        \caption{}
        \label{fig:865}
    \end{subfigure} 
     \hfill
    \begin{subfigure}[b]{\textwidth}
        \includegraphics[width=\textwidth]{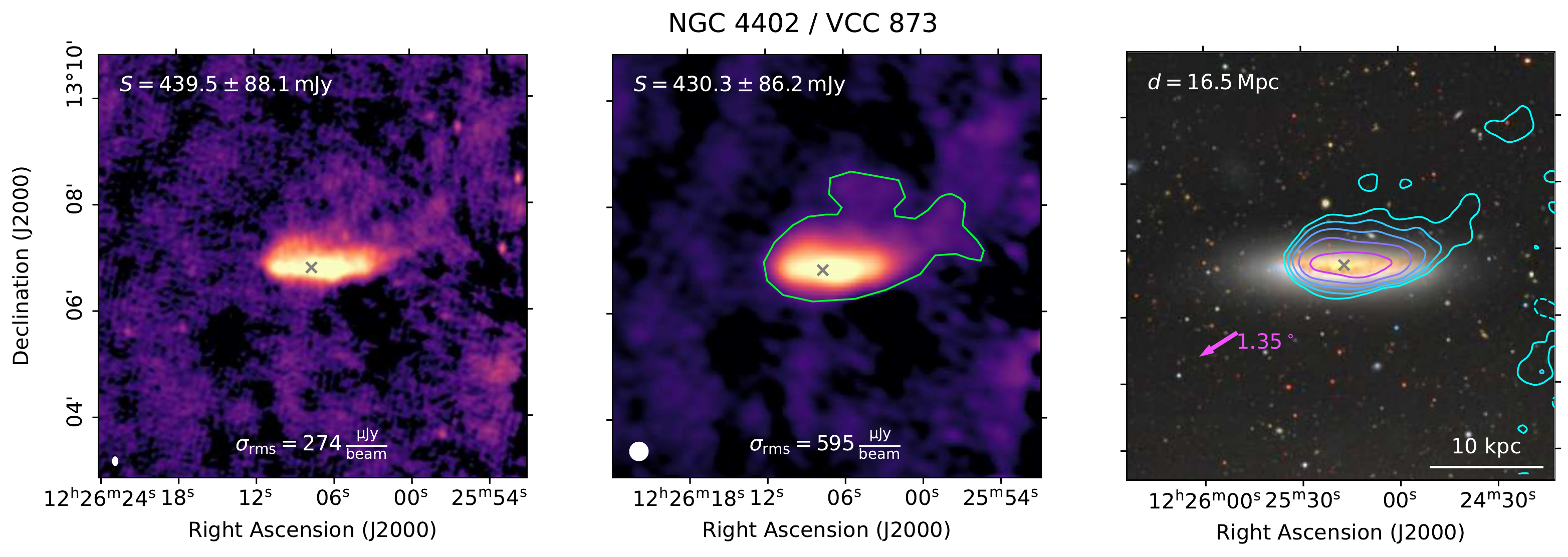}
        \caption{}
        \label{fig:873}
    \end{subfigure} 
    \caption{Same as \autoref{fig:144first}.}
\end{figure}

\begin{figure}
    \centering
    \begin{subfigure}[b]{\textwidth}
        \includegraphics[width=\textwidth]{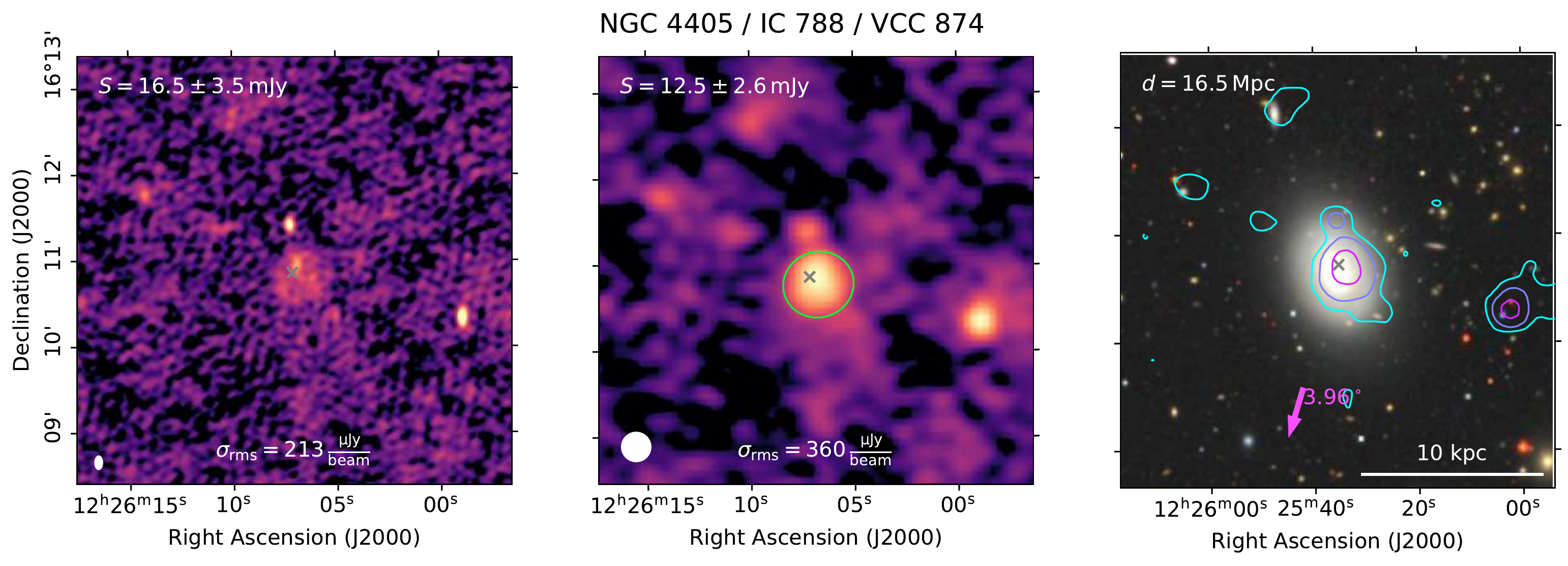}
        \caption{}
        \label{fig:874}
    \end{subfigure}
     \hfill
    \begin{subfigure}[b]{\textwidth}
        \includegraphics[width=\textwidth]{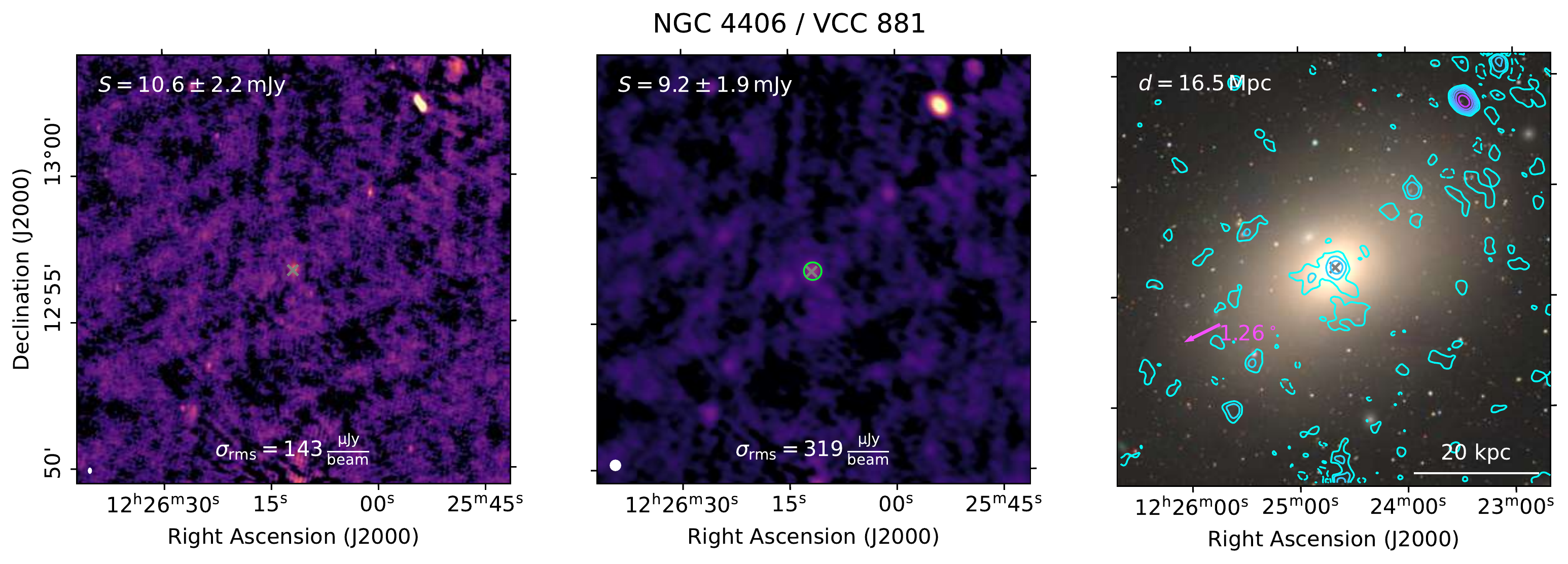}
        \caption{}
        \label{fig:881}
    \end{subfigure} 
     \hfill
    \begin{subfigure}[b]{\textwidth}
        \includegraphics[width=\textwidth]{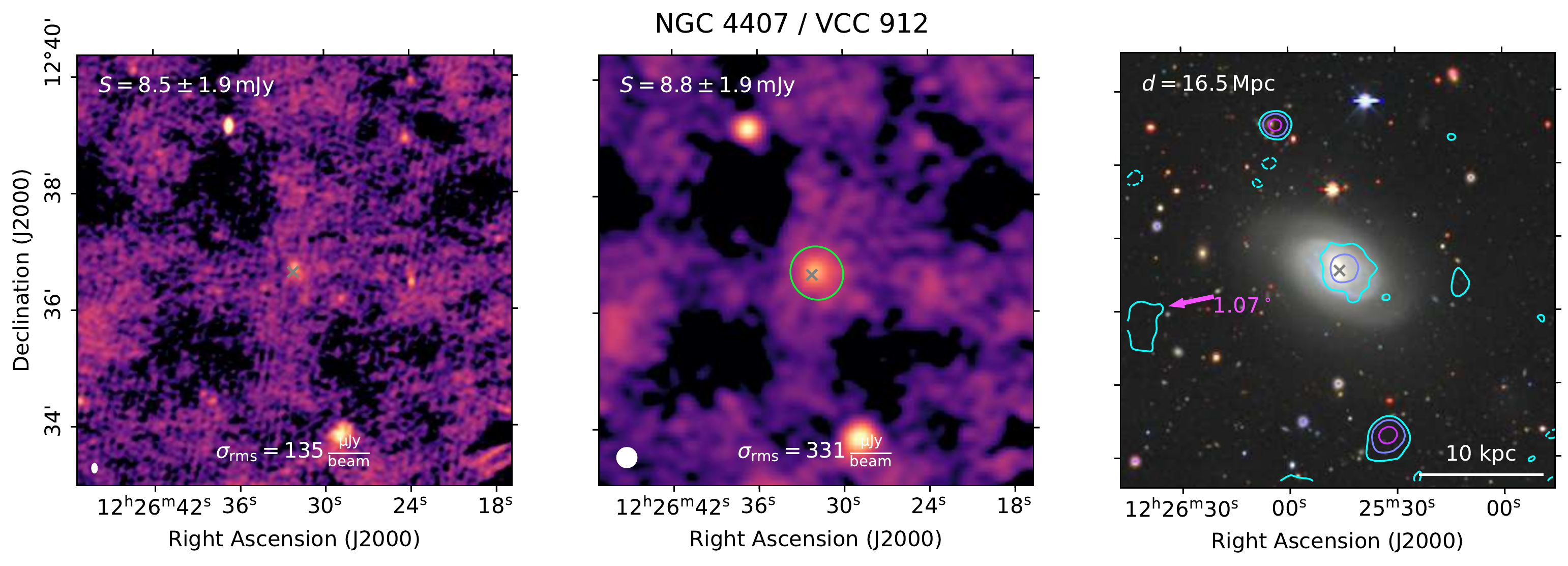}
        \caption{}
    \end{subfigure} 
    \caption{Same as \autoref{fig:144first}.}
\end{figure}

\begin{figure}
    \centering
    \begin{subfigure}[b]{\textwidth}
        \includegraphics[width=\textwidth]{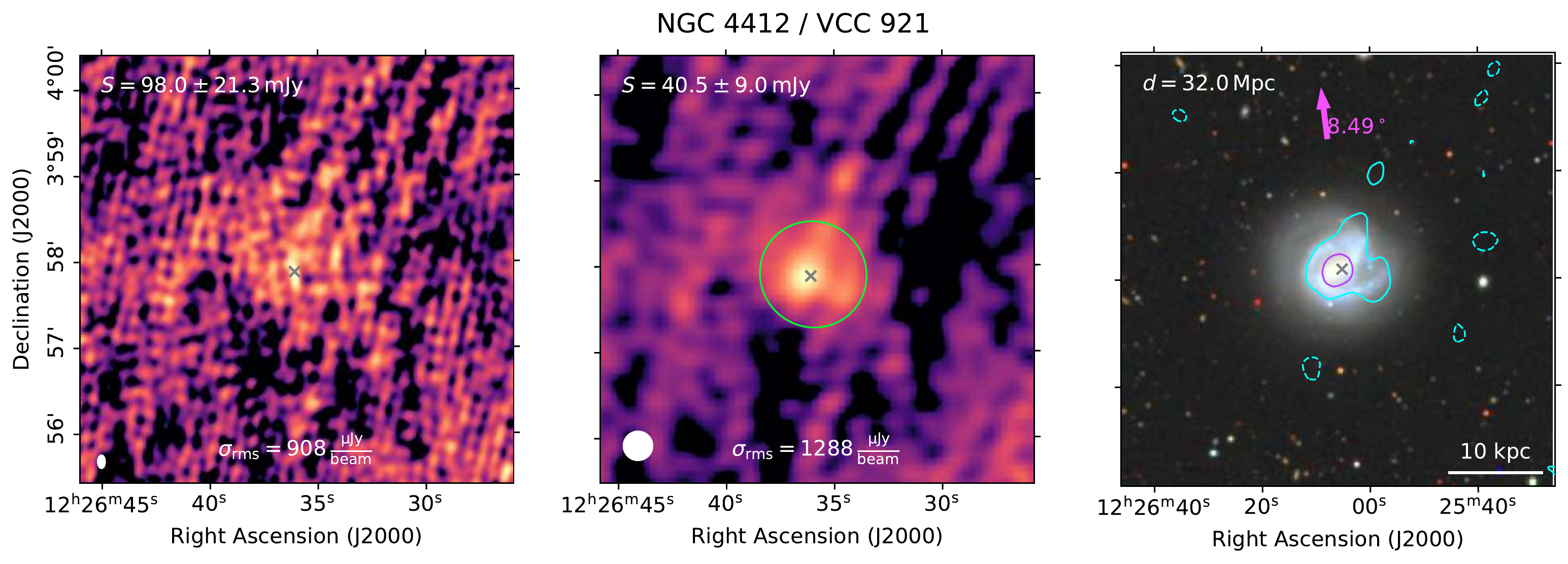}
        \caption{}
    \end{subfigure}
     \hfill
    \begin{subfigure}[b]{\textwidth}
        \includegraphics[width=\textwidth]{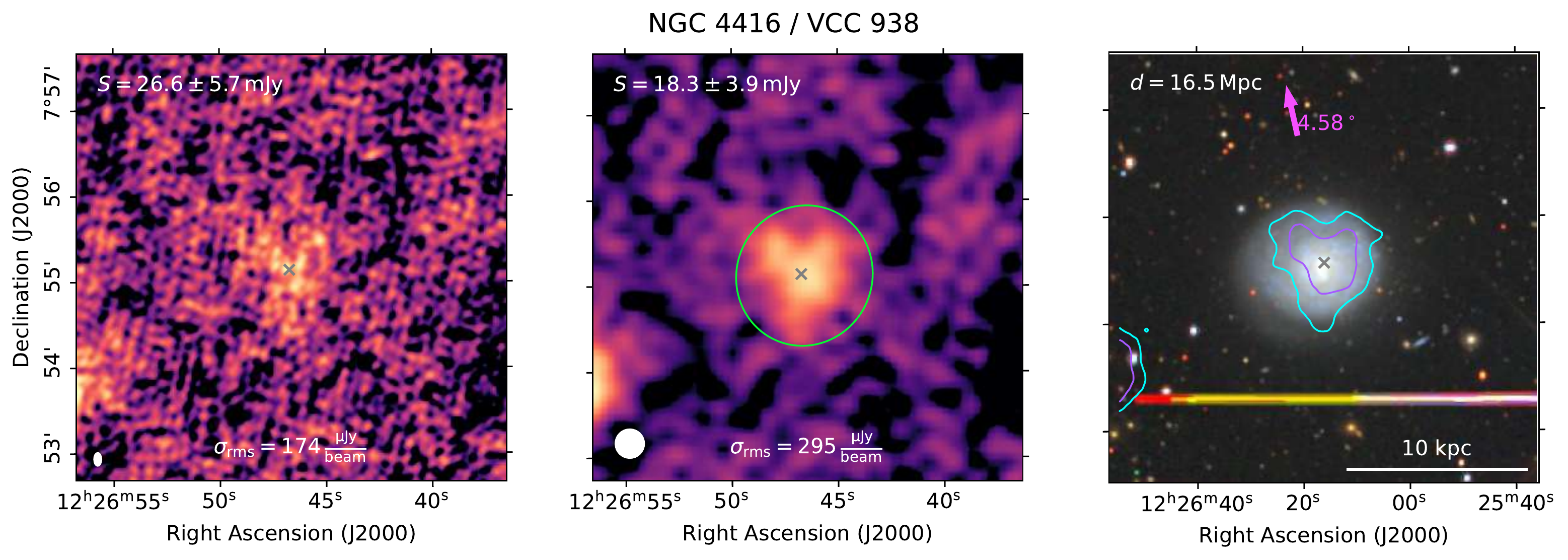}
        \caption{}
    \end{subfigure} 
     \hfill
    \begin{subfigure}[b]{\textwidth}
        \includegraphics[width=\textwidth]{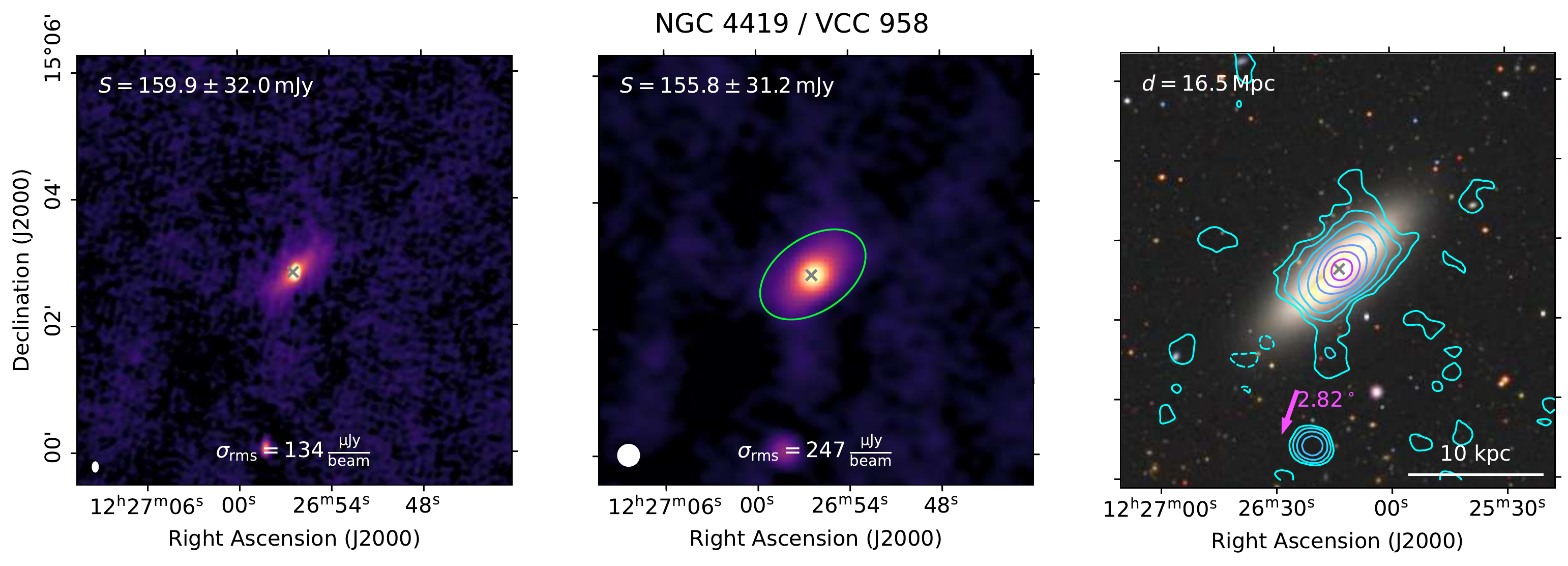}
        \caption{}
    \end{subfigure} 
    \caption{Same as \autoref{fig:144first}.}
\end{figure}

\begin{figure}
    \centering
    \begin{subfigure}[b]{\textwidth}
        \includegraphics[width=\textwidth]{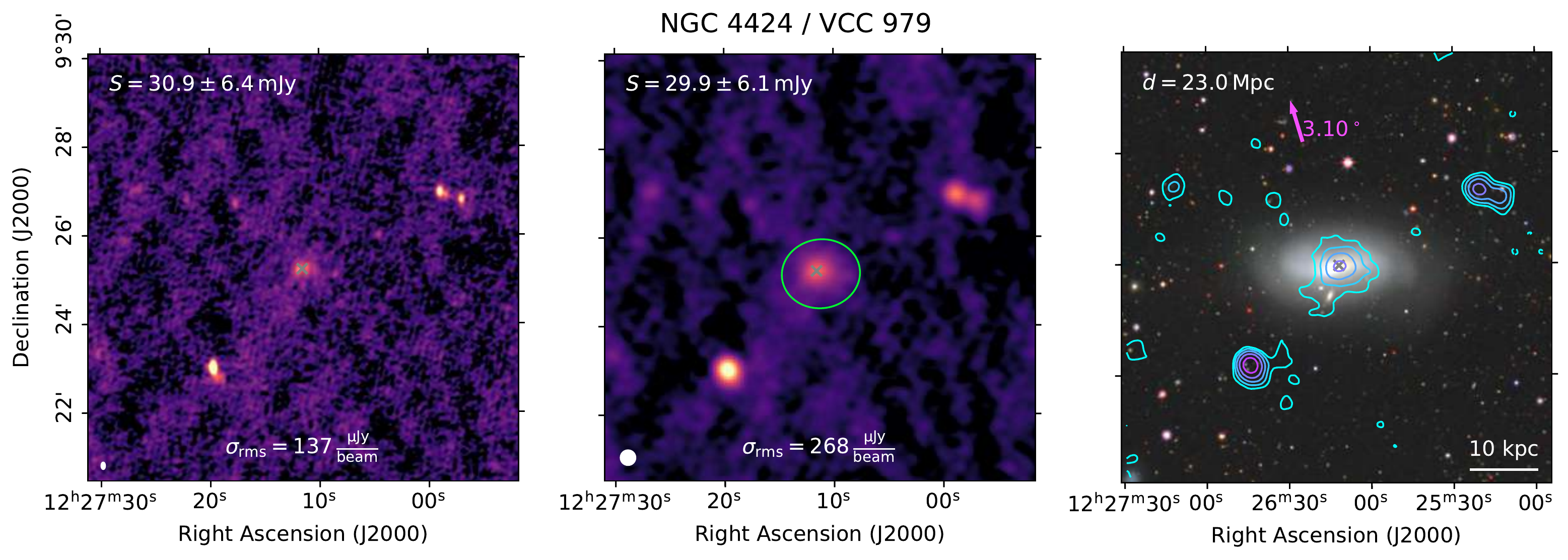}
        \caption{}
    \end{subfigure}
     \hfill
    \begin{subfigure}[b]{\textwidth}
        \includegraphics[width=\textwidth]{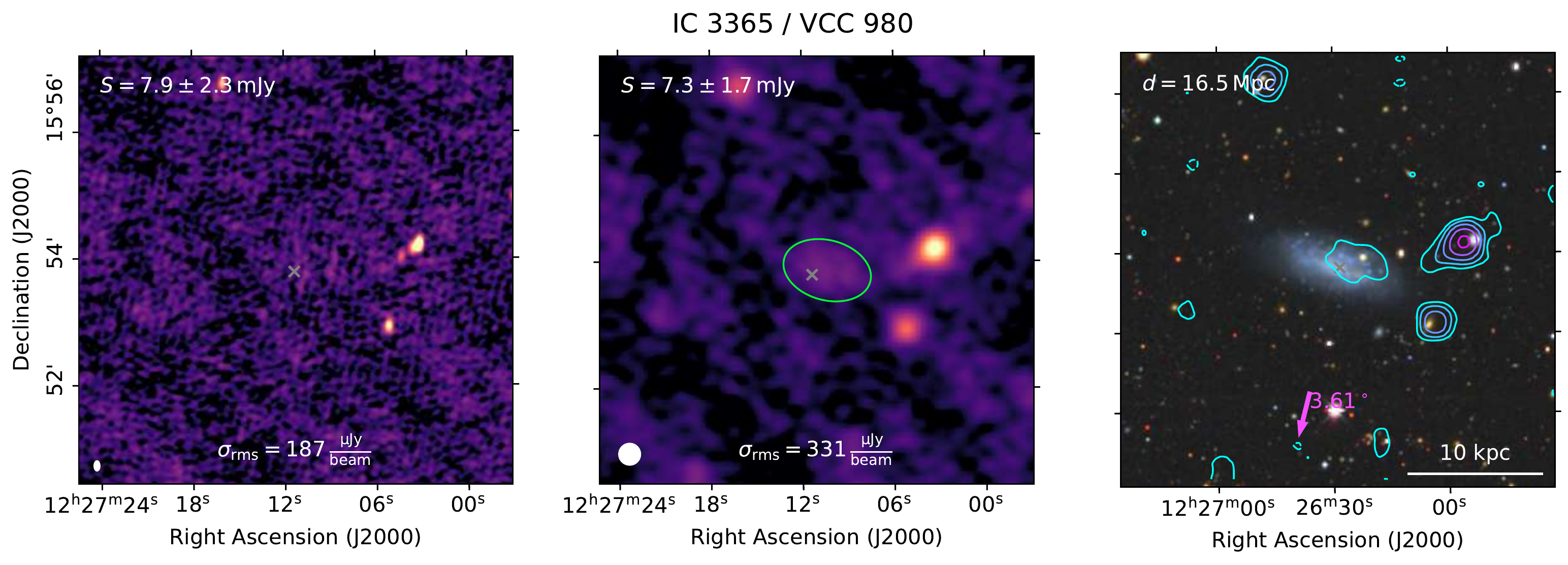}
        \caption{}
    \end{subfigure} 
     \hfill
    \begin{subfigure}[b]{\textwidth}
        \includegraphics[width=\textwidth]{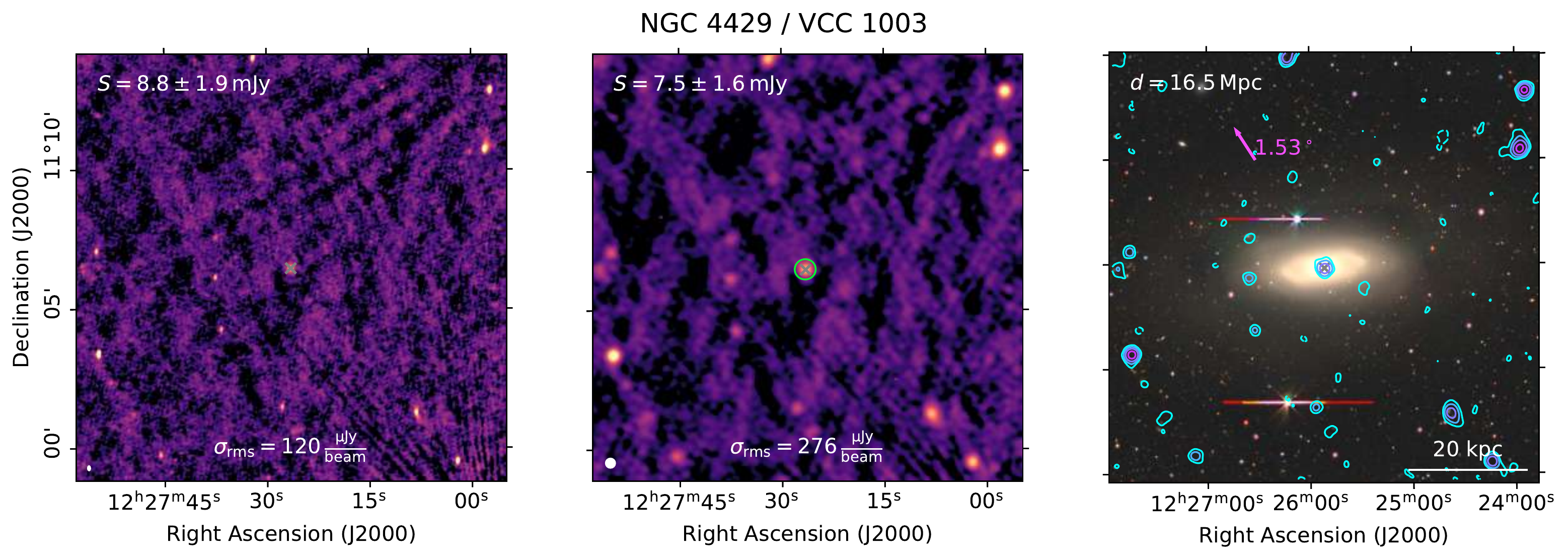}
        \caption{}
    \end{subfigure} 
    \caption{Same as \autoref{fig:144first}.}
\end{figure}

\begin{figure}
    \centering
    \begin{subfigure}[b]{\textwidth}
        \includegraphics[width=\textwidth]{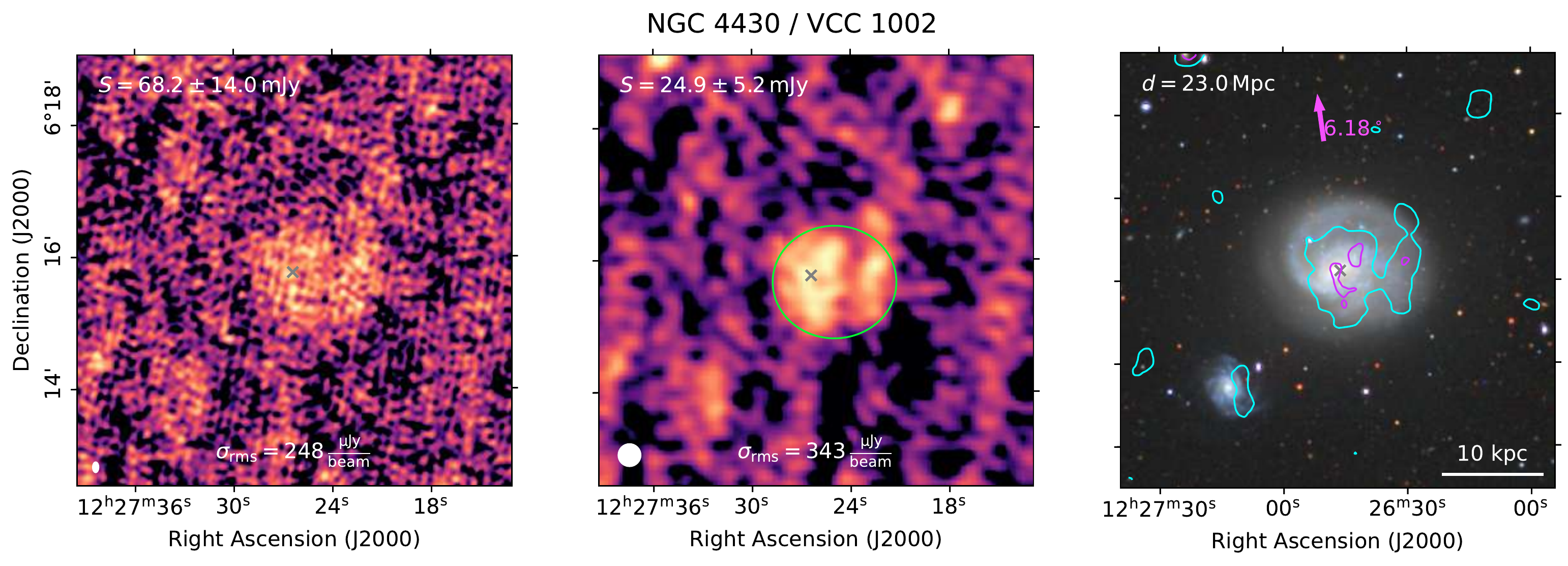}
        \caption{}
    \end{subfigure}
    \begin{subfigure}[b]{\textwidth}
        \includegraphics[width=\textwidth]{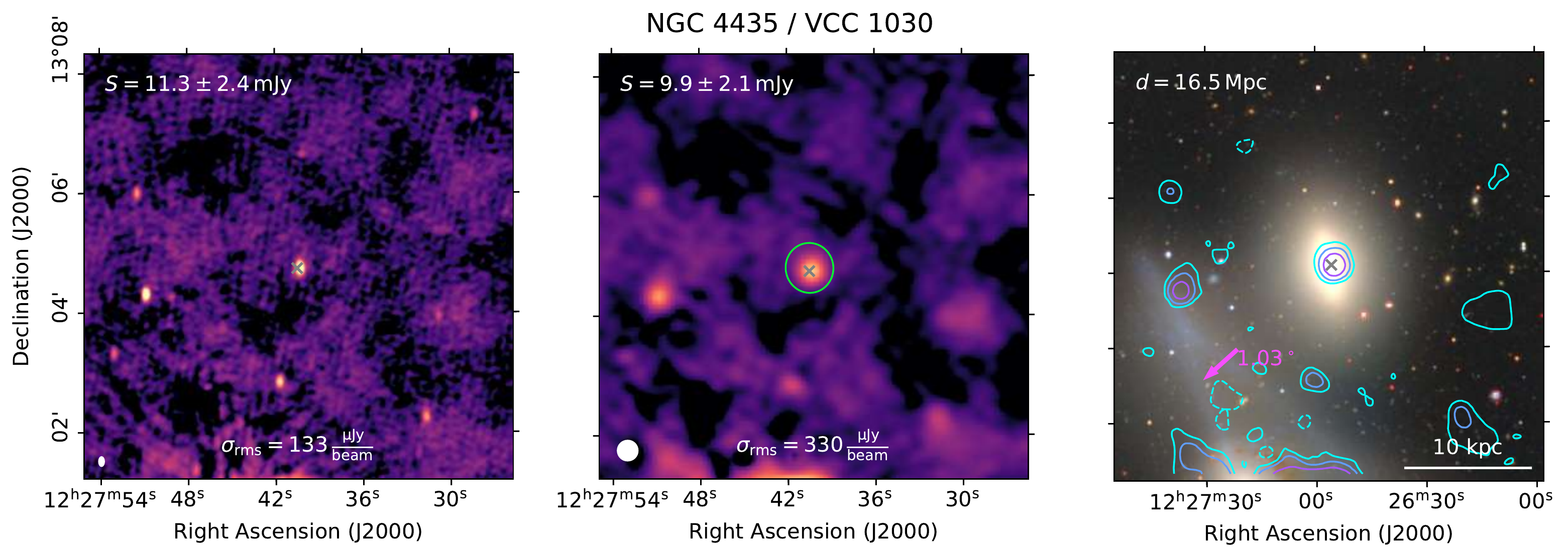}
        \caption{}
    \end{subfigure}
     \hfill
    \begin{subfigure}[b]{\textwidth}
        \includegraphics[width=\textwidth]{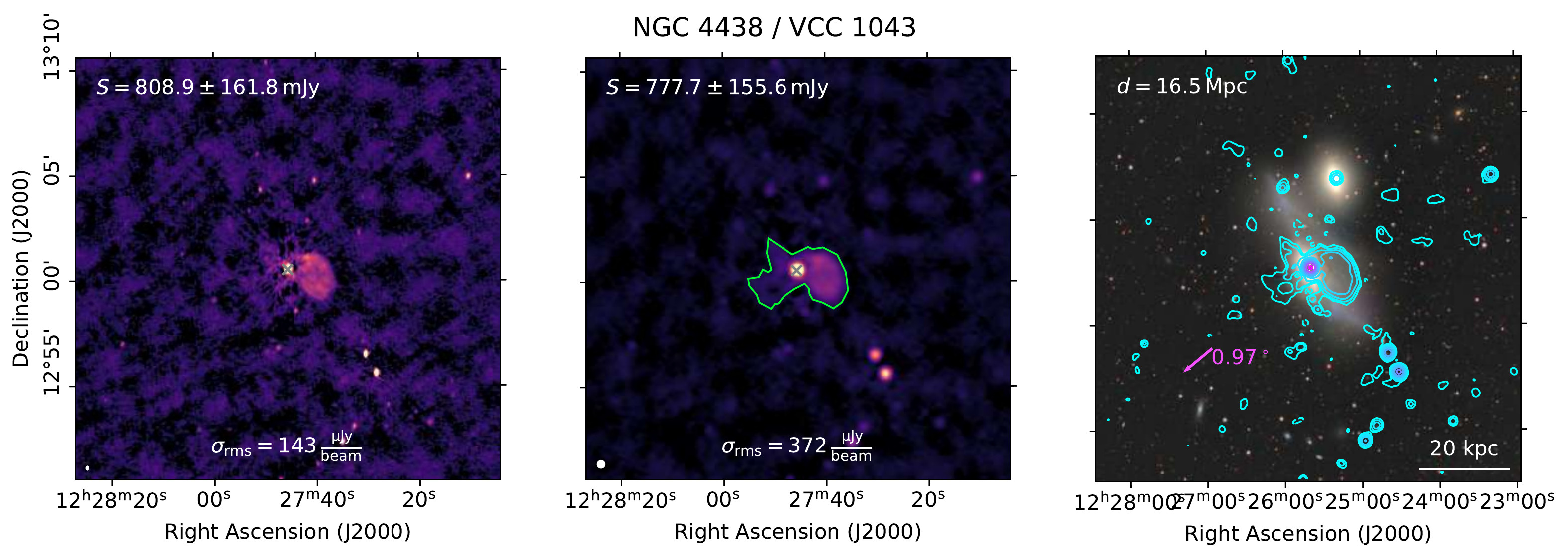}
        \caption{}
        \label{fig:1043}
    \end{subfigure} 
    \caption{Same as \autoref{fig:144first}.}
\end{figure}

\begin{figure}
    \centering
    \begin{subfigure}[b]{\textwidth}
        \includegraphics[width=\textwidth]{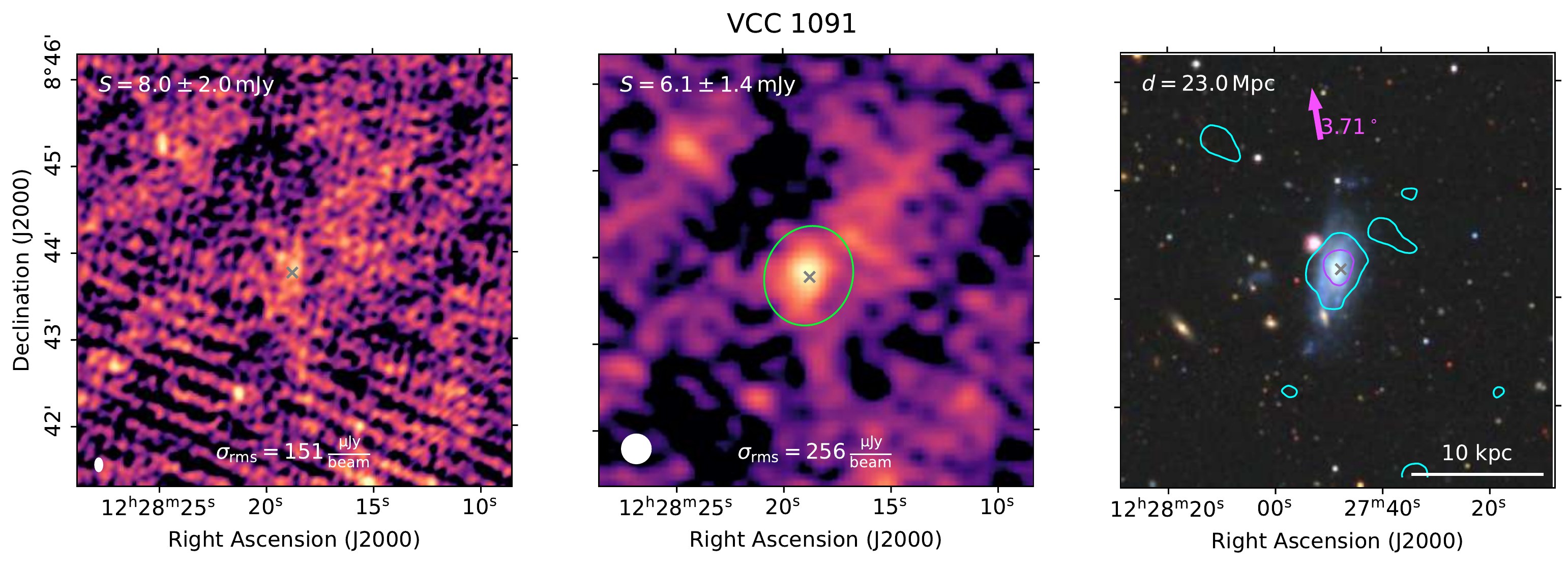}
        \caption{}
    \end{subfigure} 
    \hfill
    \begin{subfigure}[b]{\textwidth}
        \includegraphics[width=\textwidth]{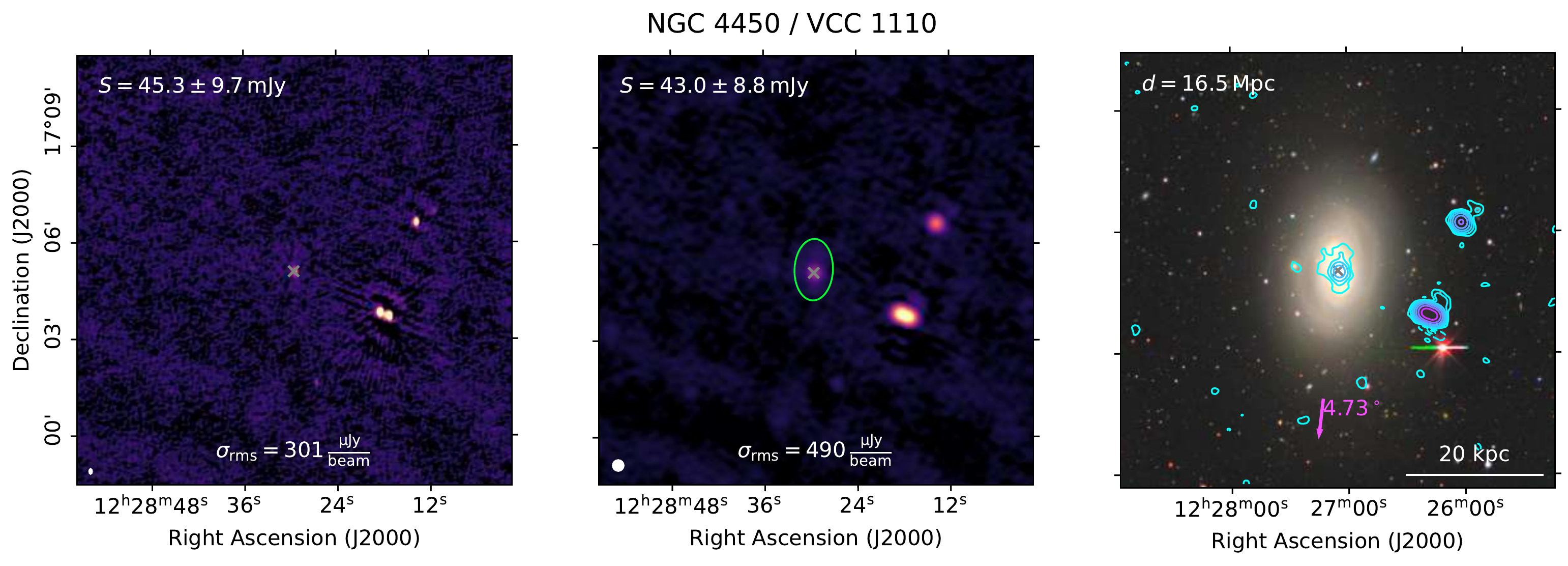}
        \caption{}
    \end{subfigure}
     \hfill
    \begin{subfigure}[b]{\textwidth}
        \includegraphics[width=\textwidth]{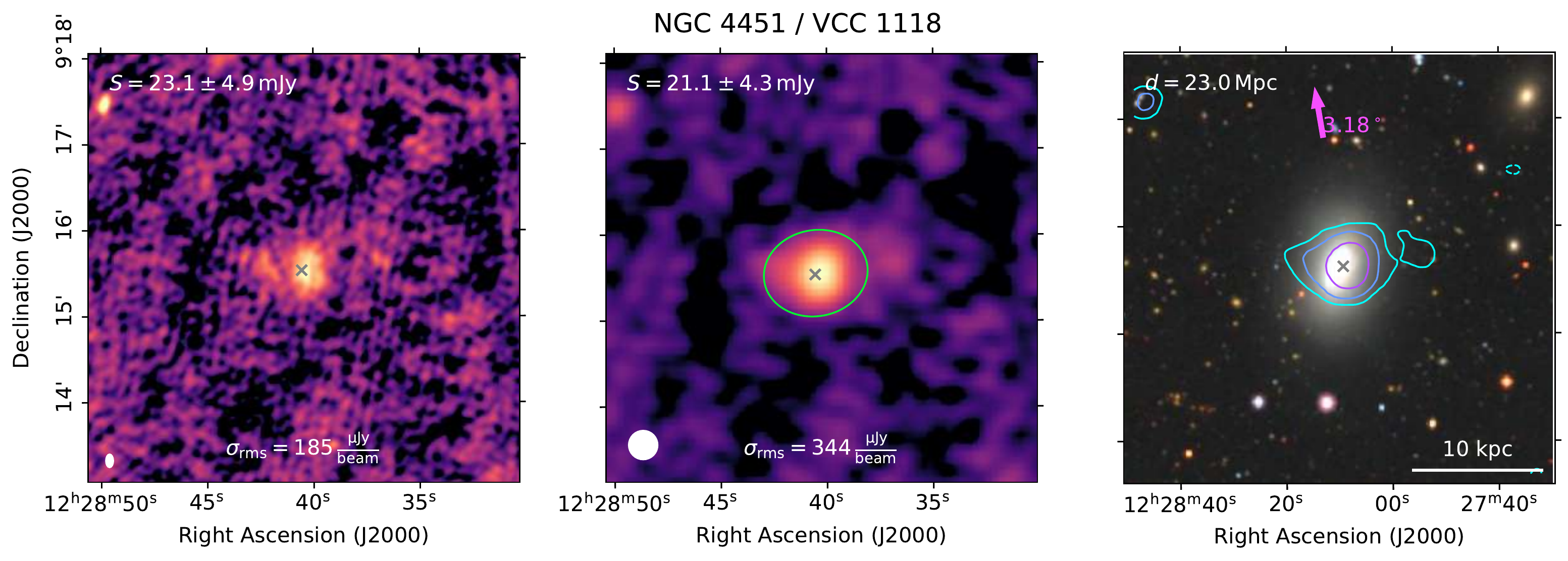}
        \caption{}
    \end{subfigure} 
    \caption{Same as \autoref{fig:144first}.}
\end{figure}

\begin{figure}
    \centering
    \begin{subfigure}[b]{\textwidth}
        \includegraphics[width=\textwidth]{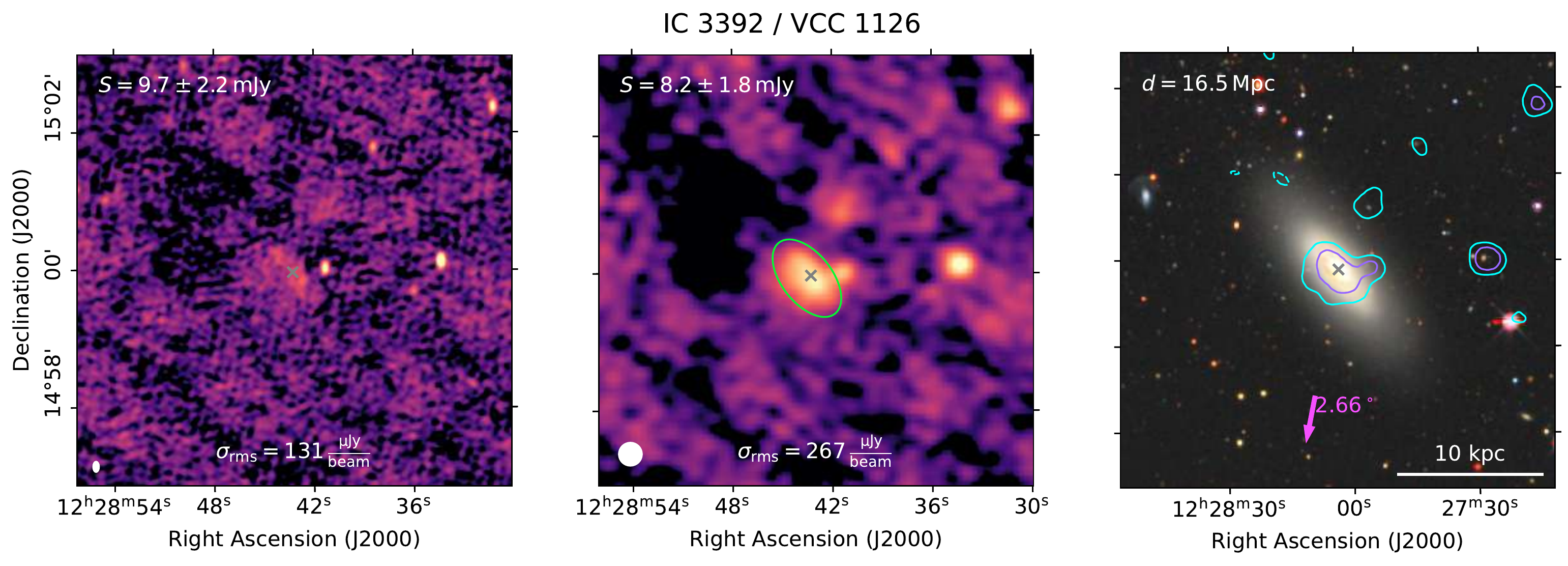}
        \caption{}
    \end{subfigure} 
     \hfill
    \begin{subfigure}[b]{\textwidth}
        \includegraphics[width=\textwidth]{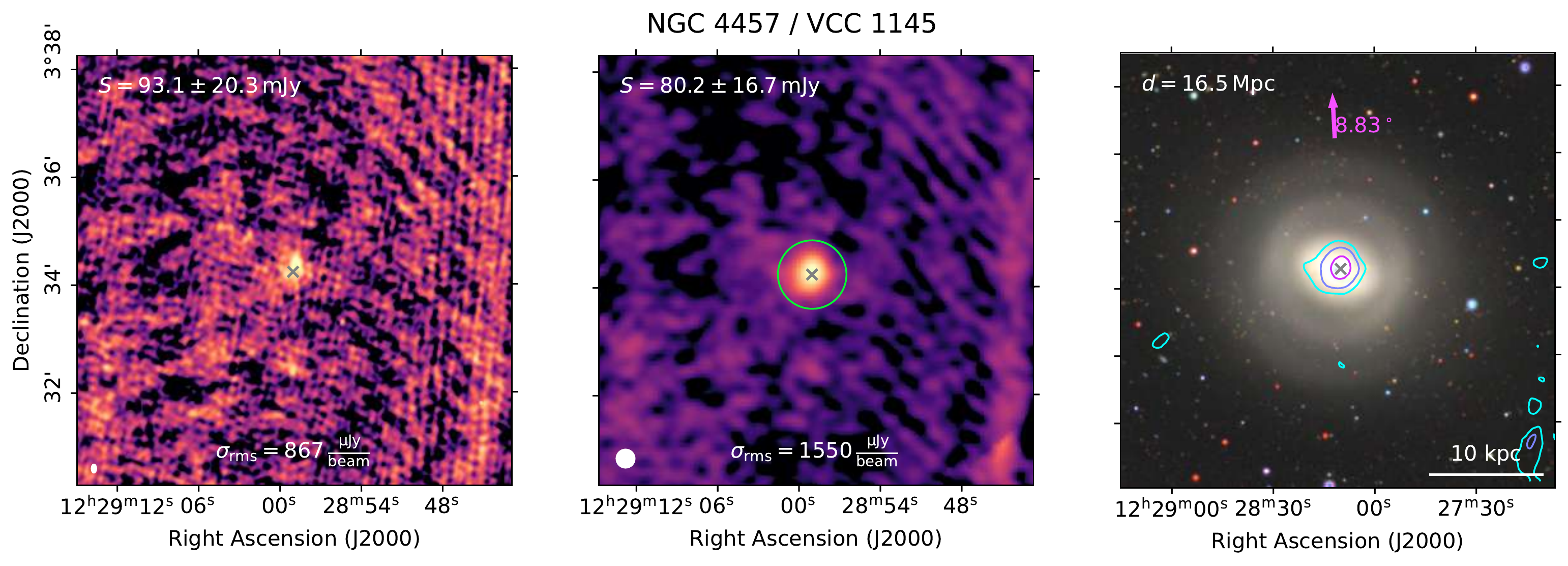}
        \caption{}
    \end{subfigure}
     \hfill
    \begin{subfigure}[b]{\textwidth}
        \includegraphics[width=\textwidth]{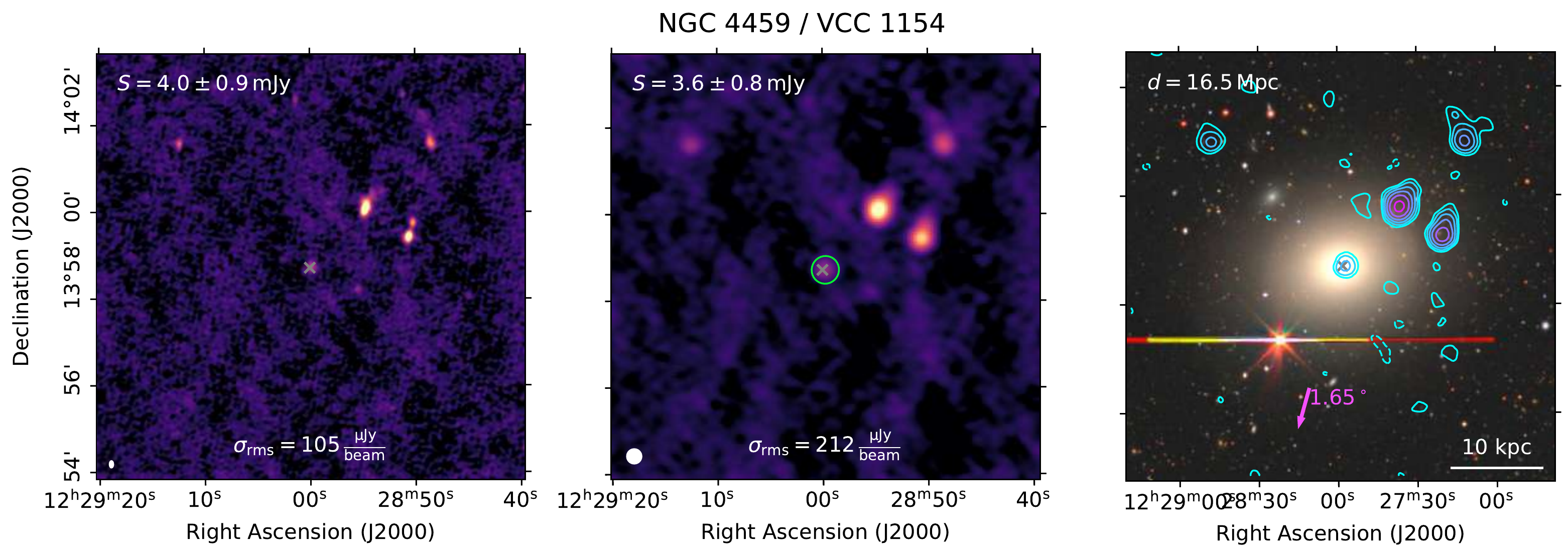}
        \caption{}
    \end{subfigure} 
    \caption{Same as \autoref{fig:144first}.}
\end{figure}

\begin{figure}
    \centering
    \begin{subfigure}[b]{\textwidth}
        \includegraphics[width=\textwidth]{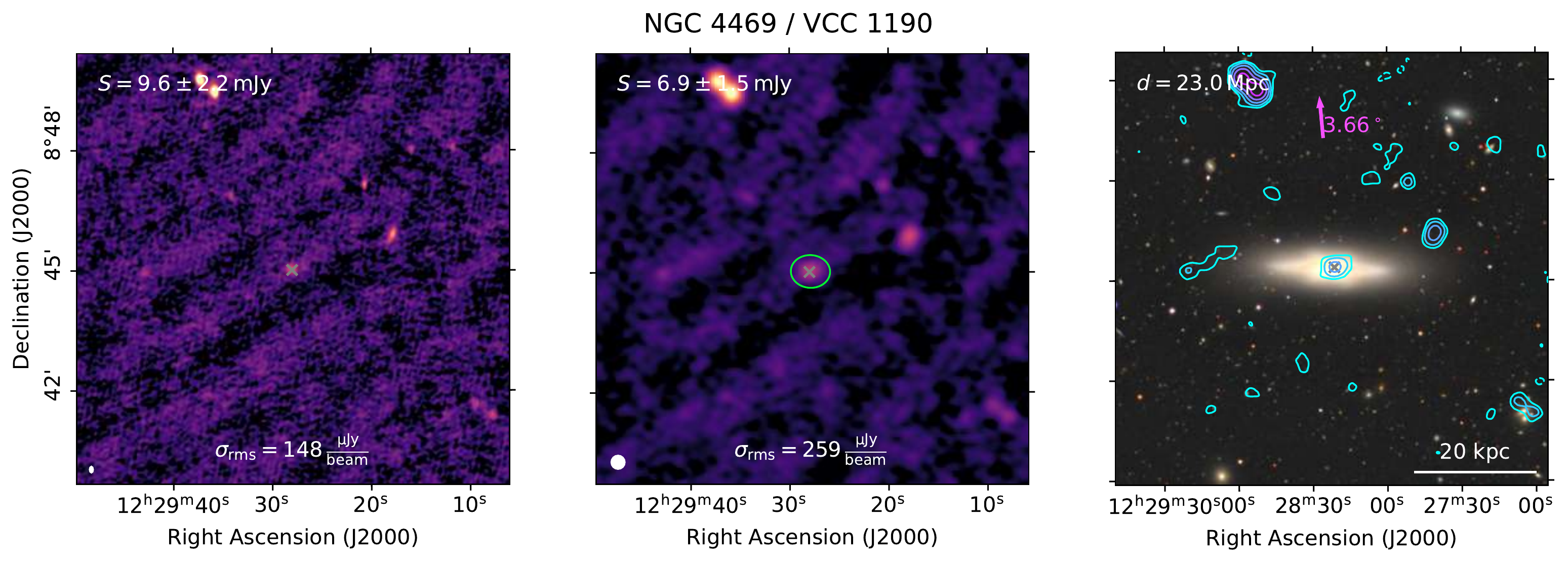}
        \caption{}
    \end{subfigure} 
     \hfill
    \begin{subfigure}[b]{\textwidth}
        \includegraphics[width=\textwidth]{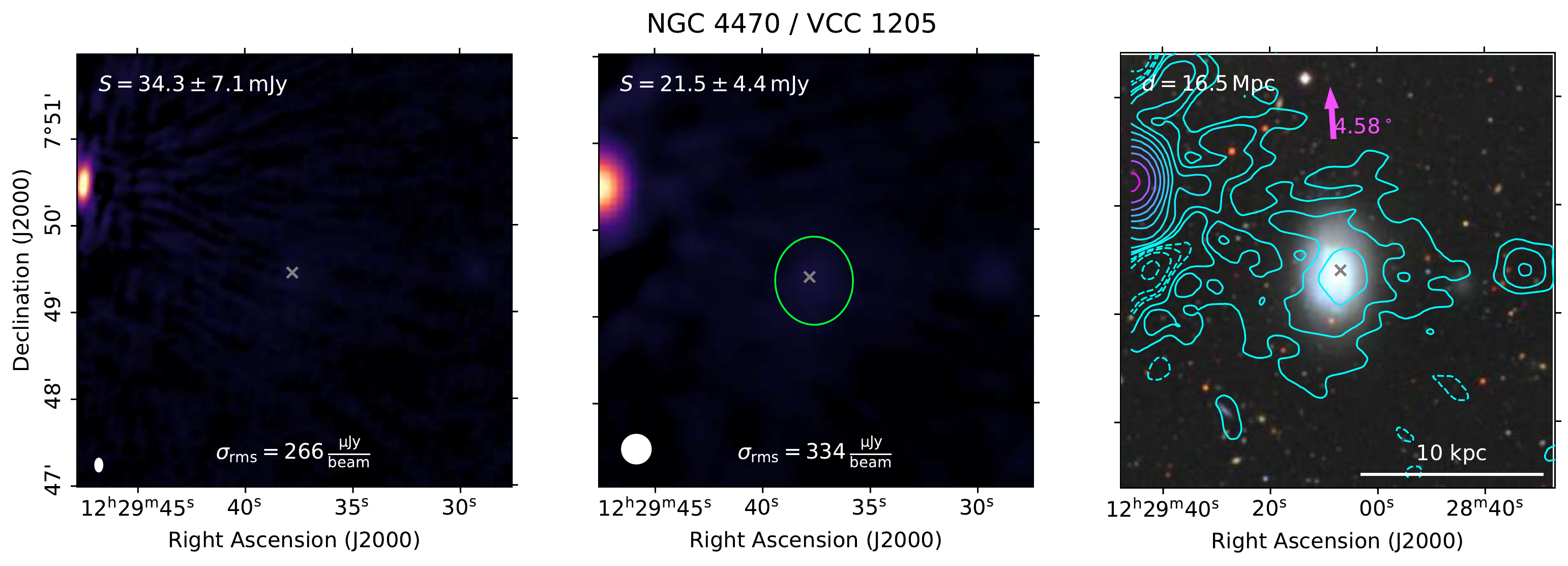}
        \caption{}
    \end{subfigure}
     \hfill
    \begin{subfigure}[b]{\textwidth}
        \includegraphics[width=\textwidth]{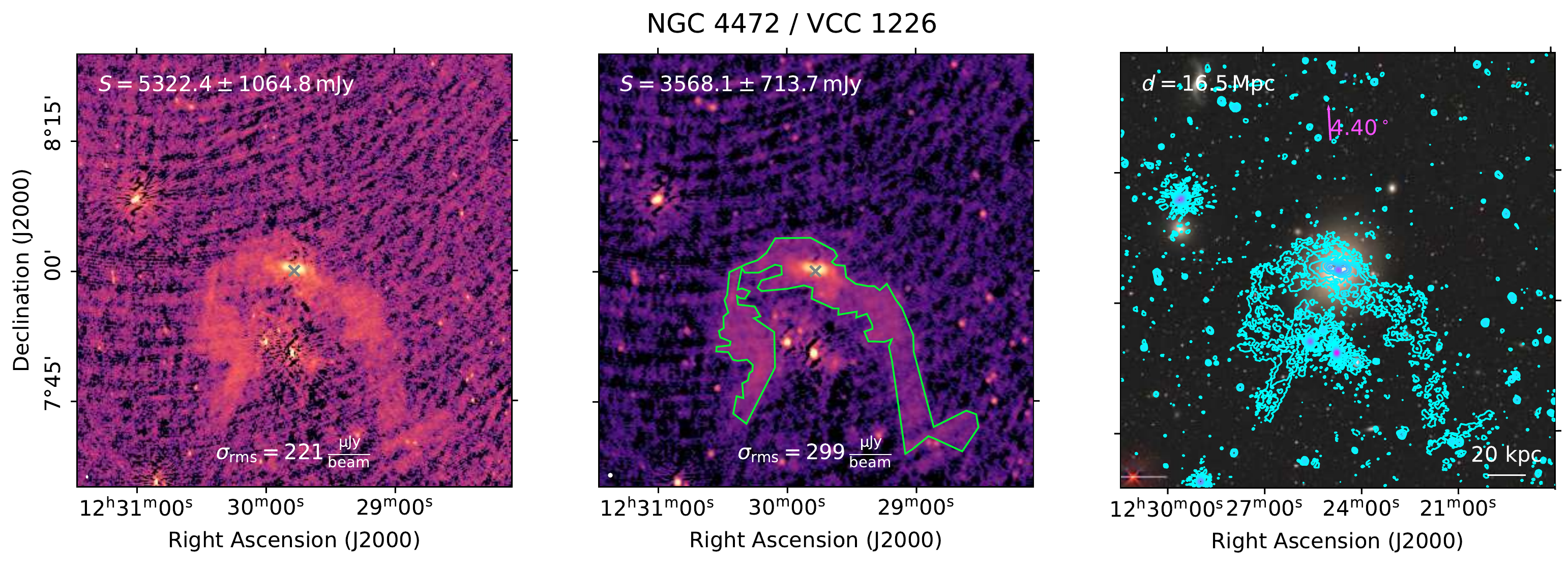}
        \caption{}
        \label{fig:1226}
    \end{subfigure} 
    \caption{Same as \autoref{fig:144first}.}
\end{figure}

\begin{figure}
    \centering
    \begin{subfigure}[b]{\textwidth}
        \includegraphics[width=\textwidth]{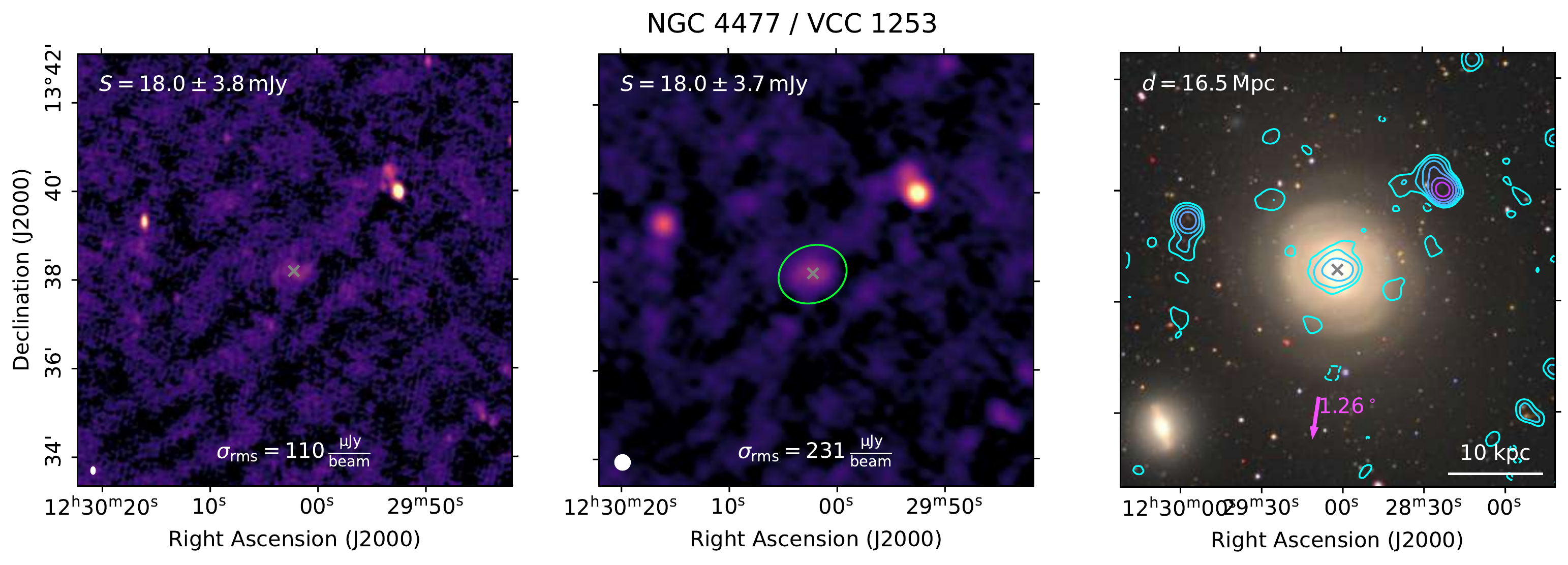}
        \caption{}
    \end{subfigure} 
     \hfill
    \begin{subfigure}[b]{\textwidth}
        \includegraphics[width=\textwidth]{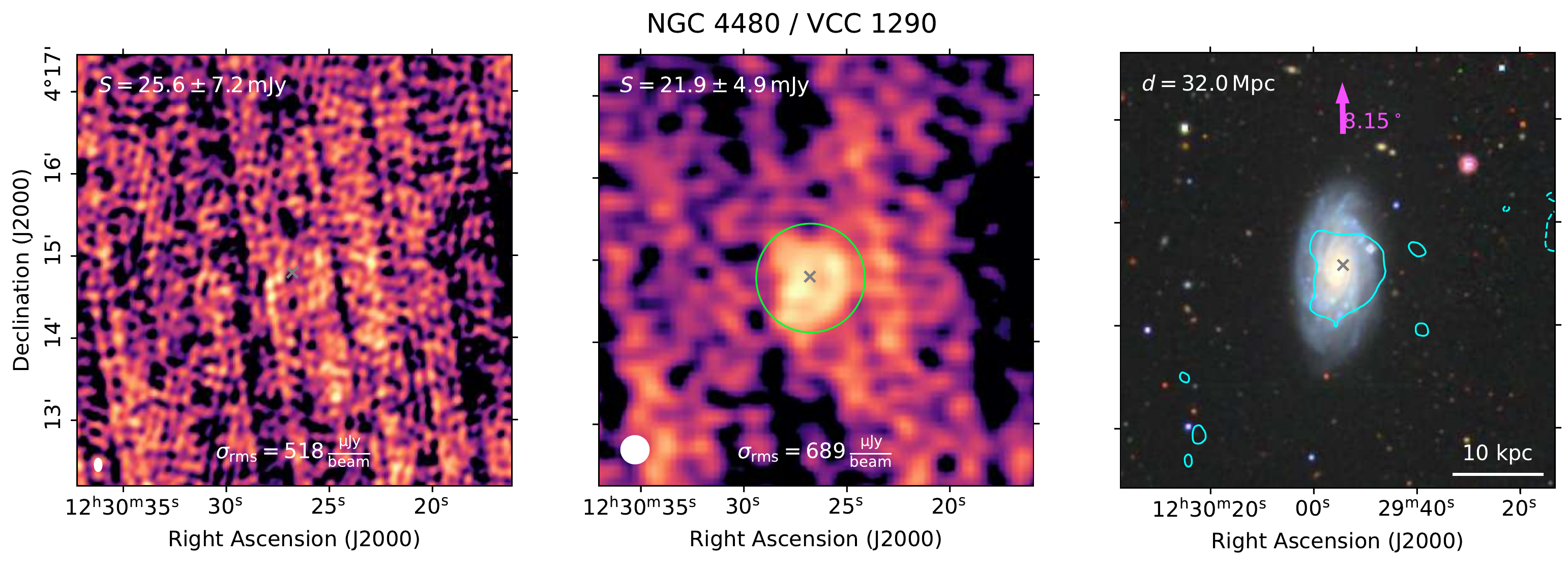}
        \caption{}
    \end{subfigure}
     \hfill
    \begin{subfigure}[b]{\textwidth}
        \includegraphics[width=\textwidth]{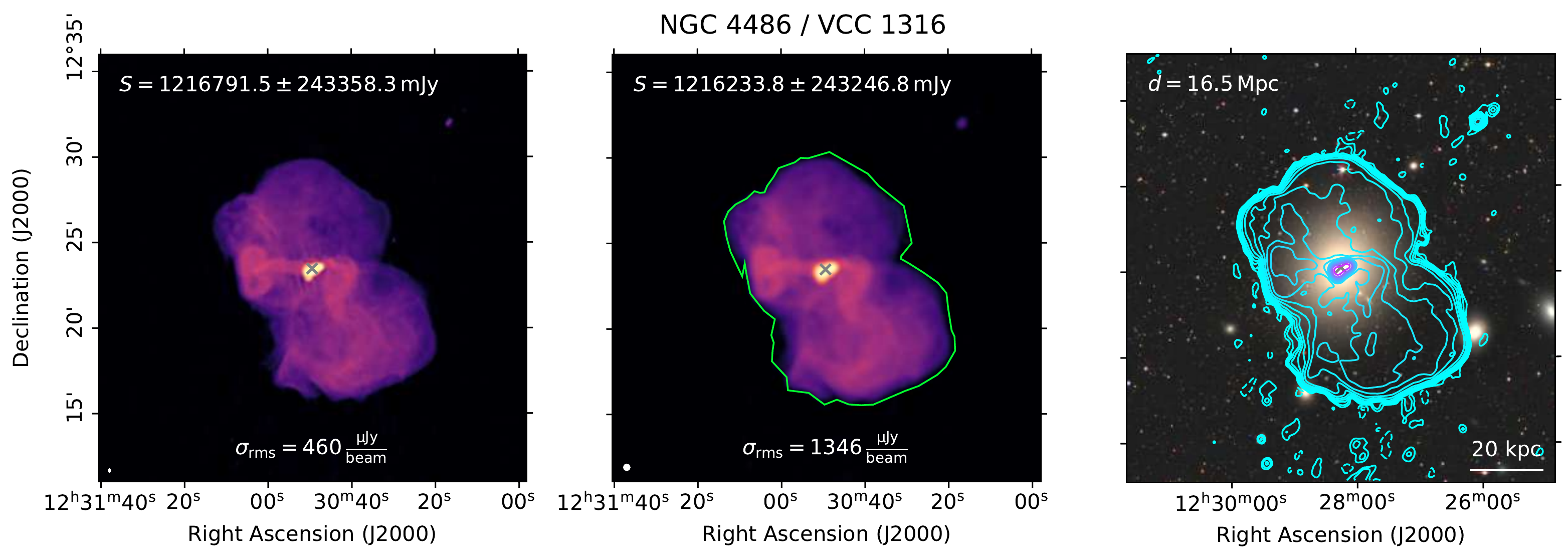}
        \caption{}
        \label{fig:1316}
    \end{subfigure} 
    \caption{Same as \autoref{fig:144first}.}
\end{figure}

\begin{figure}
    \centering
    \begin{subfigure}[b]{\textwidth}
        \includegraphics[width=\textwidth]{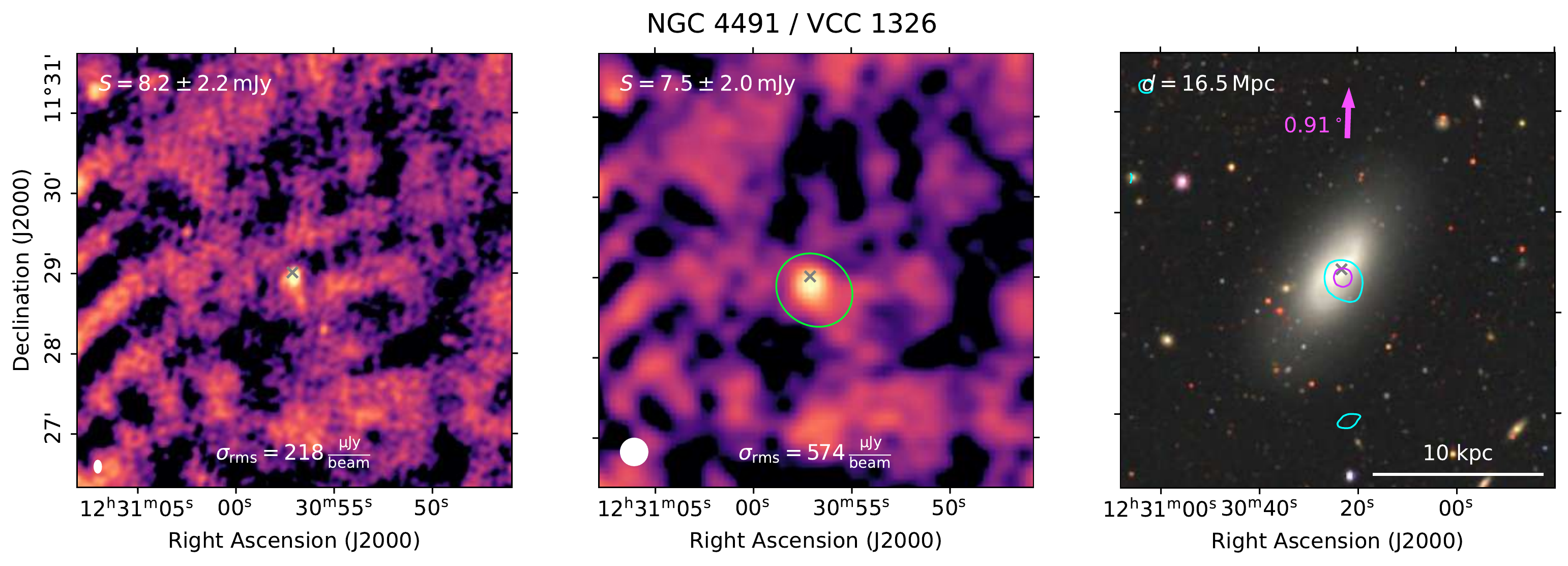}
        \caption{}
    \end{subfigure} 
     \hfill
    \begin{subfigure}[b]{\textwidth}
        \includegraphics[width=\textwidth]{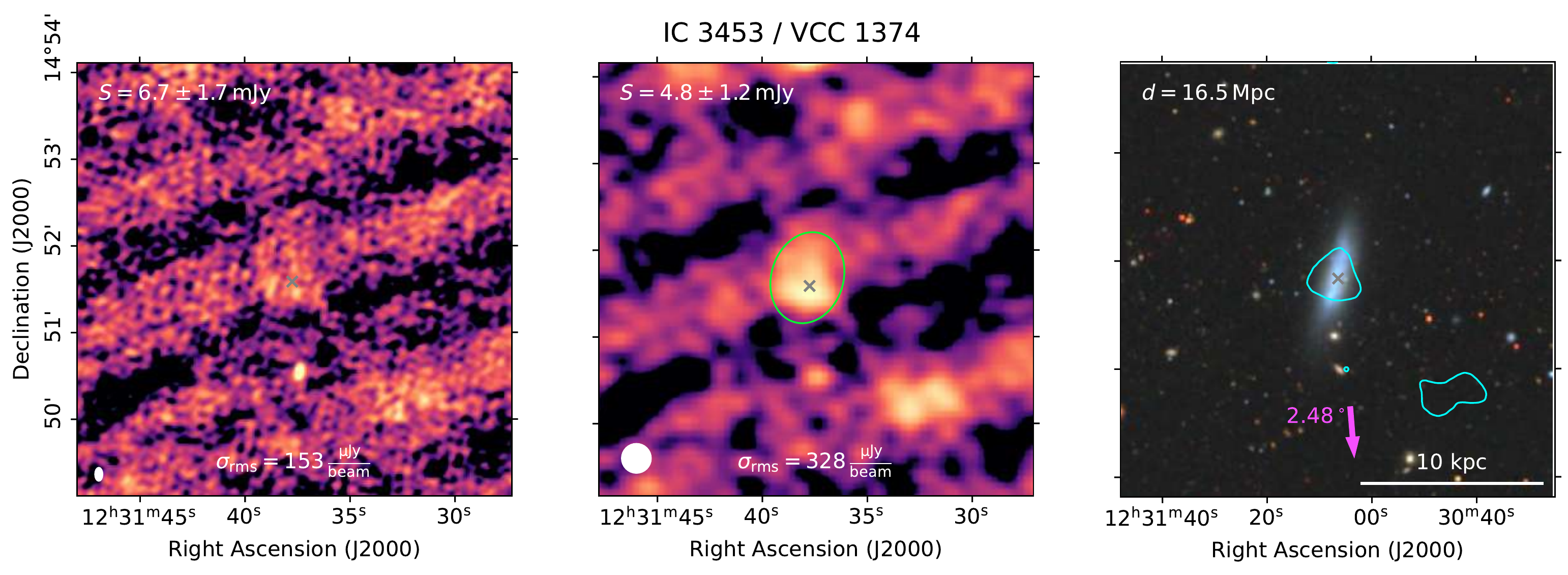}
        \caption{}
    \end{subfigure}
     \hfill
    \begin{subfigure}[b]{\textwidth}
        \includegraphics[width=\textwidth]{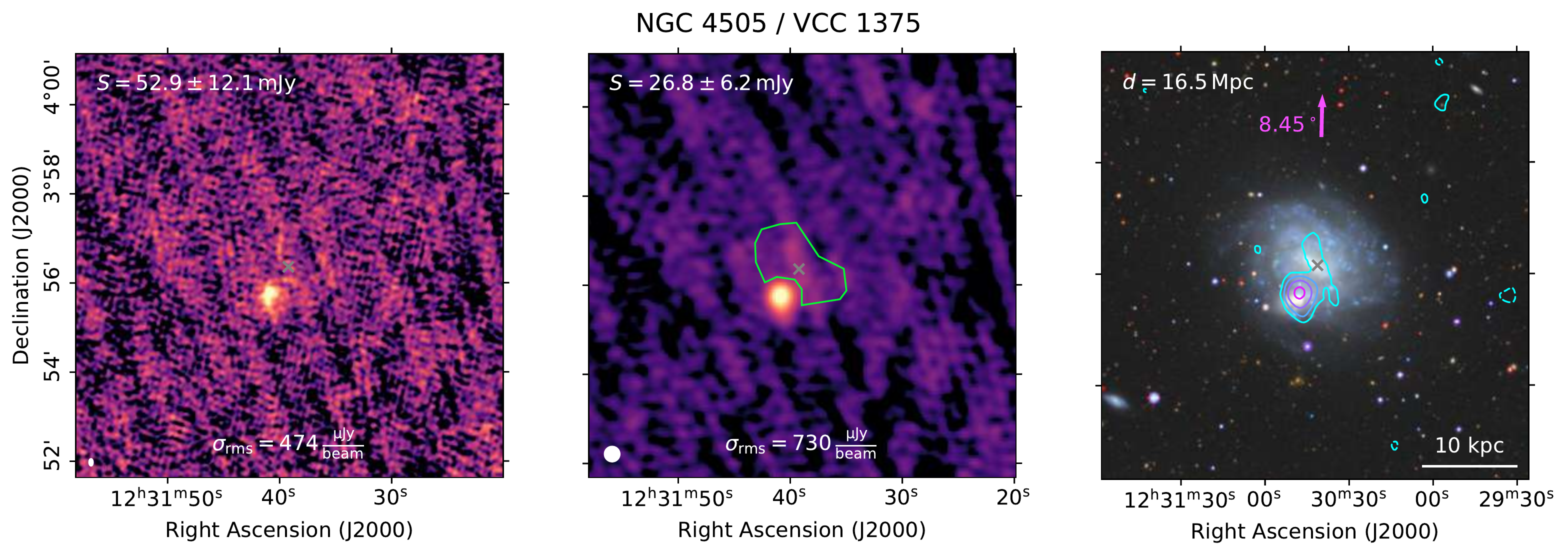}
        \caption{}
    \end{subfigure} 
    \caption{Same as \autoref{fig:144first}.}
\end{figure}

\begin{figure}
    \centering
    \begin{subfigure}[b]{\textwidth}
        \includegraphics[width=\textwidth]{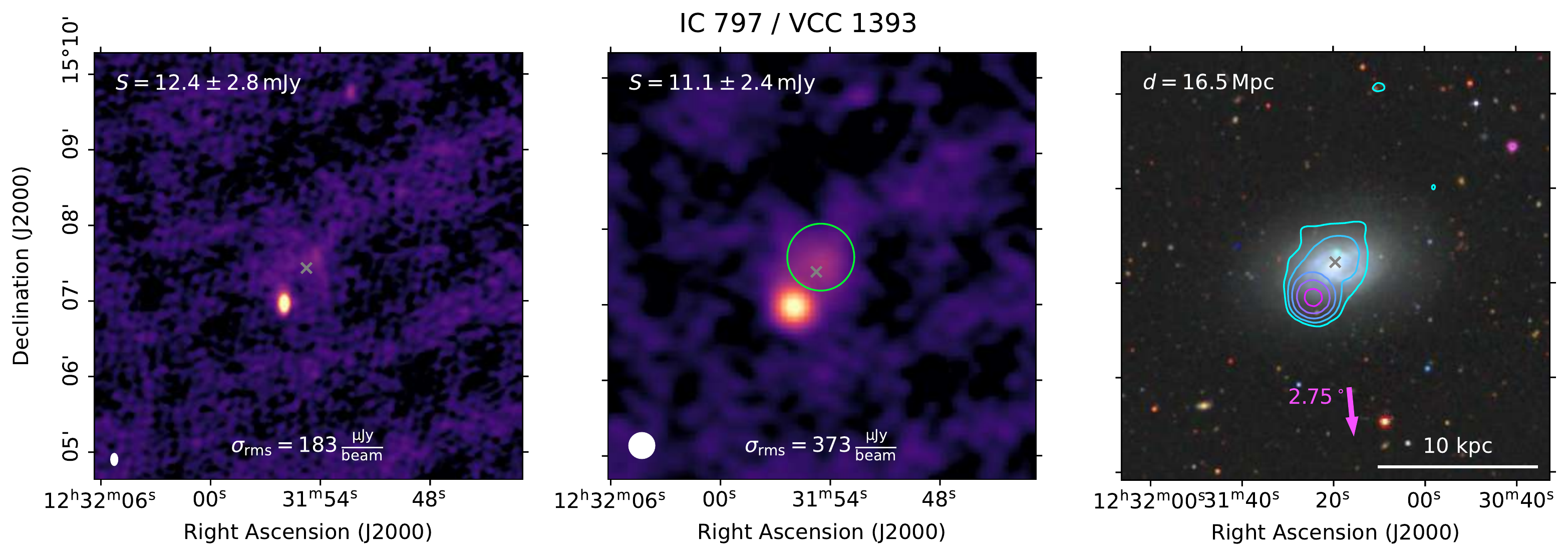}
        \caption{}
    \end{subfigure} 
     \hfill
    \begin{subfigure}[b]{\textwidth}
        \includegraphics[width=\textwidth]{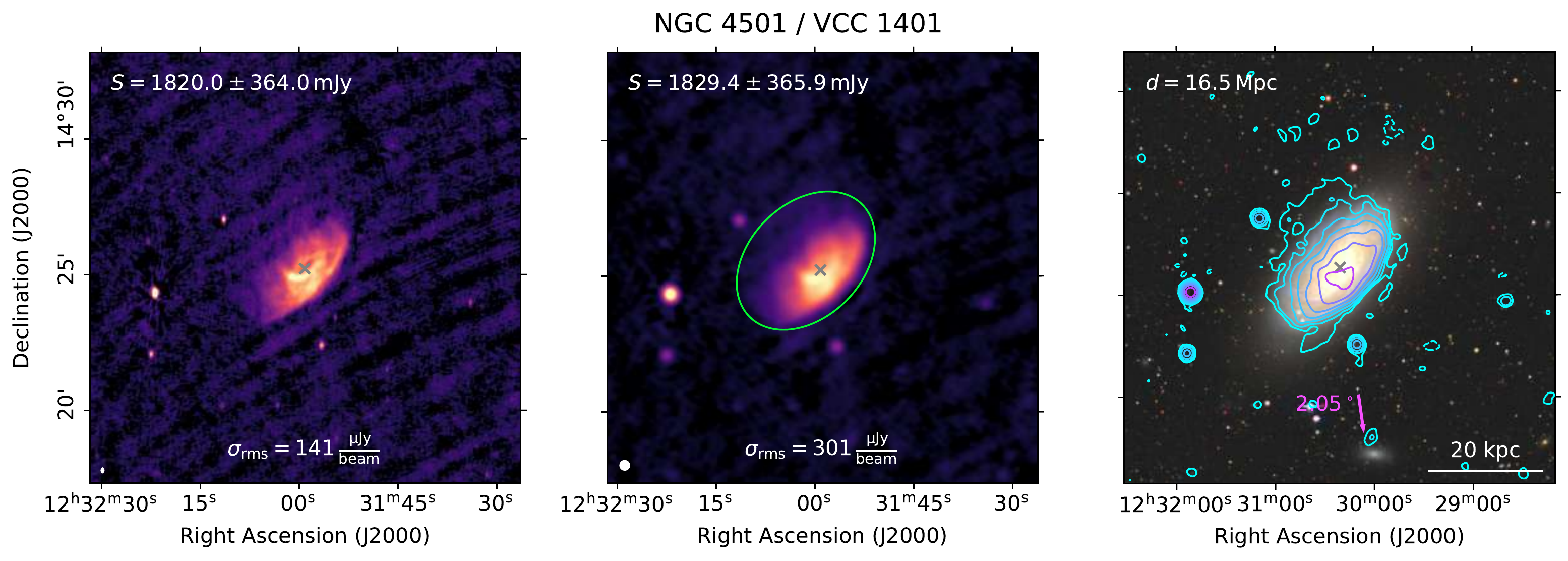}
        \caption{}
        \label{fig:1401}
    \end{subfigure}
     \hfill
    \begin{subfigure}[b]{\textwidth}
        \includegraphics[width=\textwidth]{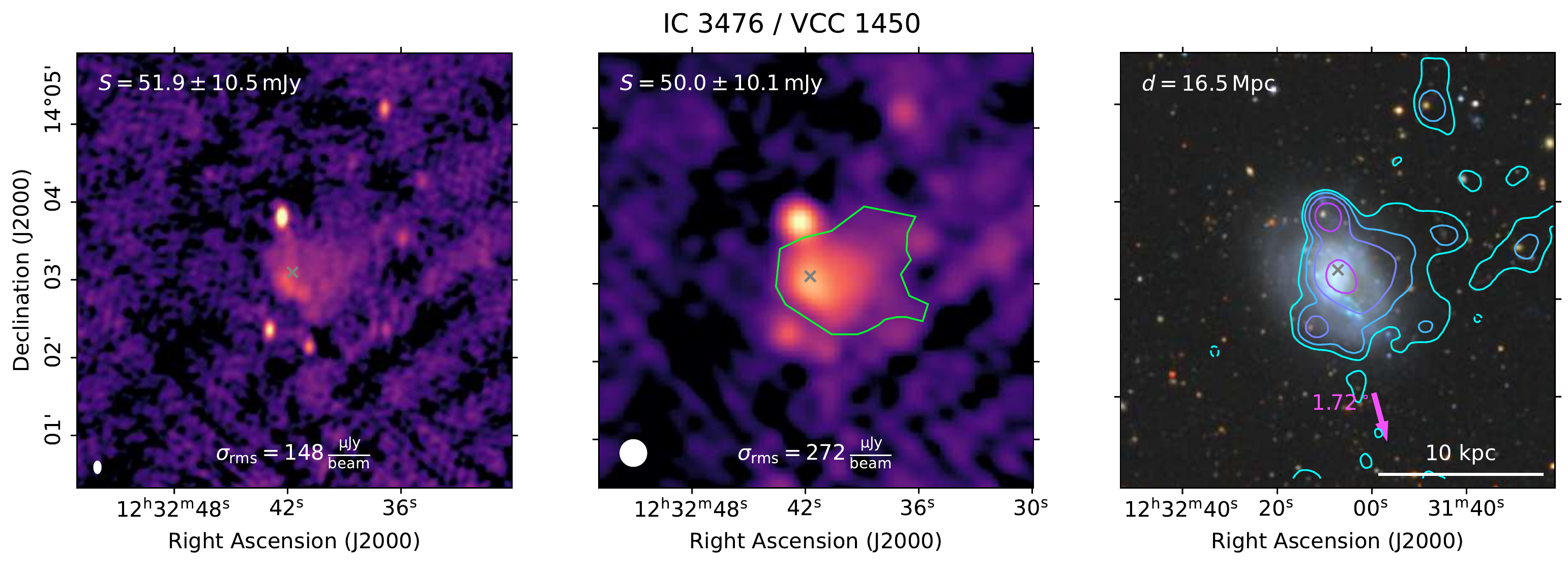}
        \caption{}
        \label{fig:1450}
    \end{subfigure} 
    \caption{Same as \autoref{fig:144first}.}
\end{figure}

\begin{figure}
    \centering
    \begin{subfigure}[b]{\textwidth}
        \includegraphics[width=\textwidth]{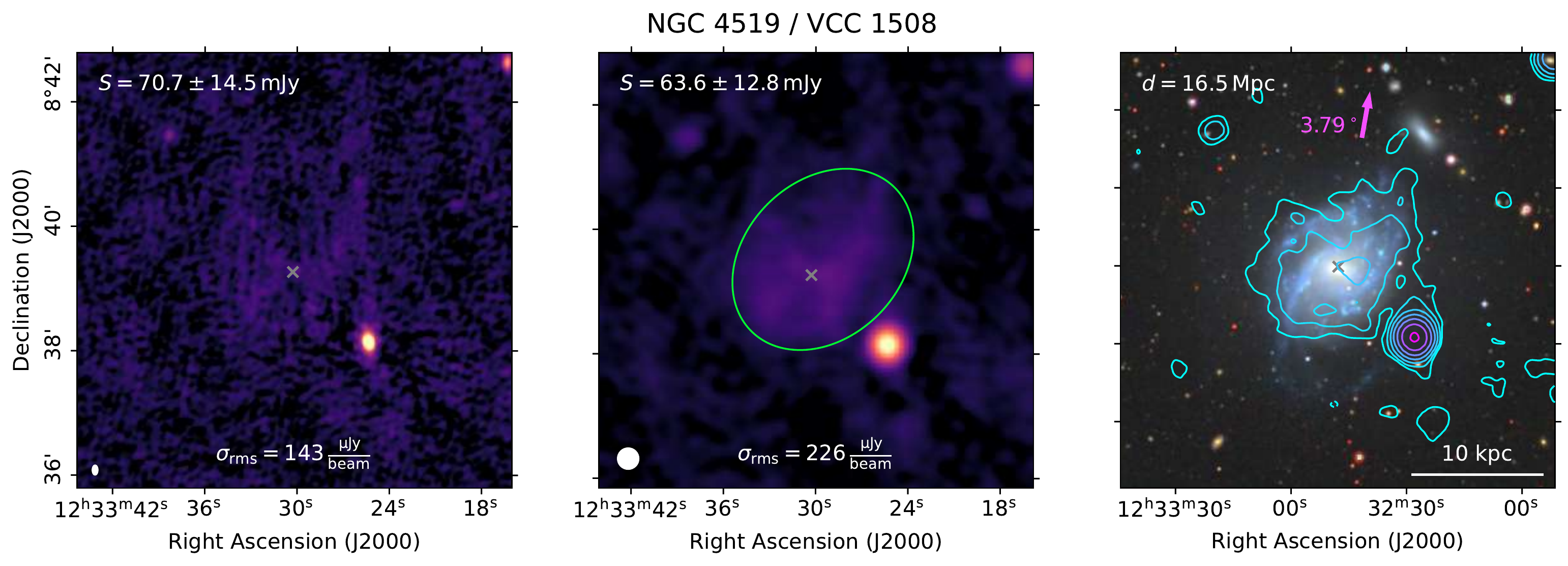}
        \caption{}
    \end{subfigure} 
     \hfill
    \begin{subfigure}[b]{\textwidth}
        \includegraphics[width=\textwidth]{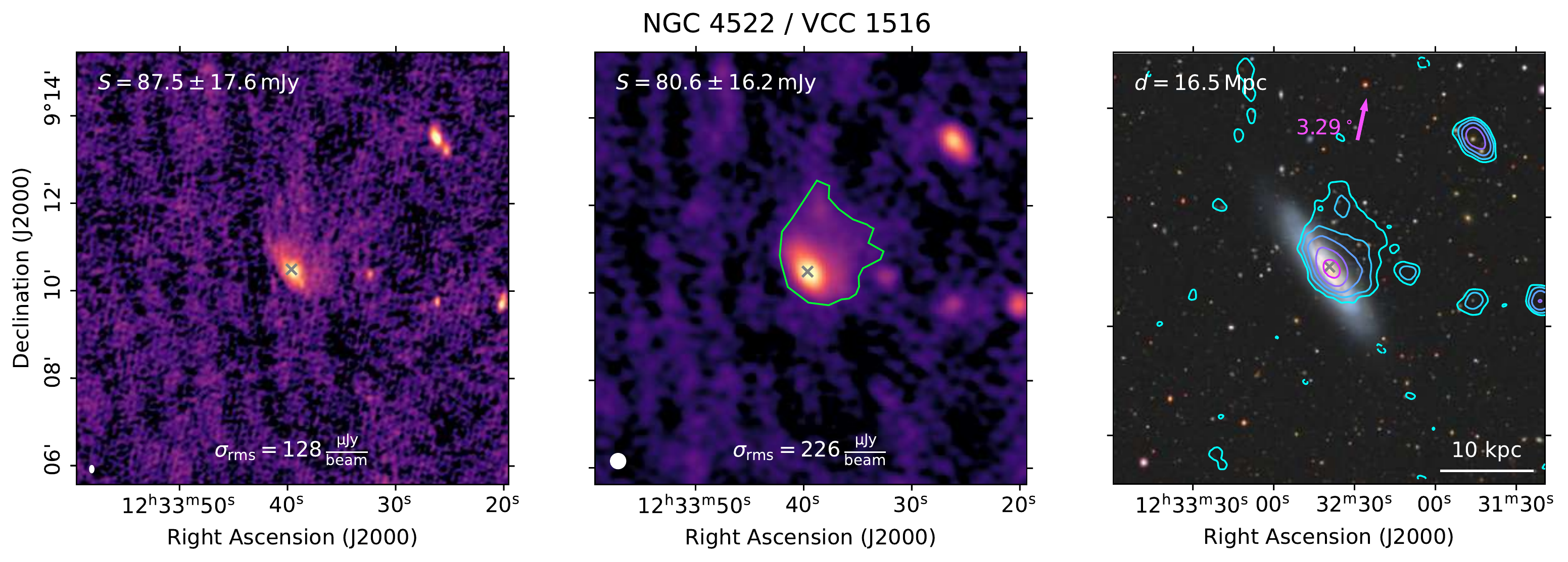}
        \caption{}
    \end{subfigure}
     \hfill
    \begin{subfigure}[b]{\textwidth}
        \includegraphics[width=\textwidth]{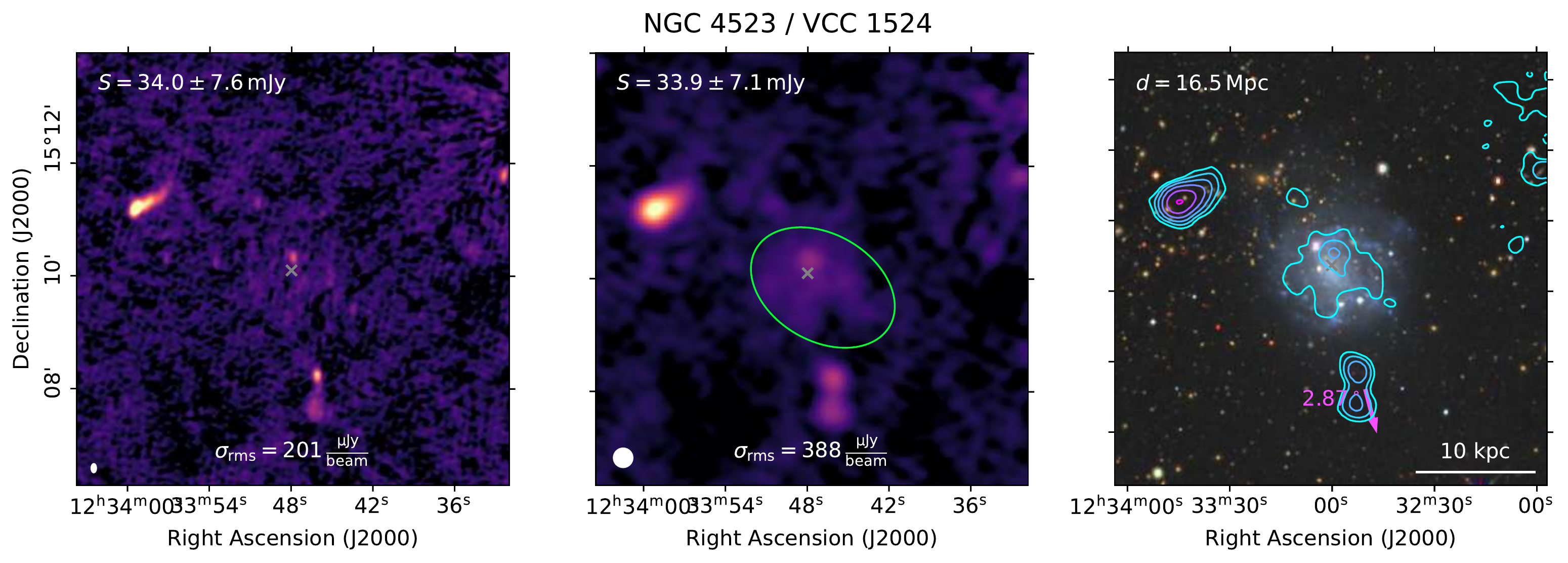}
        \caption{}
    \end{subfigure} 
    \caption{Same as \autoref{fig:144first}.}
\end{figure}

\begin{figure}
    \centering
    \begin{subfigure}[b]{\textwidth}
        \includegraphics[width=\textwidth]{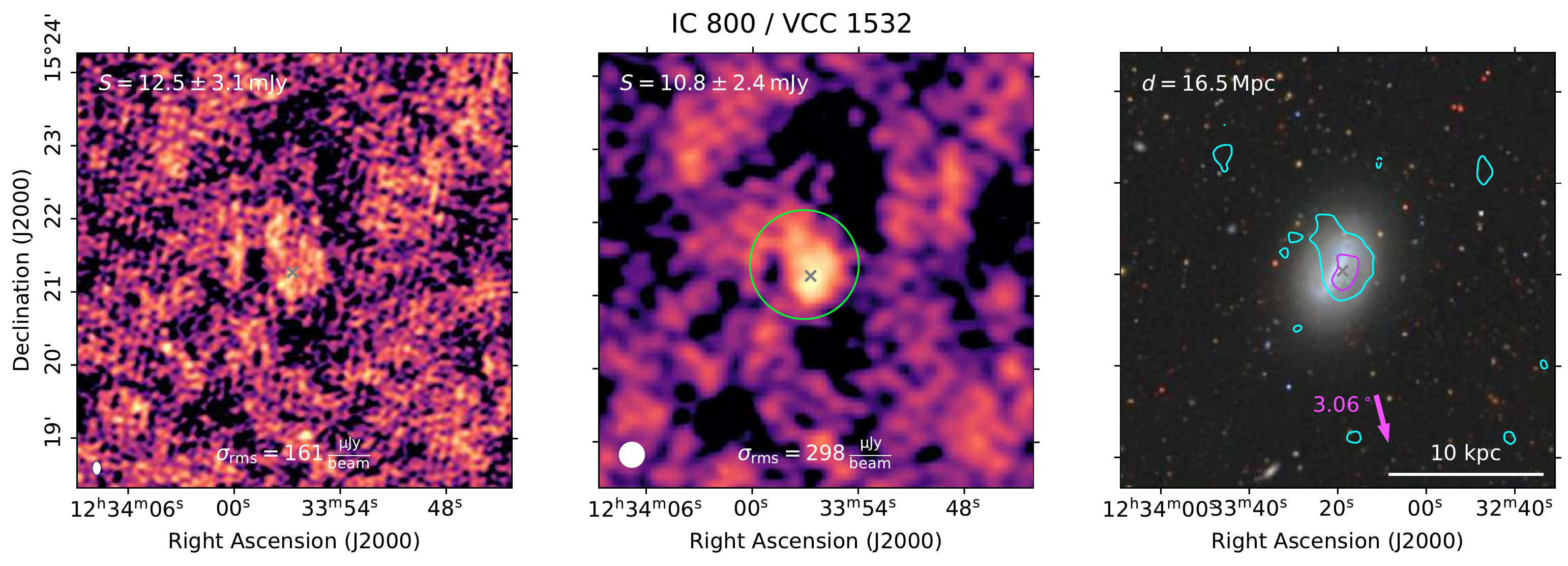}
        \caption{}
    \end{subfigure} 
     \hfill
    \begin{subfigure}[b]{\textwidth}
        \includegraphics[width=\textwidth]{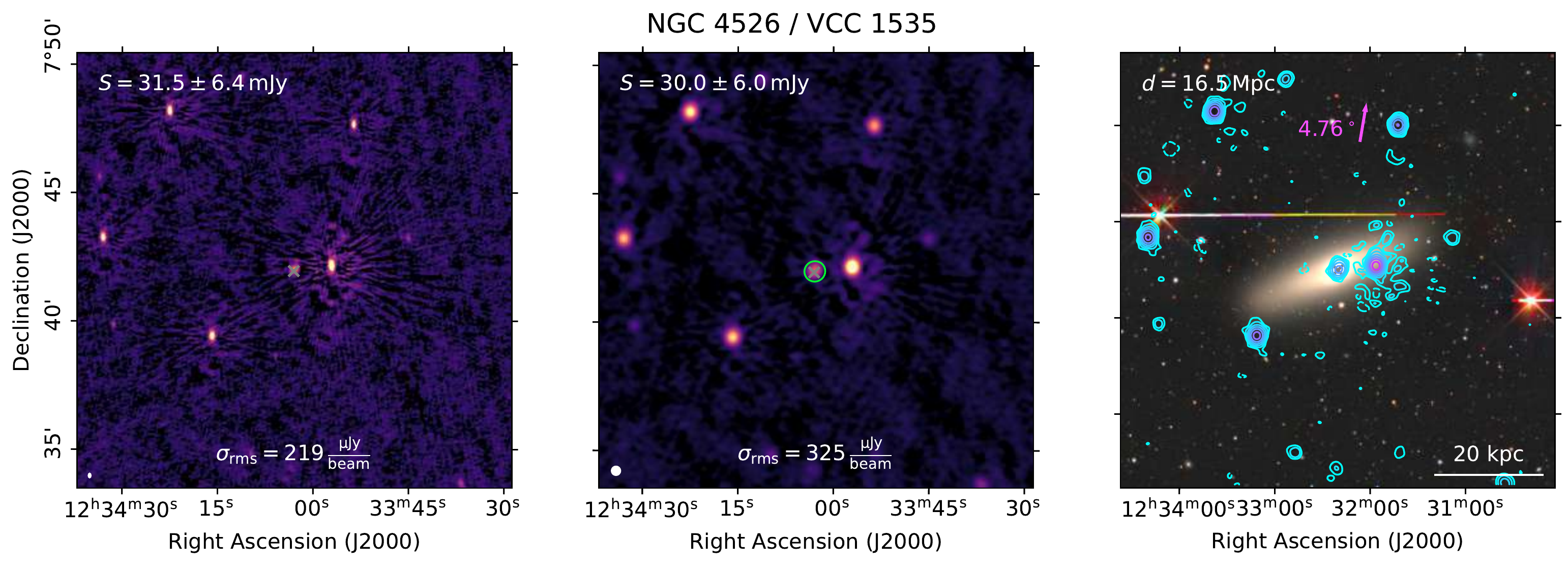}
        \caption{}
    \end{subfigure}
     \hfill
    \begin{subfigure}[b]{\textwidth}
        \includegraphics[width=\textwidth]{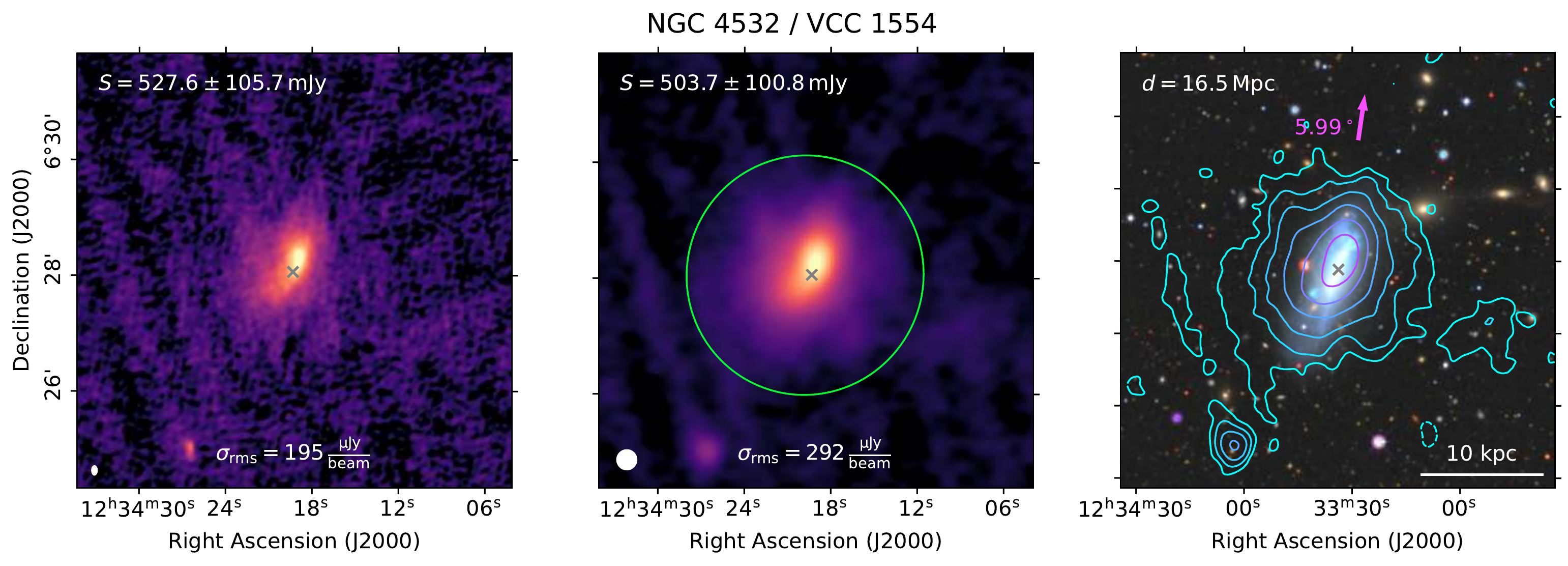}
        \caption{}
    \end{subfigure} 
    \caption{Same as \autoref{fig:144first}.}
\end{figure}

\begin{figure}
    \centering
    \begin{subfigure}[b]{\textwidth}
        \includegraphics[width=\textwidth]{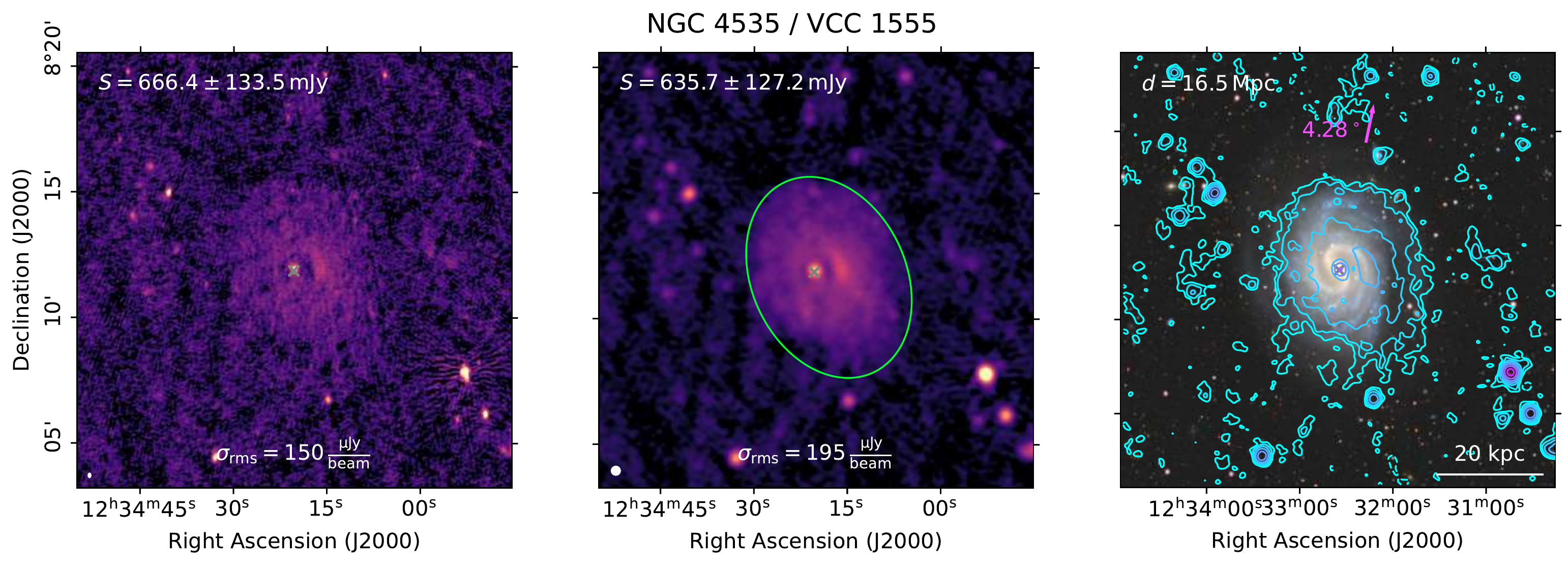}
        \caption{}
    \end{subfigure} 
     \hfill
    \begin{subfigure}[b]{\textwidth}
        \includegraphics[width=\textwidth]{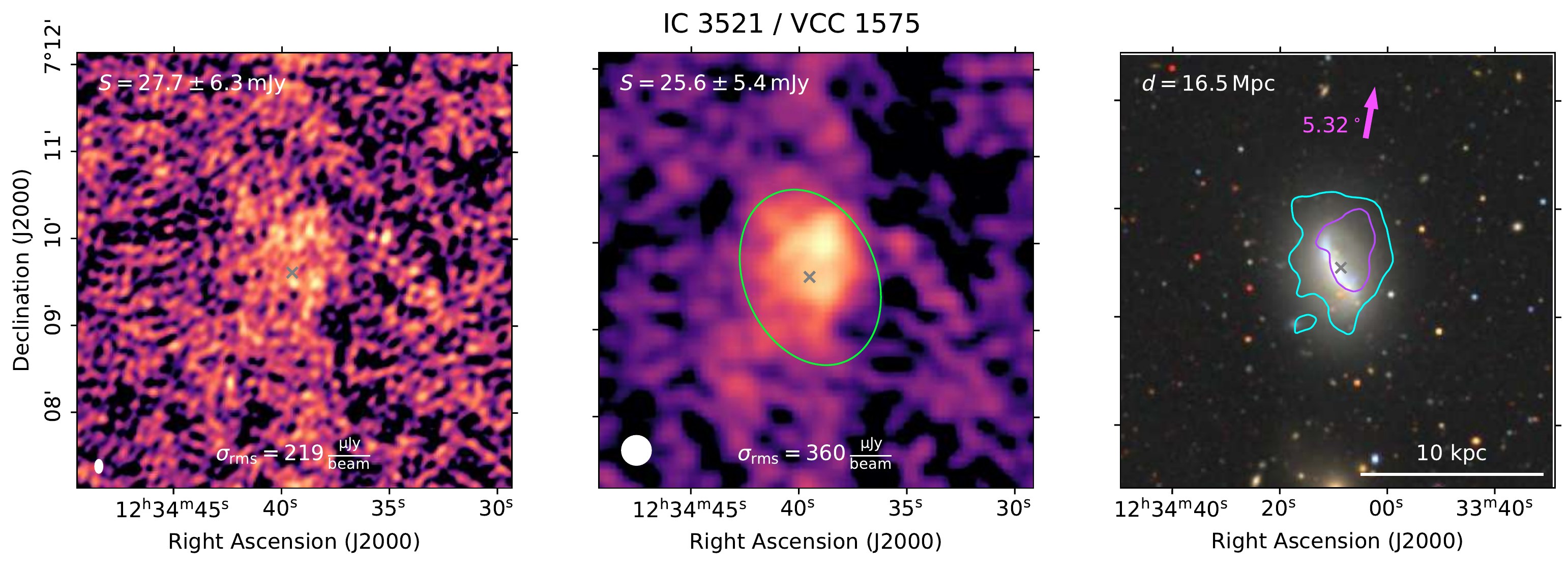}
        \caption{}
    \end{subfigure}
     \hfill
    \begin{subfigure}[b]{\textwidth}
        \includegraphics[width=\textwidth]{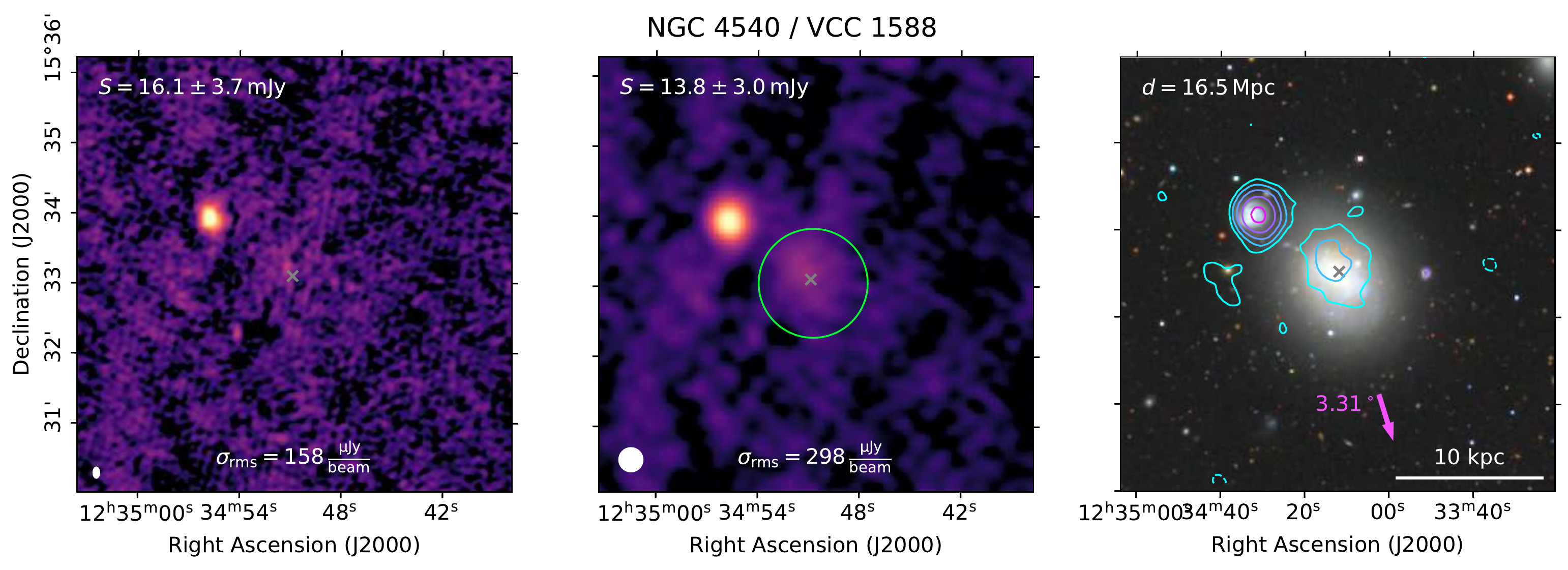}
        \caption{}
    \end{subfigure} 
    \caption{Same as \autoref{fig:144first}.}
\end{figure}

\begin{figure}
    \centering
    \begin{subfigure}[b]{\textwidth}
        \includegraphics[width=\textwidth]{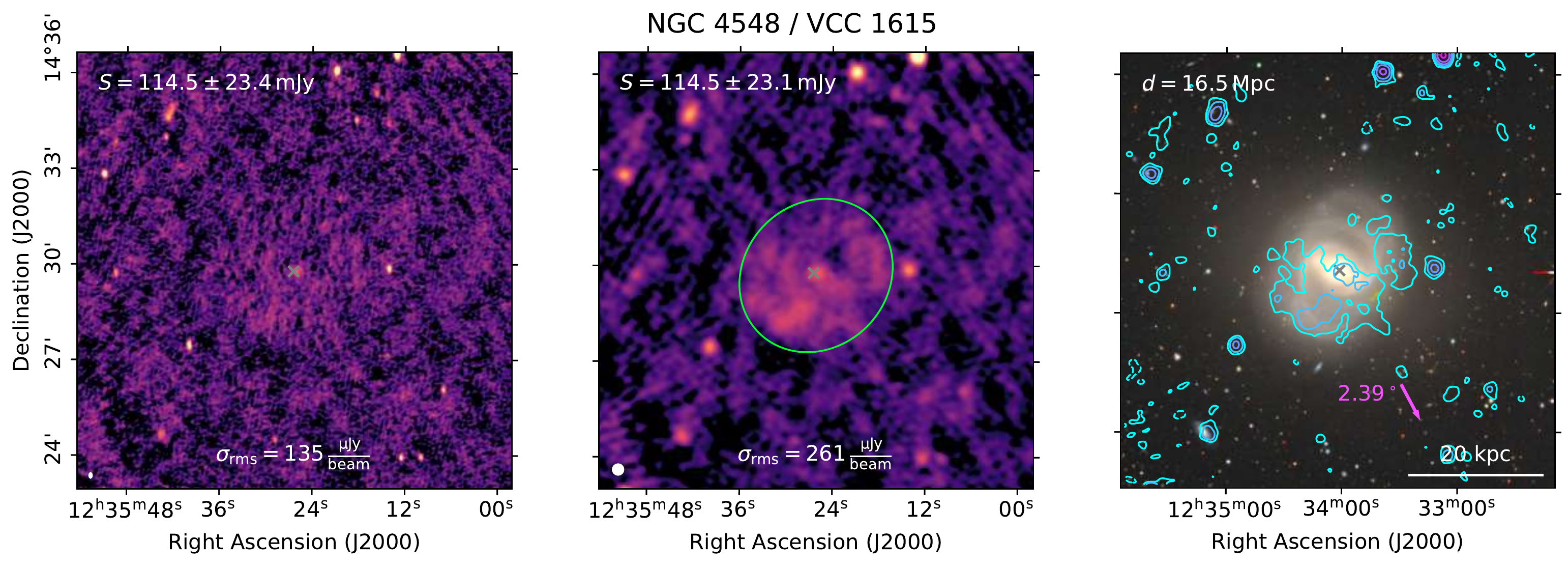}
        \caption{}
    \end{subfigure} 
     \hfill
    \begin{subfigure}[b]{\textwidth}
        \includegraphics[width=\textwidth]{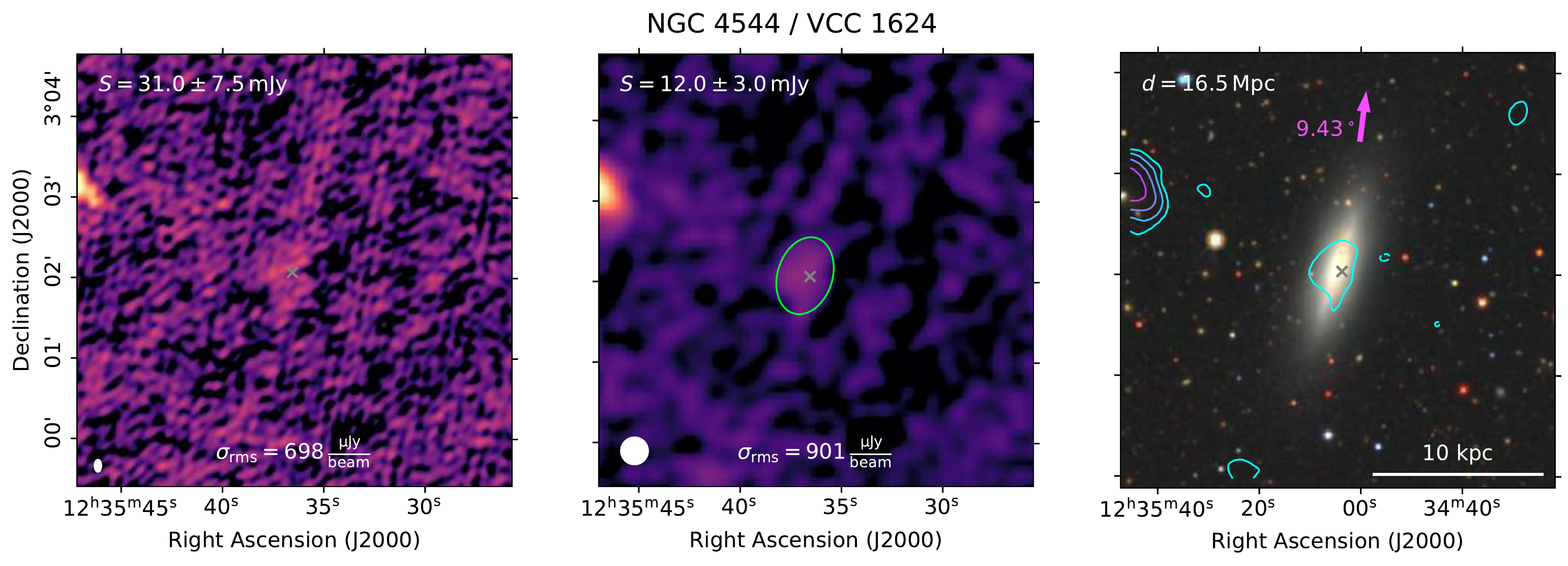}
        \caption{}
    \end{subfigure}
     \hfill
    \begin{subfigure}[b]{\textwidth}
        \includegraphics[width=\textwidth]{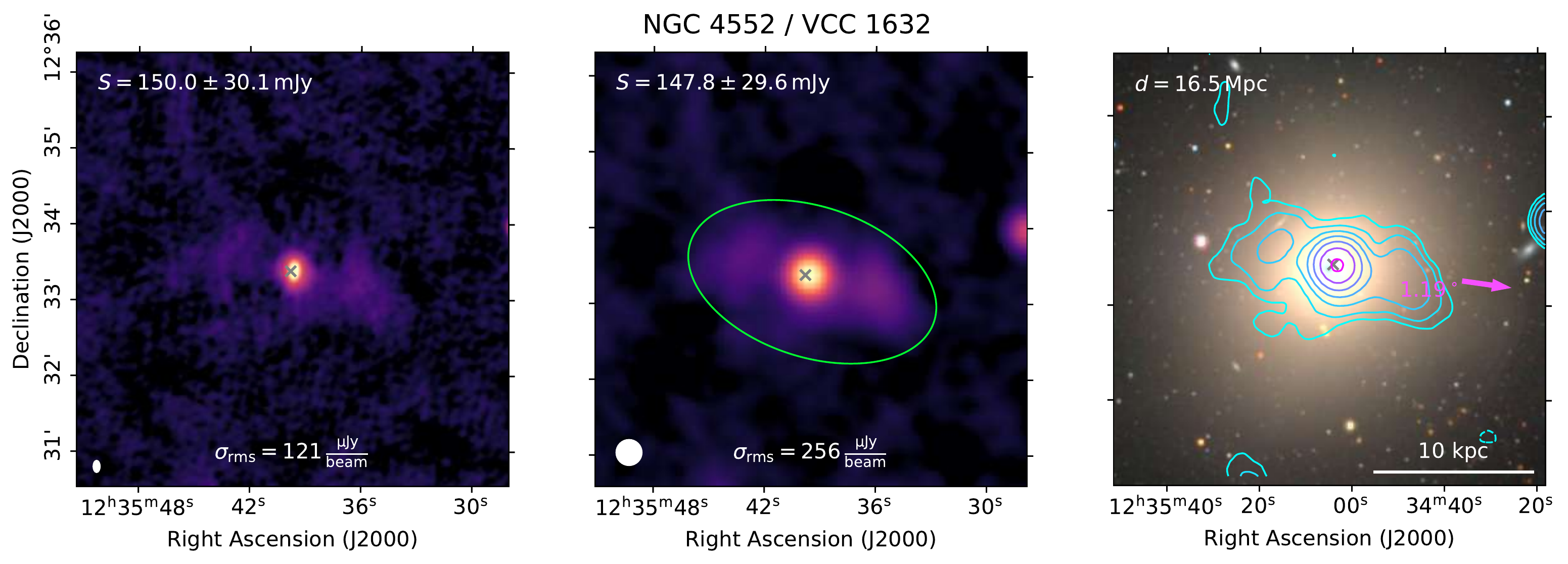}
        \caption{}
        \label{fig:1632}
    \end{subfigure} 
    \caption{Same as \autoref{fig:144first}.}
\end{figure}

\begin{figure}
    \centering
    \begin{subfigure}[b]{\textwidth}
        \includegraphics[width=\textwidth]{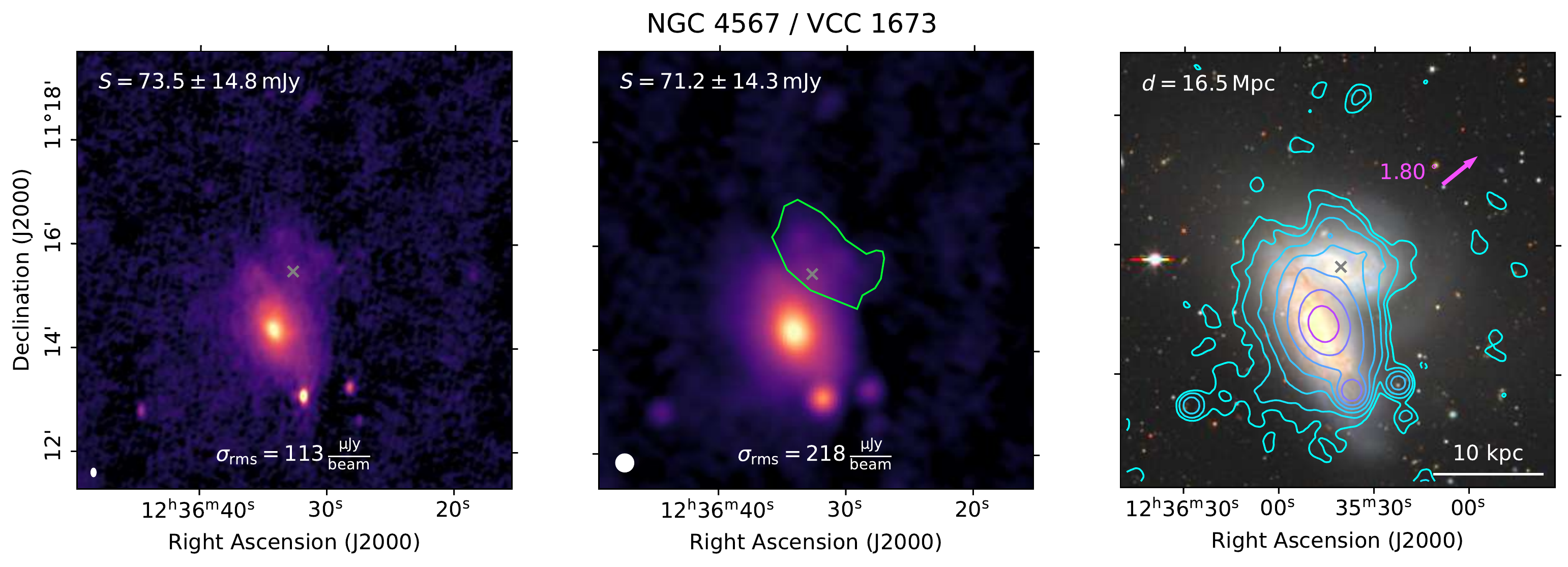}
        \caption{}
    \end{subfigure} 
     \hfill
    \begin{subfigure}[b]{\textwidth}
        \includegraphics[width=\textwidth]{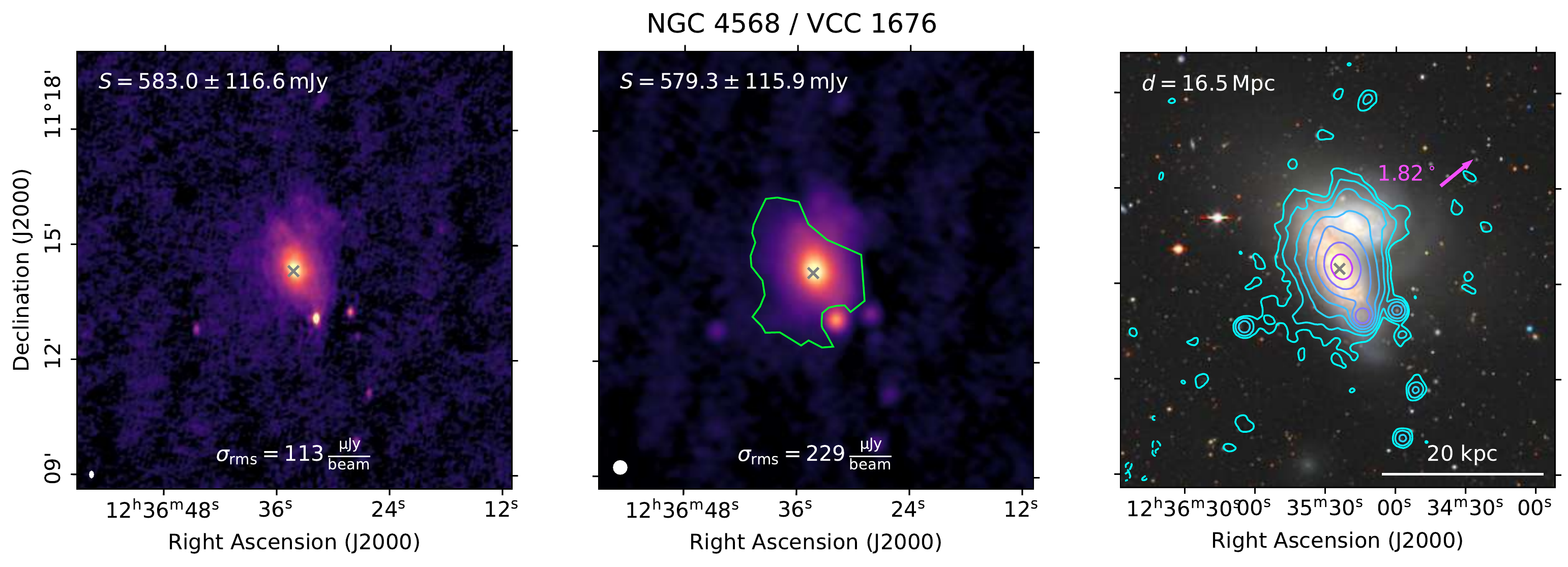}
        \caption{}
    \end{subfigure}
     \hfill
    \begin{subfigure}[b]{\textwidth}
        \includegraphics[width=\textwidth]{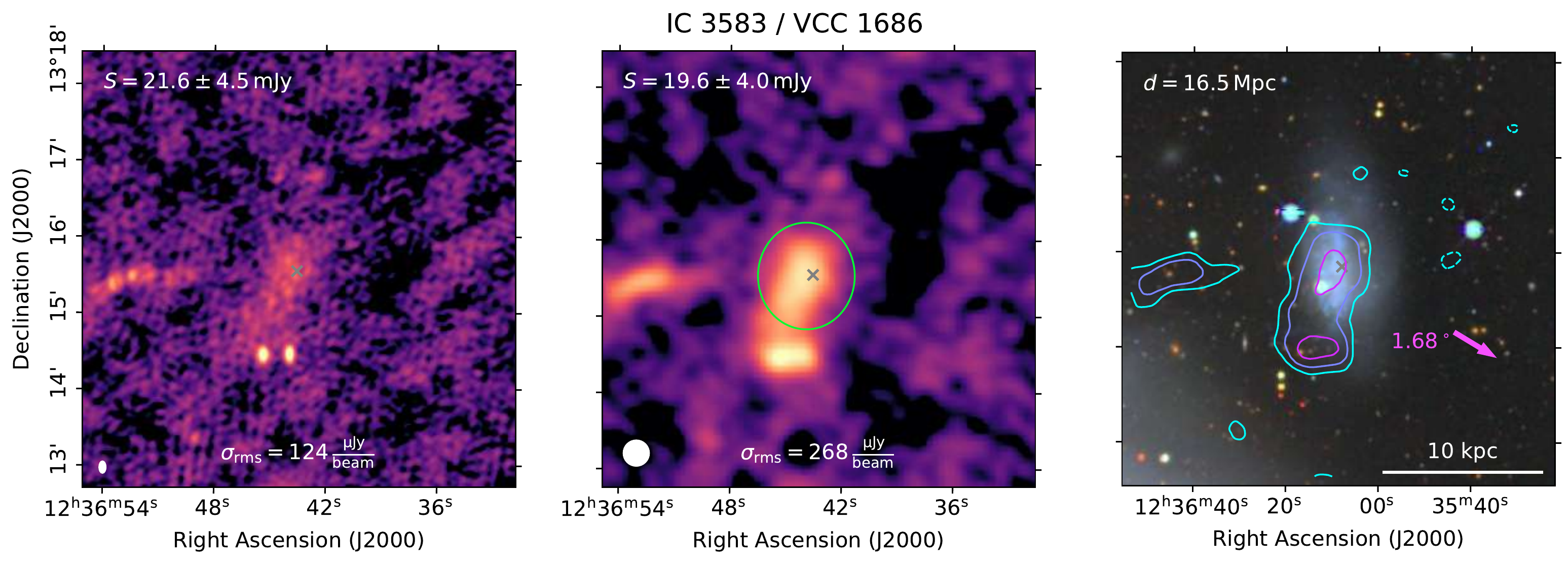}
        \caption{}
    \end{subfigure} 
    \caption{Same as \autoref{fig:144first}.}
\end{figure}

\begin{figure}
    \centering
    \begin{subfigure}[b]{\textwidth}
        \includegraphics[width=\textwidth]{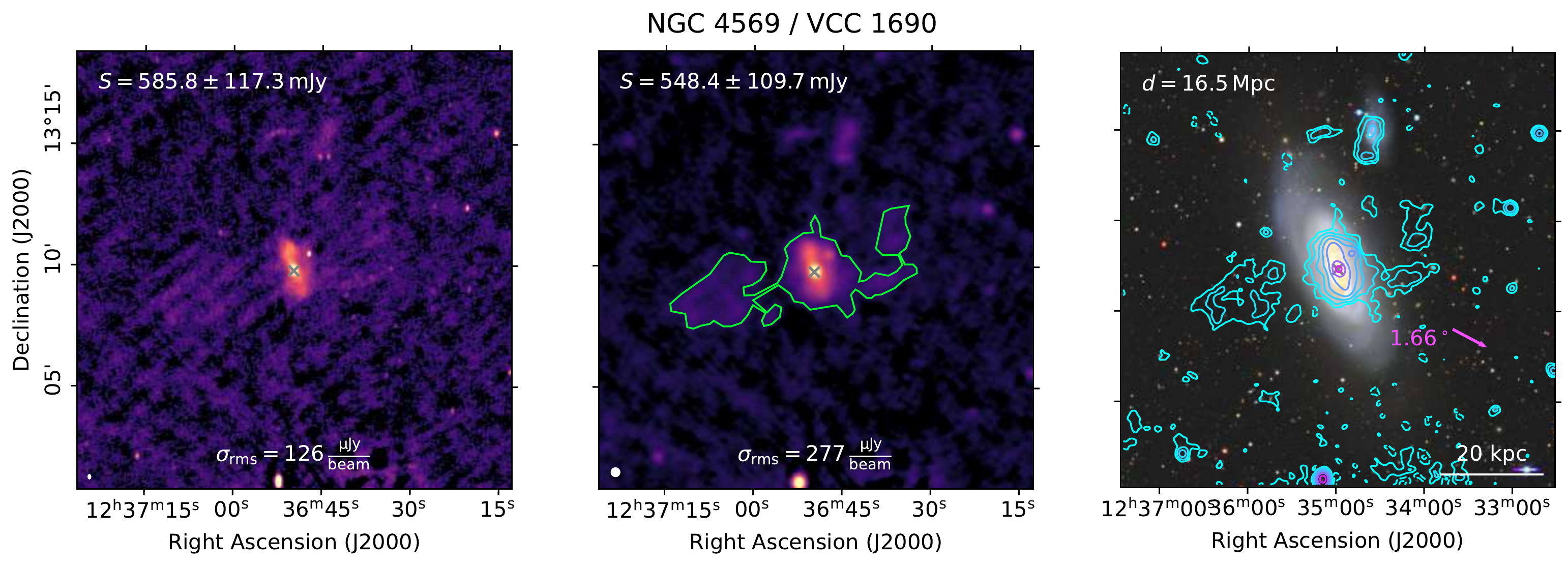}
        \caption{}
        \label{fig:1690}
    \end{subfigure} 
     \hfill
    \begin{subfigure}[b]{\textwidth}
        \includegraphics[width=\textwidth]{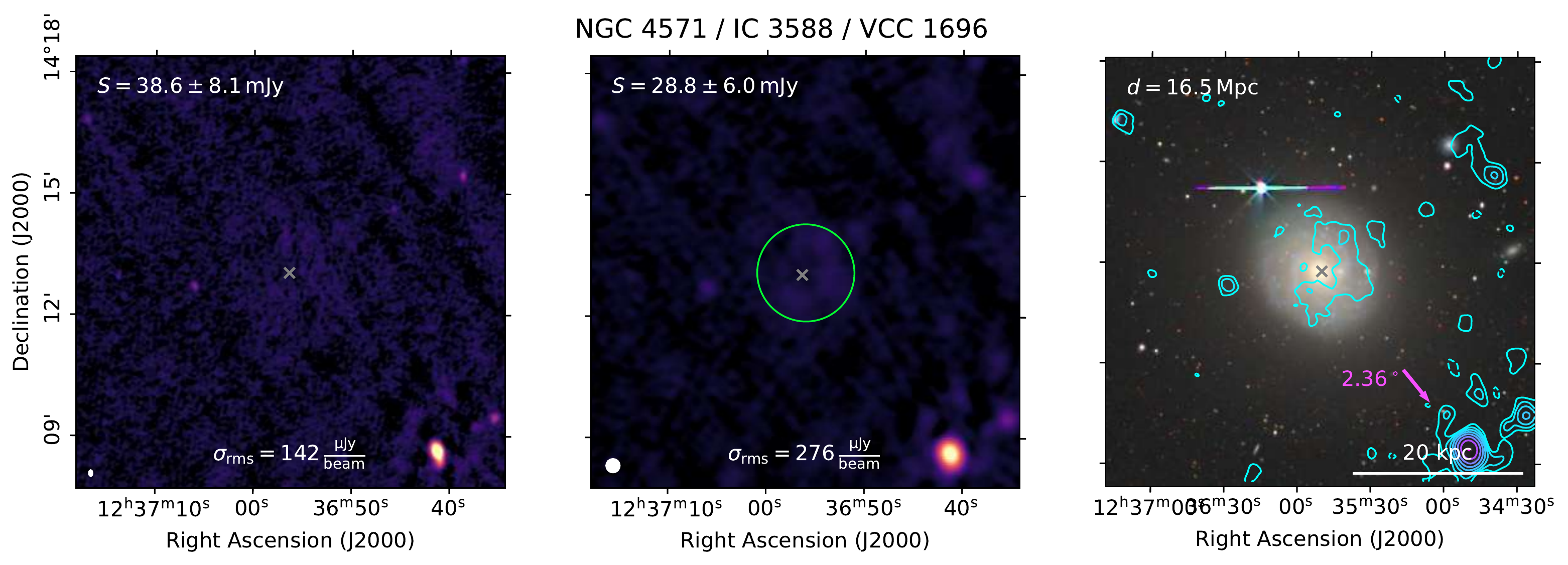}
        \caption{}
    \end{subfigure}
     \hfill
    \begin{subfigure}[b]{\textwidth}
        \includegraphics[width=\textwidth]{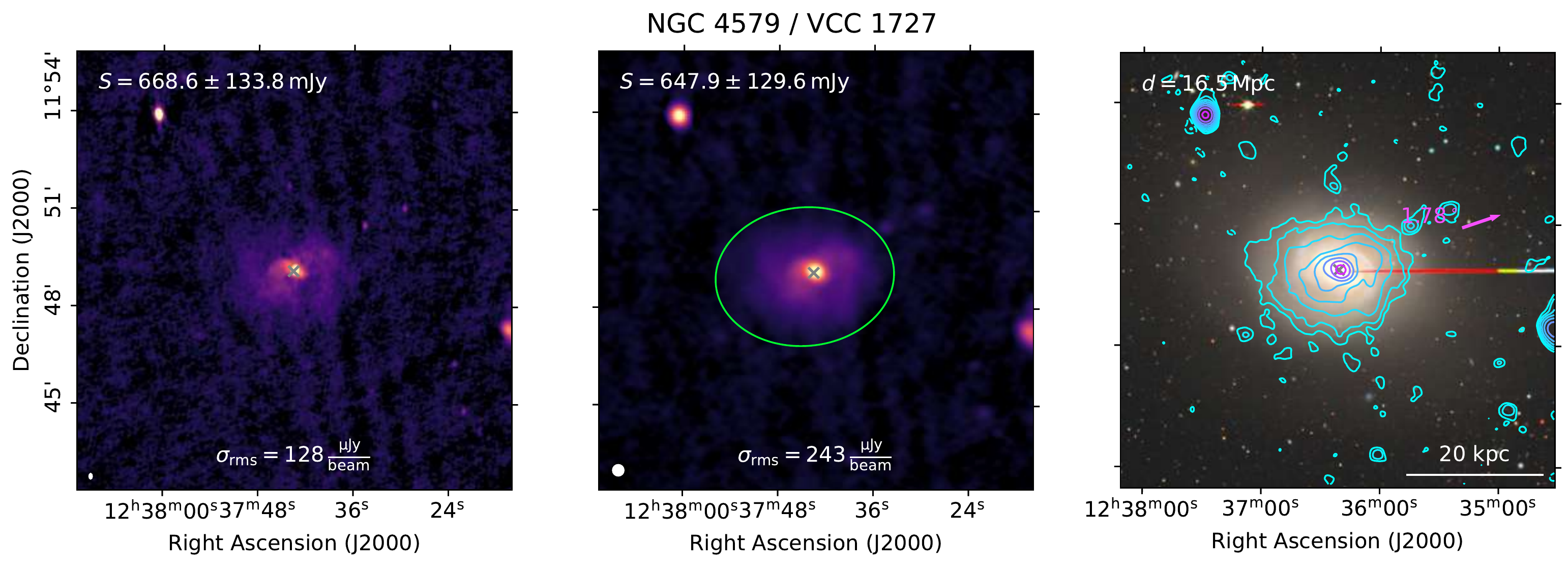}
        \caption{}
        \label{fig:1727}
    \end{subfigure} 
    \caption{Same as \autoref{fig:144first}.}
\end{figure}

\begin{figure}
    \centering
    \begin{subfigure}[b]{\textwidth}
        \includegraphics[width=\textwidth]{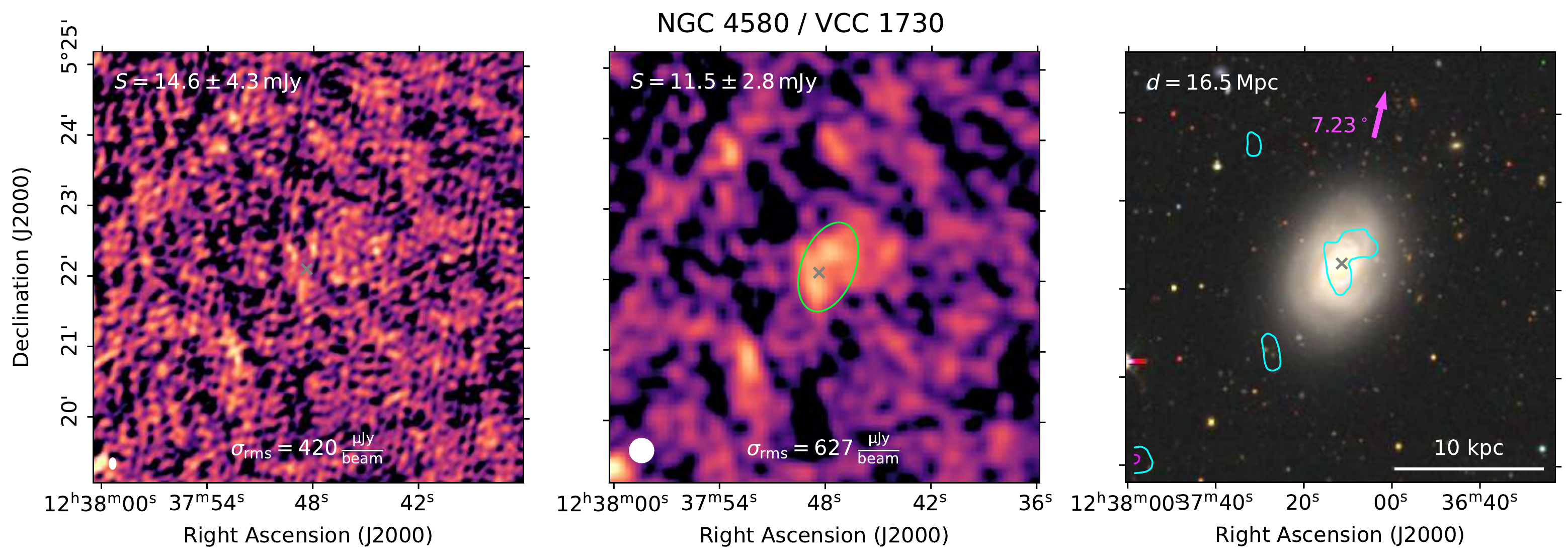}
        \caption{}
    \end{subfigure} 
     \hfill
    \begin{subfigure}[b]{\textwidth}
        \includegraphics[width=\textwidth]{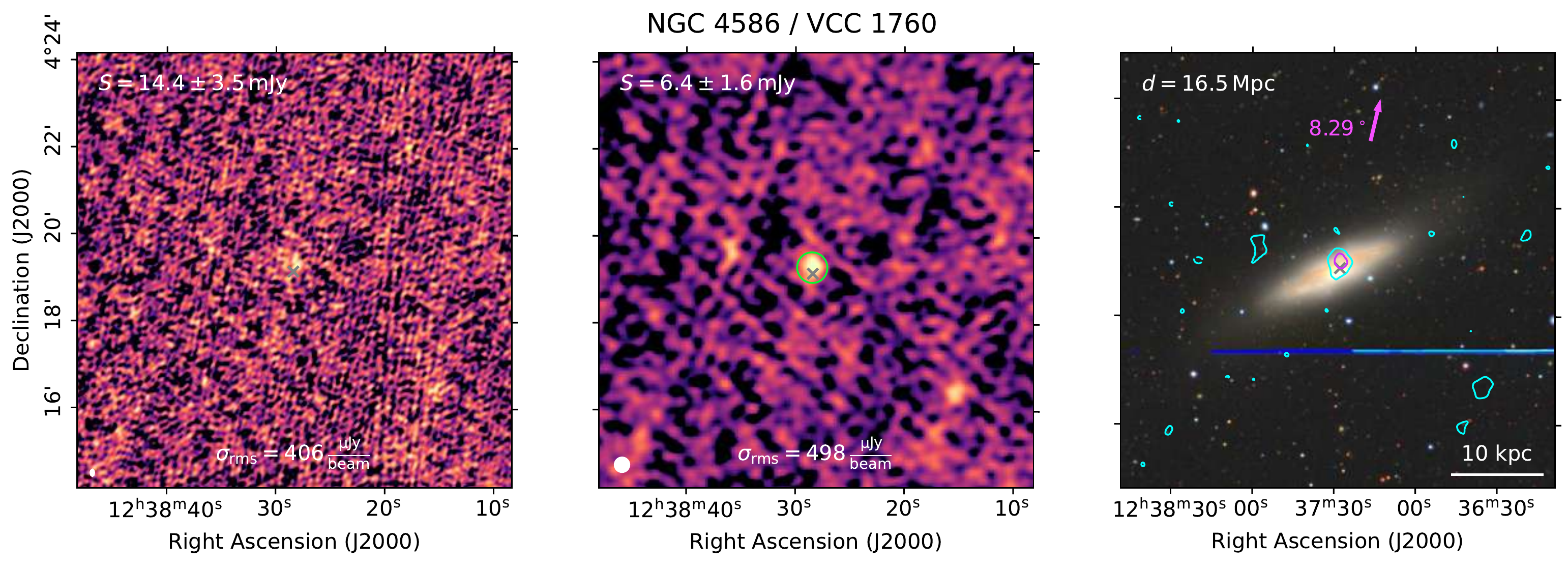}
        \caption{}
    \end{subfigure}
     \hfill
    \begin{subfigure}[b]{\textwidth}
        \includegraphics[width=\textwidth]{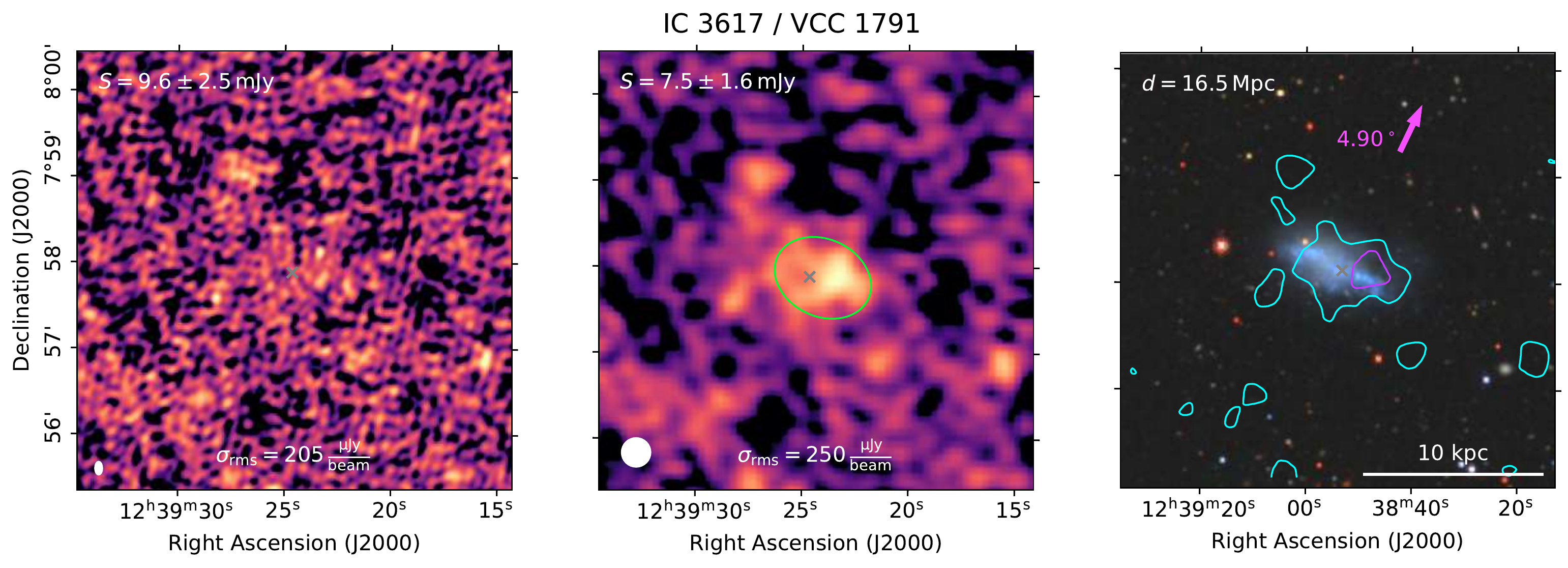}
        \caption{}
    \end{subfigure} 
    \caption{Same as \autoref{fig:144first}.}
\end{figure}

\begin{figure}
    \centering
    \begin{subfigure}[b]{\textwidth}
        \includegraphics[width=\textwidth]{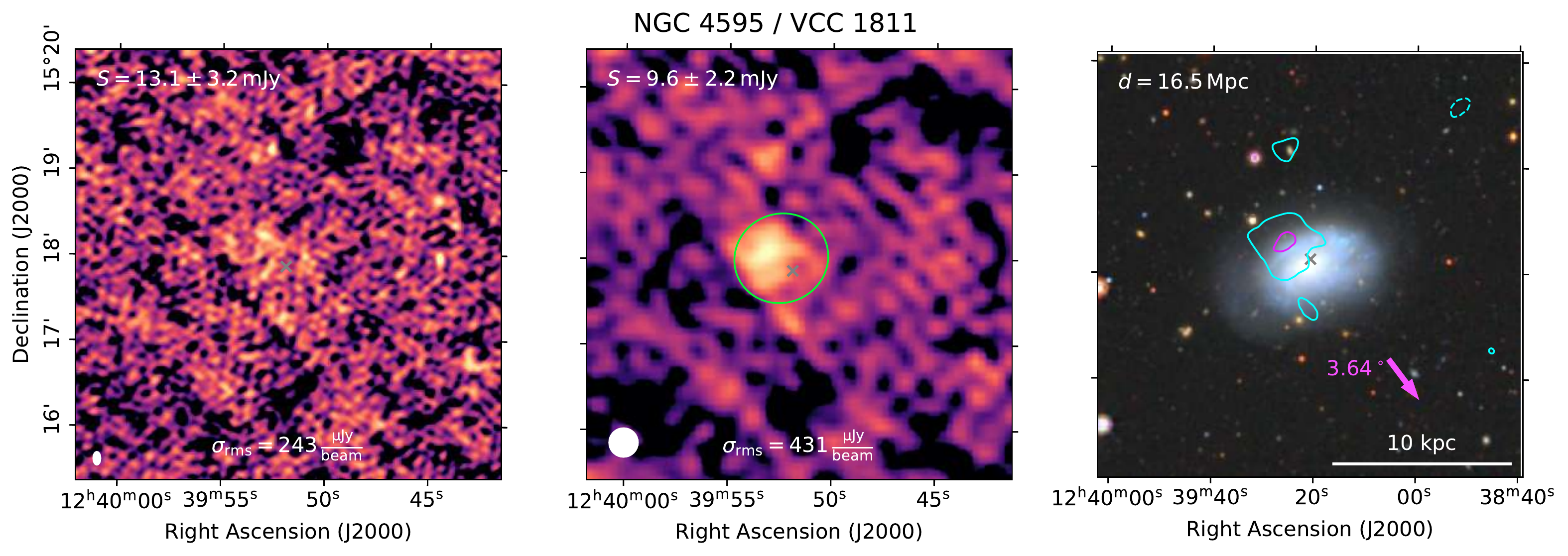}
        \caption{}
    \end{subfigure} 
     \hfill
    \begin{subfigure}[b]{\textwidth}
        \includegraphics[width=\textwidth]{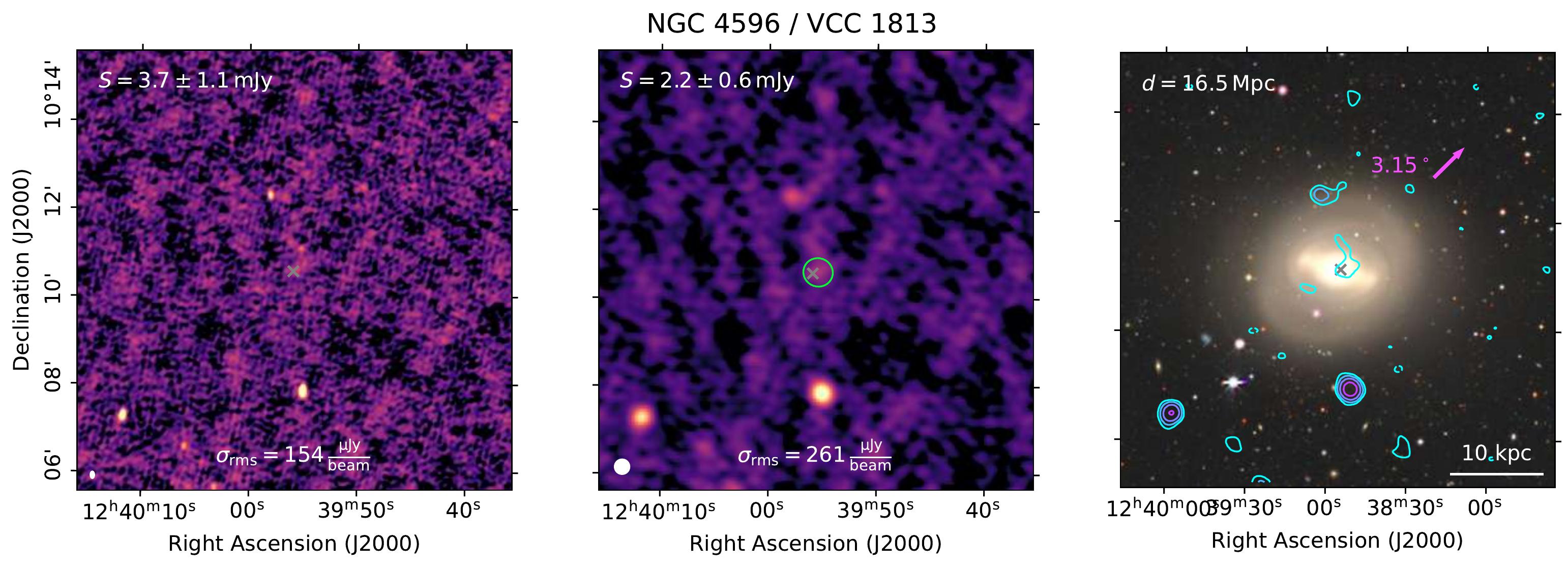}
        \caption{}
    \end{subfigure}
     \hfill
    \begin{subfigure}[b]{\textwidth}
        \includegraphics[width=\textwidth]{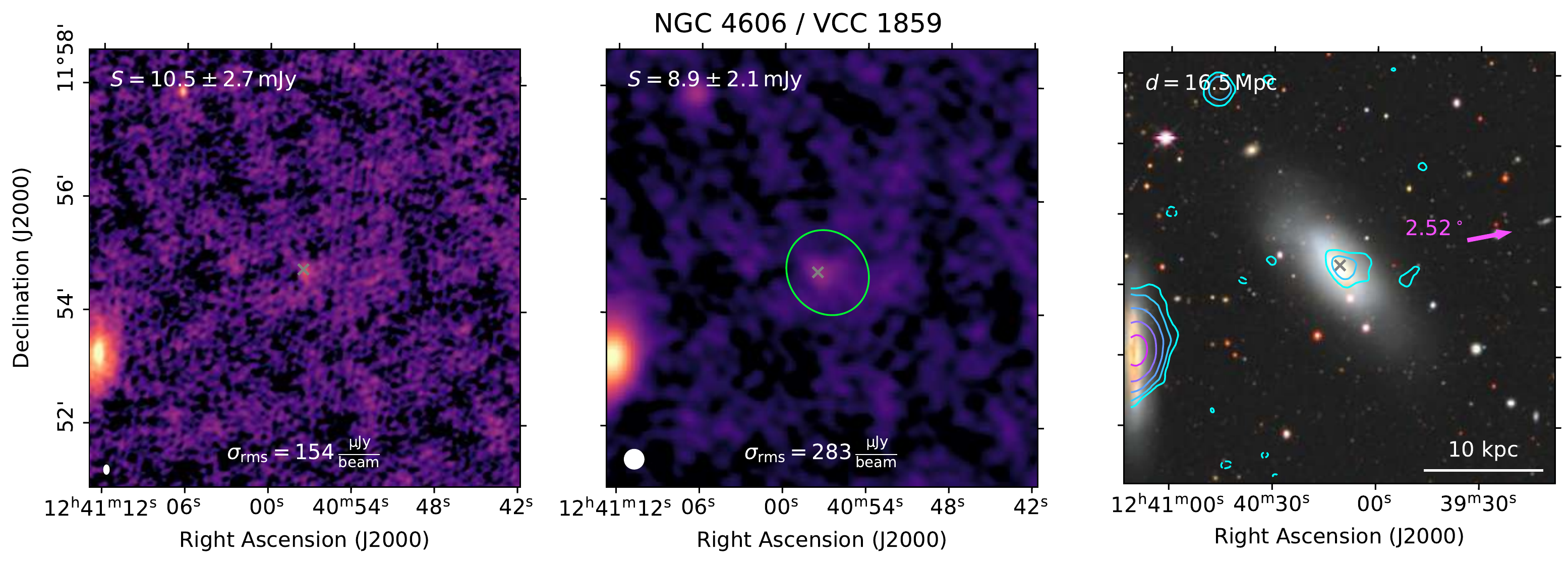}
        \caption{}
    \end{subfigure} 
    \caption{Same as \autoref{fig:144first}.}
\end{figure}

\begin{figure}
    \centering
    \begin{subfigure}[b]{\textwidth}
        \includegraphics[width=\textwidth]{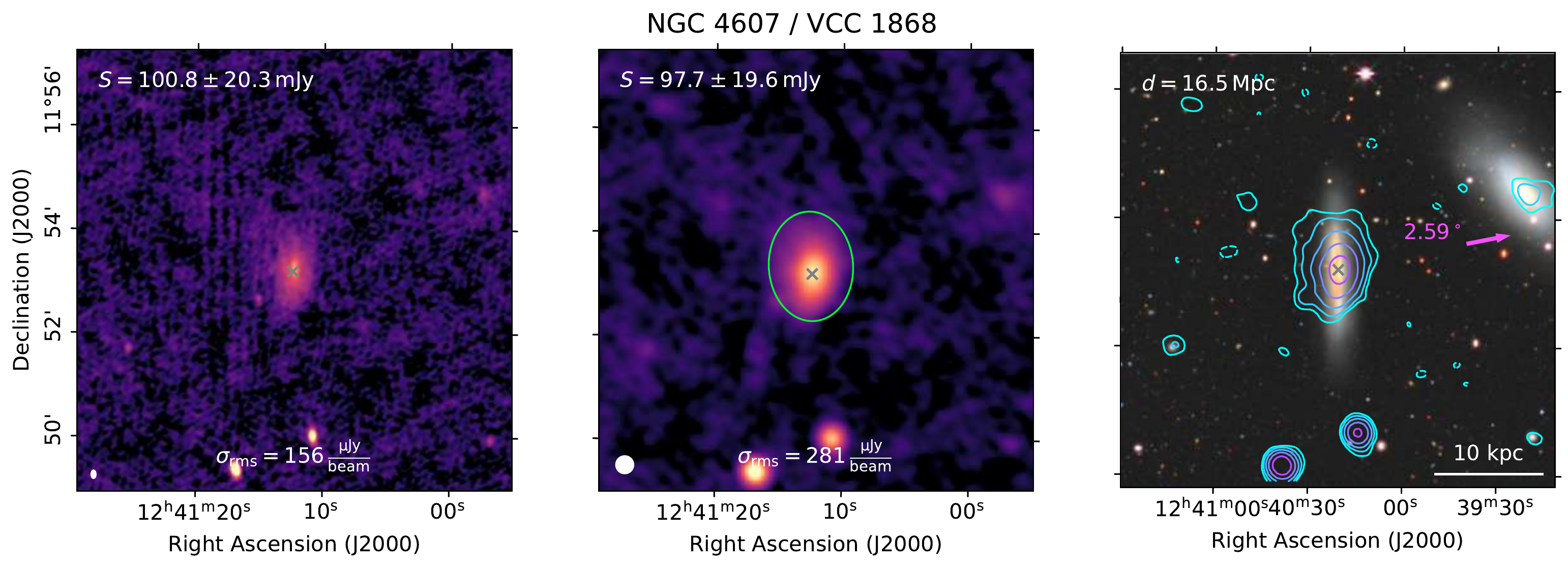}
        \caption{}
    \end{subfigure} 
     \hfill
    \begin{subfigure}[b]{\textwidth}
        \includegraphics[width=\textwidth]{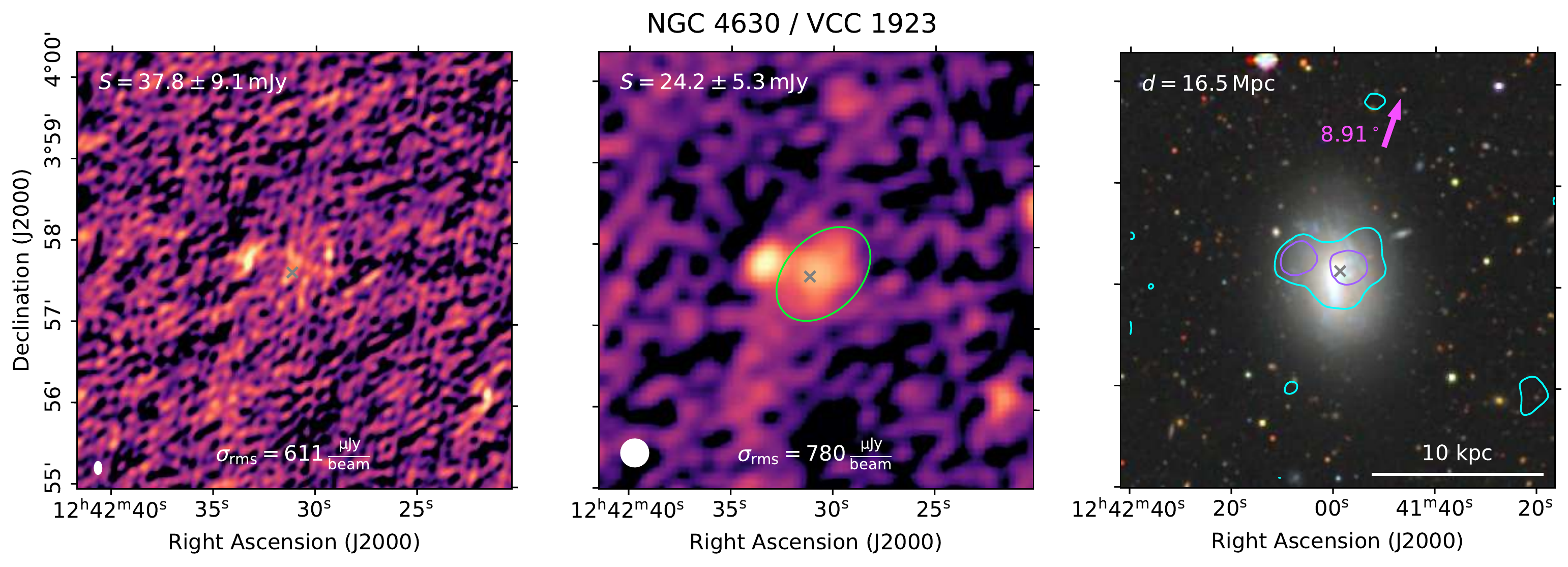}
        \caption{}
    \end{subfigure}
     \hfill
    \begin{subfigure}[b]{\textwidth}
        \includegraphics[width=\textwidth]{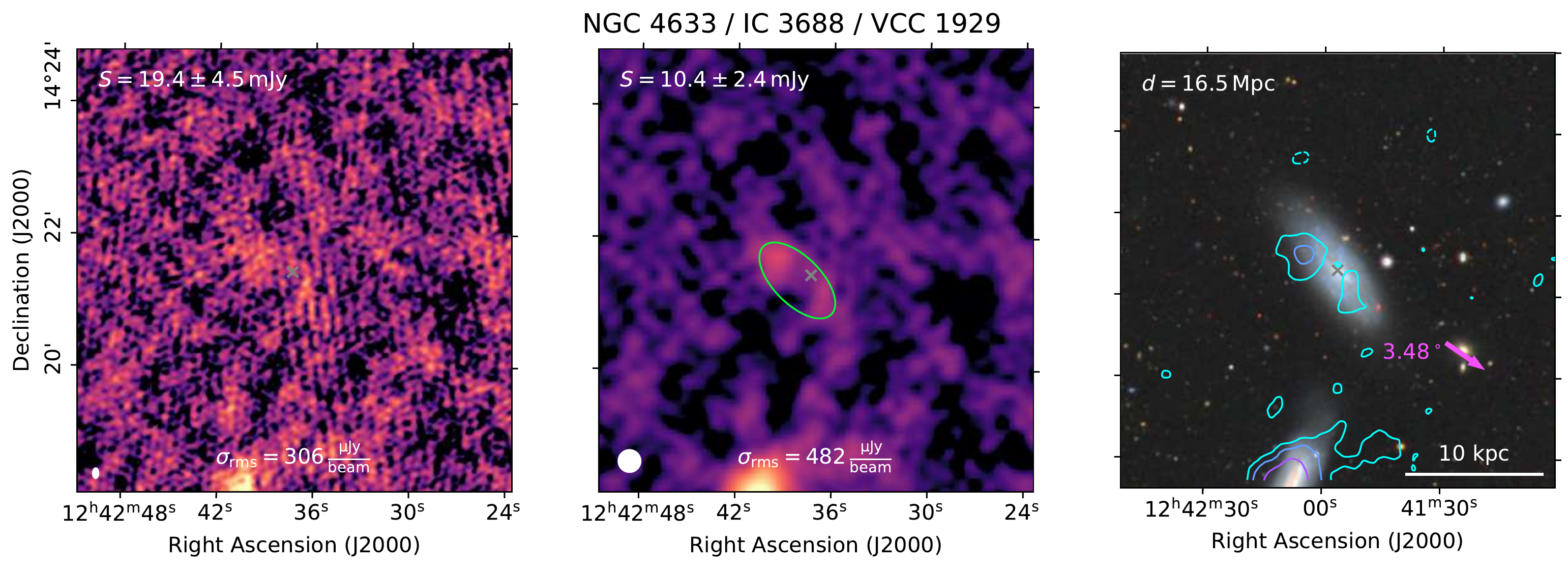}
        \caption{}
    \end{subfigure} 
    \caption{Same as \autoref{fig:144first}.}
\end{figure}

\begin{figure}
    \centering
    \begin{subfigure}[b]{\textwidth}
        \includegraphics[width=\textwidth]{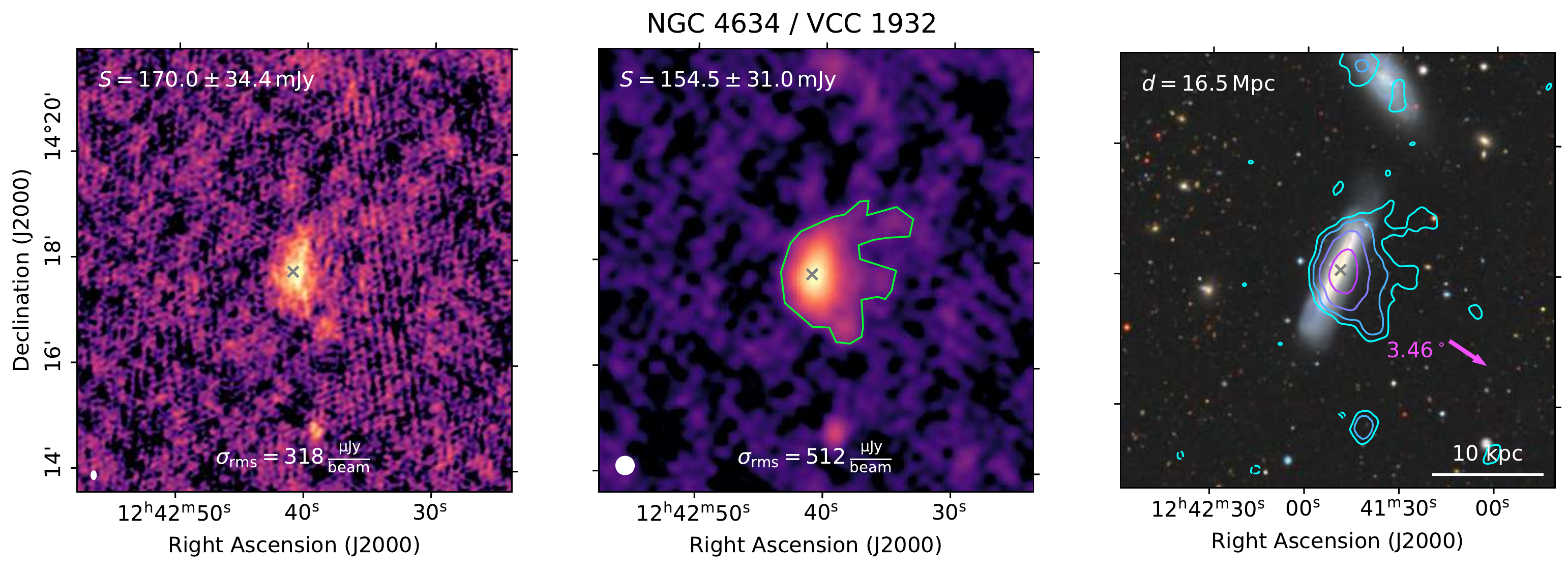}
        \caption{}
    \end{subfigure} 
     \hfill
    \begin{subfigure}[b]{\textwidth}
        \includegraphics[width=\textwidth]{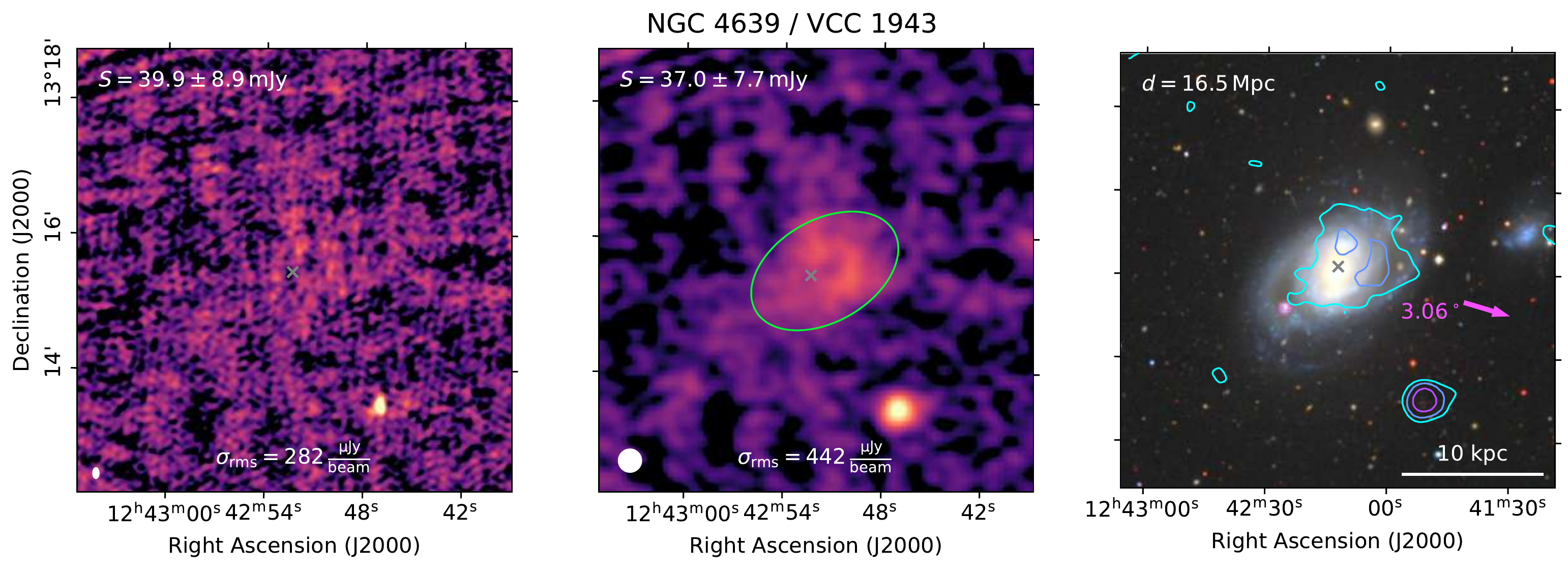}
        \caption{}
    \end{subfigure}
     \hfill
    \begin{subfigure}[b]{\textwidth}
        \includegraphics[width=\textwidth]{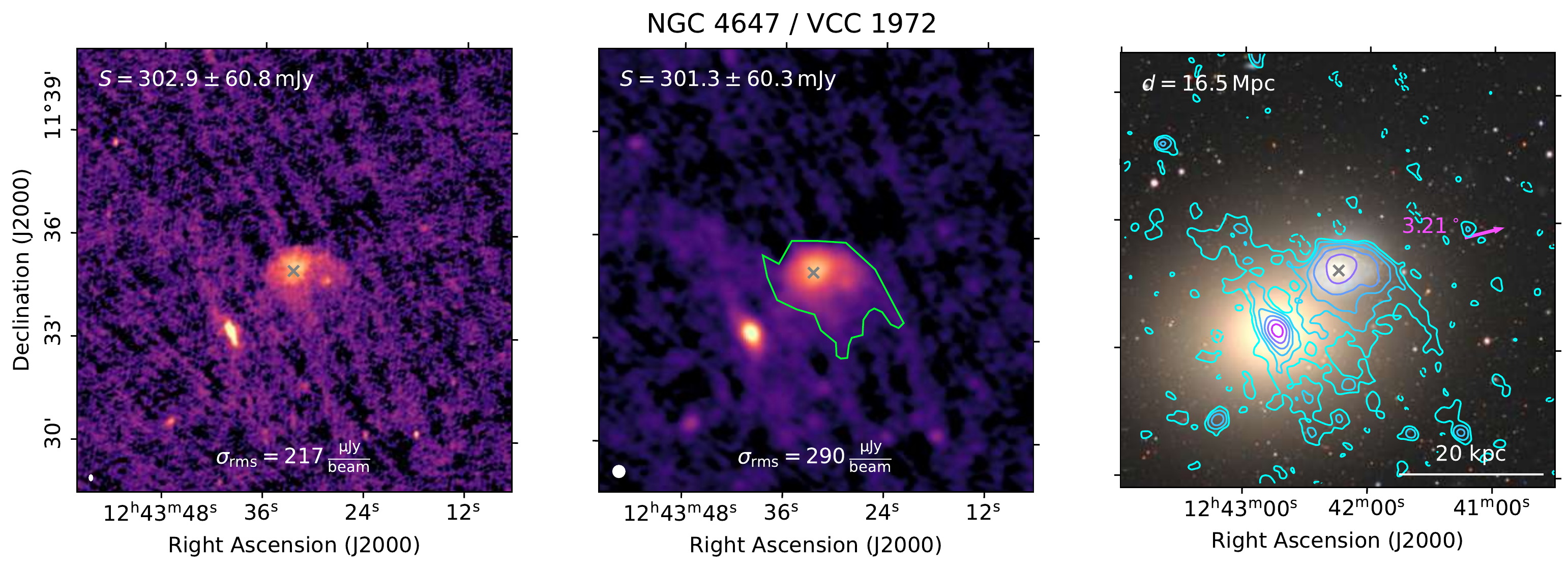}
        \caption{}
        \label{fig:1972}
    \end{subfigure} 
    \caption{Same as \autoref{fig:144first}.}
\end{figure}

\begin{figure}
    \centering
    \begin{subfigure}[b]{\textwidth}
        \includegraphics[width=\textwidth]{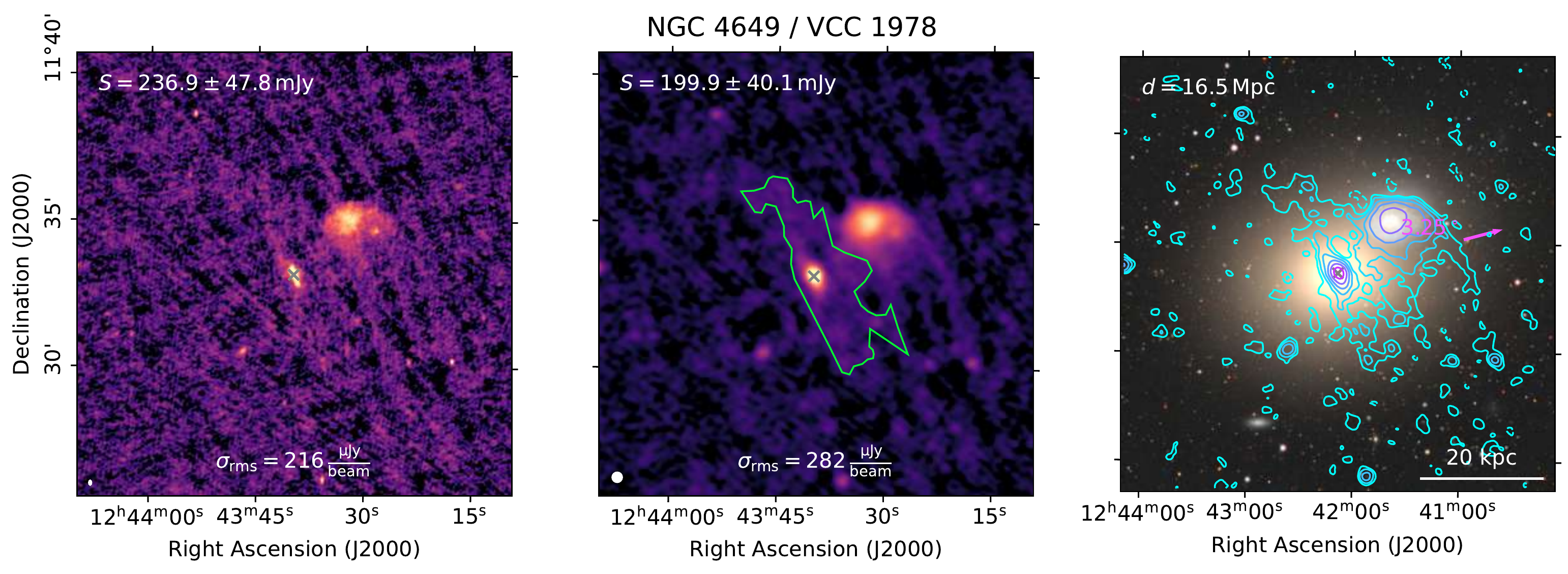}
        \caption{}
        \label{fig:1978}
    \end{subfigure} 
     \hfill
    \begin{subfigure}[b]{\textwidth}
        \includegraphics[width=\textwidth]{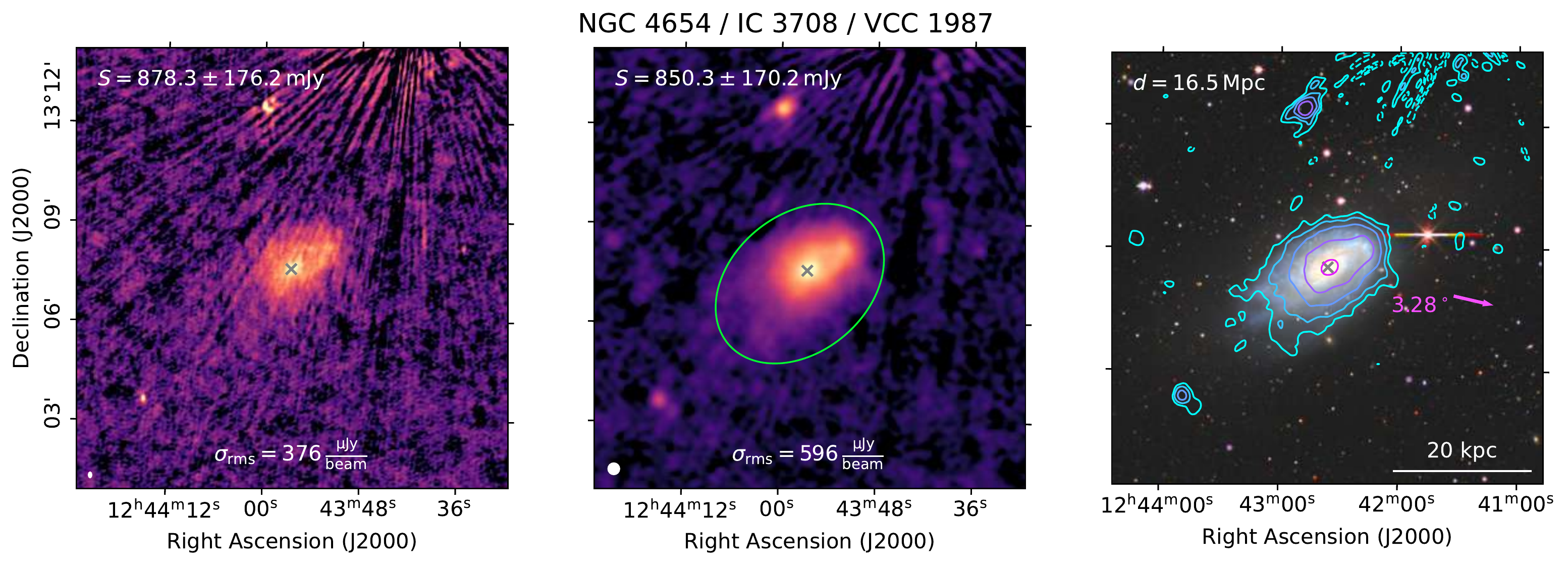}
        \caption{}
    \end{subfigure}
     \hfill
    \begin{subfigure}[b]{\textwidth}
        \includegraphics[width=\textwidth]{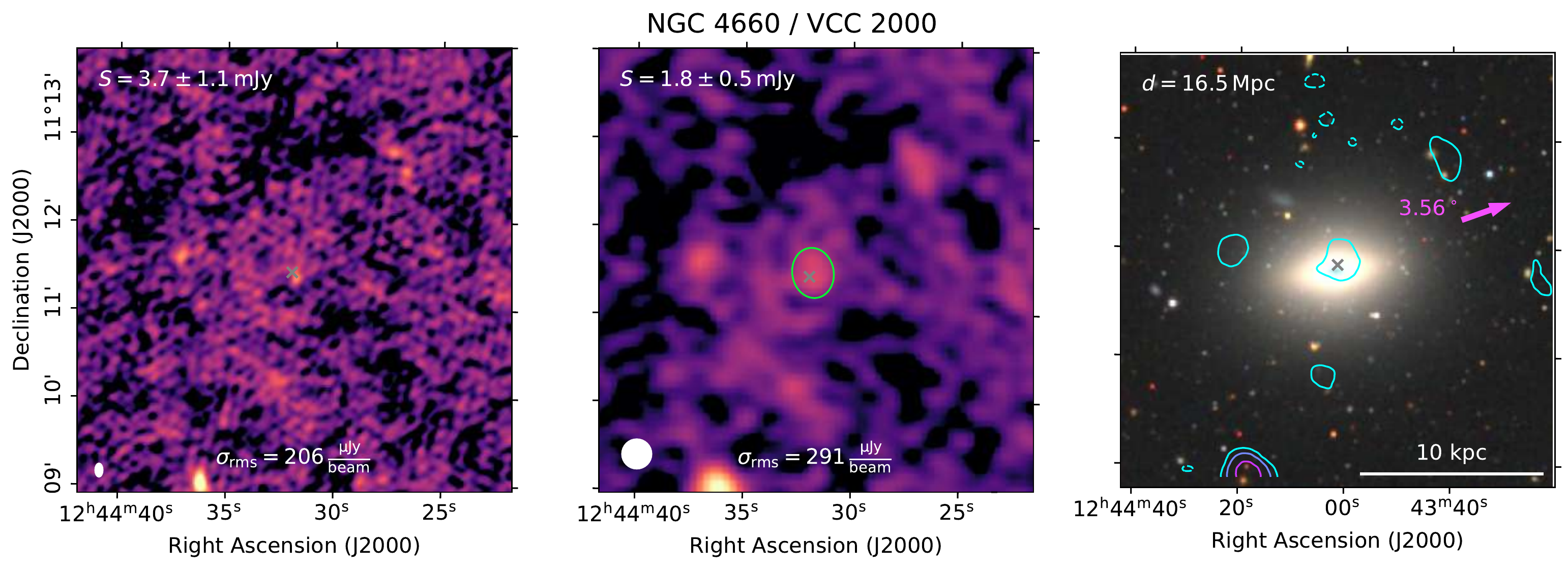}
        \caption{}
    \end{subfigure} 
    \caption{Same as \autoref{fig:144first}.}
\end{figure}

\begin{figure}
    \centering
    \begin{subfigure}[b]{\textwidth}
        \includegraphics[width=\textwidth]{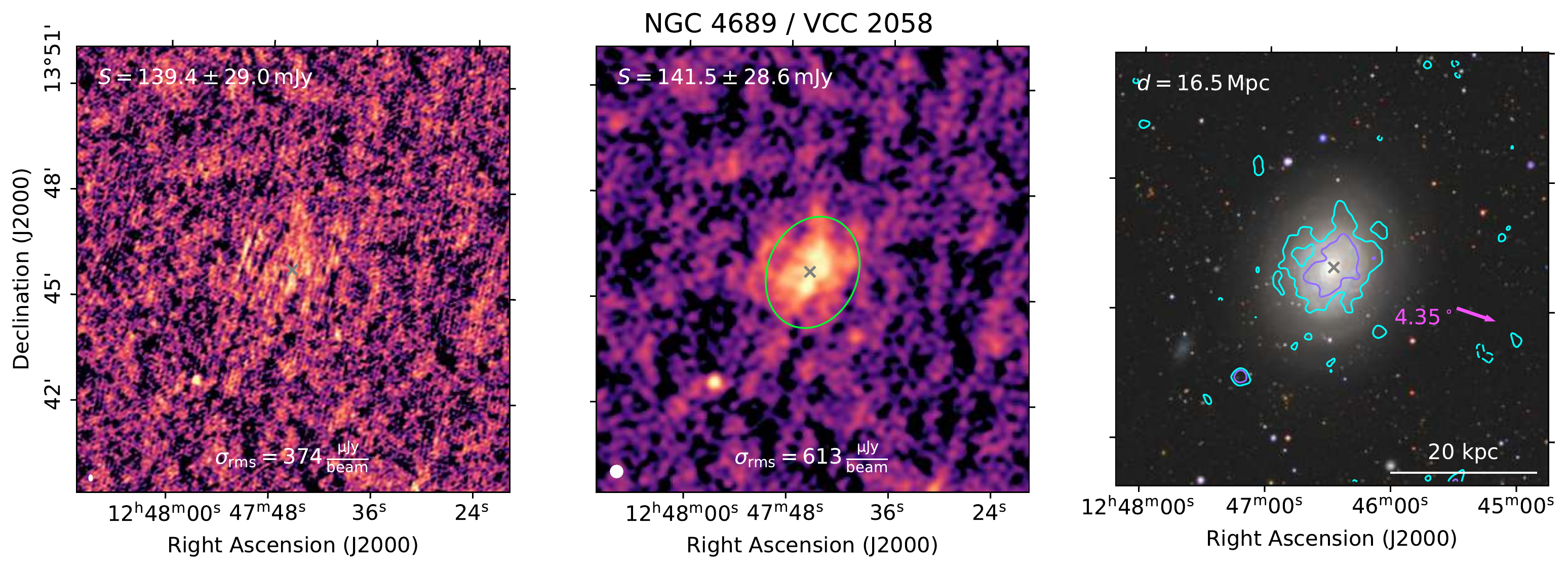}
        \caption{}
    \end{subfigure} 
     \hfill
    \begin{subfigure}[b]{\textwidth}
        \includegraphics[width=\textwidth]{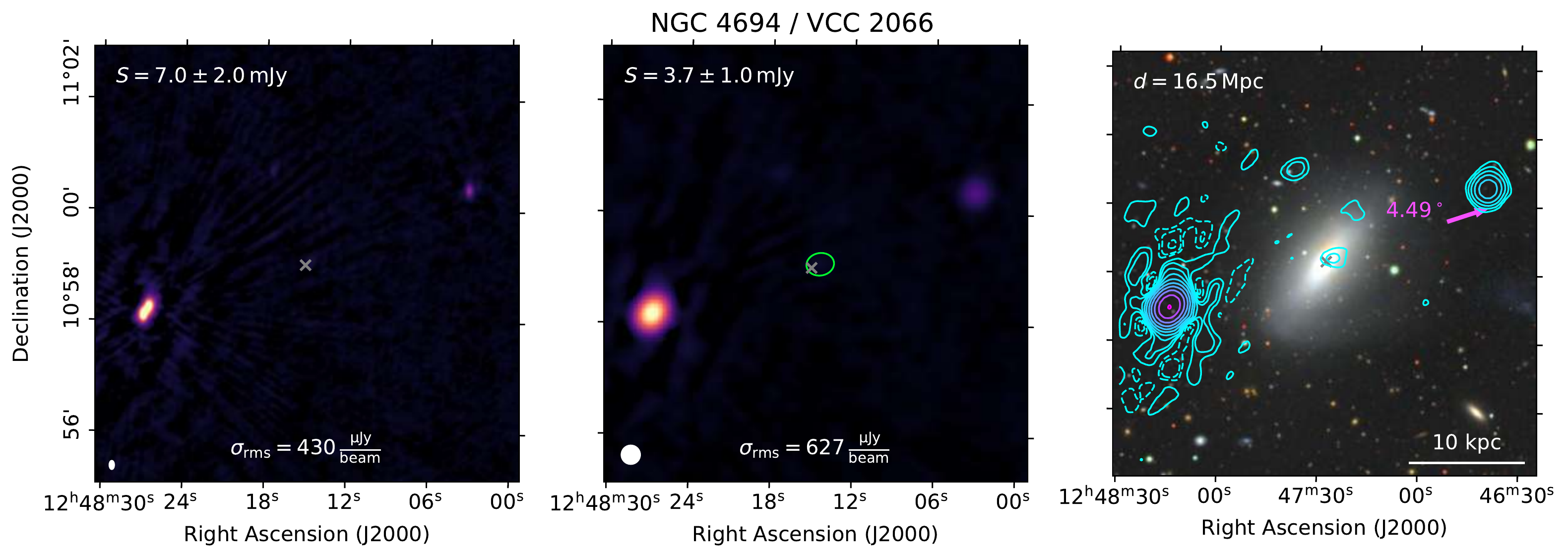}
        \caption{}
    \end{subfigure}
     \hfill
    \begin{subfigure}[b]{\textwidth}
        \includegraphics[width=\textwidth]{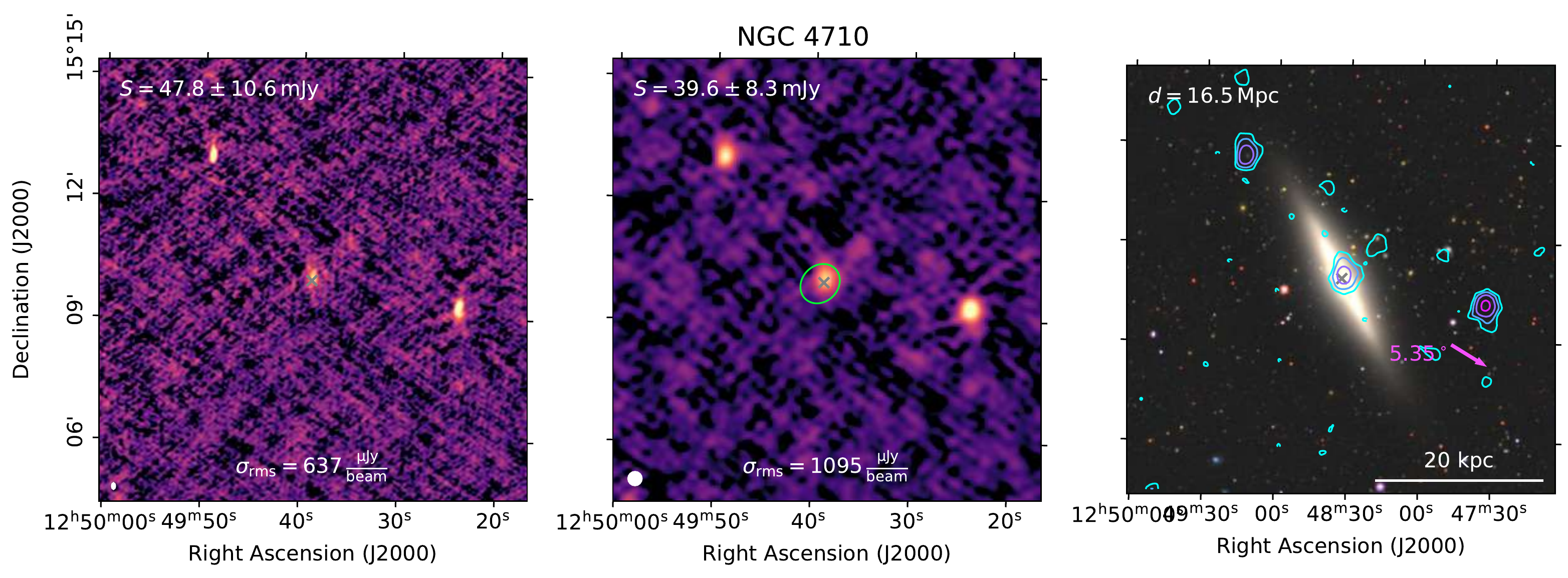}
        \caption{}
    \end{subfigure} 
    \caption{Same as \autoref{fig:144first}.}
\end{figure}

\begin{figure}
    \centering
    \begin{subfigure}[b]{\textwidth}
        \includegraphics[width=\textwidth]{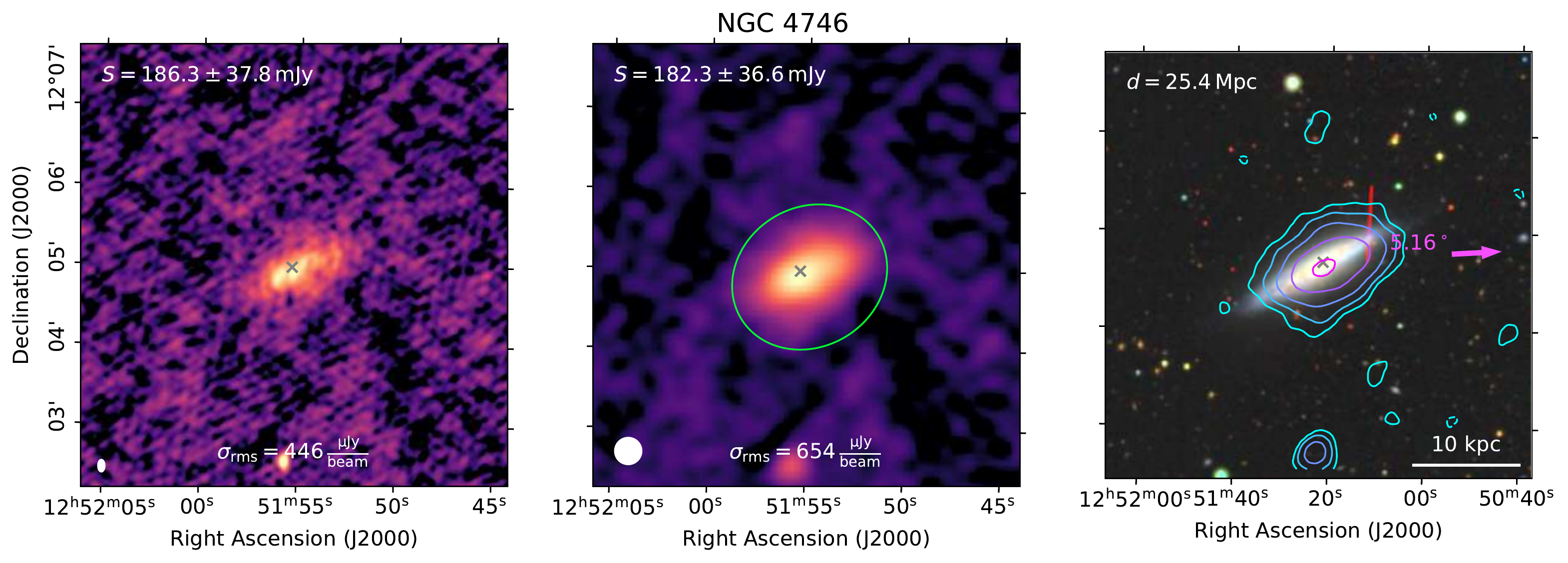}
    \end{subfigure} 
    \label{fig:4746last}
    \caption{Same as \autoref{fig:144first}.}
\end{figure}
\end{appendix}
\end{document}